\documentclass[journal=chreay,manuscript=review]{achemso}

\usepackage[version=3]{mhchem} 
\usepackage{amssymb}
\usepackage{lscape}
\usepackage{longtable}
\newcommand*{\la}{\lesssim}
\newcommand*{\ga}{\gtrsim}
\newcommand*{\bbf}{}

\author{Thomas Henning}
\email{henning@mpia.de}
\author{Dmitry Semenov}
\email{semenov@mpia.de}

\affiliation[MPIA]
{Max Planck Institute for Astronomy, K\"onigstuhl 17, D-69117 Heidelberg, Germany}

\title[Chemistry in Disks]{Chemistry in Protoplanetary Disks}

\begin{document}

 \tableofcontents

\section{Introduction}
\label{sec:intro}
Since the discovery of the first extrasolar planet around a solar-type star\cite{51Peg},
more than 900 such planets outside of our solar system have been
detected by ground- and space-based astronomical observations\cite{Udry_Santos07,2013ApJS..204...24B,Barclay_ea13}. 
Even more 
planetary candidates
discovered by the \textit{Kepler} space mission await secure identification ({\it http://exoplanet.eu}).
These exoplanets show
a wide diversity in orbital parameters, ranging from close-in Hot Jupiters and Neptunes 
to planets on very eccentric orbits, in orbital resonances, on retrograde orbits, and some located
at very large distances from their host stars. A similar diversity has been found for
the masses and radii of exoplanets and initial investigations point to a diversity 
in the chemical composition and physical structure of their atmospheres\cite{Stevenson_ea10a,Barman_ea11a,Konopacky_ea13}.

The wide range of planetary system architectures and exoplanet properties is certainly linked 
to a range of properties of their birth-places, the disk-like structures around young stars composed
of gas and dust particles\cite{2012ApJ...755...74K,2004ApJ...616..567I,2012A&A...541A..97M,2012A&A...547A.111M}.
These disks share many of the properties of the solar nebula
from which the Sun and our planetary system formed, although their masses, radial dimensions and internal
structures can be very different\cite{2011ARA&A..49...67W}. Protoplanetary disks form
through the gravitational collapse of their parental molecular cloud cores in a process regulated by the 
balance of gravitational, magnetic, gas pressure and rotational forces\cite{Stahler_Palla05}. 
After the dissipation of the protostellar
envelope and the birth of a central star (central stars), the surrounding protoplanetary disk {\bbf continues to
regulate} the inward radial transport of matter and the associated angular momentum transport outward, 
and, therefore, forms a special class of accretion disks\cite{2009apsf.book.....H}.

Thanks to the progress in infrared and radio astronomy starting in the
early 1990's, and followed by highly successful infrared space observatories such as \textit{ISO}, \textit{Spitzer}, and
\textit{Herschel}, protoplanetary disks have been found and characterized in large numbers in regions of nearby
star formation. They are discovered through their infrared and (sub-)millimeter thermal dust emission
being in excess of the radiation from the stellar photosphere of their central 
stars\cite{1989AJ.....97.1451S,1990AJ.....99..924B} 
(for a review of their properties, see \cite{2011ARA&A..49...67W}).  Dust
spectroscopy has revealed the mineralogical composition of the protoplanetary dust particles that are mostly found in the form of
amorphous silicates, crystalline forsterite, water ice, and other molecular ices.
There is strong observational evidence that the dust particles in disks can grow in size far beyond the typical submicron
sizes of interstellar dust grains\cite{2011ppcd.book..114H}.

\begin{figure}
\includegraphics[angle=0,width=6cm]{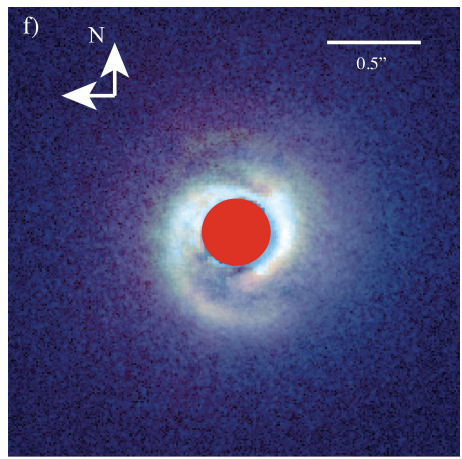}
\caption{Near-IR scattered light image of the protoplanetary disk around the Herbig~Ae star MWC~758 obtained with
the Subaru telescope by the Strategic Exploration of Exoplanets and Disks (SEEDS) collaboration.
{\bbf Reprinted with permission from Reference\cite{2013ApJ...762...48G}. Copyright 2013 American Astronomical Society.} 
\label{fig:disk_SEEDS}}
\end{figure}

In recent years, some of these disks have even been directly imaged with the \textit{Hubble Space Telescope}, ground-based
adaptive-optics assisted instruments (see Fig.~\ref{fig:disk_SEEDS}), and infrared and millimeter interferometry. The 
interferometry technique combines the
light gathered by a number of telescopes to reach higher spatial resolution than available  with the largest
single-dish telescopes. These observations have revealed a wide diversity of protoplanetary disk structures, including some
disks at advanced evolutionary phases showing inner holes or gaps devoid of emitting dust and sometimes even 
gas\cite{Hughes_ea07,Ratzka_ea07,Brown_09,Brown_ea12}.
{\bbf These spatially resolved data confirmed the earlier inference of such structures based on the analysis of spectral
energy distributions, especially provided by the {\it Spitzer} infrared space 
telescope\cite{SiciliaAguilar_ea06b,SiciliaAguilar_ea08,Furlan_ea09,Merin_ea10,Espaillat_ea10}.}
Other recent discoveries obtained with ground-based coronagraphic near-IR telescopes show large-scale, strong
asymmetries in disk structure, such as spiral arms and density ``knots''\cite{Fea04,2013ApJ...762...48G,2013arXiv1306.1768V} (see 
Fig.~\ref{fig:disk_SEEDS}).
These asymmetries are likely produced by a variety of
physical processes such as magnetohydrodynamical turbulence\cite{2012ApJ...744..144F}, grain growth beyond 
cm-sizes\cite{Hughes_ea07}, planet formation, and
gravitational instabilities\cite{Boley_ea10a}.
These spatial structures immediately
show that protoplanetary disks are not static systems, but are subject to strong dynamical changes on a timescale of
several million years\cite{2011ARA&A..49...67W}.  

The advent of sensitive infrared and (sub-)millimeter spectroscopic observations enabled the discovery of thermal 
emission and scattered
light from dust particles. In addition, a first inventory of atomic and molecular species has been provided,
ranging from molecular hydrogen to water and more complex molecules such as polycyclic aromatic hydrocarbons 
(PAHs)\cite{2006PNAS..10312249V,DGH07,Bergin_09}. At the same
time, comprehensive chemical models for protoplanetary disks have been developed by a number of research groups {\bbf (see 
Table~\ref{tab:disk_models})}, taking into
account the wide range of radiation fields (UV and/or X-rays), temperatures (10 - several 1000~K) and hydrogen number densities
(10$^4$ - 10$^{12}$ cm$^{-3}$). The combination of astronomical observations with advanced disk physical and chemical models has
provided first constraints on the thermal structure and molecular composition of protoplanetary disks orbiting young stars of
various temperatures and masses\cite{Aikawa_ea03,Pietu_ea07,Schreyer_ea08,Panic_Hogerheijde09,Oeberg_ea10a}. 
{\bbf These models have demonstrated that the chemistry in disks is mostly regulated by their
temperature and density structure, and stellar and interstellar radiation fields as well as cosmic 
rays\cite{rh00,wl00,vZea03,Red2,vDea_06,Gorti_ea09,Visser_ea09b,Henning_ea10,Owen_ea11a,Walsh_ea12}. A special feature
of protoplanetary disks is the very low temperatures in the outer midplane regions, leading to a considerable freeze-out of
molecules\cite{Dutrey_ea97,Qi_ea13a}. At the same time, chemistry, together with grain evolution, regulates the ionization 
structure of disks\cite{zetaxa,Red2,Ilgner_Nelson06,Ilgner_Nelson06a,Ilgner_Nelson06b,2011ApJ...743..152O}, and,
thereby influencing the magnetically-driven transport of mass and angular momentum\cite{MRI}. 
This means that disk chemistry and the physical
structure of disks are ultimately linked.} The impact of {\bbf radial and/or vertical} 
transport processes and dust evolution on disk chemical composition has been 
thoroughly theoretically 
investigated\cite{G02,Wehrstedt_Gail02,Boss2004,IHMM04,Willacy_ea06,Ilgner_Nelson06a,Aikawa_07,
Turner_ea07, TG07, Hersant_ea09,Heinzeller_ea11,Semenov_Wiebe11a}, 
and the predictions are being observationally confirmed.

With the \textit{Atacama Large Millimeter/Submillimeter Array 
(ALMA)\footnote{http://almascience.eso.org/about-alma/overview/alma-basics}} in Chile becoming fully operational and providing a 
giant
step in sensitivity, spectral, and spatial resolution, and with models connecting planet formation with 
planet evolution, we can expect that protoplanetary disk chemistry will become a major topic in the rapidly evolving field of 
astrochemistry.
This review will discuss these fascinating, chemically active places of planet formation and will summarize the information we 
have
obtained about the physical structure and molecular chemistry of protoplanetary disks.

{\bbf Because chemistry depends so much on temperature, density, and radiation fields in disks, and is influenced by 
disk dynamics, in Chapter~\ref{sec:phys_struc} we first provide  a comprehensive description of the physics of 
protoplanetary disks. Chapter~\ref{sec:chemistry} summarizes the fundamental chemical processes in disks and presents
an inventory of gas-phase molecules and ices. This review emphasizes the role of water in disks in 
Chapter~\ref{sec:chemistry:water} because of its relevance for planet formation and the delivery of water to Earth.
A special Chapter~\ref{sec:chemistry:deuterium} is devoted to deuterium fractionation in disks because this process may allow
to reconstruct the history of chemical and physical processes in early phases of the solar nebula and protoplanetary disks.
The formation of complex organic molecules is the focus of Chapter~\ref{sec:organics} because of its relevance for the delivery
of organic materials to terrestrial planets.}

\section{Physical Properties of Protoplanetary Disks}
\label{sec:phys_struc}

\subsection{Protoplanetary Disks as Accretion Disks}
\label{sec:phys_struc:acc_disks}
\begin{figure}
\includegraphics[angle=0,width=14cm]{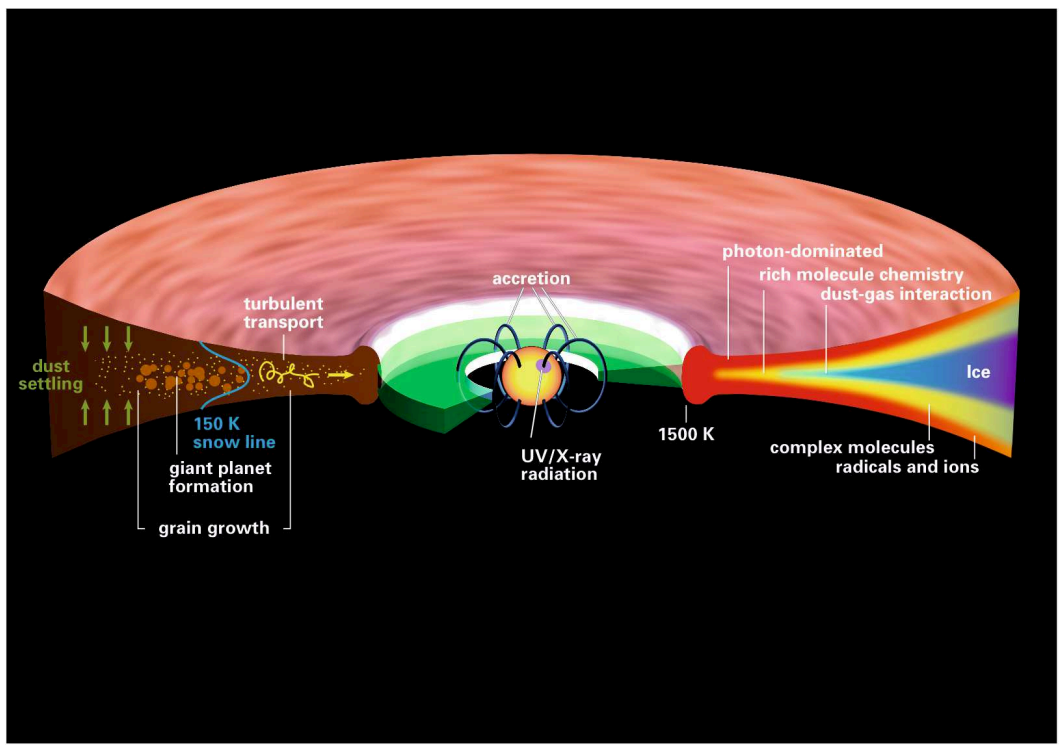}
\caption{Sketch of the physical and chemical structure of a $\sim 1-5$~Myr old protoplanetary disk around a 
Sun-like star.
\label{fig:disk_scheme}}
\end{figure}

\begin{figure}
\includegraphics[width=0.27\textwidth,angle=90]{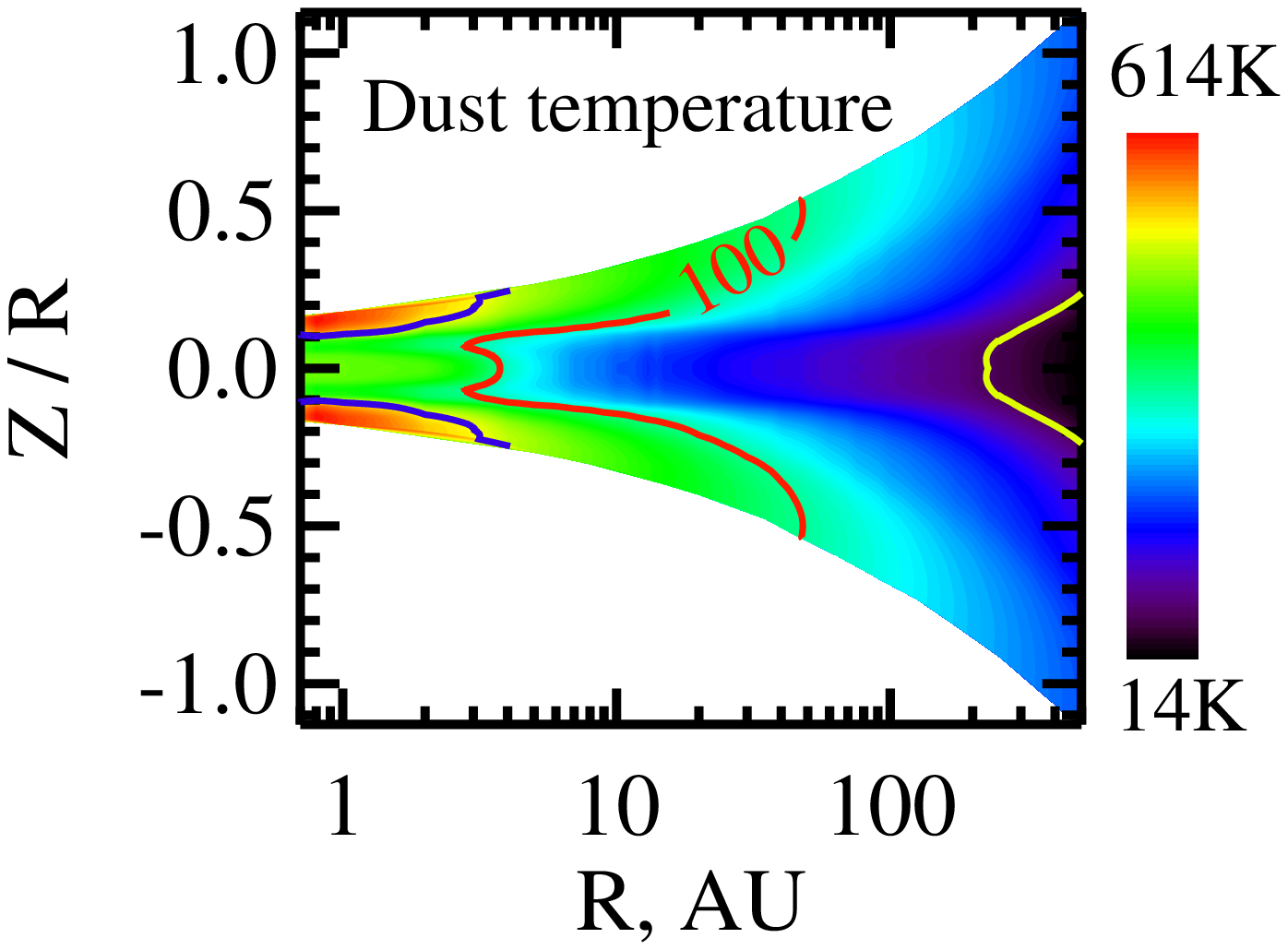}
\includegraphics[width=0.27\textwidth,angle=90]{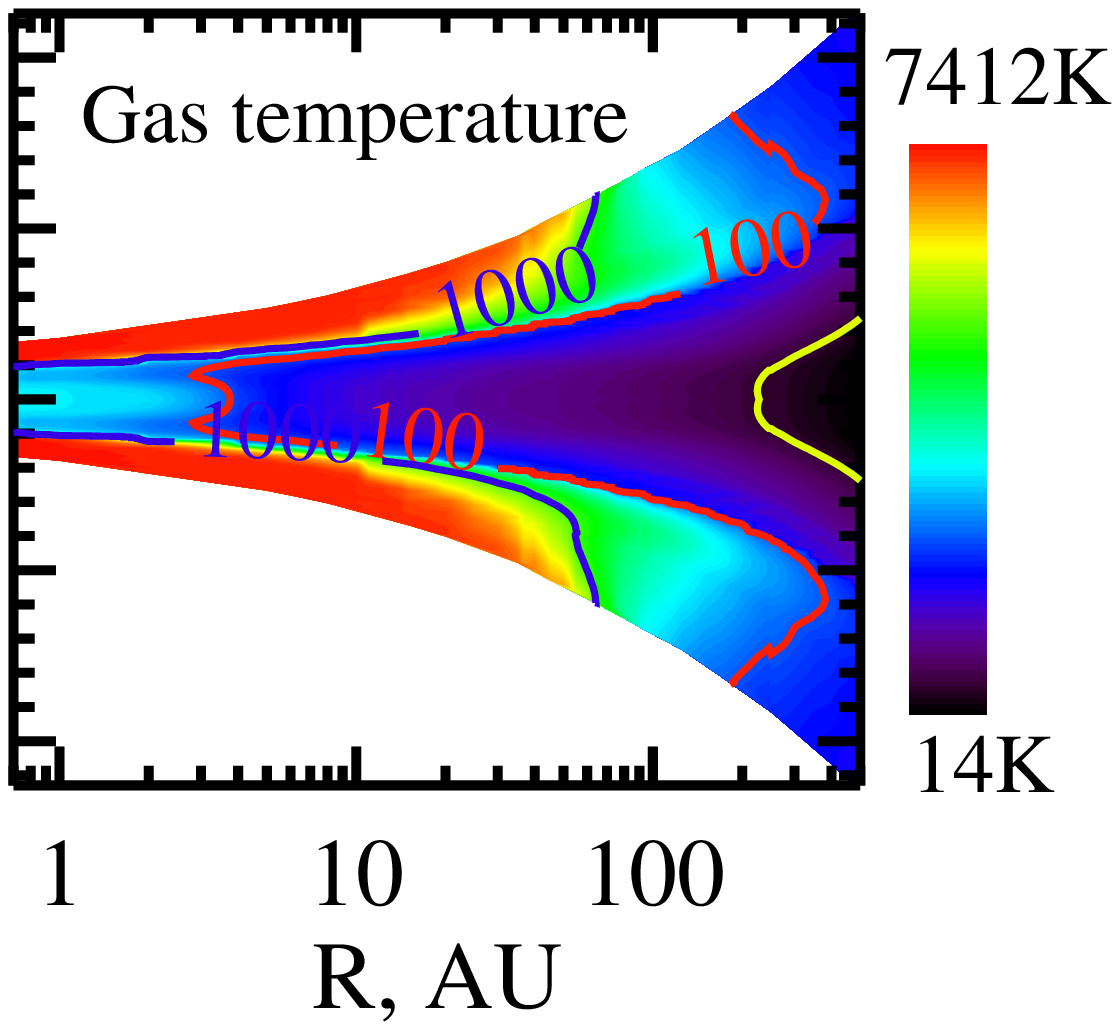}
\includegraphics[width=0.27\textwidth,angle=90]{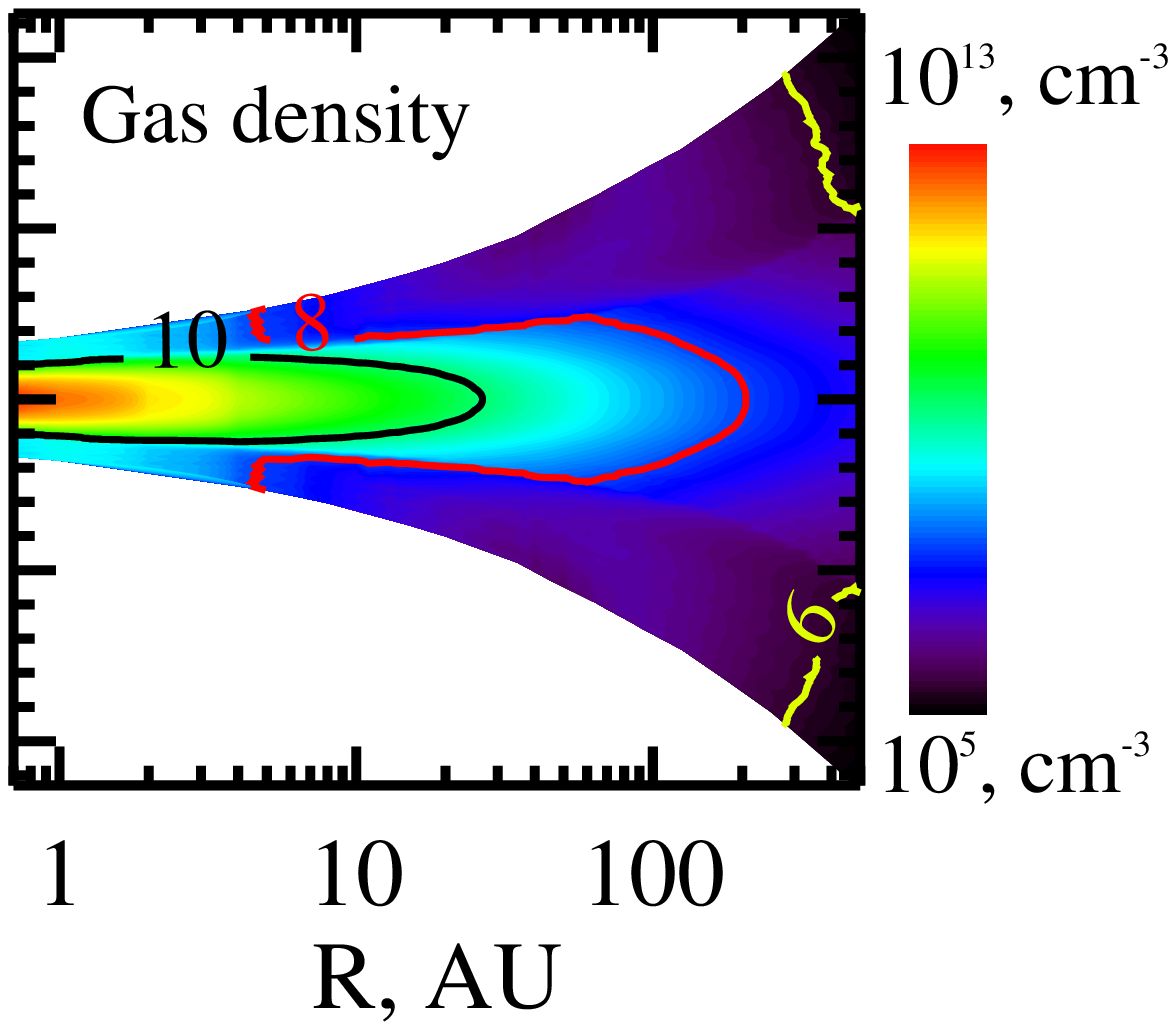}
\caption{A vertical slice through a protoplanetary disk resembling the 
DM~Tau system as calculated with the ANDES thermochemical disk model\cite{ANDES}. The 
high values are marked by red color, the low values are marked by dark
blue/black colors. 
(Left) The radial and vertical profile of the dust temperature (in Kelvin). 
(Middle) The radial and vertical profile of the gas temperature (in Kelvin). 
Note that the gas temperatures are much higher than the dust temperatures in
the disk surface layers. 
(Right) The radial and vertical profile of the gas particle density (in
cm$^{-3}$).
}
\label{fig:ANDES_disk_struc}
\end{figure}

Protoplanetary disks can be described as rotating dusty gaseous systems transporting a net amount of mass towards the
central star and the angular momentum outwards (see Fig.~\ref{fig:disk_scheme}). {\bbf These disks are characterized 
by strong radial and vertical temperature and density gradients (see Fig.~\ref{fig:ANDES_disk_struc}). High-energy stellar and 
interstellar radiation may  penetrate into the upper layers of disks,
enabling a rich molecular chemistry. In the deep and well-shielded interiors  temperatures become 
so low that molecules freeze out. The shielding is mostly  provided by micron-sized solid dust particles.
Apart from chemical evolution, the disks are characterized by
strong evolution of the initially micron-sized dust particles towards pebbles and, finally, planets.
This process has a strong impact on the physical structure of the disks, and therefore
 on the chemistry.} 

Protoplanetary disks are a special class of accretion disks.
Accretion is a mass flow
caused by the loss of potential energy due to frictional dissipation, which also leads to mechanical heating of the gas.
The velocity, temperature, and density structure of accretion disks
can be described by the conservation equations for energy, mass, and momentum. For a geometrically thin disk the time evolution of
the surface density $\Sigma$ can be expressed in the form of a non-linear diffusion equation with the viscosity $\nu$ as the
regulating parameter of the diffusion process\cite{Lynden-BellPringle74,Pringle81}.

The viscous stresses that are required for the evolution of accretion disks cannot be solely provided by the molecular viscosity
of the gas, which is many orders of magnitude too small to have any considerable effect on mass and angular momentum transport.
Instead, a ``turbulent'' viscosity has been invoked to explain the accretion behavior of protoplanetary disks. The origin of this
viscosity was  initially not known, and thus it has been often conveniently parameterized by the so-called $\alpha$ parameter:
$\nu$ = $\alpha c_{\rm s}H$,  where $c_{\rm s}$ is the sound speed and $H$ is the scale height of the 
disk\cite{ShakuraSunyaev73}. A quantity $\alpha \ll 1$
describes the regime of subsonic turbulence, $\alpha \sim 1$ -- transonic turbulence, and $\alpha \gg 1$ -- supersonic turbulence.
Typical values for $\alpha$ inferred for protoplanetary disks range between 0.001 and 
0.1\cite{2011ApJ...735..122F,2005A&A...442..703H,2011ARA&A..49...67W}.

In the case of steady-state optically thick disks with local energy dissipation the kinetic temperature $T$ decreases with
disk radius as $R^{-3/4}$. The mass accretion rate in such an $\alpha$-disk is given by
$\frac{dM(r)}{dt}\propto \nu\Sigma(r) \propto \alpha c_{\rm s}(r)H(r) \Sigma(r)$, where $\Sigma(r)$ is the surface density 
(g\,cm$^{-2}$).
The gas in disks moves on nearly circular orbits because the radial velocity component is much smaller than the angular
velocity. In most cases the mass of the central star(s) exceeds by far the mass of the disk itself, and the angular
velocity is given by the Kepler law, with $V_{\rm K} = \sqrt{GM_\star/R}$ (where $G$ is the gravitational constant and
$M_\star$ is the mass of the central star). If one assumes that the disk is isothermal in vertical direction, then the ratio $H/R$
should increase with radius $R$ and the disk has a ``flared'' geometry\cite{KenyonHartmann1987,B97}.

The physical origin of the turbulent viscosity in accretion disks, especially in protoplanetary disks, is a major topic of ongoing
astrophysical research. In ionized accretion disks a powerful magnetorotational instability (MRI) can efficiently drive
angular momentum transport and provides effective $\alpha$ values of the right order of 
magnitude\cite{MRI,Balbus_Hawley98,2011ApJ...735..122F}.
However, in deep disk interiors, where even cosmic rays are not able to penetrate efficiently, the ionization degree drops to
very low values, and dust grains become the dominant charge carriers\cite{1996ApJ...457..355G,sano,Red2}.
This can halt the MRI locally and thus results in a turbulent-inactive region (where $\alpha$-values drop well below 0.01).

These dynamically quiet ``dead'' zones together with the more turbulent disk regions can result in
very complicated density, temperature, and velocity structures in protoplanetary 
disks\cite{2012ApJ...744..144F,Dzyurkevich_ea13a,2013ApJ...764...65M}.
In very massive disks and/or the colder outer regions of disks global gravitational instabilities can occur, which are
another potential source of angular momentum transport. The effective operation of gravitational instabilities
depends strongly on the delicate balance between local cooling and heating rates, which are often difficult to 
{\bbf constrain}\cite{2003ApJ...590.1060P,2006ApJ...651..517B}.

The thermal structure of protoplanetary disks, obviously a key parameter for disk chemistry, is not only determined by the
dissipation of accretion energy. In fact, accretion heating is only dominant in the very inner, densest disk region where
planets form, while in the inner surface layers and the outer disk regions the processing of stellar and interstellar radiation 
by dust particles plays a key role\cite{2007prpl.conf..555D,Hirose_Turner11a}. {\bbf The protoplanetary disks are divided
on three different classes according to the luminosity and mass of their central stars: 
(1) disks around brown dwarfs, (2) disks around Sun-like T~Tauri stars, and (3) disks around more luminous and massive 
Herbig~Ae/Be stars.} 

{\bbf Brown dwarfs are ``failed'' stars that are not massive enough to empower hydrogen burning, and which
produce internal energy solely by gravitational contraction followed by slow, steady cooling. 
Not much is known yet with respect to chemistry in their disks, both theoretically and observationally, 
so we will not discuss it in our review. 
T~Tauri stars are young, $\la 10$~Myr, pre-main-sequence stars of the F--M spectral types, which are surrounded by gaseous 
nebulae. They have masses below $\sim 2M_{\rm Sun}$, surface temperatures similar to that of the Sun, and large radii $\ga 
2R_{\rm Sun}$.  Hydrogen burning does not start in their interiors until they are about 100~million years old, so, like brown 
dwarfs, they are still powered by gravitational contraction. T~Tauri stars are active and highly variable, with strong stellar 
winds, and intense thermal X-ray and non-thermal FUV radiation\cite{1989ARA&A..27..351B,Bergin_ea04,Preibisch_ea05}. 
Herbig Ae/Be stars are more massive ($\sim 2-8M_{\rm Sun}$), hotter ($\sim 8\,000-15\,000$~K), and more luminous ($\ga 
15-200L_{\rm Sun}$) counterparts to the T~Tauri stars of the spectral types A or B\cite{1998ARA&A..36..233W}. Being hot,
Herbig Ae/Be stars produce stronger thermal UV radiation than the T~Tauri stars. Their X-ray luminosities are in 
general lower than those of the T~Tauri stars due to the lack of efficient dynamo mechanism 
in their non-convective photospheres\cite{Guedel_Naze09,Bethell_Bergin11}.}

In strongly accreting disks the midplane temperatures can be as high as 1\,000~K {\bbf at radii of several~AU}\cite{DAea99}. 
{\bbf 
The direct irradiation from the central star can lead to the formation of inner disk walls, and thus disk 
shadows and complicated
flaring patterns\cite{2001ApJ...560..957D}. The highest dust temperatures in the disks, $T \sim 1\,500$~K, are set by the 
sublimation temperatures of the most refractory solids (e.g. corundum, Al$_2$O$_3$) and combustion-like destruction of 
carbonaceous compounds\cite{Gail_2010}.} Apart from radial temperature gradients, we can also expect a steep rise of
temperature in the vertical direction. All these factors make the
construction of realistic disk models a challenge, and, obviously, a multi-dimensional description is required.

In most of current disk physical models the disk structure is considered to be in hydrostatic equilibrium
between gravity and thermal pressure. Consequently, the vertical density distribution can be approximately described by a
Gaussian function with $\exp{(-(z/H)^2)}$, where $z$ is the vertical disk height. In the deeper interiors of the disks dust and 
gas
remain well coupled through collisions and {\bbf the kinetic temperature of the gas is equal to the dust temperature. This is not 
the case in the upper tenuous disk layers, where a detailed balance of gas heating processes, mostly 
dominated by photoelectric heating by the stellar FUV radiation, and gas cooling through CO, C, C$^+$ and 
O line emission and other cooling lines must be
calculated\cite{Kamp_Dullemond04,Gorti_Hollenbach04,Gorti_Hollenbach08,ANDES}. 
In disks where dust and gas are well mixed,
dust and gas temperatures are equal slightly above the optical depth $=1$ surface\cite{Kamp_Dullemond04}.
This corresponds to gas particle densities of about $10^6-10^7$~cm$^{-3}$\cite{ANDES}. The exact density value for the 
dust-gas coupling depends on 
the dust cross section per H atom. The collisional coupling between gas and dust may be reduced by dust settling to
the midplane, decreasing the dust-to-gas ratio. In addition, grains may coagulate, which will also lead to a reduction of the
dust cross section per hydrogen atom\cite{Gorti_Hollenbach08,Vasyunin2011,ANDES}.}

Photoelectric heating rates depend sensitively on the abundance of very small particles and polyaromatic hydrocarbons (PAHs), 
which is often
difficult to estimate\cite{Kamp_Dullemond04,Jonkheid_ea06,ANDES}.
In addition to FUV radiation, strong X-ray emission and flares are observed in T Tauri stars, which is an important additional
heating mechanism of the disk atmospheres by energetic secondary electrons\cite{Glassgold_ea04,2012ApJ...756..157G}.
At a radial distance of 1~AU from the star the gas temperature in the disk atmosphere can become as high as 5\,000-10\,000~K 
{\bbf 
(see 
Fig.~\ref{fig:ANDES_disk_struc})},
which raises
a question regarding the dynamical stability of this region. According to advanced disk models, it is plausible that the inner 
atmosphere is gradually lost, steadily reducing the disk mass and changing the global disk structure\cite{Gorti_ea09,Owen_ea11a}.

The gas accretion process is not the only driving force of disk evolution. Protoplanetary disks are also characterized by a
variety of dust evolution processes, including dust growth, fragmentation,
vertical sedimentation, and radial drift\cite{Brauer_ea08a} (for reviews see\cite{BHN00,2011ppcd.book..114H}).
The overall dust growth implies transformation of sub-micron sized particles into km-sized bodies,
which covers many orders of magnitude on a spatial scale and which is governed by a multitude of physical processes. In essence,
dust particles are assembled into cm-sized pebbles by Brownian motion, differential drift, and turbulence, followed by the dust
decoupling from the gas 
and rain down of bigger grains toward the disk midplane. The headwind exerted on these pebbles by the gas that orbits at slightly
sub-Keplerian velocity leads to rapid inward transport and loss of these pebbles (also due to mutual destructive collisions at
$\ga 10$~m\,s$^{-1}$)\cite{Weidenschilling_Cuzzi93,Brauer_ea08a,Birnstiel_ea10a}. 
Currently it is difficult to single out the most robust mechanism to overcome this so-called
``1~m-size'' barrier to continue the grain growth into the meter and kilometer regime. Plausible explanations include trapping
of solids in turbulent eddies and other long-lived over-densities in disks produced by 
turbulence\cite{2003ApJ...598.1301H,Johansen_ea09a,Birnstiel_ea10a,2012A&A...545A.134M}. When km-sized bodies are formed from the 
pebbles, they interact with each
other gravitationally and efficient growth is regulated by the few most massive first planet embryos (``oligarchic'' growth).

In the modern version of the so-called core-accretion scenario of planet 
formation\cite{1969edo..book.....S,Hayashiea85,1996Icar..124...62P} meter-sized rocks in the disk midplane become subject to
gravitational instabilities leading to the formation of km-sized planetesimals\cite{Johansen_ea07}. These planetesimals grow
through gravitationally
induced mutual collisions and coalescence to solid planet cores. Giant planets form if there is enough gas left in the disk to
gravitationally collapse onto these cores, forming their massive atmospheres at the last stage of evolution. Upon formation
planets can further interact with the turbulent gas in disks and can radially migrate and clear gaps or entire inner holes,
depending on the planet mass and the actual disk structure\cite{2007prpl.conf..655P}.
{\bbf An alternative scenario is the formation of massive clumps in outer cold regions of mostly massive 
protoplanetary disks\cite{Boss_97,Boley_ea10a}. This scenario has recently again attracted attention in order to explain
the presence of directly-imaged massive planets or brown dwarfs on wide orbits\cite{Carson_ea13a}. Detailed analysis of direct 
imaging data indicates
that sub-stellar companions formed by disk instabilities are rare and the core accretion remains the likely dominant formation
mechanism for the entire planet population\cite{Janson_ea12a}.}

\subsection{Global Parameters of Protoplanetary Disks}
\label{sec:phys_struc:parameters}
The disks masses and sizes and their radial and vertical temperature and density distributions are important quantities both for
the disk chemistry and the planet formation process.
A comprehensive overview how these quantities are actually measured through
astronomical observations and which results have been obtained is provided by Williams \& Cieza\cite{2011ARA&A..49...67W}.
Here, we will only summarize some of the most important results.

The mass of protoplanetary disks is dominated by molecular hydrogen and helium in a mass ratio defined by the cosmic
abundance of these elements. Unfortunately, the total disk masses cannot be determined directly from molecular hydrogen emission
because  H$_2$ is a homonuclear molecule without allowed electric dipole transitions and most of the disk mass is at too low
temperatures to emit in ro-vibrational magnetic quadrupole transitions. These lines occur at near- and mid-infrared wavelengths
and need excitation temperatures that are only provided in the very inner disk regions at several AU from the star
(1~AU (astronomical unit) $=$ mean distance between Sun and Earth). 
{\bbf Another complication is the fact that the emission from dust itself is often optically thick at these wavelengths and 
that one can 
only probe a very limited surface region of the disk\cite{Carmona_ea08}.}
Disk emission from the CO molecule in its rotational transitions are frequently measured
thanks to its relatively high abundance and permanent dipole moment. However, the CO/H$_2$ abundance ratio changes with radius and
vertical depth because of three main factors. First, CO gas severely freezes-out on dust grains in the cold disk midplane (where
temperatures drop below {\bbf about 20~K})\cite{Tielens_ea91,Qi_ea13a}. 
Second, in the irradiated disk atmosphere CO can be photodissociated by FUV 
radiation,
despite the ability of this molecule to self-shield itself\cite{vD88,Clayton2002,vDea_06,Lyons2007,Visser_ea09b}. Third, due to 
large concentrations of CO in disks,
$^{12}$C$^{16}$O rotational emission lines become saturated and probe the disk matter only at specific depths.
The last problem can be circumvented by using rare isotopologue lines of CO having much lower optical depths, such as
$^{12}$C$^{18}$O and even $^{12}$C$^{17}$O\cite{2006ApJ...637L.129S}.
This implies that mass estimates based on CO observations alone are highly
uncertain, at least for the colder disks around T~Tauri stars.

{\bbf An elegant way to infer disk masses would be to use the rotational lines of HD to circumvent the H$_2$ and CO problem
with disk mass determinations. However, this requires very high sensitivity at far-infrared wavelengths. 
The {\it Herschel} observatory provided the first such detection in the disk around the nearby star 
TW~Hya\cite{2013Natur.493..644B}. This observation demonstrated that the disk mass in this system is not as small as 
previously thought and is at least $0.05M_\odot$.}

The thermal emission of the dust particles can be relatively easily measured because of their much higher opacities. Under the
assumption of optically thin disk emission, which is best fulfilled for protoplanetary disks at (sub-)millimeter wavelengths,
and with a specified dust
opacity, which is often poorly constrained, the measured flux values can be directly converted into dust masses. 
{\bbf The dust disk mass $M_{\rm dust}$ is given by

\begin{equation}
 M_{\rm dust} = F_{\rm 1.3 mm}D^2/\kappa({\rm 1.3mm})B_{1.3mm}(T_{\rm dust}). 
\end{equation}

Here, $F_{\rm 1.3 mm}$ is the flux at 1.3~mm wavelength, $D$ is the distance to the object, $B_{1.3mm}$ is the Planck
function at the dust temperature $T_{\rm dust}$, and $\kappa({\rm 1.3mm})$ is the mass absorption coefficient per gram of dust.
One} should
note that the measured dust emission is only sensitive to grain sizes up to millimeters and the mass reservoir in larger
``boulders'' cannot be constrained this way. Under the assumption of a canonical gas-to-dust mass ratio of 100, total disk masses
can then be derived. With these caveats in mind,
inferred median disk masses of $5\,10^{-3}$ solar masses and a median disk-to-star mass ratio of 0.5\% has been obtained from
surveys of the Taurus-Auriga and $\rho$~Ophiuchus star-forming 
regions\cite{Andrews_Williams05,2007ApJ...671.1800A,Guilloteau_ea11a}. {\bbf In a recent study\cite{Andrews_ea13a} an
inherently linear disk mass vs. stellar mass scaling relation was found, 
but with a considerable dispersion at any given stellar mass, probably
reflecting the combined effect of evolution, different dust properties, and temperatures.}
A considerable fraction of these disks has masses above 10 Jupiter masses, a minimum value of mass needed to form the solar system
within the orbit of Neptune at $\sim 30$~AU. 
Other factors that influence the mass of disks through orbital interaction are close-in binaries or planetary systems.
As stars rarely form in isolation, the proximity to very massive stars may lead to premature gas photoevaporation
or disk disruption through dynamical interactions between cluster members.

{\bbf Accretion rates in protoplanetary disks, mostly measured close to the star, cover a wide range of values from $10^{-9}$ to
$10^{-12}~M_\odot$~yr$^{-1}$. They depend on a variety of parameters, with the most important being the stellar 
mass\cite{Muzerolle_ea03, 
Natta_ea06,Fang_ea09} and the stellar age and evolutionary status of the disk\cite{SiciliaAguilar_ea06b,2010A&A...510A..72F}.}

The disk radii are often difficult to determine precisely because of the vanishing emission at their cold and
low-density outer edges.
Submillimeter interferometric studies with high sensitivity and spatial resolution provided disk radii between 10~AU and
1\,000~AU (with a typical value of $\sim 200$~AU). 
{\bbf An interesting observational finding has been that the sizes of the gas disks, as determined by CO emission, 
are significantly larger than the sizes of the dust continuum images\cite{Pietu_ea07,Andrews_Wilner_ea12a}.}
These studies also provided estimates for
the radial surface density distribution, assuming a power law for regions not too close to the outer disk edge.
Towards the outer disk edge radial surface density tends to fall off exponentially\cite{Guilloteau_ea11a}.
The obtained power law is relatively flat with a power law index close to $-0.9$.

The radial gas temperature distribution in the outer disk seems to follow a power law with a power law exponent between
$-0.4$ and $-0.7$\cite{GD98,Pietu_ea07,2008A&A...491..219P}.
These temperature distributions are similar to what has been obtained from the modeling of the continuum spectral energy
distributions of disks\cite{2008A&A...491..219P,Andrews_Williams05}.

Multi-line studies in the low-lying rotational transitions ($J=1-0$ up to $6-5$) of CO isotopologues are presently providing a
first insight in the vertical temperature structure of the outer regions of protoplanetary 
disks\cite{Qi_ea06,Pietu_ea07,2012arXiv1205.6573A}.
These lines originate from different vertical layers of the disks depending on the location of the regions where their optical
depths
is close to 1. A rather surprising result of these studies was the discovery of a significant fraction of cold CO gas that has
temperatures below the freeze-out temperature of CO of about 20~K\cite{DDG03,Pietu_ea07}.

Infrared surveys of large populations of disks in stellar clusters of various ages have shown that the disk frequency is a
function of cluster age and steadily decreases from young star-forming regions with ages of 1~Myr to older regions of about 
10~Myr\cite{disc_fraction} with a median disk lifetime of several Myr.
Even shorter lifetimes were estimated for the presence of gas in inner disk regions,
based on measurements of the gas accretion rates\cite{2010A&A...510A..72F}. The search for colder gas in somewhat
older systems has been largely unsuccessful with a few exceptions\cite{2011ApJ...740L...7M}.
This implies that giant planet formation has to occur over the relatively short timescale of a few million years,
much smaller compared to the age of our solar system
of 4.567 billion years, as measured by radioactive dating of various meteoritic samples\cite{2007M&PS...42.1321A}.
{\bbf Our present insight into the gas-phase composition of protoplanetary disks indicates that
timescales of key chemical processes have to be shorter than $\sim 1$~Myr\cite{Semenov_Wiebe11a}.}

\section{Material Inventory and Fundamental Chemical Processes in Disks}
\label{sec:chemistry}

\subsection{Inventory of Gas-Phase Molecules}
\label{sec:chemistry:gas}
The detection of molecular line emission at infrared wavelengths requires relatively high spectral resolution in order
to isolate the
weak molecular lines from the bright dust continuum emission of the disks. In addition, the protoplanetary dust disks are 
optically thick at
infrared wavelengths and line emission can only be observed from the tenuous warm surface layers. This is different for the
(sub-)millimeter wavelength range where the dust disks are optically thin and molecular line emission can be observed throughout
the entire {\bbf disk, although the inner dust disk may still be optically thick. We should note that the emission of 
many molecules originates from above the very cold midplane, where freeze-out of gaseous
species onto grain surfaces occurs}. The sensitivity and spatial resolution of the present (sub-)millimeter facilities
limits most of the observations
to the very outer regions of disks beyond $\approx 30-100$~AU and to the most massive disks around nearby objects such as TW~Hya,
DM~Tau and MWC~480. On the other hand, infrared spectroscopy can trace both {\bbf molecular lines as well as characteristic 
absorption or emission features of solids. It} is also sensitive to
molecules without strong dipole moments, including PAHs with their forest of infrared emission lines.
{\bbf The different wavelength regimes correspond roughly to different temperatures and thus distinct disk regions,
keeping in mind that the temperature is decreasing from the inner to the outer disk. We now discuss the results starting
with the longest (sub-)millimeter wavelengths and finishing with the near-infrared wavelengths.}

\subsubsection{Results from (Sub-)millimeter Spectroscopy}
\label{sec:submm_results}
(Sub-)millimeter spectroscopy with single-dish and interferometric facilities has provided a first inventory of molecules in the
outer regions of disks. Two major programs, ``Chemistry in Disks'' at the IRAM 30-m telescope and the Plateau de Bure
Interferometer\cite{Dutrey_ea07,Schreyer_ea08,Henning_ea10,Dutrey_ea11a,Chapillon_ea12a,Chapillon_ea12b,Guilloteau_ea12a}
and ``DISCS'' at the SMA interferometer on Hawaii\cite{Oeberg_ea10a,Oeberg_ea11a,2011ApJ...743..152O} have provided 
most of the molecular line data on disk chemistry.

Apart from CO with its main isotopologues
($^{13}$CO and C$^{18}$O), {\bbf a handful of relatively simple polyatomic} molecules (HD, H$_2$CO, CS, C$_2$H, c-C$_3$H$_2$, 
HCN, 
HNC, CN, DCN) and
molecular ions (N$_2$H$^+$, HCO$^+$, DCO$^+$, H$_2$D$^+$) have been discovered by a variety of facilities.
The various molecules trace different physical and
chemical processes in the disks (see Table~\ref{table:lines}). The abundances of the discovered molecules relative to molecular
hydrogen range between $\sim 10^{-10}-10^{-4}$. Here we note that astronomical observations provide line intensities, which have
to be converted into column densities, assuming a specific temperature and density structure of the disk. This makes the analysis
of molecular emission
lines a challenging, non-trivial task, with resulting quantities usually uncertain by a factor of several.

\begin{table}
  \caption{Molecular Lines as Tracers of Physical and Chemical Conditions in Disks}
  \footnotesize
  \label{table:lines}
  \begin{tabular}{llll}
    \hline
    Molecule & Transitions & Quantity & Wavelengths \\
    \hline
    $^{12}$CO, $^{13}$CO           & Rotational and Ro-     & Temperature       & IR, FIR, sub-mm/mm\\
                                   & vibrational            &        &               \\
    H$_2$                          & Ro-vibrational            & Temperature       & IR                \\
    CS, H$_2$CO, HC$_3$N           & Rotational             & Density    & sub-mm/mm          \\
    HCO$^+$, N$_2$H$^+$, CH$^+$     & Rotational             & Ionization        & sub-mm/mm, FIR          \\
    CN, HCN, HNC                   & Rotational             & Photochemistry    & sub-mm/mm          \\
   H$_2$O, OH       (inner disc)  &  Rotational               & Temperature       & IR         \\
                                   &                        & Photochemistry    &            \\
    H$_2$O, OH       (outer disc)  & Rotational                & Photodesorption   & FIR       \\
    Complex organics (outer disc)  & Rotational             & Grain surface processes & sub-mm/mm   \\
    Complex organics (inner disc)  & (Ro-)vibrational       & High-T chemistry  & IR         \\
                                   &                        & Photochemistry         &           \\
    HD, DCO$^+$, DCN, H$_2$D$^+$   & Rotational             & Deuteration       & FIR, sub-mm/mm \\
    CS, HC$_3$N                    & Rotational              & Turbulence        & sub-mm/mm\\
    \hline
  \end{tabular}

\end{table}

Recently, the first heavier organic molecule, the cyanoacetylene HC$_3$N, was discovered in the disks around GO~Tau and 
MWC~480\cite{Chapillon_ea12b}. {\bbf Heavier polyatomic molecules remain  undetected because of their low abundances,
weak line intensities due to energy partitioning into a multitude of levels, and the limitations 
in sensitivity of the present-day facilities. The situation will change when the {\it Atacama Large Millimeter/Submillimeter 
Array} (ALMA) becomes fully operational by the end of 2013, bringing unprecedented sensitivity and spatial and spectral 
resolution. This will allow searches for complex species at low spatial resolution (to maximize sensitivity) and 
detection of strong molecular emission from inner regions of protoplanetary disks at high resolution ($r \gtrsim 10-20$~AU).
The first discovery of a somewhat heavier molecule with ALMA was the detection of cyclopropenylidene, c-C$_3$H$_2$,  in the disk
around TW~Hya\cite{2013arXiv1302.0251Q}.}

A general result of these molecular line studies is evidence for the depletion of molecules relative to molecular abundances
observed in the interstellar medium\cite{Dutrey_ea97,Kastner_ea97} (see Fig.~\ref{fig:disk_vs_TMC1}). This is caused primarily by 
two effects. First, freeze-out
(or ``depletion'') of
molecules onto dust grains occur in cold disk midplanes, where later some of these ices may become incorporated in icy bodies
such as comets. Second, molecules are destroyed by photodissociation in disk atmospheres. Indeed, photodissociation products
like CN and the elevated CN/HCN ratios point to the presence of photon-dominated regions at disk surfaces.

\subsubsection{Results from Far-Infrared Spectroscopy}
\label{sec:fir_results}
The \textit{Herschel} far-infrared observatory with its spectrometer instruments \textit{PACS}
(with a moderate spectral resolution of 1\,000--4\,000) and \textit{HIFI} (with a high spectral resolution up to 10$^6$) has
provided a flood of interesting molecular data on disk chemistry. 
{\bbf Apart from the discovery of cold and warm hydroxyl and water in 
a number of disks\cite{2012A&A...544L...9F,Meeus_ea12,2012A&A...538L...3R,Fedele_ea13a} and the recent detection of cold water 
vapor in the TW~Hya disk\cite{2011Sci...334..338H} (see Section~\ref{sec:chemistry:water})}, 
the observatory revealed warm CO emission in a large number of disks around
more massive, hotter Herbig~Ae/Be stars\cite{2010A&A...518L.129S,Meeus_ea12,Meeus_ea13a}.
The wavelength range of the \textit{PACS} instrument covers mid- to high-lying rotational J transitions (and CO transitions up
to J=31-30 could be discovered), demonstrating the presence of gas with temperatures between 100-1\,000~K\cite{Meeus_ea13a}.

In addition, the oxygen fine structure line at 63$\mu$m, and much less frequently the [OI] line at
145$\mu$m, have been observed in a number of protoplanetary disks with {\it Herschel}\cite{Meeus_ea12}.
The fine structure line of
neutral atomic carbon at 158$\mu$m has also been detected, albeit much less frequently than was anticipated
from preliminary modeling\cite{Bruderer_ea12}.
A surprise was the discovery of CH$^+$ in the disks surrounding the two Herbig~Ae
stars HD~100546\cite{Thi_ea11a} and HD~97048\cite{Meeus_ea12}.
The main formation pathway for this ion is the endothermic reaction (with an activation energy of $\sim 4\,500$~K)
between C$^+$  and molecular hydrogen, implying that CH$^+$ is tracing warm gas in the inner disk atmosphere.
An alternative explanation is a steady fragmentation of PAHs by intense stellar high-energy radiation, releasing
aromatic rings and its ``debris''.

\subsubsection{Results from Mid- and Near-Infrared Spectroscopy}
\label{sec:nir_results}
\begin{figure}
\includegraphics[angle=0,width=12cm]{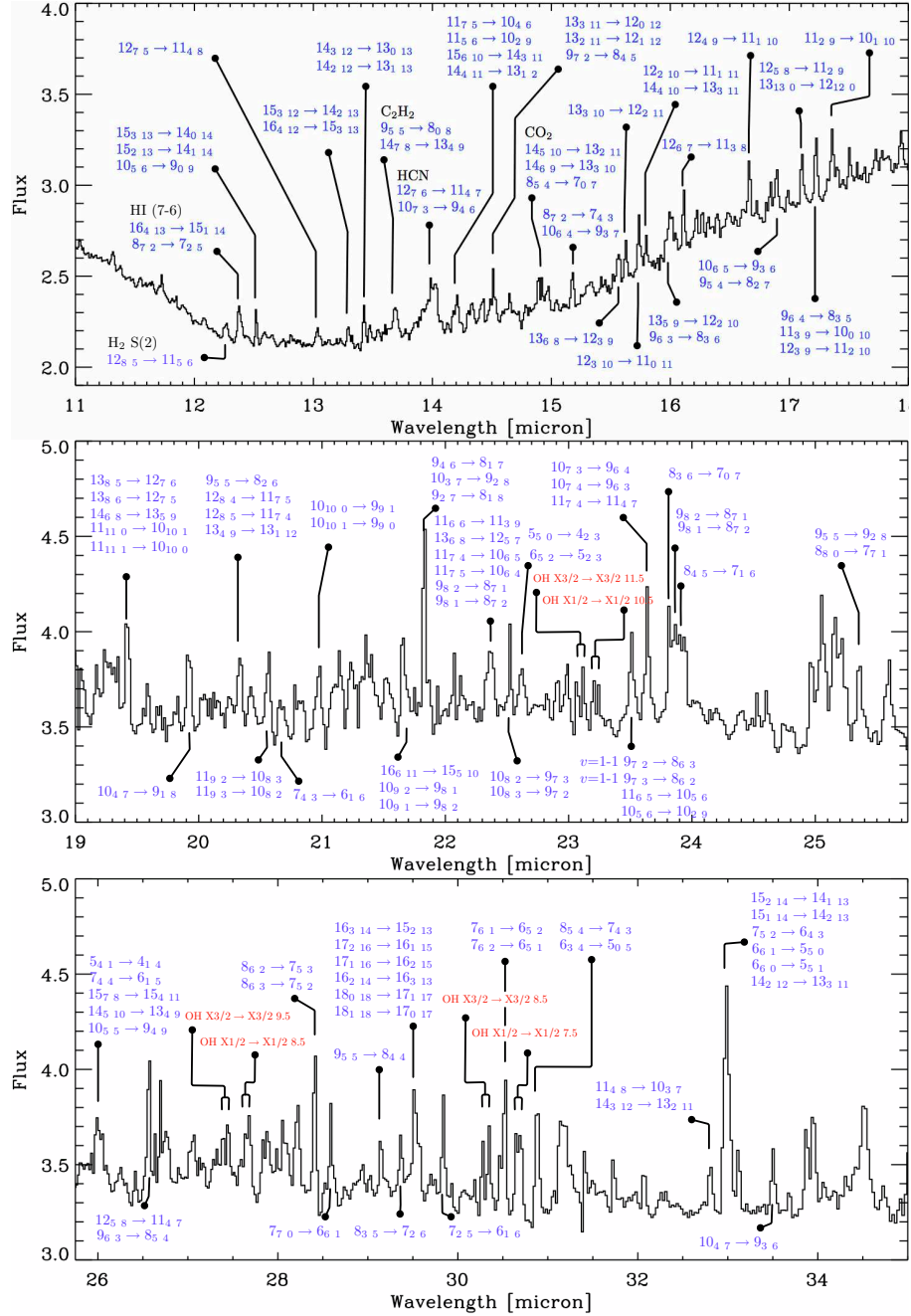}
\caption{Selected major molecular lines in the {\it Spitzer}
high-resolution spectrum of a disk around the T~Tauri star RNO~90. The transitions refer to
the rotational quantum numbers $J_{K_aK_c}$ in the ground vibrational state of water.
Rotational transitions with odd sum of $K_a$ and  $K_c$ correspond to ortho nuclear spin configurations.
{\bbf Reprinted with permission from Reference\cite{2010ApJ...720..887P}. Copyright 2010 American Astronomical Society.} 
\label{fig:water_disk}}
\end{figure}

Despite the relatively low spectral resolution provided by the infrared spectrometer on board the \textit{Spitzer} observatory
(resolution of $90-600$ over the wavelength range from $5.3$ to $38\mu$m), this mission provided extremely interesting
constraints on disk chemistry in the planet-forming zones.
Apart from the {\bbf first} discovery of a forest of emission lines from {\bbf hot} H$_2$O and OH in a
number of disks (see Fig.~\ref{fig:water_disk}), CO$_2$ and organic molecules such as HCN and C$_2$H$_2$ could be 
discovered\cite{Carr_Najita08,Salyk_ea08,2009ApJ...696..143P,2010ApJ...722L.173P,2011ApJ...733..102C,Salyk_ea11a,
2012ApJ...747...92M}.
Based on a comparison between disks around Sun-like stars and
cool stars/brown dwarfs a significant underabundance
of HCN relative to C$_2$H$_2$ was found in the disk surface of cool stars\cite{2009ApJ...696..143P}. This difference between the 
two classes of objects indicates
different chemical regimes due to large differences in the UV irradiation of their disks. Additionally, strong vibration-rotation
absorption bands of CO$_2$, C$_2$H$_2$ and HCN could be discovered in disk systems seen 
edge-on\cite{Lahuis_ea06,2007ApJ...660.1572G}.

{\bbf The high temperatures between a few 100~K and a few 1\,000~K and  relatively high densities of $>10^8$~cm$^{-3}$
in the very inner regions of protoplanetary disks are}  appropriate for excitation of rotational-vibrational molecular
transitions. Indeed, rotational-vibrational emission of CO at $4.7\mu$m has been frequently observed in disks around Herbig~Ae
stars and T~Tauri stars\cite{Bea03,Pontoppidan_ea08}. An analysis of the
excitation conditions and velocity profiles suggests that the lines originate from a range of radii from about
$0.1$~AU out to $1-2$~AU\cite{2003ApJ...589..931N}. In a number of objects, CO overtone emission at $2.3\mu$m
has been observed and traces hot and very dense gas close to the star\cite{2007prpl.conf..507N}.
In addition, high-resolution
near-infrared spectroscopy (with spectral resolution between 25\,000 and 96\,000) with the CRIRES instrument at the
\textit{Very Large Telescope} in Chile and the NIRSPEC instrument at the \textit{Keck} 
telescope revealed the presence of H$_2$O, OH, HCN and C$_2$H$_2$ in the very inner disk regions\cite{2012ApJ...747...92M}.

The presence of the larger PAH molecules in disk surface layers would have important implications for their gas temperature
through photoelectric heating and chemistry on their surfaces, including the formation of molecular 
hydrogen\cite{2012ApJ...752....3T}.
Infrared emission from PAHs has mostly been observed from the disks around Herbig~Ae stars\cite{2010ApJ...718..558A}, with many
non-detections toward the T~Tauri disks\cite{2007A&A...476..279G,2010A&A...511A...6S}. The
overall PAH emission strength is generally higher in targets with a flared disk geometry, {\bbf pointing to the importance
of the radiation field. The relative differences in the IR emission features are mainly caused by chemical differences, 
especially the ratio of aromatic to aliphatic components induced by the stellar UV radiation field\cite{2010ApJ...718..558A}.}

\begin{figure}
\includegraphics[angle=0,width=15cm]{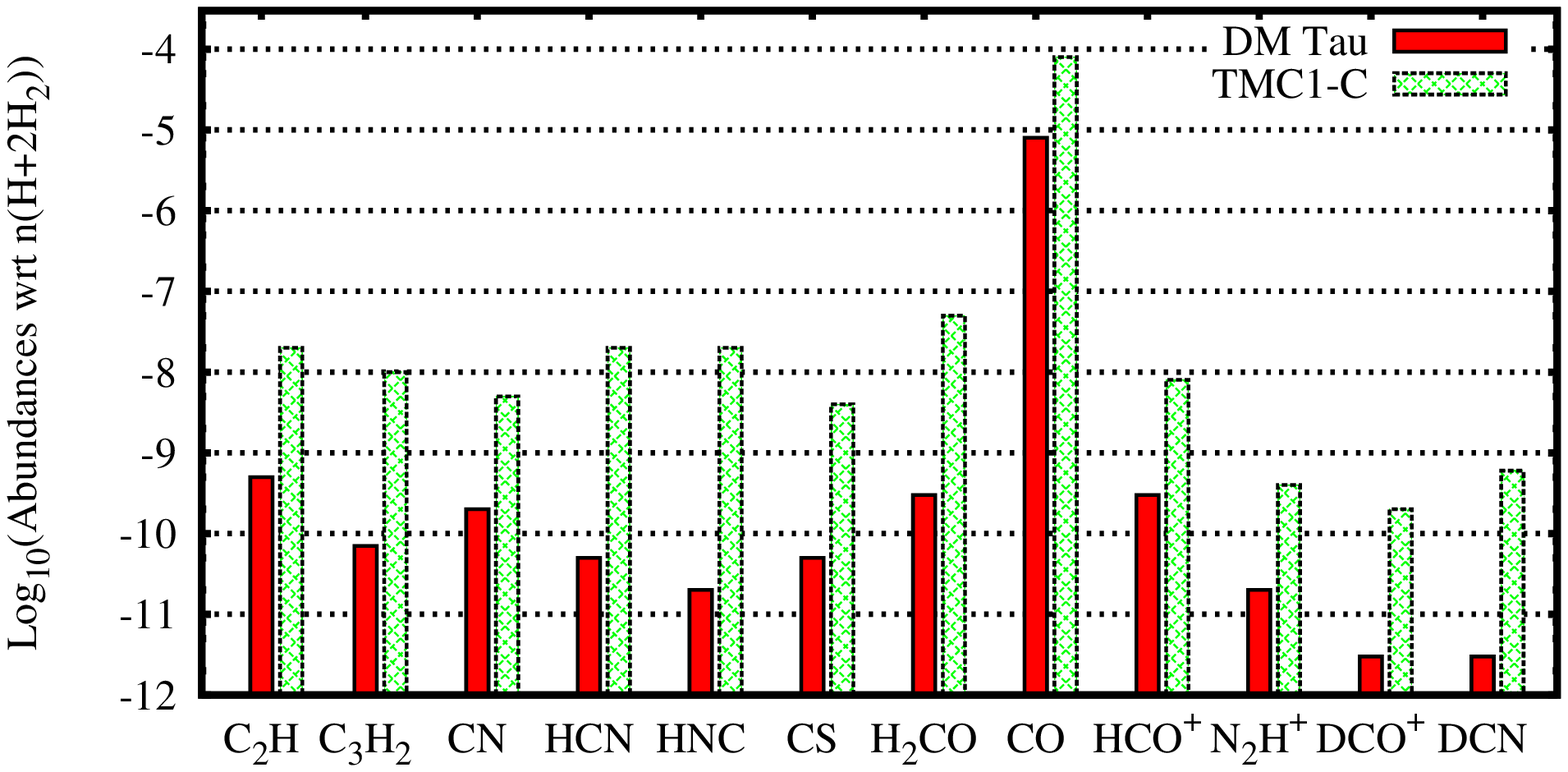}
\caption{Observed fractional abundances (wrt the amount of hydrogen nuclei) 
in the disk around the Sun-like T~Tauri star DM~Tau and the
prestellar core 
TMC1-CP. {\bbf Data derived from 
References\cite{2000A&A...356.1039T,2002A&A...381.1026R,2006A&A...448L...5G,Dutrey_ea07,2008A&A...492..703C,Chapillon_ea11, 
Dutrey_ea11a,
2011ApJ...743..152O,Semenov_Wiebe11a,Chapillon_ea12b,Oeberg_ea12a}.} 
\label{fig:disk_vs_TMC1}}
\end{figure}

\subsection{Refractory Grains and Molecular Ices}
\label{sec:chemistry:ices}
\begin{figure}
\includegraphics[angle=0,width=16cm]{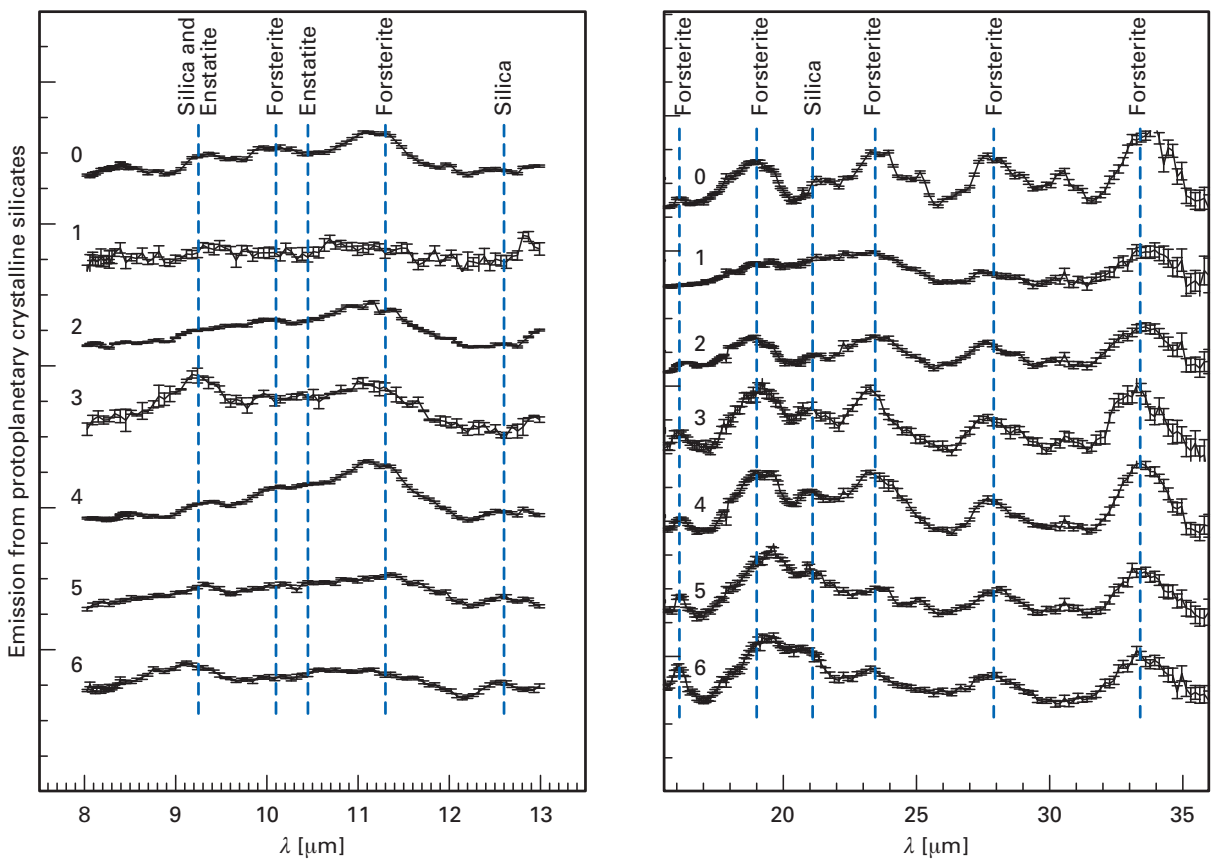}
\caption{The emission bands of crystalline silicates as observed with {\it Spitzer}
in the spectra of several T~Tauri stars. The amorphous silicate and PAH features are removed from the spectra
to enhance the contrast. {\bbf Data derived from Reference\cite{2011ppcd.book..114H}}.\label{fig:disk_silicates}}
\end{figure}

Ground-based infrared observations in the $8-13\mu$m atmospheric window and spectroscopic data from the
\textit{Infrared Space Observatory}, the \textit{Spitzer} satellite and the \textit{Herschel} space mission have
provided a very rich collection of infrared spectra of protoplanetary disks around young stars.
The mid-infrared spectra of T~Tauri stars and Herbig~Ae/Be stars are dominated by emission features produced by
vibrational resonances in amorphous and crystalline silicates, see Figure~\ref{fig:disk_silicates}
(see \cite{Natta_ea07,2011ppcd.book..114H} for reviews).
The emission of the silicate dust particles comes from the optically thin warm
surface layer of disks with typical temperatures above 100~K. The comparison between calculated absorption cross sections and
fluxes based on experimentally determined optical properties and the observed silicate emission features led to the conclusion
that a mixture of amorphous silicates with olivine and pyroxene stoichiometry, crystalline forsterite and enstatite and in some
cases silica can best explain the observed spectra\cite{Juhasz_ea10a}. A comprehensive study of
high-quality \textit{Spitzer} spectra of Herbig~Ae/Be stars indicates that porous iron-poor amorphous silicates are responsible
for the observed $10\mu$m features produced by the Si-O stretching mode. {\bbf In addition, the analysis of the strength
and shape of the $10\mu$m silicate feature has provided strong evidence for considerable grain growth to micron-sized particles,
which are much larger than the ``pristine'' submicron-sized dust grains of the interstellar medium 
(see\cite{Natta_ea07,2011ppcd.book..114H} for a review). The solid-state 
infrared  bands become flatter and finally disappear when the sizes of the emitting grains become comparable to the 
wavelength, 
i.e., for the $10\mu$m silicate feature this occurs for grains bigger than several 
microns\cite{2005A&A...437..189V,Juhasz_ea10a,2011ppcd.book..114H}.
The observed {\bbf anti-}correlation between the size of the amorphous grains and disk flaring points to the combined 
effect of coagulation and sedimentation\cite{Bouwman_ea08,2009A&A...497..379M,Juhasz_ea10a}. Coagulation leads to depletion 
of growing dust grains from the extended disk atmosphere, as they are gravitationally settling toward the midplane. 
Consequently, the surface where disks become optically thick moves along and the disk vertical structure becomes flatter.
Evidence for much larger grains up to centimeter sizes comes from the analysis of dust emission at millimeter and even centimeter
wavelengths\cite{Testi_ea03,Wilner_ea05,Rodmann_ea06,Perez_ea12}.}

Sharp bands of crystalline silicates can be observed in nearly all disk spectra (see Fig.~\ref{fig:disk_silicates}). The 
fractional abundances of crystalline
silicates cover values between $\approx 1\%$ and $30\%$. The observed bands are best explained by emission from forsterite and
enstatite particles. The forsterite-to-enstatite mass ratio changes with location, with lower values in the inner disks and
higher values in the outer disks. The analysis of the $69\mu$m feature produced by forsterite particles clearly shows that the
particles are nearly iron-free\cite{2010A&A...518L.129S,Sturm_ea13a}. The presence of crystalline silicates in
protoplanetary disks is a finding that is 
in strong contrast to the properties of dust in molecular clouds and the diffuse interstellar medium,
where crystalline silicates are not found. The presence of crystalline silicates in disks can only be explained by strong thermal
processing in protoplanetary disks either through thermal annealing and condensation in the inner 
regions of disks\cite{Gail_04} or shock heating at several astronomical units from the central star\cite{HD_02}. None of these 
theories is
without problems and fully consistent with the observational constraints. An interesting {\bbf piece of information to address} 
the puzzle of crystal formation in disks was the observation of {\em in-situ} crystal formation through the annealing of dust in 
the surface layers of the protoplanetary disk around the eruptive star EX~Lupi\cite{Abraham_ea09,2012ApJ...744..118J}.

Protoplanetary disks should certainly contain other solid phases such as Fe and FeS grains as well as carbonaceous 
particles\cite{Pea94,RP_opacities}. However, such particles have not been discovered by infrared spectroscopy
so far either because they do not show intrinsic infrared bands, they are too large in size to show strong features, or they are
simply not abundant enough. In a few disks around Herbig~Ae/Be stars infrared features at $3.43$ and $3.53\mu$m have been
detected\cite{2006A&A...457..171A}. These features were identified as the vibrational modes
of hydrogen-terminated facets of nano-diamonds\cite{1999ApJ...521L.133G}.

In the rare situation of disks seen edge-on evidence for the presence of molecular ices can be found through their absorption
features\cite{Terada_ea07a,2012A&A...538A..57A}.
An interesting example of such an object is the source
CRBR~2422.8-3423\cite{2002A&A...394L..27T,Pontoppidan_ea05},
where at least part of the H$_2$O and CO$_2$ absorption features are apparently produced in the disk.
In addition, a feature at $6.85\mu$m, tentatively attributed to
NH$_4^+$, shows evidence for {\bbf grain} heating to $\approx 50$~K and is certainly produced in the disk around this object.

Librational features of crystalline water ice at $44$ and $63\mu$m have the potential to provide important information about the
frozen water reservoir. So far, these features have only be detected in very few objects\cite{1999A&A...345..181M}, 
but the \textit{Herschel} mission should find the $63\mu$m feature in some additional disk 
sources\cite{McClure_ea12}.

\subsection{Main Chemical Processes}
\label{sec:chemistry:reactions}
As discussed in Chapter~\ref{sec:phys_struc} protoplanetary disks are characterized by strong vertical and radial temperature and 
density gradients together with vastly different radiation fields at various disk locations. These locally different disk 
properties
imply a rich and diverse disk chemistry, including photochemistry, molecular-ion reactions, neutral-neutral reactions,
gas-grain surface interactions, and grain surface reactions. A summary of relevant reactions is provided in
Table~\ref{table:reactions}. 

{\bbf Based on the radially decreasing temperature, disk} chemistry can be roughly divided in inner disk chemistry {\bbf 
($\la 20$~AU) and chemistry in the
outer disk regions beyond 20~AU}. Observationally the products 
of inner disk chemistry are best characterized by infrared
spectroscopy, whereas the outer disk is the domain of (sub-)millimeter observations. 

\begin{table}
\caption{Chemical reactions active in disks}
\label{table:reactions}
\footnotesize
\centering
\begin{tabular}{llcccc}
\hline
{\bbf Process} & {\bbf Example} & {\bbf Midplane} & {\bbf Molecular}   & {\bbf Atmosphere}   & {\bbf Inner} \\
              &               &                & {\bbf layer}       &                    & {\bbf zone} \\
              &               & $r > 20$~AU    & $r > 20$~AU    & $r > 20$~AU    & $r < 20$~AU    \\
\hline
{\bbf Bond formation} & & & & & \\
Radiative association  & C$^+$ + H$_2$ $\rightarrow$ CH$_2^+$ + h$\nu$          & X         &  X    & X   &  X  \\
Surface formation      & H + H$\|$gr    $\rightarrow$ H$_2$ + gr    &  X         & X    & 0   &  0 \\
Three-body             & H + H + H $\rightarrow$ H$_2$ + H          &  0         & 0    & 0   & X \\
\hline
{\bbf Bond destruction} & & & & & \\
Photodissociation      & CO + h$\nu$ $\rightarrow$ C + O          &  0         & X    & X   & X \\
Dissociation by CRP    & H$_2$ + CRP $\rightarrow$ H + H                & X           & X    & 0    & 0 \\
Dissociation by X-rays & ---                                      & 0           & X    & X    & X \\
Dissociative           & H$_3$O$^+$ + e$^-$ $\rightarrow$ H$_2$O + H       & X          & X    & X   & X  \\
recombination          & & & & & \\
\hline
{\bbf Bond restructuring} & & & & & \\
Neutral-neutral        & O + CH$_3$ $\rightarrow$ H$_2$CO + H                & X         & X     & 0   & X \\
Ion-molecule           & H$_3^+$ + CO $\rightarrow$ HCO$^+$ + H$_2$      &  X         & X     & X   & X \\
Charge transfer        & He$^+$ + H$_2$O $\rightarrow$ He + H$_2$O$^+$     &  X         & X     & X   & X \\
\hline
{\bbf Unchanged bond}   & & & & & \\
Photoionization        & C + h$\nu$ $\rightarrow$ C$^+$ + e$^-$ & 0       & X     & X   & X \\
Ionization by CRP      & C + CRP $\rightarrow$ C$^+$ + e$^-$           & X       & X     & 0   & 0 \\
Ionization by X-rays   & ---                                          & 0       & X     & X   & X \\
\hline
\end{tabular}
\end{table}

Inner disks are characterized by their high temperatures (from about 100~K to 5000~K) and high densities up to 10$^{12}$
cm$^{-3}$ (and more). At these high temperatures and densities in the disk, chemistry approaches a quasi-equilibrium. 
In the absence of intense sources of ionizing radiation neutral-neutral reactions with
barriers ($\ga 100-1\,000$~K or $\ga 0.2$~eV) start playing an important role in the densest warm disk inner 
regions\cite{Harada_ea10}. 
{\bbf Thus the inner disk chemistry comes closer to conditions known for
``terrestrial'' chemistry, driven by 3-body collisions, albeit characteristic timescales of the disk chemical
processes are usually much longer\cite{Semenov_Wiebe11a}. }
At the very high densities{, \bbf >10$^{12}$
cm$^{-3}$\cite{Aea99},} 3-body reactions become important in inner disks, such as the 
formation of molecular hydrogen by collisions of two hydrogen atoms and another particle that takes away the
excess of energy of formation.

At the high temperatures {\bbf and} densities of inner disks molecules should be abundant in the gas phase until they are 
destroyed
by thermal dissociation ($T \ga 2\,500-3\,500$~K). Chemical models of the inner disk 
chemistry\cite{WKMH98,Mea02,IHMM04,Woods_Willacy07,Bethell_Bergin09,2011ApJ...743..147N} predict high abundances of H$_2$O and CO 
vapor at 1~AU and the
presence of N-bearing molecules (NH$_3$, HCN, HNC) and a variety of hydrocarbons (e.g. CH$_4$ and C$_2$H$_2$).
Radiation fields for higher-mass stars may lead to the destruction of water and the formation of OH 
molecules in inner disk atmospheres\cite{2011ApJ...732..106F}.

{\bbf In contrast to ``terrestrial'' chemistry typical for very inner dense disk regions, 
high-energy radiation and cosmic rays are key drivers of 
outer disk 
chemistry\cite{ah1999,Aea02,vZea03,Gorti_Hollenbach04,2006PNAS..10312249V,vDea_06,Fogel_ea11,Vasyunin2011,Walsh_ea12,ANDES}.
The ionizing radiation leads to the production of various ions, including H$_3^+$. The  proton transfer processes from
ions to other neutrals drive rapid ion-molecule chemistry\cite{HerbstKlemperer73,Watson_74a,Woodall_ea07,KIDA}. 
The ion-molecule processes are mostly barrierless and thus effective even at temperatures well below $100$~K,
particularly in reactions involving long-range Coulomb attraction between an ion and a polarizable molecule.}

{\bbf Another important feature of disk chemistry is the freeze-out of molecules at low temperatures in the outer 
disks. These molecules are then no longer available for gas-phase chemistry. 
The ices on dust surfaces may remain chemically active.}
These ices no longer exist at the high temperatures of the inner disks due to sublimation, where dust grains are bare solids. 
The sublimation of water ice occurs at about 150~K and defines the so-called
``water snow line'', which is located at $2-3$~AU in the early solar 
nebula\cite{2000M&PS...35.1309M,Lecar_ea06,Encrenaz_08,Min_ea11}. 
The CO ``snow line'' is located at a larger
radial distance of about 20~AU (beyond the orbit of Uranus), where the gas temperature drops below 20~K.
{\bbf We should note that the positions of  the ``snow lines'' evolve with evolutionary stage of the disks and are a function
of the luminosity of the central star\cite{Qi_ea13a,Zhang_ea13a}.}

In the outer disk ($ r \ga 20$~AU) we can distinguish three chemically different regimes depending on the vertical disk
location\cite{ah1999}. In the disk surface layers stellar UV radiation and the interstellar radiation field ionize
and dissociate molecules and drive ion-molecule chemistry. In this photon-dominated region photochemistry is particularly
important and depends strongly on the strength and shape of the radiation field. T~Tauri stars emit intense non-thermal UV
radiation from the accretion shock, often in a pronounced Lyman $\alpha$ line\cite{Herczeg_ea02,Schindhelm_ea12}, 
while the hotter Herbig~Ae/Be stars produce large
amounts of thermal UV emission. The integrated flux of the stellar UV radiation at 100~AU can be higher by a factor of
$100-1\,000$ for a T~Tauri disk\cite{Bergin_ea04} and 10$^5$ for a Herbig~Ae disk\cite{Semenov_ea05,Jonkheid_ea06}, respectively, 
compared to the  interstellar radiation field\cite{G}. Photodissociation operates very differently for different molecules and is 
a sensitive function of the actual radiation field. As an example Lyman $\alpha$ photons will 
selectively dissociate HCN and H$_2$O, while other molecules such as CO and H$_2$ are practically unaffected\cite{vDea_06}. Many 
of the other important molecules such as CO and H$_2$ are dissociated by FUV radiation at wavelengths
between 91.2 and 110~nm\cite{1988ApJ...334..771V,vDea_06}. Since the dissociative
destruction of the abundant H$_2$ and CO
molecules operates through photoabsorption at discrete wavelengths isotopically-selective photodissociation based on various
degrees of self-shielding is possible\cite{DB96,Visser_ea09b}. Selective photodissociation of CO by the interstellar
radiation field can also play an important role at the far outer edges of disks as has been demonstrated for the DM~Tau
disk\cite{DDG03}.

The keV X-ray radiation is another important energy source for disk chemistry. Unlike the stellar UV luminosities,
the stellar X-ray luminosities decline from T~Tauri to Herbig~Ae/Be stars. The representative median values for T~Tauri
stars are $\log (L_{\rm X}/L_{\rm bol}) \approx -3.5$ (where $L_{\rm bol}$
is the total bolometric luminosity of the stars), which gives $L_{\rm X} \approx 3\,10^{29}$~erg\,s$^{-1}$
(with an uncertainty of an order of
magnitude)\cite{Preibisch_ea05,Getman_ea09}. This radiation is generated by coronal activity similar to our Sun
but $\ga 1\,000$ times stronger, which is driven by magnetic fields generated by an dynamo mechanism
in convective stellar interiors. In contrast, Herbig~Ae stars have weak surface magnetic fields due
to their non-convective interiors, and, consequently, their X-ray
luminosities are $\ga 10$ times lower than those of T~Tauri stars\cite{Guedel_Naze09}.
The X-ray emitting source is often {\bbf posited} high above the stellar photosphere, at distances of several stellar radii,
and thus X-ray photons reach the disk atmosphere at an oblique angle and are able to penetrate deeper into the  disk 
compared to the stellar FUV photons\cite{zetaxa}. Also, having average energies of several keV, the X-ray photons are able to 
penetrate
through higher gas columns of $\sim 0.1-1$~g\,cm$^{-2}$ compared to the FUV photons ($<0.01$~g\,cm$^{-2}$).
The unique role of X-rays in disk chemistry is their ability to ionize He (with a huge ionization potential of 24.6~eV),
producing chemically active He$^+$. Due to its high electron affinity, ionized helium is able to destroy the tightly bound
CO molecules (and other gas species), replenishing elemental carbon and oxygen back to the gas. This process
drives a rapid and rich gas-phase hydrocarbon chemistry and enriches overall gas molecular complexity.

Adjacent to the disk surface is a warm molecular layer ($T \approx 30-70$~K), where the CO molecule is protected from
freeze-out. This region is partly shielded from stellar and interstellar UV/X-ray radiation allowing a rich molecular chemistry.
Water is still frozen onto dust grains, removing most of the oxygen from the gas phase. This implies relatively high gas C/O 
ratios
close or even larger than 1, leading to a carbon-based chemistry. {\bbf Finally, UV photons may drive photodesorption in less 
opaque regions with experimentally determined rates available for CO, H$_2$O, CH$_4$, and 
NH$_3$\cite{Oeberg_ea07,Oeberg_ea09a,Oeberg_ea09b,Fayolle_ea11a}.}

The third layer is located deep in the interior of the disk close to the midplane. This layer is completely shielded from
high-energy radiation (apart from cosmic ray particles and locally produced energetic particles due to decay of short-lived
radionuclides). The temperature drops below 20~K; freeze-out of molecules and hydrogenation reactions on grain surfaces
dominate the chemistry. {\bbf The freeze-out timescale $t_{\rm freeze}$ for the standard gas-to-dust mass ratio of 100 and a 
sticking probability of 1 can be roughly estimated by the following expression:
\begin{equation}
 t_{\rm freeze} \approx \frac{10^{9} {\rm cm}^{-3}}{n_{\rm H}}\frac{a_d}{0.1\mu{\rm m}}~{\rm years},
\end{equation}
where $n_{\rm H}$ is the gas particle density (in cm$^{-3}$) and $a_d$ is the typical grain radius (in $\mu$m).
The freeze-out timescales in the cold midplanes of protoplanetary disks (with typical densities of $\sim 10^6-10^8$~cm$^{-3}$) 
are of the order of {\bbf 10 to 10$^3$} years, indicating that most of the material in the gas phase is frozen-out within a 
fraction of the disk lifetime.} 
The most important desorption process for
volatiles such as CO, N$_2$, and CH$_4$ is thermal desorption. In addition, cosmic ray and X-ray spot heating may 
release mantle material back to the gas phase\cite{Leger_ea85,HH93,Najita_ea01}.

\subsection{Status of Chemical Models for Protoplanetary Disks}
\label{sec:chemistry:models}
In this Section we provide an overview of the status of chemical models of protoplanetary disks, but do not discuss in detail
chemical models tuned to the specific conditions of the solar nebula.
 
Historically, the first chemical models were developed for studies of the chemical composition of planets
and primitive bodies in the solar system\cite{cameron1995}. These models were usually restricted to the inner,
planet-forming midplane region of the nebula ($r \la 10-30$~AU). In these models ``dark'' conditions were assumed,
with cosmic rays and the decay of short-living radionuclides as the only ionizing sources. These warm, dense and dark
conditions imply that chemical equilibrium is likely to be reached within several million years of the
nebula's lifetime. Consequently, thermodynamical equilibrium was usually assumed in the chemical models of the early 
{\bbf inner} solar nebula.
The primary focus of research of these nebular models was sublimation and condensation of various types of minerals as a
function of distance, and the impact of nebular dynamics and evolution on these 
processes\cite{Grossman1972,Morfill_Voelk84,Wood_Hashimoto93,Prinn_93}. {\bbf We should note however that even
for the chemical models of the inner solar nebula the assumption of chemical equilibrium may not be appropriate
because radiation is certainly more important than previously assumed.}
 
For example, comprehensive studies of dust evolutionary processes and the interaction of high-energy radiation with gas and dust,
together with radial advective transport, were performed by the Heidelberg ITA 
group\cite{Duschl_ea96,bauer,FG97,Finocchi_Gail97,Gail98,G01,G02,Wehrstedt_Gail02,Gail_04,Keller_Gail04}.
Initially in these studies ice mantle evaporation and accumulation, and dust evaporation and destruction were included in the 
models.
Then, ionization of nebular matter by cosmic rays, radionuclides and UV photons was investigated.
Later, various annealing and combustion processes for carbonaceous dust and metamorphosis of silicate dust were considered.
Similar studies that investigated the evolution of more volatile materials, in particular,
water ice, and isotopic fractionation, were conducted\cite{Boss2004,Lyons_Young05,Ciesla_Cuzzi06,Lyons2007,Ciesla_09}.
In \cite{TG07} a self-consistent chemo-dynamical model of the solar nebula was presented, which coupled 2D-hydrodynamics
with extended gas-phase neutral-neutral chemistry and the consideration of dynamical transport processes.
A series of studies were devoted to explain the origin of complex organic materials found in carbonaceous meteorites, which show
extreme hydrogen and nitrogen isotopic anomalies (see Section~\ref{sec:organics}).
The importance of  non-equilibrium chemistry in the outer, more distant or more UV-irradiated parts of the solar nebula
was also recognized, albeit not widely considered\cite{Geiss_Reeves81,Petaev_Wood98,Fegley00}.

In contrast to the chemical models of the solar nebula, chemical models of protoplanetary disks are mostly based on
detailed chemical kinetics models, including a multitude of processes with hundreds and thousands of chemical reactions (see
Table~\ref{tab:disk_models}). This makes extensive disk chemical modeling a computationally intensive task, and thus
many modern astrochemical models are decoupled from disk 
dynamics\cite{Aea96,Aikawa_ea97a,WKMH98,wl00,Aea02,Mea02,vZea03,Kamp_Dullemond04,Semenov_ea05,Aikawa_ea06,Agundez_ea08,Walsh_ea10,
Bruderer_ea12}. The underlying disk physical structure is often based on a steady-state, 1+1D $\alpha$-model in vertical
hydrostatic equilibrium\cite{DAea98,DAea99,Dutrey_ea07}, mostly with uniformly distributed, single-sized dust grains
or power law size distributions with various minimum and maximum cutoff grain sizes\cite{Weingartner_Draine01}.
While in the past equilibrium between dust and gas temperatures throughout the disk was typically assumed, this
simplistic assumption tends to be relaxed in recent models, where chemistry in disk atmospheres and gas thermal balance are
calculated\cite{Kamp_Dullemond04,Woitke_ea09,ANDES}.
This however requires detailed calculations of gas thermal balance, including various heating and cooling processes such as
photoelectric heating of gas by FUV-irradiated dust and PAHs, gas-grain collisions, dust heating and cooling, fine-structure line
cooling via atomic species, etc., which is a non-trivial undertaking.

In contrast to the solar nebula models, most of the disk chemical models include photoprocessing and a detailed treatment of the
far-UV radiative transfer, either in an 1+1D plane-parallel\cite{Nomura_ea07a} or a full 2D approximation\cite{vZea03}, also
including resonant scattering of Ly$\alpha$ photons\cite{Bethell_Bergin11}. Almost every disk model includes ionization by
cosmic ray particles and many include ionization due to decay of short-living radionuclides. Disk models
targeted to investigate chemistry in irradiated disk atmospheres and the molecular layer also include X-ray radiative 
transport\cite{zetaxa,Glassgold_ea05,Henning_ea10,Aresu_ea10a}. In several studies deuterium chemistry was added to the chemical
models\cite{AH99b,ah2001,Ceccarelli_Dominik05,Willacy_07,Willacy_Woods09}.

A recent advancement in studies of disk chemistry is the treatment of grain evolution. Grain coagulation, fragmentation,
sedimentation, turbulent stirring and radial transport are all important processes to be taken into account. Grain growth
depletes the upper disk layers of small grains and hence reduces the opacity of disk matter, allowing the far-UV radiation
to penetrate more efficiently into the disk, and {\bbf to heat and} dissociate molecules {\bbf deeper in the disk}.
Also, larger grains populating the disk midplane may delay the depletion of gaseous
species because of the reduced total surface area\cite{Aikawa_ea06,Vasyunin2011,Fogel_ea11,ANDES}.

In addition to laminar disk models, a number of chemo-dynamical models of protoplanetary disks were developed.
For instance, chemical evolution in protoplanetary disks with 1D radial advective mass transport was 
studied\cite{Aea99,Woods_Willacy07,Woods_Willacy08,Willacy_Woods09,Nomura_ea09}. The transformation of a parental molecular cloud 
into a
protoplanetary disk was studied using a 2D hydrodynamical code with 2D advection flow coupled to gas-grain 
chemistry\cite{Visser_ea09b,Visser_ea11a}. Another class of chemo-dynamical disk models is based on turbulent diffusive mixing, 
which
smears out chemical gradients, and is modeled in 1D, 2D, or even full 3D
\cite{Ilgner_Nelson06,Ilgner_Nelson06a,Ilgner_Nelson06b,Ilgner_ea08,Semenov_ea06,Aikawa_07,Semenov_Wiebe11a,Willacy_ea06,
Hersant_ea09,Turner_ea06,Turner_ea07} (see also Section~\ref{sec:chemistry:dynamics}). In several disk models both advective and
turbulent transport was considered\cite{IHMM04,Heinzeller_ea11}, while \cite{Gorti_ea09} studied photoevaporation of disks
and loss of the gas due to the stellar far-UV and X-ray radiation.

All those disk models are mainly axisymmetric models where mass is transported locally.
Such models may be inappropriate during the early phases of disk evolution when disks are massive enough to trigger
gravitational instabilities\cite{Vorobyov_Basu06}. Gravitational instabilities produce
transient non-axisymmetric structures in the form of spiral waves, density clumps, etc.\cite{Boley_ea10a,Vorobyov_Basu10a}, which 
might explain asymmetries observed in some protoplanetary
disks\cite{2013ApJ...762...48G}. More importantly, gravitational instabilities lead to efficient
mass transport and angular momentum redistribution, which could be characterized by a relatively high viscosity parameter $\alpha
\sim 1$. A first study where these effects were considered along with time-dependent chemistry
was recently performed\cite{Ilee_ea11a}.

In the coming years, we can expect that sophisticated multi-dimensional magneto-hydrodynamical models will be coupled with
time-dependent chemistry. First steps in this direction have been made\cite{Turner_ea07}, where a simple chemical model has
been coupled to a local 3D MHD simulation. Another research direction is to add line radiation transfer to the disk chemical 
models in
order to make predictions of line intensities and spectra\cite{Pavlyuchenkov_ea07a,Semenov_ea08,Kamp_ea11a}.
{\bbf In addition, a detailed 2D/3D treatment of the X-ray and UV radiation transfer with scattering is an essential ingredient 
for realistic 
disk chemical models\cite{IG99,vZea03}.} An accurately calculated UV spectrum including Ly$\alpha$ scattering is
required to calculate photodissociation and photoionization rates, and shielding factors for CO, H$_2$, and 
H$_2$O\cite{2012ApJ...756..157G,Bethell_Bergin09,Bethell_Bergin11}. In heavily irradiated disk atmospheres many species will 
exist in excited 
(ro-)vibrational states, which may then react differently with other species and require addition of state-to-state processes in
the models\cite{Pierce_AHearn10a}. In the outer, cold disk regions addition of nuclear-spin-dependent chemical reactions
involving ortho- and para-states of key species is required. Last but not least, a better {\bbf understanding} of surface 
processes,
including non-thermal desorption, chemisorption, high-energy processing of ices, diffusion through the ice mantle, and related 
factors have to be considered.

\begin{landscape}

\begingroup
\footnotesize
\begin{longtable}{rlllllllllll}
\caption{Chemical models of protoplanetary disks.} \label{tab:disk_models}\\

\hline
Year  &  Ref. & Disk       &  Viscosity & \multicolumn{3}{c}{High-energy} & Gas     & Dust  & \multicolumn{3}{c}{Chemistry} \\
      &       & structure  &            &  \multicolumn{3}{c}{radiation}  & thermal & grain & Reactions & Time- & Dynamics \\
      &       &            &            &   UV & X-rays & CRP             & balance & sizes &           & dep.      & \\
\hline
\endfirsthead
\multicolumn{12}{c}
{\tablename\ \thetable\ -- \textit{Continued from previous page}} \\
\hline
Year  &  Ref. & Structure  &  Viscosity & \multicolumn{3}{c}{High-energy} & Gas     & Dust  & \multicolumn{3}{c}{Chemistry} \\
      &       &            &            &  \multicolumn{3}{c}{radiation}  & thermal & grain & Reactions & Time- & Dynamics \\
      &       &            &            &   UV & X-rays & CRP             & balance & sizes &           & dep.      & \\
\hline
\endhead
\hline \multicolumn{12}{r}{\textit{Continued on next page}} \\
\endfoot
\hline
\endlastfoot
$1996$ & Aikawa et al.\cite{Aea96,Aikawa_ea97a} & MMSN\cite{Hayashi_81}, steady & passive & no & no & $10^{-17}$~s$^{-1}$ & no
& $0.1\mu$m & gas-grain & yes & no \\
$\ge 1998$ & Aikawa et al.\cite{Aikawa_ea98,Aea99,AH99b} & 1+1D\cite{Lynden-BellPringle74}, steady & $\alpha=0.01$ & no & no &
$10^{-17}$~s$^{-1}$ & no & $0.1\mu$m & gas-grain & yes & no \\
& & & & & & & & & gas-grain, D & & no \\
& & & & & & & & & & & rad. advection \\
1998 & Willacy et al.\cite{WKMH98} & 1+1D, steady & $\alpha=0.01$ & no & no & $10^{-17}$~s$^{-1}$ & no & uniform &
gas-grain & yes & no \\
$\ge 1999$ & Aikawa \& Herbst\cite{ah1999,ah2001} & MMSN\cite{Hayashi_81}, steady & passive & 1+1D & 1+1D\cite{zetaxa}
& $10^{-17}$~s$^{-1}$ & no & $0.1\mu$m & gas-grain & yes & no \\
2000 & Willacy \& Langer\cite{wl00} & 1+1D\cite{CG97}, steady & passive & 1+1D & no & $10^{-17}$~s$^{-1}$ & no & uniform &
gas-grain & yes & no \\
2001 & Kamp \& van Zadelhoff\cite{Kamp_Zadelhoff01} & (1+)1D\cite{Kamp_Bertoldi00}, steady & passive & 1+1D & no &
no & yes & power law & gas-phase & no & no \\
2002 & Aikawa et al.\cite{Aea02} & 1+1D\cite{DAea98,DAea99}, steady & $\alpha=0.01$ & 1+1D & no & $10^{-17}$~s$^{-1}$ & no &
$0.1\mu$m & gas-grain & yes & no \\
2002 & Markwick et al.\cite{Mea02,Millar_ea03} & 1+1D, steady & $\alpha=0.01$ & 1+1D & 3-layer & $10^{-17}$~s$^{-1}$ & no &
uniform & gas-grain & yes & no \\
2003 & van Zadelhoff et al.\cite{vZea03} & 1+1D\cite{DAea98,DAea99}, steady & $\alpha=0.01$ & 2D & no & $10^{-17}$~s$^{-1}$ & no
& $0.1\mu$m & gas-grain & yes & no \\
2004 & Glassgold et al.\cite{Glassgold_ea04,Glassgold_ea07,Meijerink_ea08a,Glassgold_ea09} & 1+1D\cite{DAea99}, steady &
$\alpha=0.01-$, & no & 1+1D\cite{zetaxa} & no & yes & power law & gas-phase & yes & no \\
& & & $2.0$ & & & & & &  & &  \\
$\ge 2004$ & Gorti \& Hollenbach\cite{Gorti_Hollenbach04,Gorti_Hollenbach08} & 1+1D, steady & passive & 1+1D &
1+1D & 1D & yes & power law & gas-phase & yes & no\\
2004 & Ilgner et al.\cite{IHMM04} & 1+1D, steady & $\alpha=0.005$, & 1+1D & 3-layer\cite{Mea02} & 1D & no &
uniform & gas-grain & yes & rad. advection, \\
& & & $0.01$ & & & & & & & & vert. mixing, \\
2004 & Jonkheid et al.\cite{Jonkheid_ea04,Jonkheid_ea06} & 1+1D\cite{DAea99}, steady & $\alpha=0.01$ & 1+1D & no &
$5\,10^{-17}$~s$^{-1}$ & yes & power law, & gas-phase & yes & no \\
& & & & & & & & evol. $m_{\rm dust}/m_{\rm gas}$ & & & \\
2004 & Kamp \& Dullemond\cite{Kamp_Dullemond04} & 1+1D\cite{Dullemond_ea02}, steady & passive & 1+1D & no &
yes & yes & $0.1\mu$m & gas-phase & no & no \\
2004 & Semenov et al.\cite{Red2} & 1+1D\cite{DAea99}, steady & $\alpha=0.01$ & 1+1D & 1+1D\cite{zetaxa} & 1D & no & $0.1\mu$m &
gas-grain & yes & no \\
2005 & Ceccarelli \& Dominik\cite{Ceccarelli_Dominik05,Dominik_ea05a} & 2D\cite{DD04}, steady & passive & no &
no & $3\,10^{-19}-$ & no & $0.1\mu$m & gas-phase, D & no & no \\
& & & & &  & $3\,10^{-15}$~s$^{-1}$ & & &  & & \\
$\ge 2005$ & Nomura et al.\cite{Nomura_Millar05,Nomura_ea07a} & 1+1D, steady & $\alpha=0.01$ & 1+1D & 1+1D & 1D
& yes & power law\cite{Weingartner_Draine01} & gas-grain & yes & no \\
2006 & Aikawa \& Nomura\cite{Aikawa_ea06} & 1+1D\cite{Nomura_Millar05}, steady & $\alpha=0.01$ & 1+1D & no & $10^{-17}$~s$^{-1}$ &
yes & power law, & gas-grain & yes & no \\
& & & & & & & & $10~\mu{\rm m}-$ & & \\
& & & & & & & & $10$~cm & & \\
$\ge 2006$ & Ilgner \& Nelson\cite{Ilgner_Nelson06,Ilgner_Nelson06a,Ilgner_Nelson06b,Ilgner_ea08} & 1+1D, steady &
$\alpha=0.01$ & 1+1D & 1+1D, & 1D & no & uniform & gas-phase, & yes & vert. mixing \\
& & & & & flares & & & & gas-grain & & \\
2006 & Semenov et al.\cite{Semenov_ea06} & 1+1D\cite{DAea99}, steady & $\alpha=0.01$ & 1+1D & 1+1D\cite{zetaxa} & 1D & no &
$0.1\mu$m & gas-grain & yes & 2D mixing \\
2006 & Willacy et al.\cite{Willacy_ea06} & 1+1D, steady & $\alpha=0.01$ & 1+1D & no & $10^{-17}$~s$^{-1}$ & no &
power law & gas-grain & yes & vert. mixing \\
2007 & Aikawa\cite{Aikawa_07} & 1+1D\cite{Nomura_Millar05}, steady & $\alpha=0.01$ & 1+1D & no & $10^{-17}$~s$^{-1}$ &
no & MRN\cite{MRN} law, & gas-grain & yes & vert. mixing \\
& & & & &  & & & $a_{\rm max} = 10\mu$m, &  & & \\
& & & & &  & & & 1~mm &  & & \\
$\ge 2007$ & Dutrey et al.\cite{Dutrey_ea07,Schreyer_ea08,Semenov_ea10} & 1+1D, steady & $\alpha=0.01-$ & 1+1D &
2D\cite{zetaxa,Glassgold_ea05} & 1D & no & $0.1\mu$m & gas-grain & yes & no \\
& & & $0.03$ & & & & & & & &  \\
2007 & Jonkheid et al.\cite{Jonkheid_ea07} & 2D\cite{DD04}, steady & passive & 2D & no &
$5\,10^{-17}$~s$^{-1}$ & yes & power law, & gas-phase & yes & no \\
& & & & & & & & evol. $m_{\rm dust}/m_{\rm gas}$ & & & \\
2007 & Turner et al.\cite{Turner_ea06,Turner_ea07} & 3D MHD, & 3D MHD & no & no & 1D & no
& $0.1\mu$m & gas-phase, & yes & 3D mixing \\
& & shear.-box & & &  & & & & reduced\cite{OD74} & & \\
2007 & Willacy\cite{Willacy_07} & 1+1D\cite{DCH01}, steady & $\alpha=0.01$ & 1+1D & no & $10^{-17}$~s$^{-1}$ & no &
power law & gas-grain, D & yes & no \\
$\ge 2007$ & Willacy \& Woods\cite{Woods_Willacy07,Willacy_Woods09,Woods_Willacy08} & 1+1D\cite{DCH01}, steady & $\alpha=0.01$, &
2D\cite{rh00} & 1+1D\cite{Gorti_Hollenbach04} & 1D & yes & power law & gas-grain, D & yes & rad. advection \\
& & & $0.025$ & & & & & & & &  \\
& & &  & & & & & & gas-grain, $^{13}C$ & &  \\
2008 & Ag{\'u}ndez et al.\cite{Agundez_ea08} & 1+1D\cite{DAea98,DAea99}, steady & $\alpha=0.01$ & 1+1D & no & $10^{-17}$~s$^{-1}$
& no & MRN\cite{MRN} law & gas-phase & yes & no \\
2008 & Chapillon et al.\cite{Chapillon_ea08,Chapillon_ea10} & (1+)1D, steady & passive & 1+1D & no &
$10^{-17}$~s$^{-1}$ & no & power law & gas-phase & yes & no \\
& & & & & & & & evol. $m_{\rm dust}/m_{\rm gas}$ & & & \\
$2008$ & Vasyunin et al.\cite{Vasyunin_ea08} & 1+1D\cite{DAea99}, steady & $\alpha=0.01$ & 1+1D & 1+1D\cite{zetaxa} &
$10^{-17}$~s$^{-1}$ & no & $0.1\mu$m & gas-grain & yes & no \\
$2009$ & Gorti et al.\cite{Gorti_ea09} & 1+1D\cite{Lynden-BellPringle74}, evolv. & $\alpha=0.01$, & 1+1D &
1+1D & 1D & yes & $0.3\mu$m & gas-phase & yes & no\\
& & & $0.001$, & & & & & & & &  \\
& & & $0.1$ & & & & & & & &  \\
$2009$ & Hersant et al.\cite{Hersant_ea09} & 2-layer & passive & 1+1D & no & 1D & no &
$0.1\mu$m & gas-grain & yes & vert. mixing \\
$2009$ & Nomura et al.\cite{Nomura_ea09} & 1+1D, steady & $\alpha=0.01$ & 1+1D & 1+1D\cite{Nomura_ea07a} & 1D
& yes & power law\cite{Weingartner_Draine01} & gas-grain & yes & rad. advection \\
$\ge 2009$ & Visser et al.\cite{Visser_ea09,Visser_ea11a} & 2D, evol. & $\alpha=0.01$ & 1+1D & no & $5\,10^{-17}$~s$^{-1}$
& no & $0.1\mu$m & gas-grain & yes & 2D advection \\
$\ge 2009$ & Woitke et al.\cite{Woitke_ea09,Aresu_ea10a,Kamp_ea11a,Thi_ea11a,Woitke_ea11a,Meijerink_ea12a}
& 2D~\cite{Pinte_ea06}, steady & $\alpha$ & 2D & 2D & 1D & yes & power law & gas-grain & no & no \\
2010 & Henning et al.\cite{Henning_ea10} & 1+1D, steady & $\alpha=0.01-$ & 1+1D & 2D\cite{zetaxa} & 1D &
no & $0.1\mu$m & gas-grain & yes & no \\
& & & $0.03$ & & & & & & & &  \\
$\ge 2010$ & Walsh et al.\cite{Walsh_ea10,Walsh_ea12} & 1+1D, steady & $\alpha=0.01$ & 1+1D & 1+1D\cite{Nomura_ea07a} & 1D &
yes & power law\cite{Weingartner_Draine01} & gas-grain & yes & no \\
$2011$ & Cleeves et al.\cite{Cleeves_ea11a} & 1+1D, steady & passive & 2D\cite{Bethell_Bergin11}, & 1+1D\cite{zetaxa} &
$10^{-17}$~s$^{-1}$ & no & power law\cite{Weingartner_Draine01} & gas-grain & yes & no \\
& & & & Ly$\alpha$ & & & &  & & &  \\
$\ge 2011$ & Dutrey et al.\cite{Dutrey_ea11a,Chapillon_ea12a,Chapillon_ea12b,Guilloteau_ea12a} &
2-layer\cite{Hersant_ea09} & passive & 1+1D & no & 1D & no & $0.1\mu$m & gas-grain & yes & no \\
$2011$ & Fogel et al.\cite{Fogel_ea11} & 1+1D, steady & $\alpha=0.01$ & 2D\cite{Bethell_Bergin11}, & 1+1D\cite{zetaxa} & 1D & no
& power law, & gas-grain & yes & no \\
& & & & Ly$\alpha$ & & & & settling & & &  \\
2011 & Ilee et al.\cite{Ilee_ea11a} & 3D HD & grav. & no & no & 1D & no & $0.1\mu$m & gas-grain & yes & 3D advection \\
& & & instab. & &  & & & &  & & \\
2011 & Heinzeller et al.\cite{Heinzeller_ea11} & 1+1D, steady & $\alpha=0.01$ & 1+1D & 1+1D\cite{Nomura_ea07a} & 1D & yes &
power law\cite{Weingartner_Draine01} & gas-grain & yes & rad. advection, \\
& & & & & & & & & & & vert. mixing, \\
& & & & & & & & & & & vert. disk wind \\
2011 & Najita et al.\cite{2011ApJ...743..147N} & 1+1D\cite{DAea99}, steady & $\alpha=0.01$ & no & 1+1D\cite{zetaxa} & no & yes &
power law & gas-phase & no & no \\
2011 & Semenov \& Wiebe\cite{Semenov_Wiebe11a} & 1+1D, steady & $\alpha=0.01$ & 1+1D & 2D\cite{Glassgold_ea04} & 1D &
no & $0.1\mu$m & gas-grain & yes & 2D mixing \\
$2011$ & Vasyunin et al.\cite{Vasyunin2011} & 1+1D, steady & $\alpha=0.01$ & 1+1D & 1+1D\cite{zetaxa} & 1D & no & 2D
evol.\cite{Brauer_ea08a,Birnstiel_ea10a} & gas-grain & yes & no \\
2012 & Bruderer et al.\cite{Bruderer_ea12} & 2D\cite{Bjorkman_Wood01}, steady & passive & 2D & 1+1D\cite{Bruderer_ea09a} &
1D & yes & MRN\cite{MRN} law & gas-grain & no & no \\
$2013$ & Akimkin et al.\cite{ANDES} & 1+1D, steady & $\alpha=0.01$ & 1+1D & 2D\cite{Glassgold_ea05} & 1D
& yes & 2D evol.\cite{Birnstiel_ea10a,Birnstiel_ea12} & gas-grain & yes & no \\
\hline
\end{longtable}
\endgroup

\end{landscape}

\subsection{Chemistry and Dynamics}
\label{sec:chemistry:dynamics}
An isotopic analysis of refractory condensates in the most primitive, unaltered chondrites (a class of stony meteorites) shows 
strong evidence that the
elemental composition of the inner part of the solar nebula within about 0.1 to several AU {\bbf was 
homogenized during several million years, but not completely\cite{Boss2004,Ciesla_09}. 
A plausible scenario for such mixing is dynamical transport due to turbulent diffusion and angular momentum
removal, combined with a repeated sequence of evaporation and re-condensation of solids in the very inner hot
nebular region.
One of the most notable exceptions is oxygen, which shows anomalous isotopic signatures between  all its 3 isotopes at the bulk 
\% 
level\cite{Lyons_Young05,Yurimoto_ea07}. The most likely candidates for the production of the oxygen isotopic anomalies
are chemical mass-independent fractionation and photochemical self-shielding effects, combined with transport
processes. Also, only a small fraction of presolar grains survived the dynamically violent and thermally hazardous 
process of the solar nebula formation and are routinely found in primitive meteorites\cite{2003ARA&A..41..241D}.
Thus the thermally-reprocessed solids in the inner solar nebula do not retain much memories of
the pristine interstellar chemical composition\cite{Palme_01,Trieloff_Palme06}. In contrast,
pristine ices were likely able to survive the formation of the solar nebula in its outer cold 
regions\cite{Ehrenfreund_Charnley00,Visser_ea09}, and later
were incorporated into comets.}

Another indication that dynamics of the solar nebula was important for its chemical
evolution is the rich variety of organic compounds found in carbonaceous meteorites, including amino acids.
It has been suggested that complex organics formed just prior or during the formation of planets
in heavily irradiated, warm regions of the nebula\cite{Ehrenfreund_Charnley00,Busemann_ea06,Pizzarello_ea06,Herbst_vanDishoeck09}.
Combustion and pyrolysis of hydrocarbons and PAHs at high ($\sim 1\,000$~K) temperatures and high densities
followed by radial outward transport have been proposed as a mechanism by which kerogene-like (mainly aromatic)
carbonaceous materials found in meteoritic and cometary samples were synthesized\cite{Morgan_ea91}.
{\bbf A route to a second-generation production of complex organics may be aqueous alteration inside the parent bodies of
carbonaceous meteorites, although its role is debated\cite{Ehrenfreund_Charnley00,Pizzarello_ea06}.}

The presence of crystalline silicates annealed at temperatures above 800~K in cometary samples
collected by the {\it Stardust} mission\cite{2012M&PS...47..660W} may be further evidence for efficient, large-scale transport
of solids in the solar nebula. Evidence for the presence of crystalline silicates in comets have also been found in their infrared
spectra\cite{Crovisier1997,2006ApJ...651.1256K}.
Various transport processes have been suggested, e.g. turbulent mixing,
accretion flows, stellar/disk winds, radiation pressure, or aerodynamic sorting (or a combination of 
all)\cite{2007A&A...466L...9M,2007prpl.conf..815W,2009M&PS...44.1663C,2012ApJ...760L..23W}.
Alternative mechanisms of the {\it in situ} production of the crystalline silicates by heating events include 
frequent electric discharges in a weakly ionized plasma\cite{2012ApJ...761...58H}, outbursts of a central 
star\cite{Abraham_ea09}, or shocks\cite{HD_02}.
The omni-presence of high-temperature crystalline silicates observed in other protoplanetary disks and
radial variations of the crystallinity fraction advocate for robustness of this crystallization 
process\cite{2005A&A...437..189V,Juhasz_ea10a,2011ppcd.book..114H,2011ApJ...734...51O,2012MNRAS.420.2603R}.

The turbulent nature of protoplanetary accretion disks, as discussed in Section~\ref{sec:phys_struc:acc_disks},
leads to non-thermal gas motions and radial transport of gas. As turbulence is a 3D-phenomenon driven by
a magneto-rotational instability in a rotating gaseous disk, it requires computationally
intensive modeling that has only recently become manageable (assuming a simplified chemical 
structure)\cite{2011ApJ...735..122F,2012ApJ...744..144F}. These global MHD simulations show that advection has no specified
direction in various disk regions, and in each location goes both inward and outward.
The corresponding turbulent velocity of the gas $V_{\rm turb}$ depends on the viscosity parameter $\alpha$
and scales with it somewhere between linear and square-root dependence: $\alpha  <  V_{\rm turb}/c_{\rm s} < \sqrt{\alpha}$,
here $c_{\rm s}$ is the sound speed.
The calculated $\alpha$-viscosity stresses have values in a range of $10^{-4}$ and $10^{-2}$ in the midplane and the molecular 
layer,
respectively, and rise steeply to transonic values of $\sim 0.5$ in the disk atmosphere. Similar values of micro-turbulent
velocities in the disk of DM~Tau have been derived\cite{Guilloteau_ea12a}, using the heavy 
molecule CS to
distinguish between Keplerian, thermal and non-thermal line components. Hughes et al.~(2011)\cite{Hughes_ea11a} have used the CO
molecule and
estimated turbulent velocities in disks around the Sun-like star TW~Hya and the hotter, twice as massive star HD~163296, with
about 10\% and 40\% of the sound speed, respectively.

Given this large body of empirical evidence, for decades modelers have investigated the impact of various dynamical
processes on the chemical evolution of gas and solids in the solar nebula and protoplanetary disks.
Models of the early solar nebula with radial transport by advective flows have been 
developed\cite{Morfill_Voelk84,G01,Boss2004,Keller_Gail04}, with a simple 1D analytical disk model and passive tracers.
The 2D radial mixing of gaseous and solid water in the inner nebula has been studied\cite{Cyr_ea98}.
Bauer et al.\cite{bauer} have
investigated the influence of radial transport on the gas-phase C-, H-, N-, O-chemistry driven
by dust destruction and evaporation of ices. The major result is that radial
transport enriches the outer, $r>10$~AU regions with methane and acetylene produced by oxidation of carbon dust
at $r\la1$~AU, as was observed in comets Hyakutake and Hale-Bopp.
In a recent study of that kind, Tscharnuter \& Gail~(2007)\cite{TG07} have considered
a 2D disk chemo-hydrodynamical model in which global circulation flow patterns exist, transporting disk matter
outward in the disk midplane and inward in elevated disk layers. They found that gas-phase species
produced by warm neutral-neutral chemistry in the inner region can be  transported into the cold outer region and
freeze out onto dust grain surfaces. {\bbf We should note that global 3D MHD simulations do not show any evidence
for such meridional circulation patterns\cite{2011ApJ...735..122F,2012ApJ...744..144F}, which would imply that 
the meridional  transport of molecules does not occur.}

Studies of the chemistry coupled to the dynamics in protoplanetary disks is a relatively recent research field.
Aikawa et al.~(1999)\cite{Aea99} have used an isothermal hydrostatic $\alpha$-disk model\cite{Lynden-BellPringle74} with an 
accretion rate of
$\dot{M}= 10^{-8}\,M_\odot$\,yr$^{-1}$, and modeled chemical evolution with inward accretion.
They found that within 3~Myr one can transport a parcel of gas from 400~AU to 10~AU.
This inward transport enhances concentrations of heavy hydrocarbons at $\la20$~AU,
whereas methane remains a dominant hydrocarbon in the outer disk region.
This also leads to the simultaneous existence of  reduced (e.g. CH$_4$) and
oxidized (e.g. CO$_2$) ices, similar to what is observed in comets.

Ilgner et al.~(2004)\cite{IHMM04} have used a steady $\alpha$-disk model and considered  1D vertical mixing
using a turbulent eddy turnover description\cite{Xie_ea95} and a 1D Lagrangian description for advective
transport (from 10~AU to 1~AU). They found that mixing lower vertical abundance gradients, and
that local changes in species concentrations due to mixing can be radially transported by advection.
They have concluded that diffusion does not affect disk regions dominated by gas-grain kinetics,
while it enhances abundances of simple gas-phase species like O, SO, SO$_2$, CS, etc.
Later, Ilgner \& Nelson~(2006)\cite{Ilgner_Nelson06b} have studied the
ionization structure of inner disks ($r<10$~AU), considering
vertical mixing and other effects like X-ray flares, various elemental compositions, etc. 
They found that mixing has no profound effect on electron concentration if metals are absent in the gas since recombination
timescales are faster than dynamical timescales. However, when $T\ga 200$~K and alkali atoms are present in the gas,
chemistry of ionization becomes sensitive to transport, such that diffusion may reduce the
size of the turbulent-inactive disk ``dead'' zone.

Willacy et al.~(2006)\cite{Willacy_ea06} have attempted to  systematically study the impact of disk viscosity on
evolution of various
chemical families. They used a steady-state $\alpha$-disk model similar to that of Ilgner et al.~(2004) and considered
1D-vertical mixing in the outer disk region with $r>100$~AU. They found that vertical transport can increase column
densities (vertically integrated concentrations) by up to 2 orders of magnitude for some complex species.
Still, the layered disk structure was largely preserved even in the presence of vertical mixing.
Semenov et al.~(2006)\cite{Semenov_ea06} and Aikawa~(2007)\cite{Aikawa_07} have found that turbulence can
transport gaseous CO
from the molecular layer down towards the cold midplane where it otherwise remains frozen out, which
may explain the large amount of cold ($\la15$~K) CO gas detected in the disk of DM~Tau\cite{DDG03}.
Hersant et al.~(2009)\cite{Hersant_ea09} have studied various mechanisms to retain gas-phase CO in very cold disk regions,
including vertical mixing. They found that  photodesorption in upper, less obscured molecular layer greatly increases
the gas-phase CO concentration, whereas the role of vertical mixing is less important.

Later, in Woods \& Willacy~(2007)
\cite{Woods_Willacy07} the formation and destruction of benzene in turbulent protoplanetary
disks at $r\la 35$~AU has been investigated. These authors found that radial transport allows efficient synthesis of benzene at
$\la 3$~AU, mostly due to ion-molecule reactions between C$_3$H$_3$ and
C$_3$H$_4^+$ followed by grain dissociative recombination. The resulting concentration of C$_6$H$_6$ at
larger radii of $10-30$~AU is increased by  turbulent diffusion up to 2 orders of magnitude.
In a similar study, Nomura et al.~(2009)\cite{Nomura_ea09} have considered inner disk model with radial advection
($\la 12$~AU). They found that the molecular concentrations are sensitive to the transport speed, such that
in some cases gaseous molecules are able to reach the outer, cooler disk regions where they should be depleted.
This increases the production of many complex or surface-produced species such as methanol, ammonia,
hydrogen sulfide, acetylene, etc.

Heinzeller et al.~(2011)\cite{Heinzeller_ea11} have studied the chemical evolution of a protoplanetary disk with
radial viscous accretion, vertical mixing, and a vertical disk wind (in the atmosphere). They used a steady-state disk
model with $\alpha=0.01$ and $\dot{M}= 10^{-8}\,M_\odot$\,yr$^{-1}$. They found that mixing lowers concentration
gradients, enhancing abundances of NH$_3$, CH$_3$OH, C$_2$H$_2$ and sulfur-containing species. They concluded
that the disk wind has a negligible effect on  chemistry, while the radial accretion changes molecular abundances
in the midplane, and the vertical turbulent mixing enhances abundances in the intermediate molecular layer.

A detailed study of the effect of 2D radial-vertical mixing on gas-grain chemistry in a protoplanetary
disk has been performed\cite{Semenov_Wiebe11a}. These authors used the $\alpha$-model of a
$\sim5$~Myr DM~Tau-like
disk coupled to the large-scale gas-grain chemical code ``ALCHEMIC''\cite{Semenov_ea10}. To account for
production of complex molecules, an extended set of surface processes was added.
A constant value of the viscosity parameter $\alpha=0.01$ was assumed, and the diffusion coefficient was calculated as
\begin{equation}
D_{\rm turb}(r,z) = \nu(r,z)/Sc = \alpha c_{\rm s}(r) H(r) / Sc,
\end{equation}
where $Sc$ is the Schmidt number describing the efficiency of turbulent diffusivity\cite{ShakuraSunyaev73,SchraeplerHenning04}.

\begin{table}
\caption{Detectable tracers of turbulent mixing.}
\centering
  \label{tab:tracers}
  \begin{tabular}{ll}\hline
Insensitive & Hypersensitive\\ \hline
CO          &  Heavy hydrocarbons (e.g.,C$_6$H$_6$) \\
H$_2$O ice   &  C$_2$S \\
       &  C$_3$S \\
       &  CO$_2$ \\
       &  O$_2$ \\
      &  SO \\
      &  SO$_2$\\
      &  OCN \\
      &  Complex organics (e.g., HCOOH)\\
\hline
 \end{tabular}
\end{table}

In this study it was shown that the higher the ratio of the characteristic chemical timescale to the turbulent transport 
timescale
for a given species, the higher the probability that its concentration is affected by dynamics. Consequently, turbulent 
transport changes the abundances of many gas-phase species and particularly ices.
Vertical mixing is more important than radial mixing, mainly because radial temperature and density gradients in disks
are weaker than vertical ones. The simple molecules not responding to dynamical transport include
C$_2$H, C$^+$, CH$_4$, CN, CO, HCN, HNC, H$_2$CO, OH, as well as water and ammonia ices.
The  species sensitive to transport are carbon chains and other heavy species,
in particular sulfur-bearing and complex organic molecules (COMs) frozen
onto the dust grains. This is because mixing allows ice-coated grains to be  steadily transported into
warmer inner disk regions where  efficient surface recombination reactions proceed.
In the warm molecular layer these complex ices evaporate and return to the gas phase.
It was reconfirmed that mixing does not completely smear out the vertical layered structure of protoplanetary disks
(see also \cite{Aea99,Willacy_ea06,Heinzeller_ea11}).

Finally, several promising, particularly sensitive detectable tracers of
dynamical processes in protoplanetary disks were identified. These are the ratios of concentrations of CO$_2$,
O$_2$, SO, SO$_2$, C$_2$S, C$_3$S and organic molecules to that of CO and water ice (see Table~\ref{tab:tracers}).

\section{Water and Hydroxyl Molecules in Protoplanetary Disks}
\label{sec:chemistry:water}

The chemistry of water in protoplanetary disks and the conversion of water vapor into molecular ice
are directly linked to the origin of water on Earth and the formation of giant planets beyond the snow line.
This line defines the location in the disk where water vapor freezes out to molecular ice. The position
of the snow line depends on the pressure and temperature in the disk and, therefore, on the disk mass and
heating processes. For the solar nebula it is generally assumed that the snow line was located somewhere
between 2 and 3~AU from the Sun. The planets Jupiter, Saturn, Uranus, and Neptune all formed beyond the snow line {\bbf (see 
Section~\ref{sec:chemistry:reactions})} and are enriched in volatiles relative to the Sun\cite{2007prpl.conf..591L}. 
{\bbf In general, giant planets in other planetary systems are thought to form beyond the snow line\cite{Kennedy_Kenyon08}.}
It is very likely that the volatiles were trapped in molecular ices when accreted onto the planets.
The term ``volatiles'' comprises all material with low melting and condensation temperatures (gases or
molecular ices) in contrast to ``refractory'' materials such as metallic iron and silicates. Water and carbon monoxide are
by far the most abundant simple ``volatile'' molecules, with NH$_3$, CH$_4$, and CO$_2$ as other important ices.

Because of a large reservoir of volatiles in the outer solar nebula, the two giant planets Jupiter and Saturn could first form
massive solid cores and then gravitationally attract giant H-He gas envelopes. The two outermost planets in the solar system, 
Uranus
and Neptune, are largely composed of molecular ices such as water, methane, and ammonia, and are, therefore called the ice giants.
Water, CO, and CO$_2$ are the most abundant molecular ices in comets with small admixtures of methane and ammonia and
other minor components (e.g. C$_2$H$_6$)\cite{2011IAUS..280..261B}.

Two main processes have been discussed as explanation for the source of water on Earth: (i) Delivery of hydrous silicate
grains to Earth and outgassing of volatiles by volcanoes - wet proto-Earth formation\cite{2005M&PS...40..519D} and
(ii) Delivery of water by certain classes of comets and asteroids - dry proto-Earth formation\cite{2000M&PS...35.1309M}.
Arguments against delivery of water by hydrous silicates come from the low water content in anhydrous meteoritic
silicates and the non-detection of spectral features of hydrous silicates in infrared disk spectra. A key to our
understanding of the origin of water on Earth is its D/H ratio with a mean ocean water value of $1.55\times10^{-4}$,
 which is much larger than the primordial value of the interstellar medium ($1.5\times10^{-5}$), see Fig.~\ref{fig:water_DtoH}.
This enhanced D/H ratio points to mass fractionation during chemical reactions at low temperatures (see
Section~\ref{sec:chemistry:deuterium}). Measurements of D/H ratios for asteroids and comets indicate
that a mixture of these bodies delivered water to Earth\cite{2003SSRv..106..155M,2011Natur.478..218H}.

{\bbf We should note that most of the water, at least in the outer regions of protoplanetary disks, is frozen out on dust
grain surfaces\cite{Dominik_ea05a,Bergin_ea10a}. There is growing evidence that the water ices may already have been 
formed in the prestellar and protostellar phases. Visser et al.~(2009)\cite{Visser_ea09} found that water remains in the solid 
phase everywhere during
the infall and disk formation phases, and evaporates within $\sim 10$~AU of the star. In contrast, pure CO ice evaporates
during the infall phase and is reformed in those parts of the disk that cool below the CO freeze-out temperature of $\sim 20$~K.
These authors  also found that mixed CO:H$_2$O ices will keep some solid pristine CO above this temperature
 threshold and may explain the presence of CO in comets.}

Hot water ($\approx 500-1\,000$~K) has been discovered in the inner regions of protoplanetary disks ($0.1-10$~AU) around young
solar-mass T~Tauri stars by ground-based mid-infrared spectroscopy and infrared observations with the
\textit{Spitzer} Space 
Observatory\cite{Carr_Najita08,Salyk_ea08,Pontoppidan_ea08,2010ApJ...720..887P,2010ApJ...722L.173P,2011ApJ...743..147N,
2011ApJ...734...27T,2012ApJ...745...90B}, see Fig.~\ref{fig:water_disk}. Water is a highly asymmetric molecule with a rich
infrared spectrum produced by
transitions between energy levels characterized by 3 quantum numbers J, K$_A$, and K$_B$. The infrared emission of the
water molecules comes predominantly from a warm disk surface layer.

Water exists in two spin states: ortho-H$_2$O with total nuclear spin of unity and para-H$_2$O with total nuclear spin of
zero. Either direct radiative or collisional transitions between ortho- and para-H$_2$O energy levels are not allowed, and
the ortho-to-para ratio can only be modified by chemistry by proton exchange. The
ortho-to-para ratio in the high-temperature limit is regulated by spin statistics resulting in a ratio 3:1.
Water formation on cold grain surfaces may allow lower ortho-to-para ratios through equilibration at the dust 
temperature\cite{2011Sci...334..338H}.

In contrast to the observations of rich water-dominated infrared spectra from disks around T~Tauri stars, water has not been
detected in the inner disks around the more massive and luminous Herbig~Ae/Be stars\cite{2011ApJ...732..106F}. A likely
explanation of
this observational finding is the photodissociation of hot water by the stronger ultraviolet radiation field of these stars. In
fact, it has been demonstrated that the H$_2$O emission from the disk around the young eruptive star EX~Lupi is variable as
a likely consequence of the changing ultraviolet radiation field\cite{2012ApJ...745...90B}.

Cooler water emission (at temperatures of $\approx 100-500$~K) has been discovered with the
\textit{Herschel} Space Observatory in the far-infared
spectra of disks around T Tauri stars\cite{2012A&A...538L...3R,Fedele_ea13a} and in a small
number of Herbig~Ae/Be stars by line stacking\cite{2012A&A...544L...9F,Fedele_ea13a}. {\bbf It has been directly detected 
in the disk around the Herbig~Ae star HD~163296, which is enshrouded by a flat disk with settled dust\cite{Meeus_ea12}.}
Overall, disks around Herbig Ae/Be stars have higher
OH/H$_2$O abundance ratios across the inner disks than T~Tauri stars.

Above a temperature of $\approx 300$~K water should be the dominant oxygen-containing component in the 
gas, assuming solar-elemental abundances\cite{1978A&A....67..323E,1996ApJ...456..250K}. Water is formed rapidly by two 
neutral-neutral reactions in the warm gas:\\

O + H$_2$ -> OH + H\\

OH + H$_2$ -> H$_2$O + H.\\

An extensive discussion of the relevant rate constants for these reactions can be found 
here\cite{1978A&A....67..323E,1987ApJ...317..423W,1996ApJ...456..250K}. At
temperatures
lower than a few 100~K, ion-neutral reactions become an important channel for water production. The key molecular ion H$_3^+$ will
react with atomic oxygen, producing H$_3$O$^+$ which recombines via a number of dissociative reaction
channels, among them a reaction to H and H$_2$O. In the midplane of the outer disks water is frozen out on dust grains. The water
molecules can be released by UV photodesorption\cite{Dominik_ea05a} in intermediate layers of the disk and water
gets photodissociated in the upper disk atmospheres. The complicated interplay between grain evolution, grain surface chemistry
and freeze-out, photodesorption and photodissociation and radial and vertical mixing processes will regulate the abundance
of water in its different phases in the outer disk\cite{Vasyunin2011,Fogel_ea11,ANDES,2013arXiv1308.1772K}.

The discovery of the
ground-state rotational emission of both spin isomers of water from the outer disk around the star TW~Hya with \textit{Herschel's}
high-resolution spectrometer \textit{HIFI}\cite{2011Sci...334..338H}
provided strong constraints on the water vapor abundance
in the outer disk with a layer of maximum water abundance of $\approx 10^{-7}$ relative to H$_2$. Above this layer water gets
photodissociated, below it freezes. The observations allowed the determination of the ortho-to-para ratio (OPR). The derived
value of $0.77 \pm 0.07$ is much lower than the equilibrium value of 3 and indicates that grain surface reactions and
photodesorption play an important role in producing the observed water vapor. {\bbf This value is also lower than the
OPR H$_2$O ratios for solar system comets, ranging from $2-3$\cite{Mumma_ea87,Woodward_ea07,BM_ea09,Shinnaka_ea12,Bonev_ea13a}. 
If one interprets the OPR values in terms of a spin
temperature, as often done in the literature, and equates this temperature with the physical temperature of the dust on
which water ice has formed, this would indicate that water in TW~Hya has formed at lower temperatures ($10-20$~K) 
than in comets ($\sim 50$~K)\cite{2011Sci...334..338H}.}

\section{Deuterium Chemistry in Disks}
\label{sec:chemistry:deuterium}


Among the most important findings related to the chemical evolution in the early solar nebula is isotopic fractionation.
Many cometary and meteoritic materials show anomalous isotopic enrichment in such elements as oxygen, carbon, nitrogen,
and sulfur\cite{1998M&PS...33..603A,Clayton_93,Palme_01,Clayton2007,Yurimoto_ea07}.
In particular, amino acids found in carbonaceous meteorites\cite{Glavin_ea10,2012M&PS..tmp..201E} and recently discovered
glycine in cometary samples returned by the {\it Stardust} mission\cite{Elsila_ea09} show extraterrestrial isotopic signatures.
The composition of terrestrial minerals, water, and rocks on Earth retains an imprint of the rate differences at which 
fractionation
has occurred.

\begin{figure}
\includegraphics[angle=90,width=15cm]{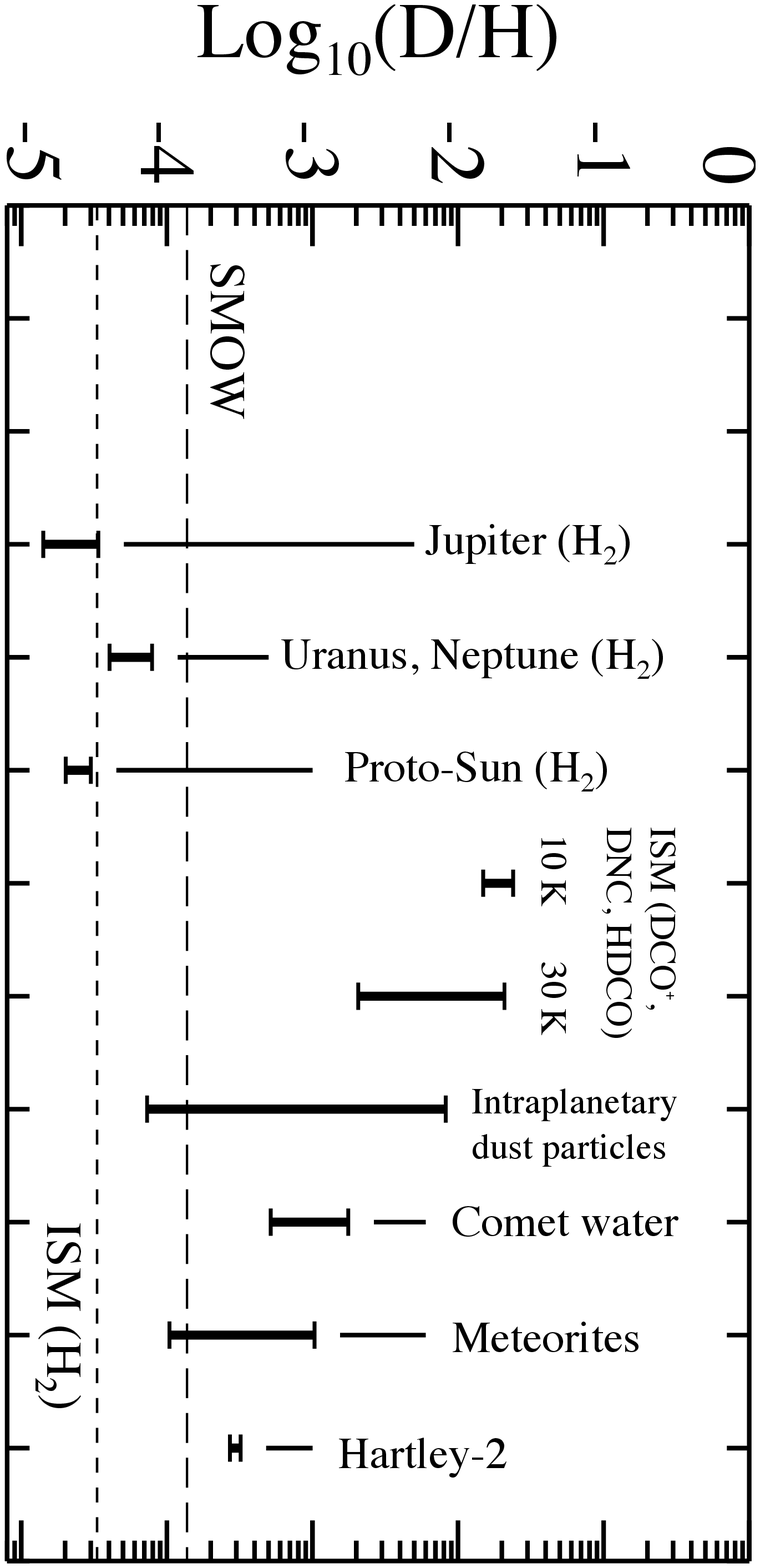}
\caption{The observed D/H ratio for H$_2$O in various astrophysical environments ranging from the interstellar medium
to comets and the Earth's ocean water. {\bbf The ``ISM (H$_2$)'' (dashed line) represents the HD/H$_2$ in dense cold ISM.
``SMOW'' (long dashed line) denotes the HDO/H$_2$O ratio in the Vienna Standard Mean Ocean Water model.
Data derived from 
Reference\cite{Bergin_09}}.\label{fig:water_DtoH}}
\end{figure}

{\bbf Deuterium} fractionation is a non-equilibrium process sensitive to temperature and thermal history of an
environment. Formed right after the Big Bang, with an initial elemental ratio of $D/H \approx 3\,10^{-5}$, only half of this
deuterium has survived till today, mostly in form of the HD molecule (the other half was burned in stellar 
interiors)\cite{2012MNRAS.425.2477P,2006ApJ...647.1106L}. However, at low temperatures, $T\la 20-80$~K, part of deuterium from 
HD can be
redistributed into other molecules by ion-molecule and surface chemistry, resulting in higher molecular D/H ratios than
the value of $1.5\,10^{-5}$. In Figure~\ref{fig:water_DtoH} we show D/H ratios measured for several
molecules (H$_2$, H$_2$O, HCN, HCO$^+$, H$_2$CO etc.) in planets of the solar system, comets, asteroids, interplanetary
dust particles and the Standard Mean Ocean Water {\bbf (SMOW)}. The D/H ratio for H$_2$ in the {\bbf diffuse} interstellar 
medium
(ISM) of $\approx 3\,10^{-5}$ is also indicated. As can be clearly seen, all gas and icy giants as well as
the (proto-)Sun have the cosmic elemental D/H ratios for H$_2$ similar to the ISM H$_2$ D/H value. These objects have
warm interiors and high densities to re-distribute deuterium among chemical species at nearly equilibrium conditions.
In contrast, cold ISM environments like prestellar cores or warm protostellar envelopes have D/H ratios for
DCO$^+$, DCN, HDCO, etc. that are higher than the cosmic value by factors of $\sim 
100-300$\cite{2002P&SS...50.1173R,2003ApJ...585L..55B,2006ApJ...645.1198V}.
Part of this parental, highly
deuterium-enriched cloud core matter ends up in a protoplanetary disk where it is incorporated into meteorites, asteroids, and
comets.
The D/H ratios measured in meteorites and observed in long-periodic comets from the Oort cloud are also enhanced
by factors of at least $10-30$ compared to the elemental D/H ratio\cite{2000prpl.conf.1159I},
which is lower than the molecular ISM values.
As discussed in Section~\ref{sec:chemistry:water}, water has likely been delivered to Earth by bodies from more distant
parts of the solar system as the planetesimals out of which Earth has formed were presumably dry. Not surprisingly, the Earth 
water
bears signature of this exogenous delivery scenario by showing about a ten times enhanced D/H ratio in ocean's water, $\approx
1.6\,10^{-4}$, compared to the {\bbf elemental} ISM D/H value. As the Oort-family long-period comets all show 2-3 times higher
HDO/H$_2$O ratios, whereas carbonaceous chondrites have about the Earth HDO/H$_2$O value, it was argued that water came to
Earth with asteroids, not comets. However, the recent discovery of nearly Earth's ocean water D/H signature in a Jupiter-family,
short-period comet Hartley-2\cite{2011Natur.478..218H} has reignited this discussion.

What happens with deuterated species in transition from the cold conditions of the ISM to the warm conditions in the
planet-forming region can be best understood by studying deuterium chemistry in protoplanetary disks.
Several deuterated molecules have been detected in protoplanetary disks, with abundances much higher
than the cosmic elemental D/H ratio of $1.5\,10^{-5}$. The measured D/H ratios of DCN and DCO$^+$ in the disk around TW~Hya
are about $1-10\%$\cite{Qi_ea08}. {\"O}berg et al.~(2012)\cite{Oeberg_ea12a} have found that the DCO$^+$ column density
increases outward, whereas DCN is more centrally peaked. This suggests different fractionation pathways, with DCN forming at
higher
temperatures than DCO$^+$. The derived DCN/HCN ratio is similar to that of DCO$^+$/HCO$^+$, $\sim 0.017$.
A smaller DCO$^+$/HCO$^+$ value, $4\,10^{-3}$, has been measured in the disk of DM~Tau\cite{2006A&A...448L...5G}.

The observed overabundance of deuterated species in the  cold ISM and protoplanetary disks
is due to isotopic deuterium fractionation. {\bbf This process is related} to the zero-point vibrational energy
difference for the isotopically substituted molecules, implying a temperature barrier for a backward reaction at
low temperatures. The main isotope exchange reaction involves isotopologues of H$_3^+$ and H$_2$, and is
effective at $T \la 10-30$~K: H$_3^+$ + HD $\leftrightarrows$ H$_2$D$^+$ + H$_2$ + 232~K\cite{Millarea89,GHR_02}, followed by 
similar reactions with higher D-isotopologues.
This leads to accumulation of H$_2$D$^+$, which further transfers D into other molecules by ion-molecule
reactions\cite{RobertsMillar00,Robertsea03,2011arXiv1110.2644A}.
For example, the dominant formation pathway to produce DCO$^+$ is via
ion-molecule reactions of CO with H$_2$D$^+$. In disks it results in an DCO$^+$ to
HCO$^+$ ratio that should increase with radius due to the outward
decrease of temperature, as actually observed.
Deuterium fractionation initiated by the H$_3^+$ isotopologues is particularly
effective at low temperatures in disk midplanes, where CO and other molecules that destroy the H$_3^+$ 
isotopologues
are severely frozen out.

On the other hand, the H$_3^+$ isotopologues can dissociatively recombine with electrons, producing a flux of
atomic H and D. In the cold, dense regions such as outer disk midplanes H and D atoms can stick to dust grains and reacts with
ices such as CO, O, C, and form multi-deuterated complex species (isotopologues of e.g. H$_2$CO, CH$_3$OH, H$_2$O).
Moreover, laboratory experiments have demonstrated that on dust surfaces a substitution of a proton by deuteron
in H-bearing species can occur, accelerating deuterium enrichment of complex 
ices\cite{Nagaoka_ea05,2005IAUS..231..415W,2006AIPC..855...86H,2006AIPC..855..107H}.

Recently, it has been realized that the ortho/para ratio of H$_{2}$ and other species can
lower the pace of deuterium fractionation\cite{2006A&A...449..621F,2009A&A...494..623P}. {\bbf The internal energy of ortho-H$_2$ 
is higher than that of para-H$_2$,
which helps to overcome the backward reaction of deuterium enrichment.} For 
example, the backward
endothermic reaction between ortho-H$_{2}$ and H$_{2}$D$^{+}$ can proceed far more
rapidly at low temperatures of 10~K than the corresponding reaction
involving its para form. Consequently, it results in a lower degree of deuterium fractionation in a medium
having a sufficient amount of ortho-H$_{2}$\cite{2006A&A...449..621F}.

Other important fractionation reactions are effective at higher
temperatures (up to $70-80$~K), and are particularly relevant for inner, warmer disk regions:
CH$_3^+$ + HD $\leftrightarrows$ CH$_2$D$^+$ + H$_2$ + 390~K\cite{Asvany_ea04} and C$_2$H$_2^+$ + HD $\leftrightarrows$ 
C$_2$HD$^+$ + H$_2$ + 550~K\cite{Herbst_ea87}. Both reactions produce DCN by the following ion-
molecule reaction: N + CH$_2$D$^+$ $\rightarrow$
DCN$^+$, followed by protonation by H$_2$ and dissociative recombination\cite{Roueff_ea07}.
Compared to DCO$^+$, DCN can thus be formed at warmer temperatures and reside closer to
the central star in protoplanetary disks, exactly as observed in the TW~Hya disk\cite{Oeberg_ea12a}.

These observational findings in protoplanetary disks have been {\bbf compared to} theoretical studies.
Aikawa \& Herbst~(1999)\cite{AH99b} have studied deuterium chemistry in the outer regions of protoplanetary disks with an 1D 
accretion flow, using a
collapse model to set up the initial molecular concentrations. They have found that the
molecular D/H ratios are enhanced wrt the protosolar values, and that the ratios at $\sim 30$~AU agree
reasonably well with the D/H ratios observed in comets. This advocates for {\it in situ} deuterium fractionation
in the solar nebula, so that comets may not necessarily be composed of primordial, unprocessed interstellar matter.
Willacy~(2007)\cite{Willacy_07} and Willacy \& Woods~(2009)\cite{Willacy_Woods09} have studied deuterium chemistry
in outer and inner disk regions, respectively. They found that the D/H ratios observed
in comets may  partly originate from the parental molecular cloud and partly be produced in the disk. They concluded that the D/H
ratios
of gaseous species are more sensitive to deuterium fractionation processes in disks due to rapid ion-molecule
chemistry compared to the deuterated ices, whose D/H values are regulated by slow surface chemistry and
are imprints of the cold conditions of the prestellar cloud.

{\bbf The previous discussion shows that fractionation processes occur both in the prestellar cloud phase and in the 
protoplanetary disk phase. This complicates the use of D/H ratios as a clean ``tool'' to distinguish between the contribution
from prestellar chemistry and disk chemistry to the molecular composition of disks.}

%

\section{Complex Organic Molecules}
\label{sec:organics}

One of the most exciting questions in astrochemistry are (1) the exact
form in which carbon-based materials exist under space conditions\cite{1998Sci...282.2204H}, and (2) how and which
prebiotic life-building blocks can survive the process of star and planet formation.
In the ISM solid carbonaceous compounds can be identified through numerous aromatic
(C$-$C) and aliphatic (C$-$H) stretching and bending
modes at infrared wavelengths\cite{2007prpl.conf..801A,2010pcim.book.....T,1999ApJ...519..687S} and a strong ultraviolet
resonance\cite{1988ApJ...328..734F,1998ApJ...498..486S,2011ApJ...742....2S},
and are believed to take the form of
hydrogenated amorphous carbon (nanoparticles), with various fraction of H as well as
sp$^2$ and sp$^3$ carbon atoms\cite{2012ApJ...761..115D,1999ApJ...519..687S}, or occur as large 
PAHs\cite{1998Sci...282.2204H,2011ApJ...742....2S}.
About $1-10\%$ (depending on far-UV/X-ray irradiation) of all cosmic carbon is
locked in PAHs\cite{1989ApJS...71..733A,2008ARA&A..46..289T}.
A small, $\la 3-4\%$ fraction of carbon condenses out as (nano-)diamond particles that are
found as presolar grains in meteorites\cite{1998M&PS...33..603A}. Also, some elemental C is locked in tightly bound
silicon carbide (SiC) grains, which are mostly of stardust origin\cite{1987Natur.330..728B,2001AcSpe..57..815H}.

Many different amino acids and other complex organics (in insolvable and solvable forms) were present
in the early solar system, as found through detailed mass-spectrometry and mineralogical and petrological analysis
of primitive carbonaceous chondrites\cite{Glavin_ea10} and interplanetary dust particles\cite{2012M&PS...47..525M}.
Analysis of the cometary dust sampled by the {\it Giotto} spacecraft in the Halley comet showed the presence
of the so-called organic ``CHON'' particles (which are large molecules composed of multiple C, H, O, and N atoms\cite{JCK88}). 
The recent identification of glycine in the
{\it Stardust} cometary dust samples\cite{Elsila_ea09} provide strong evidence that comets may have an
organically rich composition, a fact which is also supported by their extremely low albedos. These findings
were also interpreted as indication that initial carbonaceous materials of the ISM could have been
almost entirely chemically reprocessed into complex organics prior or during the formation of the inner solar system,
within several million years\cite{Pea94}. Indeed, the discovery of highly deuterated amino acids in meteoritic materials
(as well as other isotopic anomalies in C, N, and O) implies that their {\bbf (at least, initial)} synthesis
occurred under very cold and dark conditions characteristic of dense prestellar molecular clouds
\cite{2005GeCoA..69..599P,Herbst_vanDishoeck09,2012A&ARv..20...56C}.
This hypothesis is supported by the detection of complex organic molecules
(COMs) such as methanol (CH$_3$OH), acetaldehyde (CH$_3$CHO), dimethyl ether (CH$_3$OCH$_3$), methyl formate (CH$_3$OCHO), and 
ketene (CH$_2$CO), etc. in the gas phase in cold, young, low-mass prestellar 
cores\cite{CD3OH,2012A&A...541L..12B}. {\bbf Glycolaldehyde and formamide (NH$_2$CHO), the simplest 
amide, a key species in the synthesis of amino acids and metabolic molecules, have also been recently detected in envelopes of 
solar-type protostars\cite{Hollis_ea04,2012ApJ...757L...4J,2013ApJ...763L..38K}. }

On the other hand, 
COMs could also have formed at the verge of planet formation in the heavily irradiated, warm inner regions of the
solar nebula by endothermic chemistry from simple species such as CO, N$_2$, OH, and 
H$_2$\cite{Ehrenfreund_Charnley00,Busemann_ea06}.The formation of  protoplanetary disks from their parental molecular 
clouds is associated with strong shocks that can reprocess the gas and some ices\cite{Hassel_04}.
The newly produced COMs can either be trapped in the ices and become 
incorporated in comets or return into the gas phase by thermal desorption, UV-photodesorption, or heating triggered by cosmic ray 
particles and X-rays\cite{dHendecourtea82,Leger_ea85,ShalabeiaGreenberg94,Garrod_ea07}.
Also, the Fischer-Tropsch catalysis converts CO and H$_2$ into hydrocarbons at appropriate temperatures ($T\ga 1\,000$~K)
in the presence of metallic surfaces. In a similar manner,  the catalytic Haber-Bosch synthesis produces ammonia 
from N$_2$ and H$_2$\cite{2003AsBio...3..291H}. The overall efficiency of both these processes depend on
the properties of the metallic surfaces (e.g., poisoning by other materials, refractory ice coatings, topology) and the fraction
of metallic iron and nickel incorporated in silicates and left in their metallic (or oxidized) forms.
The appropriate conditions for such synthesis must have existed in the very inner, sub-AU, accretion-heated regions
of the early solar nebula and may exist in other actively accreting protoplanetary disks\cite{NB94}.
Also, in this hot region PAHs are gradually destroyed by neutral-neutral reactions with barriers with H, OH and O,
resulting in high concentrations of acetylene (C$_2$H$_2$) and, later, CO, CO$_2$ and CH$_4$\cite{Kress_ea10a}.

{\bbf The most plausible scenario of the synthesis of COMs is that the first-generation, simpler organic molecules
have been synthesized already during the pre-disk, cold cloud phase, followed by production of second-generation, more
complex organics inside warm, irradiated disk regions. The further growth in their complexity could have been enabled by aqueous
alteration inside large, radiogenically-heated asteroids.}


Despite the variety of interstellar COMs, only formaldehyde (H$_2$CO), {\bbf C$_2$H, C$_3$H$_2$, HCN, HNC, and HC$_3$N} 
and a few other non-organic species have been detected and spatially resolved
with (sub-)millimeter interferometers in the outer regions ($r \ga 50-200$~AU) of several nearby protoplanetary 
disks\cite{Dutrey_ea97,Kastner_ea97,Aikawa_ea03,Qi_ea03,Dutrey_ea07,Henning_ea10,Chapillon_ea12a,Chapillon_ea12b,
2013arXiv1302.0251Q}.
The ground-based search for simple gas-phase organic species such as methanol and
formic acid in disks has so far been unsuccessful. On the other hand, simple organic ices, like HCOOH ice,
have been identified in the {\it Spitzer} infrared spectra of the envelopes of low-mass
Class~I/II objects\cite{Zasowski_ea08}. The main reasons for the lack of detections of COMs in protoplanetary disks
are the low masses and small sizes of these objects, and the severe depletion of these heavy species onto dust surfaces {\bbf in 
outer cold regions of the disks} (see
Fig.~\ref{fig:disk_vs_TMC1}). Formaldehyde and
formic acid are just precursors for the synthesis of other complex organic ices via slow photoprocessing, forming
reactive radicals in the icy mantles, which can further recombine with each other at appropriate conditions
\cite{Garrod_ea06,2011ApJ...730...69E,2011ApJ...734...78J,2012ApJ...758...37K}.
Dynamical transport from cold to warm/irradiated regions in protoplanetary
disks can also lead to efficient desorption of heavy ices (see Section~\ref{sec:chemistry:dynamics} above).
A rich organic chemistry occurring in the inner warm disk regions has been confirmed with infrared spectroscopy by {\it Spitzer}.
Detected species include CO, CO$_2$, C$_2$H$_2$, CN, and HCN, which reside in the warm gas, $T \ga 
200-1\,000$~K\cite{Lahuis_ea06,2007prpl.conf..507N,Carr_Najita08,Salyk_ea08,2009ApJ...696..143P,2011ApJ...733..102C,Salyk_ea11a,
2011ApJ...734...27T,2012ApJ...747...92M}.

\begin{figure}
\includegraphics[width=0.535\textwidth]{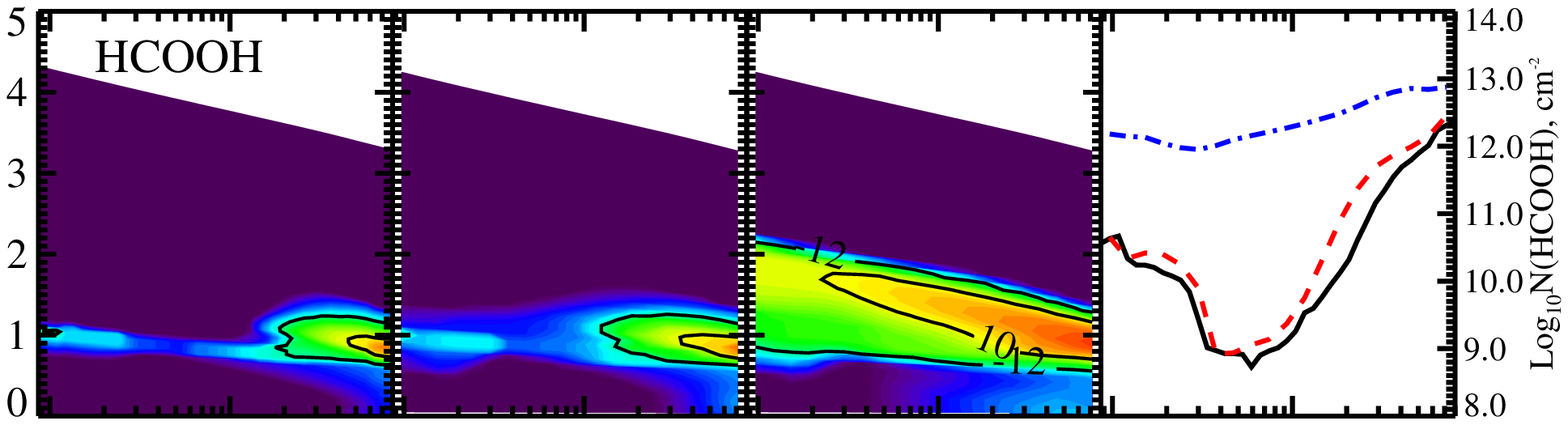}\\
\includegraphics[width=0.535\textwidth]{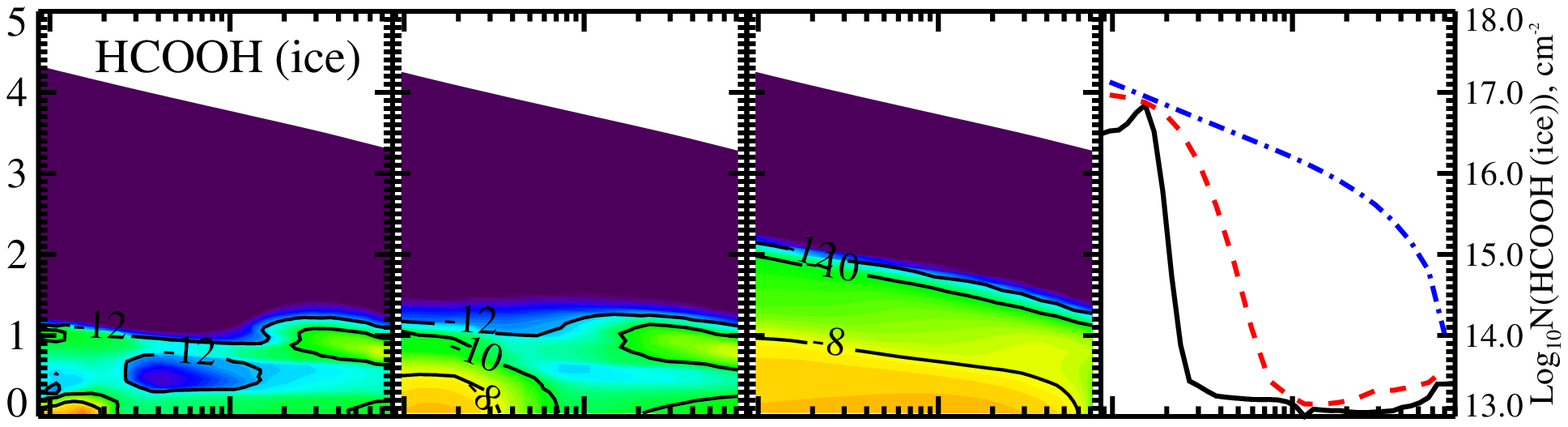}\\
\includegraphics[width=0.535\textwidth]{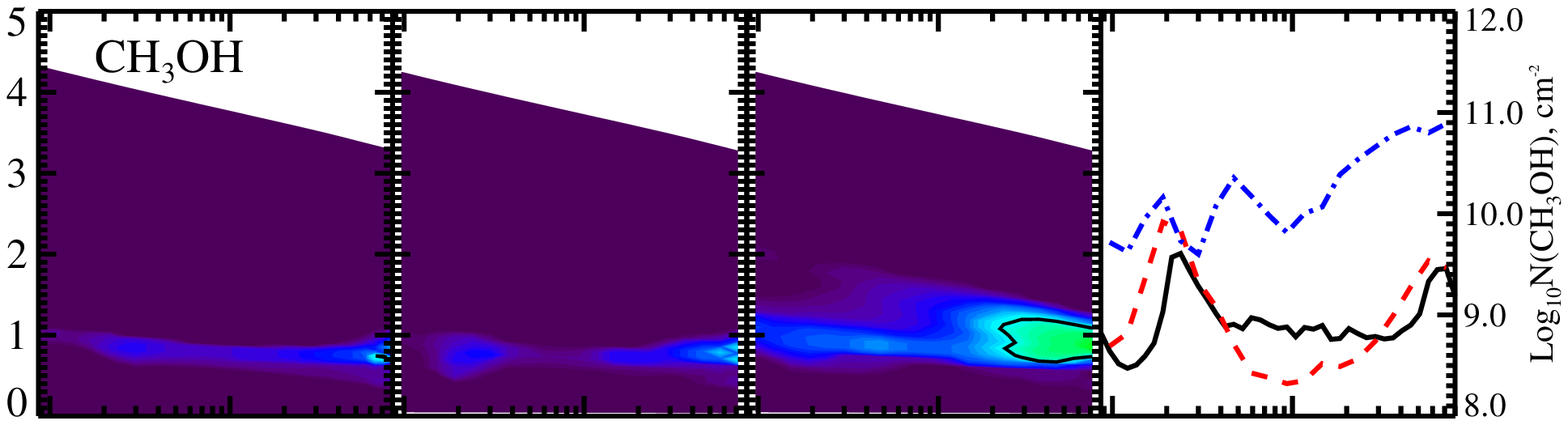}\\
\includegraphics[width=0.535\textwidth]{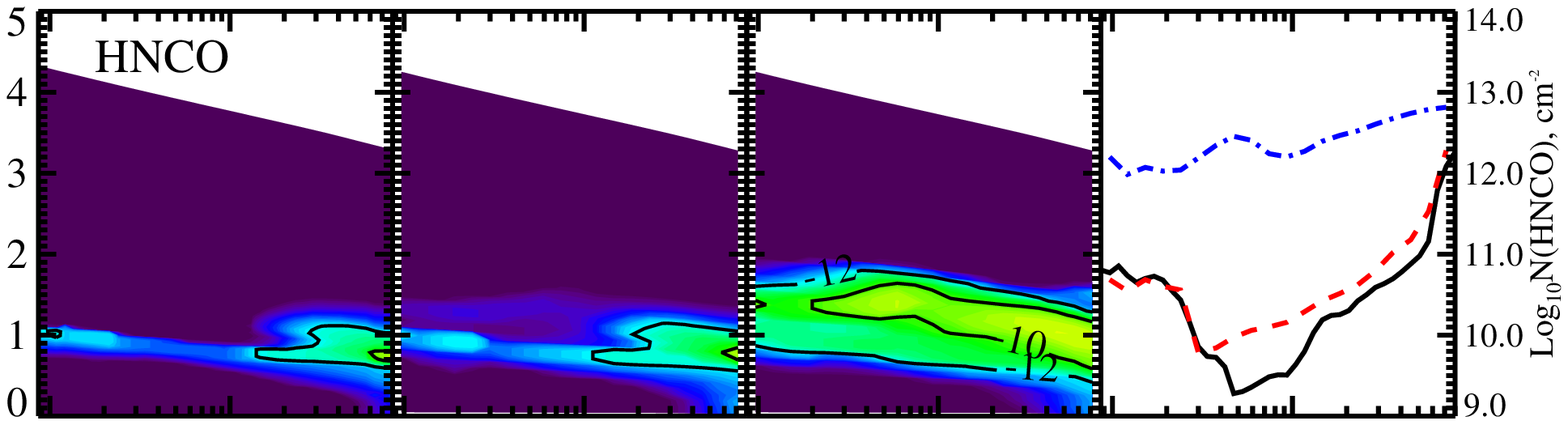}\\
\includegraphics[width=0.535\textwidth]{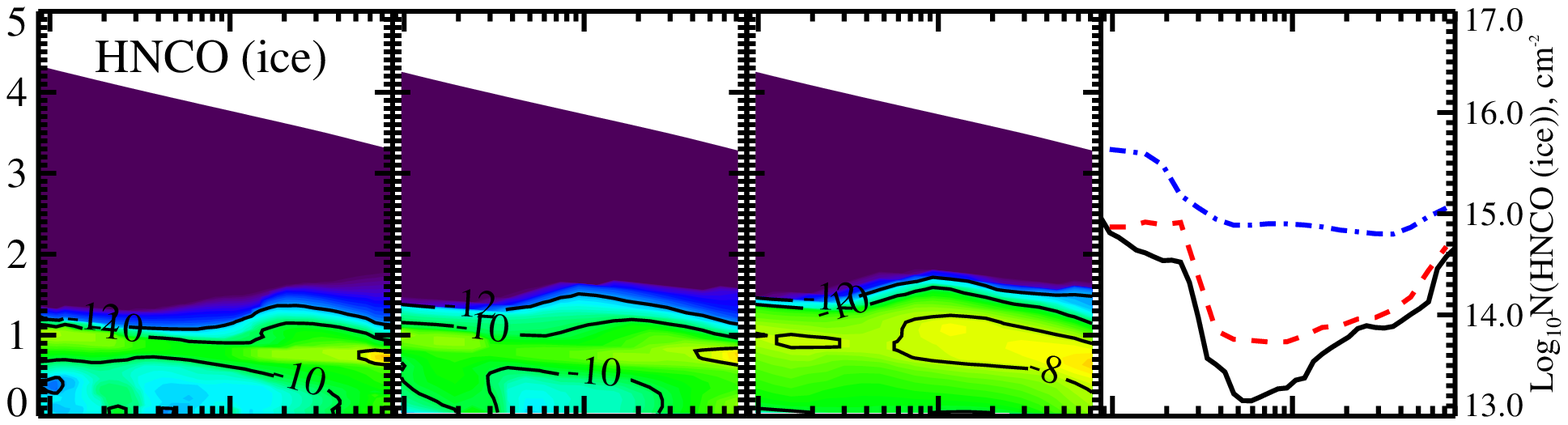}\\
\includegraphics[width=0.535\textwidth]{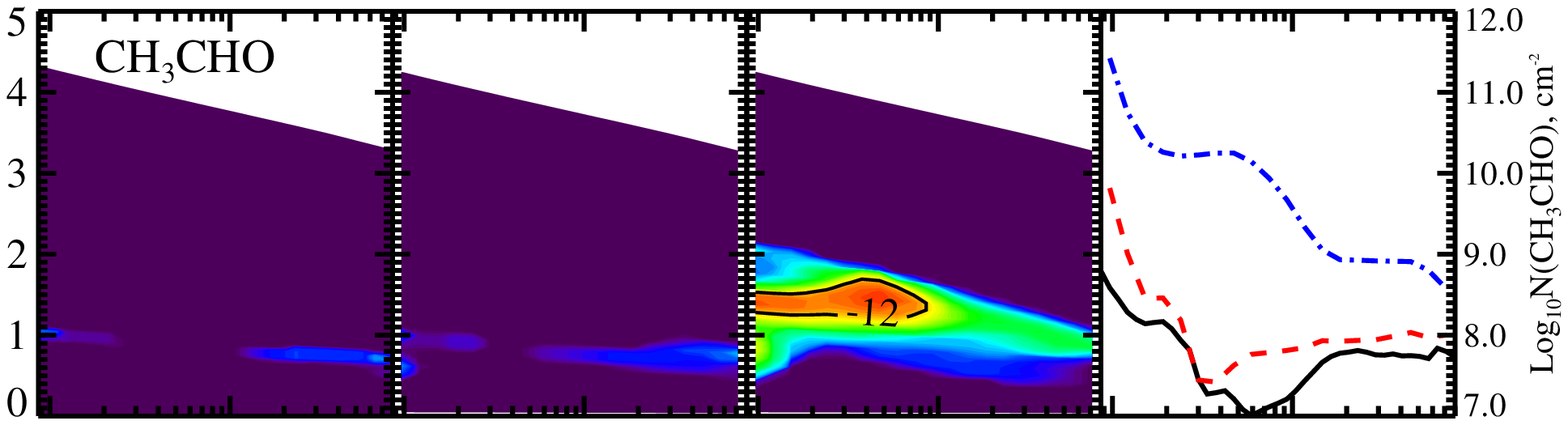}\\
\includegraphics[width=0.535\textwidth]{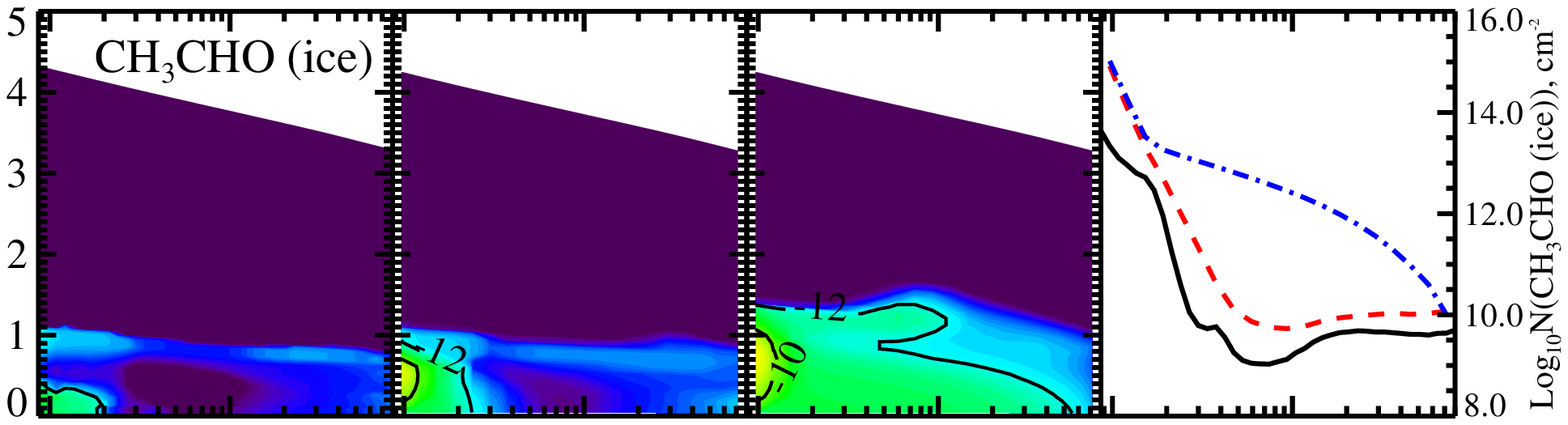}\\
\includegraphics[width=0.535\textwidth]{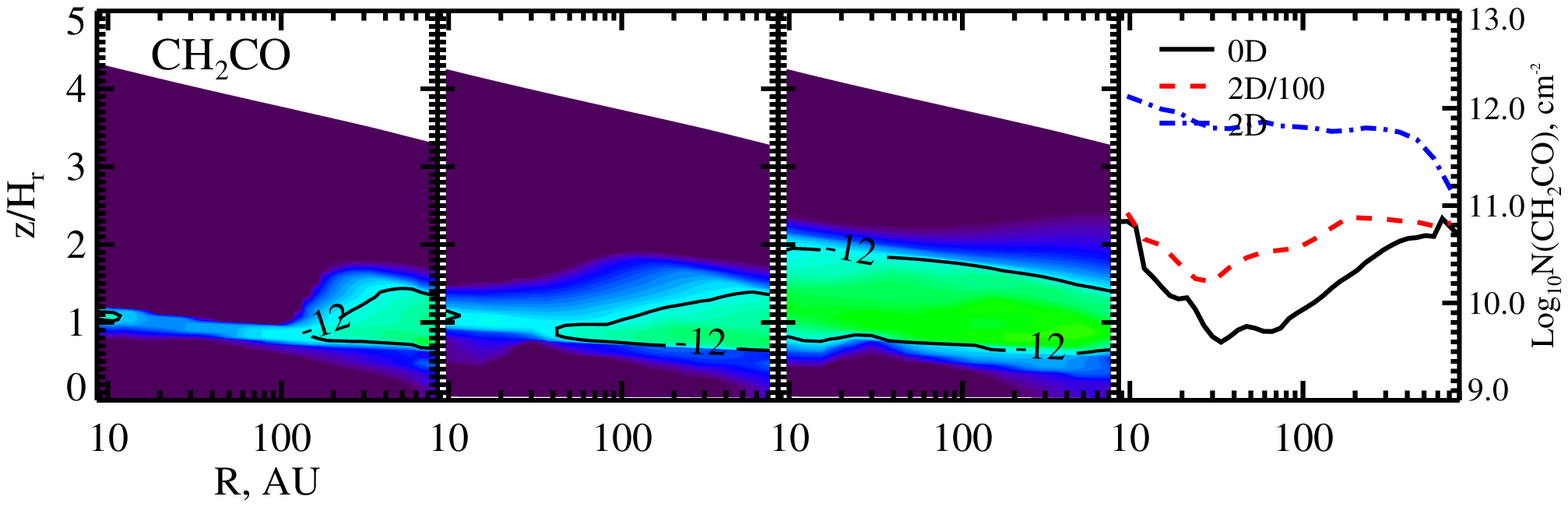}
\caption{Radial and vertical variations of abundances and column densities of complex organic
molecules in the DM~Tau disk at 5~Myr. (Top to bottom, left columns) The $\log$ of relative abundances
(with respect to the total number of hydrogen nuclei) of HCOOH, HCOOH ice, CH$_3$OH,
HNCO, HNCO ice, CH$_3$CHO, CH$_3$CHO ice, and CH$_2$CO.  Results for
the three disk models are shown: (1) laminar chemistry, 2) the slow 2D-mixing
chemistry ($Sc=100$), and (3) the fast 2D-mixing chemistry with $Sc=1$.  
The corresponding vertical column densities (abundances integrated in the vertical direction) are shown in the right prt of the 
Figure. {\bbf Reprinted with permission from Reference\cite{Semenov_Wiebe11a}. Copyright 2011 American Astronomical Society.}}
\label{fig:complex}
\end{figure}

The importance of dynamical transport for the synthesis of COMs in the early solar nebula has been studied by
\cite{Semenov_Wiebe11a} (see Section~\ref{sec:chemistry:dynamics} for a brief description of their model).
The main results are presented in Fig.~\ref{fig:complex}. The plot shows distributions of absolute concentrations and
vertically-integrated column densities of HCOOH, HCOOH ice, CH$_3$OH, HNCO, HNCO ice, CH$_3$CHO, CH$_3$CHO ice, and CH$_2$CO
calculated with a quiescent and a rapid turbulent mixing model of the solar nebula. In the dynamically-quiescent model
the abundance distributions of gaseous species have a layered structure, with very narrow layers of peak concentrations
located at $\approx 0.8-1$ pressure scale heights. Abundances of
complex ices reach peak values at either the bottom of the molecular layer (HNCO ice, HCOOH ice) or in the inner warm midplane
(CH$_3$CHO ice, HCOOH ice). The overall pattern is easily explained by photoevaporation of heavy COM ices, which in the gas
phase become susceptible to ionizing far-UV/X-ray radiation and are rapidly photodissociated.
On the other hand, dynamical transport facilitates the synthesis of COMs by transporting icy grains toward warm or irradiated
disk regions, where heavy reactive radicals are formed upon CRP/X-ray irradiation of ices, followed by surface recombination
when they become mobile at $T\ga30-50$~K.

We list the most important reactions for the evolution of HCOOH, HCOOH ice,
CH$_3$OH, HNCO, HNCO ice, CH$_3$CHO, CH$_3$CHO ice, and CH$_2$CO in the solar nebula and protoplanetary disk midplanes and
molecular layers in
Table~\ref{tab:key_reac_organics}.
The chemical evolution of formic acid (HCOOH) begins with the dissociative recombination of
CH$_3$O$_2^+$, which is produced by radiative association of HCO$^+$ and H$_2$O,
and by ion-molecule reaction of methane with ionized molecular oxygen.
HCOOH is destroyed by photodissociation and photoionization, and removed from the gas due to depletion onto dust grains at
$T\la 100$~K. {\bbf Then} HCOOH ice forms, which is destroyed by far-UV photons generated in the disk by CRP (or the attenuated
stellar far-UV photons in the disk molecular layer).
The surface and gas-phase formation of HCOOH through the neutral-neutral reaction of OH and H$_2$CO
is only a minor channel.
Gas-phase methanol is synthesized by surface recombination of H and CH$_2$OH
as well as frozen OH and CH$_3$ (at $T\ga 30$~K) followed by evaporation of the methanol ice.
The main removal pathways for CH$_3$OH are accretion onto the dust surfaces
in the disk regions with $T\la 100$~K, photodissociation and ionization.

{\bbf We now will discuss the chemistry of the isocyanic acid (HNCO) in detail because this molecule is a key
molecule in the chemical processes on dust surfaces\cite{vanBroekhuizen_ea2004} and an important diagnostic 
species\cite{Tideswell_ea10}.}
The production of HNCO is also dominated by surface reactions. It can reach the gas phase
either by evaporation of HNCO ice at $T\la 50-60$~K or by direct recombination of
surface H and OCN {\bbf (by chemisorption)}. 
The major destruction gas-phase pathways for HNCO are accretion onto the dust grain surfaces.
HNCO ice is produced by a surface reaction involving H and OCN and is destroyed by
far-UV photons. Similarly, the chemistry of acetaldehyde (CH$_3$CHO) begins with the surface recombination of CH$_3$ and HCO
ices followed by desorption to the gas phase at $T\la 60$~K. It is destroyed by photodissociation.
Gas-phase ethenone (CH$_2$CO) is produced via neutral-neutral reactions of atomic O with C$_2$H$_3$, and
direct surface recombination of the H and HC$_2$O ices.
Evaporation of ethenone ice starts when dust temperatures exceed $\sim100$~K.
Key removal channels include photodissociation and photoionization, and freeze-out in the disk midplane.

In the presence of dynamical transport, concentrations of HNCO and other COMs are enhanced, albeit differently.
This is related to their relatively long chemical timescales associated with slow surface synthesis, which is longer than
the dynamical timescales. HCOOH becomes more abundant since concentrations of its precursors, water
and HCO$^+$, are increased by mixing, leading to higher concentrations of CH$_3$O$^+$.
The effect is less pronounced for HNCO, as its precursor species, OCN, is not as sensitive to transport.
HNCO ice, produced in the molecular layer, is transported by diffusion to the cold midplane where it cannot be
effectively synthesized.

Concentration of gaseous acetaldehyde is greatly enhanced by mixing in the molecular layer at $r \la 100$~AU.
It is  synthesized via recombination of heavy CH$_3$ and HCO ices, which are mobile on dust surfaces only
in the very inner warm  disk region. In the fast mixing model large amounts of solid acetaldehyde
accumulate as more and more icy grains reach the warm disk regions due to transport.

To conclude, {\it in situ} studies of ``primitive'' material in the solar system have provided mounting evidence for the
{\bbf presence} of complex organic matter in the early solar nebula. Complex molecules should also be present in other 
protoplanetary
disks, but only a few key species have been detected so far due to the limited sensitivity of modern (sub-)millimeter
interferometers.
This situation will likely change dramatically with the beginning of the full operation of ALMA.
The potential detection of such complex species like dimethyl ether, formic acid, methyl
formate, etc. in protoplanetary disks with ALMA will provide solid evidence that the global chemical evolution
is regulated by disk dynamics.

\begin{table}
\caption{Key chemical processes: organic species}
\footnotesize
\label{tab:key_reac_organics}
\begin{center}
\begin{tabular}{llrrll}
\hline
{\bbf Reaction} & {$\alpha$} & {$\beta$} & {$\gamma$}\\
      & [(cm$^3$)\,s$^{-1}$] &  & [K]  \\
\hline
CH$_2$CO ice   + h$\nu_{\rm CRP}$  $\rightarrow$  CH$_2$ ice   + CO ice   & $9.15\,(2)$   & $0$   & $0$\\
CH$_2$CO ice   + h$\nu_{\rm CRP}$  $\rightarrow$  C$_2$ ice   + H$_2$O ice   & $4.07\,(2)$   & $0$   & $0$ \\
CH$_3$CHO ice   + h$\nu_{\rm CRP}$  $\rightarrow$  CH$_3$ ice   + HCO ice   & $5.25\,(2)$   & $0$   & $0$  \\
CH$_3$CHO ice   + h$\nu_{\rm CRP}$  $\rightarrow$  CH$_4$ ice   + CO ice   & $5.25\,(2)$   & $0$   & $0$\\
CH$_3$OH ice   + h$\nu_{\rm CRP}$  $\rightarrow$  CH$_3$ ice   + OH ice   & $1.50\,(3)$   & $0$   & $0$\\
CH$_3$OH ice   + h$\nu_{\rm CRP}$  $\rightarrow$  H$_2$CO ice   + H$_2$ ice   & $3.17\,(3)$   & $0$   & $0$\\
C$_2$H$_5$OH ice   + h$\nu_{\rm CRP}$  $\rightarrow$  CH$_3$CHO ice   + H$_2$ ice   & $6.85\,(2)$   & $0$   & $0$\\
HCOOH ice   + h$\nu_{\rm CRP}$  $\rightarrow$  CO$_2$ ice   + H ice   + H ice   & $6.50\,(2)$   & $0$   & $0$\\
HCOOH ice   + h$\nu_{\rm CRP}$  $\rightarrow$  HCO ice   + OH ice   & $2.49\,(2)$   & $0$   & $0$   \\
HNCO ice   + h$\nu_{\rm CRP}$  $\rightarrow$  NH ice   + CO ice   & $6.00\,(3)$   & $0$   & $0$   \\
CH$_3$OH ice   + UV  $\rightarrow$  H$_2$CO ice   + H$_2$ ice   & $0.72\,(-9)$   & $0$   & $1.72$   \\
C$_2$H$_5$OH ice   + UV  $\rightarrow$  CH$_3$CHO ice   + H$_2$ ice   & $0.13\,(-9)$   & $0$   & $2.35$ \\
HCOOH ice   + UV  $\rightarrow$  HCO ice   + OH ice   & $0.28\,(-9)$   & $0$   & $1.80$      \\
HNCO   + h$\nu_{\rm CRP}$  $\rightarrow$  NH   + CO   & $6.00\,(3)$   & $0$   & $0$     \\
CH$_3$CHO   + UV  $\rightarrow$  CH$_3$CHO$^+$   + e$^-$   & $0.26\,(-9)$   & $0$   & $2.28$   \\
CH$_3$CHO   + UV  $\rightarrow$  CH$_4$   + CO   & $0.34\,(-9)$   & $0$   & $1.52$     \\
CH$_3$CHO   + UV  $\rightarrow$  CH$_3$   + HCO   & $0.34\,(-9)$   & $0$   & $1.52$     \\
HCOOH   + UV  $\rightarrow$  HCOOH$^+$   + e$^-$   & $0.17\,(-9)$   & $0$   & $2.59$   \\
HCOOH   + UV  $\rightarrow$  HCO   + OH   & $0.28\,(-9)$   & $0$   & $1.80$     \\
CH$_2$CO   + grain  $\rightarrow$  CH$_2$CO ice   & --   &  --  &   --   \\
HCOOH   + grain  $\rightarrow$  HCOOH ice   & --   & --   & --    \\
HNCO   + grain  $\rightarrow$  HNCO ice   & --   & --   & --  \\
CH$_3$CHO   + grain  $\rightarrow$  CH$_3$CHO ice   & --   & --   & --    \\
CH$_3$OH   + grain  $\rightarrow$  CH$_3$OH ice   & --  &--   & --   \\
HNCO ice  $\rightarrow$  HNCO   & --   & --  & $2.85\,(3)$   \\
CH$_2$CO ice  $\rightarrow$  CH$_2$CO   & --   & --   & $2.20\,(3)$   \\
CH$_3$CHO ice  $\rightarrow$  CH$_3$CHO   & --   & --  & $2.87\,(3)$   \\
H ice   + CH$_2$OH ice  $\rightarrow$  CH$_3$OH ice   & --  & --   & --   \\
H ice   + HC$_2$O ice  $\rightarrow$  CH$_2$CO ice   & --   & --   & --   \\
H ice   + OCN ice  $\rightarrow$  HNCO ice   & --   & --   & --  \\
OH ice   + CH$_3$ ice  $\rightarrow$  CH$_3$OH ice   & --   & -- & --  \\
OH ice   + HCO ice  $\rightarrow$  HCOOH ice   & --   & --   & --  \\
HCOOH   + H$^+$  $\rightarrow$  HCOOH$^+$   + H   & $0.28\,(-8)$   & $-0.50$   & $0$  \\
HNCO   + H$^+$  $\rightarrow$  NH$_2$$^+$   + CO   & $0.15\,(-7)$   & $-0.50$   & $0$  \\
CH$_3$CHO   + H$_3$$^+$  $\rightarrow$  CH$_3$CH$_2$O$^+$   + H$_2$   & $0.62\,(-8)$   & $-0.50$   & $0$   \\
CH$_3$CHO   + HCO$^+$  $\rightarrow$  CH$_3$CH$_2$O$^+$   + CO   & $0.25\,(-8)$   & $-0.50$   & $0$   \\
HCOOH   + H$_3$$^+$  $\rightarrow$  HCO$^+$   + H$_2$O   + H$_2$   & $0.23\,(-8)$   & $-0.50$   & $0$\\
HCOOH   + HCO$^+$  $\rightarrow$  CH$_3$O$_2$$^+$   + CO   & $0.13\,(-8)$   & $-0.50$   & $0$   \\
O   + C$_2$H$_5$  $\rightarrow$  CH$_3$CHO   + H   & $0.13\,(-9)$   & $0$   & $0$   \\
OH   + H$_2$CO  $\rightarrow$  HCOOH   + H   & $0.20\,(-12)$   & $0$   & $0$   \\
\hline
\end{tabular}
\end{center}
\end{table}

%

\section{Conclusions}
Protoplanetary disks are amazing structures of gas and dust surrounding young stars, which are
characterized by a broad range in temperature, density, and radiation fields. They show a rich
variety of chemical processes, ranging from high-temperature neutral-neutral reactions in the
inner disk regions to ion-molecular chemistry and molecular freeze-out close to the midplane in
the outer disk regions. Grain surface reactions, thermal and photo-driven desorption as well as
deuteration processes are all part of the {\bbf diverse} chemistry in protoplanetary disks. Dust evolution,
ionization structure, and turbulent transport are closely linked processes, defining the thermal and
kinematic structure of disks.

Submillimeter and millimeter
observations provide constraints on the {\bbf radial and} vertical physical and chemical structure, and are delivering 
information about molecular abundances
{\bbf in the outer disks}.
Infrared spectroscopy both from the ground and from space {\bbf has been providing} an inventory of H$_2$O, OH,
CO, CO$_2$, and simple organic molecules in the warm planet-forming regions of disks. 
With the enormously increased sensitivity and spatial resolution provided by the \textit{Atacama Large
Millimeter/Submillimeter Array} in Chile, now beginning operations, and the infrared spectroscopic capabilities of the
\textit{James Webb Space Telescope}, to be launched towards the end of this decade, the field of disk
chemistry will become ever more rich in data. The development of theoretical modeling tools and the determination of key
reaction rates will form the basis for the comprehensive scientific exploitation of these astronomical data.

\acknowledgement
This research made use of NASA's Astrophysics Data System.
DS acknowledges support by the {\it Deutsche Forschungsgemeinschaft} through
SPP~1385: ``The first ten million years of the solar system - a
planetary materials approach'' (SE 1962/1-1 and 1-2).

\bibliography{achemso_no_last_p}

\providecommand*\mcitethebibliography{\thebibliography}
\csname @ifundefined\endcsname{endmcitethebibliography}
  {\let\endmcitethebibliography\endthebibliography}{}
\begin{mcitethebibliography}{422}
\providecommand*\natexlab[1]{#1}
\providecommand*\mciteSetBstSublistMode[1]{}
\providecommand*\mciteSetBstMaxWidthForm[2]{}
\providecommand*\mciteBstWouldAddEndPuncttrue
  {\def\EndOfBibitem{\unskip.}}
\providecommand*\mciteBstWouldAddEndPunctfalse
  {\let\EndOfBibitem\relax}
\providecommand*\mciteSetBstMidEndSepPunct[3]{}
\providecommand*\mciteSetBstSublistLabelBeginEnd[3]{}
\providecommand*\EndOfBibitem{}
\mciteSetBstSublistMode{f}
\mciteSetBstMaxWidthForm{subitem}{(\alph{mcitesubitemcount})}
\mciteSetBstSublistLabelBeginEnd
  {\mcitemaxwidthsubitemform\space}
  {\relax}
  {\relax}

\bibitem[{Mayor} and {Queloz}(1995){Mayor}, and {Queloz}]{51Peg}
{Mayor},~M.; {Queloz},~D. \emph{Nature} \textbf{1995}, \emph{378}, 355\relax
\mciteBstWouldAddEndPuncttrue
\mciteSetBstMidEndSepPunct{\mcitedefaultmidpunct}
{\mcitedefaultendpunct}{\mcitedefaultseppunct}\relax
\EndOfBibitem
\bibitem[{Udry} and {Santos}(2007){Udry}, and {Santos}]{Udry_Santos07}
{Udry},~S.; {Santos},~N.~C. \emph{Ann.~Rev.~Astron.~Astrophys.,} \textbf{2007},
  \emph{45}, 397\relax
\mciteBstWouldAddEndPuncttrue
\mciteSetBstMidEndSepPunct{\mcitedefaultmidpunct}
{\mcitedefaultendpunct}{\mcitedefaultseppunct}\relax
\EndOfBibitem
\bibitem[{Batalha} et~al.(2013){Batalha}, {Rowe}, {Bryson}, {Barclay}, {Burke},
  {Caldwell}, {Christiansen}, {Mullally}, {Thompson}, {Brown}, {Dupree},
  {Fabrycky}, {Ford}, {Fortney}, {Gilliland}, {Isaacson}, {Latham}, {Marcy},
  {Quinn}, {Ragozzine}, {Shporer}, {Borucki}, {Ciardi}, {Gautier}, {Haas},
  {Jenkins}, {Koch}, {Lissauer}, {Rapin}, {Basri}, {Boss}, {Buchhave},
  {Carter}, {Charbonneau}, {Christensen-Dalsgaard}, {Clarke}, {Cochran},
  {Demory}, {Desert}, {Devore}, {Doyle}, {Esquerdo}, {Everett}, {Fressin},
  {Geary}, {Girouard}, {Gould}, {Hall}, {Holman}, {Howard}, { Howell},
  {Ibrahim}, {Kinemuchi}, {Kjeldsen}, {Klaus}, {Li}, {Lucas}, {Meibom},
  {Morris}, {Pr{\v s}a}, {Quintana}, {Sanderfer}, {Sasselov}, {Seader},
  {Smith}, {Steffen}, {Still}, {Stumpe}, {Tarter}, {Tenenbaum}, {Torres},
  {Twicken}, {Uddin}, {Van Cleve}, {Walkowicz}, and
  {Welsh}]{2013ApJS..204...24B}
{Batalha},~N.~M. et~al.  \emph{Astrophys.~J.,~Suppl.~Ser.,} \textbf{2013},
  \emph{204}, 24\relax
\mciteBstWouldAddEndPuncttrue
\mciteSetBstMidEndSepPunct{\mcitedefaultmidpunct}
{\mcitedefaultendpunct}{\mcitedefaultseppunct}\relax
\EndOfBibitem
\bibitem[{Barclay} et~al.(2013){Barclay}, {Rowe}, {Lissauer}, {Huber},
  {Fressin}, {Howell}, {Bryson}, {Chaplin}, {D{\'e}sert}, {Lopez}, {Marcy},
  {Mullally}, {Ragozzine}, {Torres}, {Adams}, {Agol}, {Barrado}, {Basu},
  {Bedding}, {Buchhave}, {Charbonneau}, {Christiansen},
  {Christensen-Dalsgaard}, {Ciardi}, {Cochran}, {Dupree}, {Elsworth},
  {Everett}, {Fischer}, {Ford}, {Fortney}, {Geary}, {Haas}, {Handberg},
  {Hekker}, {Henze}, {Horch}, {Howard}, {Hunter}, {Isaacson}, {Jenkins},
  {Karoff}, {Kawaler}, {Kjeldsen}, {Klaus}, {Latham}, {Li}, {Lillo-Box},
  {Lund}, {Lundkvist}, {Metcalfe}, {Miglio}, { Morris}, {Quintana}, {Stello},
  {Smith}, {Still}, and {Thompson}]{Barclay_ea13}
{Barclay},~T. et~al.  \emph{Nature} \textbf{2013}, \emph{494}, 452\relax
\mciteBstWouldAddEndPuncttrue
\mciteSetBstMidEndSepPunct{\mcitedefaultmidpunct}
{\mcitedefaultendpunct}{\mcitedefaultseppunct}\relax
\EndOfBibitem
\bibitem[{Stevenson} et~al.(2010){Stevenson}, {Harrington}, {Nymeyer},
  {Madhusudhan}, {Seager}, {Bowman}, {Hardy}, {Deming}, {Rauscher}, and
  {Lust}]{Stevenson_ea10a}
{Stevenson},~K.~B.; {Harrington},~J.; {Nymeyer},~S.; {Madhusudhan},~N.;
  {Seager},~S.; {Bowman},~W.~C.; {Hardy},~R.~A.; {Deming},~D.; {Rauscher},~E.;
  {Lust},~N.~B. \emph{Nature} \textbf{2010}, \emph{464}, 1161\relax
\mciteBstWouldAddEndPuncttrue
\mciteSetBstMidEndSepPunct{\mcitedefaultmidpunct}
{\mcitedefaultendpunct}{\mcitedefaultseppunct}\relax
\EndOfBibitem
\bibitem[{Barman} et~al.(2011){Barman}, {Macintosh}, {Konopacky}, and
  {Marois}]{Barman_ea11a}
{Barman},~T.~S.; {Macintosh},~B.; {Konopacky},~Q.~M.; {Marois},~C.
  \emph{Astrophys.~J.,} \textbf{2011}, \emph{733}, 65\relax
\mciteBstWouldAddEndPuncttrue
\mciteSetBstMidEndSepPunct{\mcitedefaultmidpunct}
{\mcitedefaultendpunct}{\mcitedefaultseppunct}\relax
\EndOfBibitem
\bibitem[{Konopacky} et~al.(2013){Konopacky}, {Barman}, {Macintosh}, and
  {Marois}]{Konopacky_ea13}
{Konopacky},~Q.~M.; {Barman},~T.~S.; {Macintosh},~B.~A.; {Marois},~C.
  \emph{Science} \textbf{2013}, \emph{339}, 1398\relax
\mciteBstWouldAddEndPuncttrue
\mciteSetBstMidEndSepPunct{\mcitedefaultmidpunct}
{\mcitedefaultendpunct}{\mcitedefaultseppunct}\relax
\EndOfBibitem
\bibitem[{Kretke} and {Lin}(2012){Kretke}, and {Lin}]{2012ApJ...755...74K}
{Kretke},~K.~A.; {Lin},~D.~N.~C. \emph{Astrophys.~J.,} \textbf{2012},
  \emph{755}, 74\relax
\mciteBstWouldAddEndPuncttrue
\mciteSetBstMidEndSepPunct{\mcitedefaultmidpunct}
{\mcitedefaultendpunct}{\mcitedefaultseppunct}\relax
\EndOfBibitem
\bibitem[{Ida} and {Lin}(2004){Ida}, and {Lin}]{2004ApJ...616..567I}
{Ida},~S.; {Lin},~D.~N.~C. \emph{Astrophys.~J.,} \textbf{2004}, \emph{616},
  567\relax
\mciteBstWouldAddEndPuncttrue
\mciteSetBstMidEndSepPunct{\mcitedefaultmidpunct}
{\mcitedefaultendpunct}{\mcitedefaultseppunct}\relax
\EndOfBibitem
\bibitem[{Mordasini} et~al.(2012){Mordasini}, {Alibert}, {Benz}, {Klahr}, and
  {Henning}]{2012A&A...541A..97M}
{Mordasini},~C.; {Alibert},~Y.; {Benz},~W.; {Klahr},~H.; {Henning},~T.
  \emph{Astron.~Astrophys.,} \textbf{2012}, \emph{541}, A97\relax
\mciteBstWouldAddEndPuncttrue
\mciteSetBstMidEndSepPunct{\mcitedefaultmidpunct}
{\mcitedefaultendpunct}{\mcitedefaultseppunct}\relax
\EndOfBibitem
\bibitem[{Mordasini} et~al.(2012){Mordasini}, {Alibert}, {Klahr}, and
  {Henning}]{2012A&A...547A.111M}
{Mordasini},~C.; {Alibert},~Y.; {Klahr},~H.; {Henning},~T.
  \emph{Astron.~Astrophys.,} \textbf{2012}, \emph{547}, A111\relax
\mciteBstWouldAddEndPuncttrue
\mciteSetBstMidEndSepPunct{\mcitedefaultmidpunct}
{\mcitedefaultendpunct}{\mcitedefaultseppunct}\relax
\EndOfBibitem
\bibitem[{Williams} and {Cieza}(2011){Williams}, and
  {Cieza}]{2011ARA&A..49...67W}
{Williams},~J.~P.; {Cieza},~L.~A. \emph{Ann.~Rev.~Astron.~Astrophys.,}
  \textbf{2011}, \emph{49}, 67\relax
\mciteBstWouldAddEndPuncttrue
\mciteSetBstMidEndSepPunct{\mcitedefaultmidpunct}
{\mcitedefaultendpunct}{\mcitedefaultseppunct}\relax
\EndOfBibitem
\bibitem[{Stahler} and {Palla}(2005){Stahler}, and {Palla}]{Stahler_Palla05}
{Stahler},~S.~W.; {Palla},~F. \emph{The Formation of Stars}; Wiley-VCH Verlag
  GmbH, Weinheim, 2005\relax
\mciteBstWouldAddEndPuncttrue
\mciteSetBstMidEndSepPunct{\mcitedefaultmidpunct}
{\mcitedefaultendpunct}{\mcitedefaultseppunct}\relax
\EndOfBibitem
\bibitem[{Hartmann}(2009)]{2009apsf.book.....H}
{Hartmann},~L. \emph{Accretion Processes in Star Formation: Second Edition};
  Cambridge University Press, Cambridge, 2009\relax
\mciteBstWouldAddEndPuncttrue
\mciteSetBstMidEndSepPunct{\mcitedefaultmidpunct}
{\mcitedefaultendpunct}{\mcitedefaultseppunct}\relax
\EndOfBibitem
\bibitem[{Strom} et~al.(1989){Strom}, {Strom}, {Edwards}, {Cabrit}, and
  {Skrutskie}]{1989AJ.....97.1451S}
{Strom},~K.~M.; {Strom},~S.~E.; {Edwards},~S.; {Cabrit},~S.; {Skrutskie},~M.~F.
  \emph{Astron.~J,} \textbf{1989}, \emph{97}, 1451\relax
\mciteBstWouldAddEndPuncttrue
\mciteSetBstMidEndSepPunct{\mcitedefaultmidpunct}
{\mcitedefaultendpunct}{\mcitedefaultseppunct}\relax
\EndOfBibitem
\bibitem[{Beckwith} et~al.(1990){Beckwith}, {Sargent}, {Chini}, and
  {G{\"u}sten}]{1990AJ.....99..924B}
{Beckwith},~S.~V.~W.; {Sargent},~A.~I.; {Chini},~R.~S.; {G{\"u}sten},~R.
  \emph{Astron.~J,} \textbf{1990}, \emph{99}, 924\relax
\mciteBstWouldAddEndPuncttrue
\mciteSetBstMidEndSepPunct{\mcitedefaultmidpunct}
{\mcitedefaultendpunct}{\mcitedefaultseppunct}\relax
\EndOfBibitem
\bibitem[{Henning} and {Meeus}(2011){Henning}, and
  {Meeus}]{2011ppcd.book..114H}
{Henning},~T.; {Meeus},~G. In \emph{Physical Processes in Circumstellar Disks
  around Young Stars}; {Garcia},~P.~J.~V., Ed.; Chicago University Press,
  Chicago, 2011; p 114\relax
\mciteBstWouldAddEndPuncttrue
\mciteSetBstMidEndSepPunct{\mcitedefaultmidpunct}
{\mcitedefaultendpunct}{\mcitedefaultseppunct}\relax
\EndOfBibitem
\bibitem[{Grady} et~al.(2013){Grady}, {Muto}, {Hashimoto}, {Fukagawa},
  {Currie}, {Biller}, {Thalmann}, {Sitko}, {Russell}, {Wisniewski}, {Dong},
  {Kwon}, {Sai}, {Hornbeck}, {Schneider}, {Hines}, {Moro Mart{\'{\i}}n},
  {Feldt}, {Henning}, {Pott}, {Bonnefoy}, {Bouwman}, {Lacour}, {Mueller},
  {Juh{\'a}sz}, {Crida}, {Chauvin}, {Andrews}, {Wilner}, {Kraus}, {Dahm},
  {Robitaille}, {Jang-Condell}, {Abe}, {Akiyama}, {Brandner}, {Brandt},
  {Carson}, {Egner}, {Follette}, {Goto}, {Guyon}, {Hayano}, {Hayashi},
  {Hayashi}, {Hodapp}, {Ishii}, {Iye}, {Janson}, {Kandori}, {Knapp}, {Kudo},
  {Kusakabe}, {Kuzuhara}, {Mayama}, {McElwain}, {Matsuo}, { Miyama}, {Morino},
  {Nishimura}, {Pyo}, {Serabyn}, {Suto}, {Suzuki}, {Takami}, {Takato},
  {Terada}, {Tomono}, {Turner}, {Watanabe}, {Yamada}, {Takami}, {Usuda}, and
  {Tamura}]{2013ApJ...762...48G}
{Grady},~C.~A. et~al.  \emph{Astrophys.~J.,} \textbf{2013}, \emph{762},
  48\relax
\mciteBstWouldAddEndPuncttrue
\mciteSetBstMidEndSepPunct{\mcitedefaultmidpunct}
{\mcitedefaultendpunct}{\mcitedefaultseppunct}\relax
\EndOfBibitem
\bibitem[{Hughes} et~al.(2007){Hughes}, {Wilner}, {Calvet}, {D'Alessio},
  {Claussen}, and {Hogerheijde}]{Hughes_ea07}
{Hughes},~A.~M.; {Wilner},~D.~J.; {Calvet},~N.; {D'Alessio},~P.;
  {Claussen},~M.~J.; {Hogerheijde},~M.~R. \emph{Astrophys.~J.,} \textbf{2007},
  \emph{664}, 536\relax
\mciteBstWouldAddEndPuncttrue
\mciteSetBstMidEndSepPunct{\mcitedefaultmidpunct}
{\mcitedefaultendpunct}{\mcitedefaultseppunct}\relax
\EndOfBibitem
\bibitem[{Ratzka} et~al.(2007){Ratzka}, {Leinert}, {Henning}, {Bouwman},
  {Dullemond}, and {Jaffe}]{Ratzka_ea07}
{Ratzka},~T.; {Leinert},~C.; {Henning},~T.; {Bouwman},~J.; {Dullemond},~C.~P.;
  {Jaffe},~W. \emph{Astron.~Astrophys.,} \textbf{2007}, \emph{471}, 173\relax
\mciteBstWouldAddEndPuncttrue
\mciteSetBstMidEndSepPunct{\mcitedefaultmidpunct}
{\mcitedefaultendpunct}{\mcitedefaultseppunct}\relax
\EndOfBibitem
\bibitem[{Brown} et~al.(2009){Brown}, {Blake}, {Qi}, {Dullemond}, {Wilner}, and
  {Williams}]{Brown_09}
{Brown},~J.~M.; {Blake},~G.~A.; {Qi},~C.; {Dullemond},~C.~P.; {Wilner},~D.~J.;
  {Williams},~J.~P. \emph{Astrophys.~J.,} \textbf{2009}, \emph{704}, 496\relax
\mciteBstWouldAddEndPuncttrue
\mciteSetBstMidEndSepPunct{\mcitedefaultmidpunct}
{\mcitedefaultendpunct}{\mcitedefaultseppunct}\relax
\EndOfBibitem
\bibitem[{Brown} et~al.(2012){Brown}, {Herczeg}, {Pontoppidan}, and {van
  Dishoeck}]{Brown_ea12}
{Brown},~J.~M.; {Herczeg},~G.~J.; {Pontoppidan},~K.~M.; {van Dishoeck},~E.~F.
  \emph{Astrophys.~J.,} \textbf{2012}, \emph{744}, 116\relax
\mciteBstWouldAddEndPuncttrue
\mciteSetBstMidEndSepPunct{\mcitedefaultmidpunct}
{\mcitedefaultendpunct}{\mcitedefaultseppunct}\relax
\EndOfBibitem
\bibitem[{Sicilia-Aguilar} et~al.(2006){Sicilia-Aguilar}, {Hartmann}, {Calvet},
  {Megeath}, {Muzerolle}, {Allen}, {D'Alessio}, {Mer{\'{\i}}n}, {Stauffer},
  {Young}, and {Lada}]{SiciliaAguilar_ea06b}
{Sicilia-Aguilar},~A.; {Hartmann},~L.; {Calvet},~N.; {Megeath},~S.~T.;
  {Muzerolle},~J.; {Allen},~L.; {D'Alessio},~P.; {Mer{\'{\i}}n},~B.;
  {Stauffer},~J.; {Young},~E.; {Lada},~C. \emph{Astrophys.~J.,} \textbf{2006},
  \emph{638}, 897\relax
\mciteBstWouldAddEndPuncttrue
\mciteSetBstMidEndSepPunct{\mcitedefaultmidpunct}
{\mcitedefaultendpunct}{\mcitedefaultseppunct}\relax
\EndOfBibitem
\bibitem[{Sicilia-Aguilar} et~al.(2008){Sicilia-Aguilar}, {Henning},
  {Juh{\'a}sz}, {Bouwman}, {Garmire}, and {Garmire}]{SiciliaAguilar_ea08}
{Sicilia-Aguilar},~A.; {Henning},~T.; {Juh{\'a}sz},~A.; {Bouwman},~J.;
  {Garmire},~G.; {Garmire},~A. \emph{Astrophys.~J.,} \textbf{2008}, \emph{687},
  1145\relax
\mciteBstWouldAddEndPuncttrue
\mciteSetBstMidEndSepPunct{\mcitedefaultmidpunct}
{\mcitedefaultendpunct}{\mcitedefaultseppunct}\relax
\EndOfBibitem
\bibitem[{Furlan} et~al.(2009){Furlan}, {Watson}, {McClure}, {Manoj},
  {Espaillat}, {D'Alessio}, {Calvet}, {Kim}, {Sargent}, {Forrest}, and
  {Hartmann}]{Furlan_ea09}
{Furlan},~E.; {Watson},~D.~M.; {McClure},~M.~K.; {Manoj},~P.; {Espaillat},~C.;
  {D'Alessio},~P.; {Calvet},~N.; {Kim},~K.~H.; {Sargent},~B.~A.;
  {Forrest},~W.~J.; {Hartmann},~L. \emph{Astrophys.~J.,} \textbf{2009},
  \emph{703}, 1964\relax
\mciteBstWouldAddEndPuncttrue
\mciteSetBstMidEndSepPunct{\mcitedefaultmidpunct}
{\mcitedefaultendpunct}{\mcitedefaultseppunct}\relax
\EndOfBibitem
\bibitem[{Mer{\'{\i}}n} et~al.(2010){Mer{\'{\i}}n}, {Brown}, {Oliveira},
  {Herczeg}, {van Dishoeck}, {Bottinelli}, {Evans}, {Cieza}, {Spezzi},
  {Alcal{\'a}}, {Harvey}, {Blake}, {Bayo}, {Geers}, {Lahuis}, {Prusti},
  {Augereau}, {Olofsson}, {Walter}, and {Chiu}]{Merin_ea10}
{Mer{\'{\i}}n},~B. et~al.  \emph{Astrophys.~J.,} \textbf{2010}, \emph{718},
  1200\relax
\mciteBstWouldAddEndPuncttrue
\mciteSetBstMidEndSepPunct{\mcitedefaultmidpunct}
{\mcitedefaultendpunct}{\mcitedefaultseppunct}\relax
\EndOfBibitem
\bibitem[{Espaillat} et~al.(2010){Espaillat}, {D'Alessio}, {Hern{\'a}ndez},
  {Nagel}, {Luhman}, {Watson}, {Calvet}, {Muzerolle}, and
  {McClure}]{Espaillat_ea10}
{Espaillat},~C.; {D'Alessio},~P.; {Hern{\'a}ndez},~J.; {Nagel},~E.;
  {Luhman},~K.~L.; {Watson},~D.~M.; {Calvet},~N.; {Muzerolle},~J.;
  {McClure},~M. \emph{Astrophys.~J.,} \textbf{2010}, \emph{717}, 441\relax
\mciteBstWouldAddEndPuncttrue
\mciteSetBstMidEndSepPunct{\mcitedefaultmidpunct}
{\mcitedefaultendpunct}{\mcitedefaultseppunct}\relax
\EndOfBibitem
\bibitem[{Fukagawa} et~al.(2004){Fukagawa}, {Hayashi}, {Tamura}, {Itoh},
  {Hayashi}, {Oasa}, {Takeuchi}, {Morino}, {Murakawa}, {Oya}, {Yamashita},
  {Suto}, {Mayama}, {Naoi}, {Ishii}, {Pyo}, {Nishikawa}, {Takato}, {Usuda},
  {Ando}, {Iye}, {Miyama}, and {Kaifu}]{Fea04}
{Fukagawa},~M. et~al.  \emph{Astrophys. J. Lett.,} \textbf{2004}, \emph{605},
  L53\relax
\mciteBstWouldAddEndPuncttrue
\mciteSetBstMidEndSepPunct{\mcitedefaultmidpunct}
{\mcitedefaultendpunct}{\mcitedefaultseppunct}\relax
\EndOfBibitem
\bibitem[{van der Marel} et~al.(2013){van der Marel}, {van Dishoeck},
  {Bruderer}, {Birnstiel}, {Pinilla}, {Dullemond}, {van Kempen}, {Schmalzl},
  {Brown}, {Herczeg}, {Mathews}, and {Geers}]{2013arXiv1306.1768V}
{van der Marel},~N.; {van Dishoeck},~E.~F.; {Bruderer},~S.; {Birnstiel},~T.;
  {Pinilla},~P.; {Dullemond},~C.~P.; {van Kempen},~T.~A.; {Schmalzl},~M.;
  {Brown},~J.~M.; {Herczeg},~G.~J.; {Mathews},~G.~S.; {Geers},~V.
  \emph{Science} \textbf{2013}, \emph{340}, 1199\relax
\mciteBstWouldAddEndPuncttrue
\mciteSetBstMidEndSepPunct{\mcitedefaultmidpunct}
{\mcitedefaultendpunct}{\mcitedefaultseppunct}\relax
\EndOfBibitem
\bibitem[{Flock} et~al.(2012){Flock}, {Dzyurkevich}, {Klahr}, {Turner}, and
  {Henning}]{2012ApJ...744..144F}
{Flock},~M.; {Dzyurkevich},~N.; {Klahr},~H.; {Turner},~N.; {Henning},~T.
  \emph{Astrophys.~J.,} \textbf{2012}, \emph{744}, 144\relax
\mciteBstWouldAddEndPuncttrue
\mciteSetBstMidEndSepPunct{\mcitedefaultmidpunct}
{\mcitedefaultendpunct}{\mcitedefaultseppunct}\relax
\EndOfBibitem
\bibitem[{Boley} et~al.(2010){Boley}, {Hayfield}, {Mayer}, and
  {Durisen}]{Boley_ea10a}
{Boley},~A.~C.; {Hayfield},~T.; {Mayer},~L.; {Durisen},~R.~H. \emph{Icarus}
  \textbf{2010}, \emph{207}, 509\relax
\mciteBstWouldAddEndPuncttrue
\mciteSetBstMidEndSepPunct{\mcitedefaultmidpunct}
{\mcitedefaultendpunct}{\mcitedefaultseppunct}\relax
\EndOfBibitem
\bibitem[{van Dishoeck}(2006)]{2006PNAS..10312249V}
{van Dishoeck},~E.~F. \emph{Proc. Natl. Acad. Sci., U.S.A.,} \textbf{2006},
  \emph{103}, 12249\relax
\mciteBstWouldAddEndPuncttrue
\mciteSetBstMidEndSepPunct{\mcitedefaultmidpunct}
{\mcitedefaultendpunct}{\mcitedefaultseppunct}\relax
\EndOfBibitem
\bibitem[{Dutrey} et~al.(2007){Dutrey}, {Guilloteau}, and {Ho}]{DGH07}
{Dutrey},~A.; {Guilloteau},~S.; {Ho},~P. In \emph{Protostars and Planets V};
  {Reipurth},~B., {Jewitt},~D., {Keil},~K., Eds.; University of Arizona Press,
  Tucson, 2007; p 495\relax
\mciteBstWouldAddEndPuncttrue
\mciteSetBstMidEndSepPunct{\mcitedefaultmidpunct}
{\mcitedefaultendpunct}{\mcitedefaultseppunct}\relax
\EndOfBibitem
\bibitem[{Bergin}(2011)]{Bergin_09}
{Bergin},~E.~A. In \emph{Physical Processes in Circumstellar Disks around Young
  Stars}; {Garcia},~P.~J.~V., Ed.; Chicago University Press, Chicago, 2011;
  p~55\relax
\mciteBstWouldAddEndPuncttrue
\mciteSetBstMidEndSepPunct{\mcitedefaultmidpunct}
{\mcitedefaultendpunct}{\mcitedefaultseppunct}\relax
\EndOfBibitem
\bibitem[{Aikawa} et~al.(2003){Aikawa}, {Momose}, {Thi}, {van Zadelhoff}, {Qi},
  {Blake}, and {van Dishoeck}]{Aikawa_ea03}
{Aikawa},~Y.; {Momose},~M.; {Thi},~W.-F.; {van Zadelhoff},~G.-J.; {Qi},~C.;
  {Blake},~G.~A.; {van Dishoeck},~E.~F. \emph{Publ.~Astron.~Soc.~Jpn}
  \textbf{2003}, \emph{55}, 11\relax
\mciteBstWouldAddEndPuncttrue
\mciteSetBstMidEndSepPunct{\mcitedefaultmidpunct}
{\mcitedefaultendpunct}{\mcitedefaultseppunct}\relax
\EndOfBibitem
\bibitem[{Pi{\'e}tu} et~al.(2007){Pi{\'e}tu}, {Dutrey}, and
  {Guilloteau}]{Pietu_ea07}
{Pi{\'e}tu},~V.; {Dutrey},~A.; {Guilloteau},~S. \emph{Astron.~Astrophys.,}
  \textbf{2007}, \emph{467}, 163\relax
\mciteBstWouldAddEndPuncttrue
\mciteSetBstMidEndSepPunct{\mcitedefaultmidpunct}
{\mcitedefaultendpunct}{\mcitedefaultseppunct}\relax
\EndOfBibitem
\bibitem[{Schreyer} et~al.(2008){Schreyer}, {Guilloteau}, {Semenov}, {Bacmann},
  {Chapillon}, {Dutrey}, {Gueth}, {Henning}, {Hersant}, {Launhardt}, {Pety},
  and {Pi{\'e}tu}]{Schreyer_ea08}
{Schreyer},~K.; {Guilloteau},~S.; {Semenov},~D.; {Bacmann},~A.;
  {Chapillon},~E.; {Dutrey},~A.; {Gueth},~F.; {Henning},~T.; {Hersant},~F.;
  {Launhardt},~R.; {Pety},~J.; {Pi{\'e}tu},~V. \emph{Astron.~Astrophys.,}
  \textbf{2008}, \emph{491}, 821\relax
\mciteBstWouldAddEndPuncttrue
\mciteSetBstMidEndSepPunct{\mcitedefaultmidpunct}
{\mcitedefaultendpunct}{\mcitedefaultseppunct}\relax
\EndOfBibitem
\bibitem[{Pani{\'c}} and {Hogerheijde}(2009){Pani{\'c}}, and
  {Hogerheijde}]{Panic_Hogerheijde09}
{Pani{\'c}},~O.; {Hogerheijde},~M.~R. \emph{Astron.~Astrophys.,} \textbf{2009},
  \emph{508}, 707\relax
\mciteBstWouldAddEndPuncttrue
\mciteSetBstMidEndSepPunct{\mcitedefaultmidpunct}
{\mcitedefaultendpunct}{\mcitedefaultseppunct}\relax
\EndOfBibitem
\bibitem[{{\"O}berg} et~al.(2010){{\"O}berg}, {Qi}, {Fogel}, {Bergin},
  {Andrews}, {Espaillat}, {van Kempen}, {Wilner}, and {Pascucci}]{Oeberg_ea10a}
{{\"O}berg},~K.~I.; {Qi},~C.; {Fogel},~J.~K.~J.; {Bergin},~E.~A.;
  {Andrews},~S.~M.; {Espaillat},~C.; {van Kempen},~T.~A.; {Wilner},~D.~J.;
  {Pascucci},~I. \emph{Astrophys.~J.,} \textbf{2010}, \emph{720}, 480\relax
\mciteBstWouldAddEndPuncttrue
\mciteSetBstMidEndSepPunct{\mcitedefaultmidpunct}
{\mcitedefaultendpunct}{\mcitedefaultseppunct}\relax
\EndOfBibitem
\bibitem[{Richling} and {Yorke}(2000){Richling}, and {Yorke}]{rh00}
{Richling},~S.; {Yorke},~H.~W. \emph{Astrophys.~J.,} \textbf{2000}, \emph{539},
  258\relax
\mciteBstWouldAddEndPuncttrue
\mciteSetBstMidEndSepPunct{\mcitedefaultmidpunct}
{\mcitedefaultendpunct}{\mcitedefaultseppunct}\relax
\EndOfBibitem
\bibitem[{Willacy} and {Langer}(2000){Willacy}, and {Langer}]{wl00}
{Willacy},~K.; {Langer},~W.~D. \emph{Astrophys.~J.,} \textbf{2000}, \emph{544},
  903\relax
\mciteBstWouldAddEndPuncttrue
\mciteSetBstMidEndSepPunct{\mcitedefaultmidpunct}
{\mcitedefaultendpunct}{\mcitedefaultseppunct}\relax
\EndOfBibitem
\bibitem[{van Zadelhoff} et~al.(2003){van Zadelhoff}, {Aikawa}, {Hogerheijde},
  and {van Dishoeck}]{vZea03}
{van Zadelhoff},~G.-J.; {Aikawa},~Y.; {Hogerheijde},~M.~R.; {van
  Dishoeck},~E.~F. \emph{Astron.~Astrophys.,} \textbf{2003}, \emph{397},
  789\relax
\mciteBstWouldAddEndPuncttrue
\mciteSetBstMidEndSepPunct{\mcitedefaultmidpunct}
{\mcitedefaultendpunct}{\mcitedefaultseppunct}\relax
\EndOfBibitem
\bibitem[{Semenov} et~al.(2004){Semenov}, {Wiebe}, and {Henning}]{Red2}
{Semenov},~D.; {Wiebe},~D.; {Henning},~T. \emph{Astron.~Astrophys.,}
  \textbf{2004}, \emph{417}, 93\relax
\mciteBstWouldAddEndPuncttrue
\mciteSetBstMidEndSepPunct{\mcitedefaultmidpunct}
{\mcitedefaultendpunct}{\mcitedefaultseppunct}\relax
\EndOfBibitem
\bibitem[{van Dishoeck} et~al.(2006){van Dishoeck}, {Jonkheid}, and {van
  Hemert}]{vDea_06}
{van Dishoeck},~E.~F.; {Jonkheid},~B.; {van Hemert},~M.~C. In \emph{Chemical
  evolution of the Universe}; {Sims},~I.~R., {Williams},~D.~A., Eds.; Faraday
  discussion; Royal Society of Chemistry, Cambridge, 2006; Vol. 133; p
  231\relax
\mciteBstWouldAddEndPuncttrue
\mciteSetBstMidEndSepPunct{\mcitedefaultmidpunct}
{\mcitedefaultendpunct}{\mcitedefaultseppunct}\relax
\EndOfBibitem
\bibitem[{Gorti} et~al.(2009){Gorti}, {Dullemond}, and
  {Hollenbach}]{Gorti_ea09}
{Gorti},~U.; {Dullemond},~C.~P.; {Hollenbach},~D. \emph{Astrophys.~J.,}
  \textbf{2009}, \emph{705}, 1237\relax
\mciteBstWouldAddEndPuncttrue
\mciteSetBstMidEndSepPunct{\mcitedefaultmidpunct}
{\mcitedefaultendpunct}{\mcitedefaultseppunct}\relax
\EndOfBibitem
\bibitem[{Visser} et~al.(2009){Visser}, {van Dishoeck}, and
  {Black}]{Visser_ea09b}
{Visser},~R.; {van Dishoeck},~E.~F.; {Black},~J.~H. \emph{Astron.~Astrophys.,}
  \textbf{2009}, \emph{503}, 323\relax
\mciteBstWouldAddEndPuncttrue
\mciteSetBstMidEndSepPunct{\mcitedefaultmidpunct}
{\mcitedefaultendpunct}{\mcitedefaultseppunct}\relax
\EndOfBibitem
\bibitem[{Henning} et~al.(2010){Henning}, {Semenov}, {Guilloteau}, {Dutrey},
  {Hersant}, {Wakelam}, {Chapillon}, {Launhardt}, {Pi{\'e}tu}, and
  {Schreyer}]{Henning_ea10}
{Henning},~T.; {Semenov},~D.; {Guilloteau},~S.; {Dutrey},~A.; {Hersant},~F.;
  {Wakelam},~V.; {Chapillon},~E.; {Launhardt},~R.; {Pi{\'e}tu},~V.;
  {Schreyer},~K. \emph{Astrophys.~J.,} \textbf{2010}, \emph{714}, 1511\relax
\mciteBstWouldAddEndPuncttrue
\mciteSetBstMidEndSepPunct{\mcitedefaultmidpunct}
{\mcitedefaultendpunct}{\mcitedefaultseppunct}\relax
\EndOfBibitem
\bibitem[{Owen} et~al.(2011){Owen}, {Ercolano}, and {Clarke}]{Owen_ea11a}
{Owen},~J.~E.; {Ercolano},~B.; {Clarke},~C.~J. \emph{Mon.~Not.~R.~Astron.~Soc,}
  \textbf{2011}, \emph{412}, 13\relax
\mciteBstWouldAddEndPuncttrue
\mciteSetBstMidEndSepPunct{\mcitedefaultmidpunct}
{\mcitedefaultendpunct}{\mcitedefaultseppunct}\relax
\EndOfBibitem
\bibitem[{Walsh} et~al.(2012){Walsh}, {Nomura}, {Millar}, and
  {Aikawa}]{Walsh_ea12}
{Walsh},~C.; {Nomura},~H.; {Millar},~T.~J.; {Aikawa},~Y. \emph{Astrophys.~J.,}
  \textbf{2012}, \emph{747}, 114\relax
\mciteBstWouldAddEndPuncttrue
\mciteSetBstMidEndSepPunct{\mcitedefaultmidpunct}
{\mcitedefaultendpunct}{\mcitedefaultseppunct}\relax
\EndOfBibitem
\bibitem[{Dutrey} et~al.(1997){Dutrey}, {Guilloteau}, and
  {Guelin}]{Dutrey_ea97}
{Dutrey},~A.; {Guilloteau},~S.; {Guelin},~M. \emph{Astron.~Astrophys.,}
  \textbf{1997}, \emph{317}, L55\relax
\mciteBstWouldAddEndPuncttrue
\mciteSetBstMidEndSepPunct{\mcitedefaultmidpunct}
{\mcitedefaultendpunct}{\mcitedefaultseppunct}\relax
\EndOfBibitem
\bibitem[{Qi} et~al.(2013){Qi}, {{\"O}berg}, and {Wilner}]{Qi_ea13a}
{Qi},~C.; {{\"O}berg},~K.~I.; {Wilner},~D.~J. \emph{Astrophys.~J.,}
  \textbf{2013}, \emph{765}, 34\relax
\mciteBstWouldAddEndPuncttrue
\mciteSetBstMidEndSepPunct{\mcitedefaultmidpunct}
{\mcitedefaultendpunct}{\mcitedefaultseppunct}\relax
\EndOfBibitem
\bibitem[{Glassgold} et~al.(1997){Glassgold}, {Najita}, and {Igea}]{zetaxa}
{Glassgold},~A.~E.; {Najita},~J.; {Igea},~J. \emph{Astrophys.~J.,}
  \textbf{1997}, \emph{480}, 344\relax
\mciteBstWouldAddEndPuncttrue
\mciteSetBstMidEndSepPunct{\mcitedefaultmidpunct}
{\mcitedefaultendpunct}{\mcitedefaultseppunct}\relax
\EndOfBibitem
\bibitem[{Ilgner} and {Nelson}(2006){Ilgner}, and {Nelson}]{Ilgner_Nelson06}
{Ilgner},~M.; {Nelson},~R.~P. \emph{Astron.~Astrophys.,} \textbf{2006},
  \emph{445}, 205\relax
\mciteBstWouldAddEndPuncttrue
\mciteSetBstMidEndSepPunct{\mcitedefaultmidpunct}
{\mcitedefaultendpunct}{\mcitedefaultseppunct}\relax
\EndOfBibitem
\bibitem[{Ilgner} and {Nelson}(2006){Ilgner}, and {Nelson}]{Ilgner_Nelson06a}
{Ilgner},~M.; {Nelson},~R.~P. \emph{Astron.~Astrophys.,} \textbf{2006},
  \emph{445}, 223\relax
\mciteBstWouldAddEndPuncttrue
\mciteSetBstMidEndSepPunct{\mcitedefaultmidpunct}
{\mcitedefaultendpunct}{\mcitedefaultseppunct}\relax
\EndOfBibitem
\bibitem[{Ilgner} and {Nelson}(2006){Ilgner}, and {Nelson}]{Ilgner_Nelson06b}
{Ilgner},~M.; {Nelson},~R.~P. \emph{Astron.~Astrophys.,} \textbf{2006},
  \emph{455}, 731\relax
\mciteBstWouldAddEndPuncttrue
\mciteSetBstMidEndSepPunct{\mcitedefaultmidpunct}
{\mcitedefaultendpunct}{\mcitedefaultseppunct}\relax
\EndOfBibitem
\bibitem[{{\"O}berg} et~al.(2011){{\"O}berg}, {Qi}, {Wilner}, and
  {Andrews}]{2011ApJ...743..152O}
{{\"O}berg},~K.~I.; {Qi},~C.; {Wilner},~D.~J.; {Andrews},~S.~M.
  \emph{Astrophys.~J.,} \textbf{2011}, \emph{743}, 152\relax
\mciteBstWouldAddEndPuncttrue
\mciteSetBstMidEndSepPunct{\mcitedefaultmidpunct}
{\mcitedefaultendpunct}{\mcitedefaultseppunct}\relax
\EndOfBibitem
\bibitem[{Balbus} and {Hawley}(1991){Balbus}, and {Hawley}]{MRI}
{Balbus},~S.~A.; {Hawley},~J.~F. \emph{Astrophys.~J.,} \textbf{1991},
  \emph{376}, 214\relax
\mciteBstWouldAddEndPuncttrue
\mciteSetBstMidEndSepPunct{\mcitedefaultmidpunct}
{\mcitedefaultendpunct}{\mcitedefaultseppunct}\relax
\EndOfBibitem
\bibitem[{Gail}(2002)]{G02}
{Gail},~H.-P. \emph{Astron.~Astrophys.,} \textbf{2002}, \emph{390}, 253\relax
\mciteBstWouldAddEndPuncttrue
\mciteSetBstMidEndSepPunct{\mcitedefaultmidpunct}
{\mcitedefaultendpunct}{\mcitedefaultseppunct}\relax
\EndOfBibitem
\bibitem[{Wehrstedt} and {Gail}(2002){Wehrstedt}, and {Gail}]{Wehrstedt_Gail02}
{Wehrstedt},~M.; {Gail},~H. \emph{Astron.~Astrophys.,} \textbf{2002},
  \emph{385}, 181\relax
\mciteBstWouldAddEndPuncttrue
\mciteSetBstMidEndSepPunct{\mcitedefaultmidpunct}
{\mcitedefaultendpunct}{\mcitedefaultseppunct}\relax
\EndOfBibitem
\bibitem[{Boss}(2004)]{Boss2004}
{Boss},~A.~P. \emph{Astrophys.~J.,} \textbf{2004}, \emph{616}, 1265\relax
\mciteBstWouldAddEndPuncttrue
\mciteSetBstMidEndSepPunct{\mcitedefaultmidpunct}
{\mcitedefaultendpunct}{\mcitedefaultseppunct}\relax
\EndOfBibitem
\bibitem[{Ilgner} et~al.(2004){Ilgner}, {Henning}, {Markwick}, and
  {Millar}]{IHMM04}
{Ilgner},~M.; {Henning},~T.; {Markwick},~A.~J.; {Millar},~T.~J.
  \emph{Astron.~Astrophys.,} \textbf{2004}, \emph{415}, 643\relax
\mciteBstWouldAddEndPuncttrue
\mciteSetBstMidEndSepPunct{\mcitedefaultmidpunct}
{\mcitedefaultendpunct}{\mcitedefaultseppunct}\relax
\EndOfBibitem
\bibitem[{Willacy} et~al.(2006){Willacy}, {Langer}, {Allen}, and
  {Bryden}]{Willacy_ea06}
{Willacy},~K.; {Langer},~W.; {Allen},~M.; {Bryden},~G. \emph{Astrophys.~J.,}
  \textbf{2006}, \emph{644}, 1202\relax
\mciteBstWouldAddEndPuncttrue
\mciteSetBstMidEndSepPunct{\mcitedefaultmidpunct}
{\mcitedefaultendpunct}{\mcitedefaultseppunct}\relax
\EndOfBibitem
\bibitem[{Aikawa}(2007)]{Aikawa_07}
{Aikawa},~Y. \emph{Astrophys.~J.,} \textbf{2007}, \emph{656}, L93\relax
\mciteBstWouldAddEndPuncttrue
\mciteSetBstMidEndSepPunct{\mcitedefaultmidpunct}
{\mcitedefaultendpunct}{\mcitedefaultseppunct}\relax
\EndOfBibitem
\bibitem[{Turner} et~al.(2007){Turner}, {Sano}, and
  {Dziourkevitch}]{Turner_ea07}
{Turner},~N.~J.; {Sano},~T.; {Dziourkevitch},~N. \emph{Astrophys.~J.,}
  \textbf{2007}, \emph{659}, 729\relax
\mciteBstWouldAddEndPuncttrue
\mciteSetBstMidEndSepPunct{\mcitedefaultmidpunct}
{\mcitedefaultendpunct}{\mcitedefaultseppunct}\relax
\EndOfBibitem
\bibitem[{Tscharnuter} and {Gail}(2007){Tscharnuter}, and {Gail}]{TG07}
{Tscharnuter},~W.~M.; {Gail},~H.-P. \emph{Astron.~Astrophys.,} \textbf{2007},
  \emph{463}, 369\relax
\mciteBstWouldAddEndPuncttrue
\mciteSetBstMidEndSepPunct{\mcitedefaultmidpunct}
{\mcitedefaultendpunct}{\mcitedefaultseppunct}\relax
\EndOfBibitem
\bibitem[{Hersant} et~al.(2009){Hersant}, {Wakelam}, {Dutrey}, {Guilloteau},
  and {Herbst}]{Hersant_ea09}
{Hersant},~F.; {Wakelam},~V.; {Dutrey},~A.; {Guilloteau},~S.; {Herbst},~E.
  \emph{Astron.~Astrophys.,} \textbf{2009}, \emph{493}, L49\relax
\mciteBstWouldAddEndPuncttrue
\mciteSetBstMidEndSepPunct{\mcitedefaultmidpunct}
{\mcitedefaultendpunct}{\mcitedefaultseppunct}\relax
\EndOfBibitem
\bibitem[{Heinzeller} et~al.(2011){Heinzeller}, {Nomura}, {Walsh}, and
  {Millar}]{Heinzeller_ea11}
{Heinzeller},~D.; {Nomura},~H.; {Walsh},~C.; {Millar},~T.~J.
  \emph{Astrophys.~J.,} \textbf{2011}, \emph{731}, 115\relax
\mciteBstWouldAddEndPuncttrue
\mciteSetBstMidEndSepPunct{\mcitedefaultmidpunct}
{\mcitedefaultendpunct}{\mcitedefaultseppunct}\relax
\EndOfBibitem
\bibitem[{Semenov} and {Wiebe}(2011){Semenov}, and {Wiebe}]{Semenov_Wiebe11a}
{Semenov},~D.; {Wiebe},~D. \emph{Astrophys.~J.,} \textbf{2011}, \emph{196},
  25\relax
\mciteBstWouldAddEndPuncttrue
\mciteSetBstMidEndSepPunct{\mcitedefaultmidpunct}
{\mcitedefaultendpunct}{\mcitedefaultseppunct}\relax
\EndOfBibitem
\bibitem[{Lynden-Bell} and {Pringle}(1974){Lynden-Bell}, and
  {Pringle}]{Lynden-BellPringle74}
{Lynden-Bell},~D.; {Pringle},~J.~E. \emph{Mon.~Not.~R.~Astron.~Soc,}
  \textbf{1974}, \emph{168}, 603\relax
\mciteBstWouldAddEndPuncttrue
\mciteSetBstMidEndSepPunct{\mcitedefaultmidpunct}
{\mcitedefaultendpunct}{\mcitedefaultseppunct}\relax
\EndOfBibitem
\bibitem[{Pringle}(1981)]{Pringle81}
{Pringle},~J.~E. \emph{Ann.~Rev.~Astron.~Astrophys.,} \textbf{1981}, \emph{19},
  137\relax
\mciteBstWouldAddEndPuncttrue
\mciteSetBstMidEndSepPunct{\mcitedefaultmidpunct}
{\mcitedefaultendpunct}{\mcitedefaultseppunct}\relax
\EndOfBibitem
\bibitem[{Akimkin} et~al.(2013){Akimkin}, {Zhukovska}, {Wiebe}, {Semenov},
  {Pavlyuchenkov}, {Vasyunin}, {Birnstiel}, and {Henning}]{ANDES}
{Akimkin},~V.; {Zhukovska},~S.; {Wiebe},~D.; {Semenov},~D.;
  {Pavlyuchenkov},~Y.; {Vasyunin},~A.; {Birnstiel},~T.; {Henning},~T.
  \emph{Astrophys.~J.,} \textbf{2013}, \emph{766}, 8\relax
\mciteBstWouldAddEndPuncttrue
\mciteSetBstMidEndSepPunct{\mcitedefaultmidpunct}
{\mcitedefaultendpunct}{\mcitedefaultseppunct}\relax
\EndOfBibitem
\bibitem[{Shakura} and {Sunyaev}(1973){Shakura}, and
  {Sunyaev}]{ShakuraSunyaev73}
{Shakura},~N.~I.; {Sunyaev},~R.~A. \emph{Astron.~Astrophys.,} \textbf{1973},
  \emph{24}, 337\relax
\mciteBstWouldAddEndPuncttrue
\mciteSetBstMidEndSepPunct{\mcitedefaultmidpunct}
{\mcitedefaultendpunct}{\mcitedefaultseppunct}\relax
\EndOfBibitem
\bibitem[{Flock} et~al.(2011){Flock}, {Dzyurkevich}, {Klahr}, {Turner}, and
  {Henning}]{2011ApJ...735..122F}
{Flock},~M.; {Dzyurkevich},~N.; {Klahr},~H.; {Turner},~N.~J.; {Henning},~T.
  \emph{Astrophys.~J.,} \textbf{2011}, \emph{735}, 122\relax
\mciteBstWouldAddEndPuncttrue
\mciteSetBstMidEndSepPunct{\mcitedefaultmidpunct}
{\mcitedefaultendpunct}{\mcitedefaultseppunct}\relax
\EndOfBibitem
\bibitem[{Hueso} and {Guillot}(2005){Hueso}, and
  {Guillot}]{2005A&A...442..703H}
{Hueso},~R.; {Guillot},~T. \emph{Astron.~Astrophys.,} \textbf{2005},
  \emph{442}, 703\relax
\mciteBstWouldAddEndPuncttrue
\mciteSetBstMidEndSepPunct{\mcitedefaultmidpunct}
{\mcitedefaultendpunct}{\mcitedefaultseppunct}\relax
\EndOfBibitem
\bibitem[{Kenyon} and {Hartmann}(1987){Kenyon}, and
  {Hartmann}]{KenyonHartmann1987}
{Kenyon},~S.~J.; {Hartmann},~L. \emph{Astrophys.~J.,} \textbf{1987},
  \emph{323}, 714\relax
\mciteBstWouldAddEndPuncttrue
\mciteSetBstMidEndSepPunct{\mcitedefaultmidpunct}
{\mcitedefaultendpunct}{\mcitedefaultseppunct}\relax
\EndOfBibitem
\bibitem[{Bell} et~al.(1997){Bell}, {Cassen}, {Klahr}, and {Henning}]{B97}
{Bell},~K.~R.; {Cassen},~P.~M.; {Klahr},~H.~H.; {Henning},~T.
  \emph{Astrophys.~J.,} \textbf{1997}, \emph{486}, 372\relax
\mciteBstWouldAddEndPuncttrue
\mciteSetBstMidEndSepPunct{\mcitedefaultmidpunct}
{\mcitedefaultendpunct}{\mcitedefaultseppunct}\relax
\EndOfBibitem
\bibitem[{Balbus} and {Hawley}(1998){Balbus}, and {Hawley}]{Balbus_Hawley98}
{Balbus},~S.~A.; {Hawley},~J.~F. \emph{Rev. Mod. Phys.,} \textbf{1998},
  \emph{70}, 1\relax
\mciteBstWouldAddEndPuncttrue
\mciteSetBstMidEndSepPunct{\mcitedefaultmidpunct}
{\mcitedefaultendpunct}{\mcitedefaultseppunct}\relax
\EndOfBibitem
\bibitem[{Gammie}(1996)]{1996ApJ...457..355G}
{Gammie},~C.~F. \emph{Astrophys.~J.,} \textbf{1996}, \emph{457}, 355\relax
\mciteBstWouldAddEndPuncttrue
\mciteSetBstMidEndSepPunct{\mcitedefaultmidpunct}
{\mcitedefaultendpunct}{\mcitedefaultseppunct}\relax
\EndOfBibitem
\bibitem[{Sano} et~al.(2000){Sano}, {Miyama}, {Umebayashi}, and {Nakano}]{sano}
{Sano},~T.; {Miyama},~S.~M.; {Umebayashi},~T.; {Nakano},~T.
  \emph{Astrophys.~J.,} \textbf{2000}, \emph{543}, 486\relax
\mciteBstWouldAddEndPuncttrue
\mciteSetBstMidEndSepPunct{\mcitedefaultmidpunct}
{\mcitedefaultendpunct}{\mcitedefaultseppunct}\relax
\EndOfBibitem
\bibitem[{Dzyurkevich} et~al.(2013){Dzyurkevich}, {Turner}, {Henning}, and
  {Kley}]{Dzyurkevich_ea13a}
{Dzyurkevich},~N.; {Turner},~N.~J.; {Henning},~T.; {Kley},~W.
  \emph{Astrophys.~J.,} \textbf{2013}, \emph{765}, 114\relax
\mciteBstWouldAddEndPuncttrue
\mciteSetBstMidEndSepPunct{\mcitedefaultmidpunct}
{\mcitedefaultendpunct}{\mcitedefaultseppunct}\relax
\EndOfBibitem
\bibitem[{Mohanty} et~al.(2013){Mohanty}, {Ercolano}, and
  {Turner}]{2013ApJ...764...65M}
{Mohanty},~S.; {Ercolano},~B.; {Turner},~N.~J. \emph{Astrophys.~J.,}
  \textbf{2013}, \emph{764}, 65\relax
\mciteBstWouldAddEndPuncttrue
\mciteSetBstMidEndSepPunct{\mcitedefaultmidpunct}
{\mcitedefaultendpunct}{\mcitedefaultseppunct}\relax
\EndOfBibitem
\bibitem[{Pickett} et~al.(2003){Pickett}, {Mej{\'{\i}}a}, {Durisen}, {Cassen},
  {Berry}, and {Link}]{2003ApJ...590.1060P}
{Pickett},~B.~K.; {Mej{\'{\i}}a},~A.~C.; {Durisen},~R.~H.; {Cassen},~P.~M.;
  {Berry},~D.~K.; {Link},~R.~P. \emph{Astrophys.~J.,} \textbf{2003},
  \emph{590}, 1060\relax
\mciteBstWouldAddEndPuncttrue
\mciteSetBstMidEndSepPunct{\mcitedefaultmidpunct}
{\mcitedefaultendpunct}{\mcitedefaultseppunct}\relax
\EndOfBibitem
\bibitem[{Boley} et~al.(2006){Boley}, {Mej{\'{\i}}a}, {Durisen}, {Cai},
  {Pickett}, and {D'Alessio}]{2006ApJ...651..517B}
{Boley},~A.~C.; {Mej{\'{\i}}a},~A.~C.; {Durisen},~R.~H.; {Cai},~K.;
  {Pickett},~M.~K.; {D'Alessio},~P. \emph{Astrophys.~J.,} \textbf{2006},
  \emph{651}, 517\relax
\mciteBstWouldAddEndPuncttrue
\mciteSetBstMidEndSepPunct{\mcitedefaultmidpunct}
{\mcitedefaultendpunct}{\mcitedefaultseppunct}\relax
\EndOfBibitem
\bibitem[{Dullemond} et~al.(2007){Dullemond}, {Hollenbach}, {Kamp}, and
  {D'Alessio}]{2007prpl.conf..555D}
{Dullemond},~C.~P.; {Hollenbach},~D.; {Kamp},~I.; {D'Alessio},~P. In
  \emph{Protostars and Planets V}; {Reipurth},~B., {Jewitt},~D., {Keil},~K.,
  Eds.; University of Arizona Press, Tucson, 2007; p 555\relax
\mciteBstWouldAddEndPuncttrue
\mciteSetBstMidEndSepPunct{\mcitedefaultmidpunct}
{\mcitedefaultendpunct}{\mcitedefaultseppunct}\relax
\EndOfBibitem
\bibitem[{Hirose} and {Turner}(2011){Hirose}, and {Turner}]{Hirose_Turner11a}
{Hirose},~S.; {Turner},~N.~J. \emph{Astrophys.~J.,} \textbf{2011}, \emph{732},
  L30\relax
\mciteBstWouldAddEndPuncttrue
\mciteSetBstMidEndSepPunct{\mcitedefaultmidpunct}
{\mcitedefaultendpunct}{\mcitedefaultseppunct}\relax
\EndOfBibitem
\bibitem[{Bertout}(1989)]{1989ARA&A..27..351B}
{Bertout},~C. \emph{Ann.~Rev.~Astron.~Astrophys.,} \textbf{1989}, \emph{27},
  351\relax
\mciteBstWouldAddEndPuncttrue
\mciteSetBstMidEndSepPunct{\mcitedefaultmidpunct}
{\mcitedefaultendpunct}{\mcitedefaultseppunct}\relax
\EndOfBibitem
\bibitem[{Bergin} et~al.(2003){Bergin}, {Calvet}, {D'Alessio}, and
  {Herczeg}]{Bergin_ea04}
{Bergin},~E.; {Calvet},~N.; {D'Alessio},~P.; {Herczeg},~G.~J.
  \emph{Astrophys.~J.,} \textbf{2003}, \emph{591}, L159\relax
\mciteBstWouldAddEndPuncttrue
\mciteSetBstMidEndSepPunct{\mcitedefaultmidpunct}
{\mcitedefaultendpunct}{\mcitedefaultseppunct}\relax
\EndOfBibitem
\bibitem[{Preibisch} et~al.(2005){Preibisch}, {Kim}, {Favata}, {Feigelson},
  {Flaccomio}, {Getman}, {Micela}, {Sciortino}, {Stassun}, {Stelzer}, and
  {Zinnecker}]{Preibisch_ea05}
{Preibisch},~T.; {Kim},~Y.; {Favata},~F.; {Feigelson},~E.~D.; {Flaccomio},~E.;
  {Getman},~K.; {Micela},~G.; {Sciortino},~S.; {Stassun},~K.; {Stelzer},~B.;
  {Zinnecker},~H. \emph{Astrophys.~J.,} \textbf{2005}, \emph{160}, 401\relax
\mciteBstWouldAddEndPuncttrue
\mciteSetBstMidEndSepPunct{\mcitedefaultmidpunct}
{\mcitedefaultendpunct}{\mcitedefaultseppunct}\relax
\EndOfBibitem
\bibitem[{Waters} and {Waelkens}(1998){Waters}, and
  {Waelkens}]{1998ARA&A..36..233W}
{Waters},~L.~B.~F.~M.; {Waelkens},~C. \emph{Ann.~Rev.~Astron.~Astrophys.,}
  \textbf{1998}, \emph{36}, 233\relax
\mciteBstWouldAddEndPuncttrue
\mciteSetBstMidEndSepPunct{\mcitedefaultmidpunct}
{\mcitedefaultendpunct}{\mcitedefaultseppunct}\relax
\EndOfBibitem
\bibitem[{G{\"u}del} and {Naz{\'e}}(2009){G{\"u}del}, and
  {Naz{\'e}}]{Guedel_Naze09}
{G{\"u}del},~M.; {Naz{\'e}},~Y. \emph{Astron.~Astrophys.,} \textbf{2009},
  \emph{17}, 309\relax
\mciteBstWouldAddEndPuncttrue
\mciteSetBstMidEndSepPunct{\mcitedefaultmidpunct}
{\mcitedefaultendpunct}{\mcitedefaultseppunct}\relax
\EndOfBibitem
\bibitem[{Bethell} and {Bergin}(2011){Bethell}, and {Bergin}]{Bethell_Bergin11}
{Bethell},~T.~J.; {Bergin},~E.~A. \emph{Astrophys.~J.,} \textbf{2011},
  \emph{739}, 78\relax
\mciteBstWouldAddEndPuncttrue
\mciteSetBstMidEndSepPunct{\mcitedefaultmidpunct}
{\mcitedefaultendpunct}{\mcitedefaultseppunct}\relax
\EndOfBibitem
\bibitem[{D'Alessio} et~al.(1999){D'Alessio}, {Calvet}, {Hartmann}, {Lizano},
  and {Cant{\'o}}]{DAea99}
{D'Alessio},~P.; {Calvet},~N.; {Hartmann},~L.; {Lizano},~S.; {Cant{\'o}},~J.
  \emph{Astrophys.~J.,} \textbf{1999}, \emph{527}, 893\relax
\mciteBstWouldAddEndPuncttrue
\mciteSetBstMidEndSepPunct{\mcitedefaultmidpunct}
{\mcitedefaultendpunct}{\mcitedefaultseppunct}\relax
\EndOfBibitem
\bibitem[{Dullemond} et~al.(2001){Dullemond}, {Dominik}, and
  {Natta}]{2001ApJ...560..957D}
{Dullemond},~C.~P.; {Dominik},~C.; {Natta},~A. \emph{Astrophys.~J.,}
  \textbf{2001}, \emph{560}, 957\relax
\mciteBstWouldAddEndPuncttrue
\mciteSetBstMidEndSepPunct{\mcitedefaultmidpunct}
{\mcitedefaultendpunct}{\mcitedefaultseppunct}\relax
\EndOfBibitem
\bibitem[{Gail}(2010)]{Gail_2010}
{Gail},~H.-P. In \emph{Astromineralogy, 2nd edition.}; {Henning},~T., Ed.;
  Lecture Notes in Physics; Springer Verlag, Berlin, 2010; Vol. 815; p~61\relax
\mciteBstWouldAddEndPuncttrue
\mciteSetBstMidEndSepPunct{\mcitedefaultmidpunct}
{\mcitedefaultendpunct}{\mcitedefaultseppunct}\relax
\EndOfBibitem
\bibitem[{Kamp} and {Dullemond}(2004){Kamp}, and {Dullemond}]{Kamp_Dullemond04}
{Kamp},~I.; {Dullemond},~C.~P. \emph{Astrophys.~J.,} \textbf{2004}, \emph{615},
  991\relax
\mciteBstWouldAddEndPuncttrue
\mciteSetBstMidEndSepPunct{\mcitedefaultmidpunct}
{\mcitedefaultendpunct}{\mcitedefaultseppunct}\relax
\EndOfBibitem
\bibitem[{Gorti} and {Hollenbach}(2004){Gorti}, and
  {Hollenbach}]{Gorti_Hollenbach04}
{Gorti},~U.; {Hollenbach},~D. \emph{Astrophys.~J.,} \textbf{2004}, \emph{613},
  424\relax
\mciteBstWouldAddEndPuncttrue
\mciteSetBstMidEndSepPunct{\mcitedefaultmidpunct}
{\mcitedefaultendpunct}{\mcitedefaultseppunct}\relax
\EndOfBibitem
\bibitem[{Gorti} and {Hollenbach}(2008){Gorti}, and
  {Hollenbach}]{Gorti_Hollenbach08}
{Gorti},~U.; {Hollenbach},~D. \emph{Astrophys.~J.,} \textbf{2008}, \emph{683},
  287\relax
\mciteBstWouldAddEndPuncttrue
\mciteSetBstMidEndSepPunct{\mcitedefaultmidpunct}
{\mcitedefaultendpunct}{\mcitedefaultseppunct}\relax
\EndOfBibitem
\bibitem[{Vasyunin} et~al.(2011){Vasyunin}, {Wiebe}, {Birnstiel}, {Zhukovska},
  {Henning}, and {Dullemond}]{Vasyunin2011}
{Vasyunin},~A.~I.; {Wiebe},~D.~S.; {Birnstiel},~T.; {Zhukovska},~S.;
  {Henning},~T.; {Dullemond},~C.~P. \emph{Astrophys.~J.,} \textbf{2011},
  \emph{727}, 76\relax
\mciteBstWouldAddEndPuncttrue
\mciteSetBstMidEndSepPunct{\mcitedefaultmidpunct}
{\mcitedefaultendpunct}{\mcitedefaultseppunct}\relax
\EndOfBibitem
\bibitem[{Jonkheid} et~al.(2006){Jonkheid}, {Kamp}, {Augereau}, and {van
  Dishoeck}]{Jonkheid_ea06}
{Jonkheid},~B.; {Kamp},~I.; {Augereau},~J.-C.; {van Dishoeck},~E.~F.
  \emph{Astron.~Astrophys.,} \textbf{2006}, \emph{453}, 163\relax
\mciteBstWouldAddEndPuncttrue
\mciteSetBstMidEndSepPunct{\mcitedefaultmidpunct}
{\mcitedefaultendpunct}{\mcitedefaultseppunct}\relax
\EndOfBibitem
\bibitem[{Glassgold} et~al.(2004){Glassgold}, {Najita}, and
  {Igea}]{Glassgold_ea04}
{Glassgold},~A.~E.; {Najita},~J.; {Igea},~J. \emph{Astrophys.~J.,}
  \textbf{2004}, \emph{615}, 972\relax
\mciteBstWouldAddEndPuncttrue
\mciteSetBstMidEndSepPunct{\mcitedefaultmidpunct}
{\mcitedefaultendpunct}{\mcitedefaultseppunct}\relax
\EndOfBibitem
\bibitem[{Glassgold} et~al.(2012){Glassgold}, {Galli}, and
  {Padovani}]{2012ApJ...756..157G}
{Glassgold},~A.~E.; {Galli},~D.; {Padovani},~M. \emph{Astrophys.~J.,}
  \textbf{2012}, \emph{756}, 157\relax
\mciteBstWouldAddEndPuncttrue
\mciteSetBstMidEndSepPunct{\mcitedefaultmidpunct}
{\mcitedefaultendpunct}{\mcitedefaultseppunct}\relax
\EndOfBibitem
\bibitem[{Brauer} et~al.(2008){Brauer}, {Dullemond}, and
  {Henning}]{Brauer_ea08a}
{Brauer},~F.; {Dullemond},~C.~P.; {Henning},~T. \emph{Astron.~Astrophys.,}
  \textbf{2008}, \emph{480}, 859\relax
\mciteBstWouldAddEndPuncttrue
\mciteSetBstMidEndSepPunct{\mcitedefaultmidpunct}
{\mcitedefaultendpunct}{\mcitedefaultseppunct}\relax
\EndOfBibitem
\bibitem[{Beckwith} et~al.(2000){Beckwith}, {Henning}, and {Nakagawa}]{BHN00}
{Beckwith},~S.~V.~W.; {Henning},~T.; {Nakagawa},~Y. In \emph{Protostars and
  Planets IV}; {Mannings},~V., {Boss},~A.~P., {Russell},~S.~S., Eds.;
  University of Arizona Press, Tucson, 2000; p 533\relax
\mciteBstWouldAddEndPuncttrue
\mciteSetBstMidEndSepPunct{\mcitedefaultmidpunct}
{\mcitedefaultendpunct}{\mcitedefaultseppunct}\relax
\EndOfBibitem
\bibitem[{Weidenschilling} and {Cuzzi}(1993){Weidenschilling}, and
  {Cuzzi}]{Weidenschilling_Cuzzi93}
{Weidenschilling},~S.~J.; {Cuzzi},~J.~N. In \emph{Protostars and Planets III};
  {Levy},~E.~H., {Lunine},~J.~I., Eds.; University of Arizona Press, Tucson,
  1993; p 1031\relax
\mciteBstWouldAddEndPuncttrue
\mciteSetBstMidEndSepPunct{\mcitedefaultmidpunct}
{\mcitedefaultendpunct}{\mcitedefaultseppunct}\relax
\EndOfBibitem
\bibitem[{Birnstiel} et~al.(2010){Birnstiel}, {Dullemond}, and
  {Brauer}]{Birnstiel_ea10a}
{Birnstiel},~T.; {Dullemond},~C.~P.; {Brauer},~F. \emph{Astron.~Astrophys.,}
  \textbf{2010}, \emph{513}, A79\relax
\mciteBstWouldAddEndPuncttrue
\mciteSetBstMidEndSepPunct{\mcitedefaultmidpunct}
{\mcitedefaultendpunct}{\mcitedefaultseppunct}\relax
\EndOfBibitem
\bibitem[{Haghighipour} and {Boss}(2003){Haghighipour}, and
  {Boss}]{2003ApJ...598.1301H}
{Haghighipour},~N.; {Boss},~A.~P. \emph{Astrophys.~J.,} \textbf{2003},
  \emph{598}, 1301\relax
\mciteBstWouldAddEndPuncttrue
\mciteSetBstMidEndSepPunct{\mcitedefaultmidpunct}
{\mcitedefaultendpunct}{\mcitedefaultseppunct}\relax
\EndOfBibitem
\bibitem[{Johansen} et~al.(2011){Johansen}, {Klahr}, and
  {Henning}]{Johansen_ea09a}
{Johansen},~A.; {Klahr},~H.; {Henning},~T. \emph{Astron.~Astrophys.,}
  \textbf{2011}, \emph{529}, A62\relax
\mciteBstWouldAddEndPuncttrue
\mciteSetBstMidEndSepPunct{\mcitedefaultmidpunct}
{\mcitedefaultendpunct}{\mcitedefaultseppunct}\relax
\EndOfBibitem
\bibitem[{Meheut} et~al.(2012){Meheut}, {Meliani}, {Varniere}, and
  {Benz}]{2012A&A...545A.134M}
{Meheut},~H.; {Meliani},~Z.; {Varniere},~P.; {Benz},~W.
  \emph{Astron.~Astrophys.,} \textbf{2012}, \emph{545}, A134\relax
\mciteBstWouldAddEndPuncttrue
\mciteSetBstMidEndSepPunct{\mcitedefaultmidpunct}
{\mcitedefaultendpunct}{\mcitedefaultseppunct}\relax
\EndOfBibitem
\bibitem[{Safronov}(1969)]{1969edo..book.....S}
{Safronov},~V.~S. \emph{Evoliutsiia doplanetnogo oblaka i obrazovanie Zemli i
  planet.}; Izdatel'stvo ``Nauka'', Moskva, USSR, 1969\relax
\mciteBstWouldAddEndPuncttrue
\mciteSetBstMidEndSepPunct{\mcitedefaultmidpunct}
{\mcitedefaultendpunct}{\mcitedefaultseppunct}\relax
\EndOfBibitem
\bibitem[{Hayashi} et~al.(1985){Hayashi}, {Nakazawa}, and
  {Nakagawa}]{Hayashiea85}
{Hayashi},~C.; {Nakazawa},~K.; {Nakagawa},~Y. In \emph{Protostars and Planets
  II}; Black,~D.~C., Matthews,~M.~S., Eds.; University of Arizona Press,
  Tucson, 1985; p 1100\relax
\mciteBstWouldAddEndPuncttrue
\mciteSetBstMidEndSepPunct{\mcitedefaultmidpunct}
{\mcitedefaultendpunct}{\mcitedefaultseppunct}\relax
\EndOfBibitem
\bibitem[{Pollack} et~al.(1996){Pollack}, {Hubickyj}, {Bodenheimer},
  {Lissauer}, {Podolak}, and {Greenzweig}]{1996Icar..124...62P}
{Pollack},~J.~B.; {Hubickyj},~O.; {Bodenheimer},~P.; {Lissauer},~J.~J.;
  {Podolak},~M.; {Greenzweig},~Y. \emph{Icarus} \textbf{1996}, \emph{124},
  62\relax
\mciteBstWouldAddEndPuncttrue
\mciteSetBstMidEndSepPunct{\mcitedefaultmidpunct}
{\mcitedefaultendpunct}{\mcitedefaultseppunct}\relax
\EndOfBibitem
\bibitem[{Johansen} et~al.(2007){Johansen}, {Oishi}, {Mac Low}, {Klahr},
  {Henning}, and {Youdin}]{Johansen_ea07}
{Johansen},~A.; {Oishi},~J.~S.; {Mac Low},~M.-M.; {Klahr},~H.; {Henning},~T.;
  {Youdin},~A. \emph{Nature} \textbf{2007}, \emph{448}, 1022\relax
\mciteBstWouldAddEndPuncttrue
\mciteSetBstMidEndSepPunct{\mcitedefaultmidpunct}
{\mcitedefaultendpunct}{\mcitedefaultseppunct}\relax
\EndOfBibitem
\bibitem[{Papaloizou} et~al.(2007){Papaloizou}, {Nelson}, {Kley}, {Masset}, and
  {Artymowicz}]{2007prpl.conf..655P}
{Papaloizou},~J.~C.~B.; {Nelson},~R.~P.; {Kley},~W.; {Masset},~F.~S.;
  {Artymowicz},~P. In \emph{Protostars and Planets V}; {Reipurth},~B.,
  {Jewitt},~D., {Keil},~K., Eds.; University of Arizona Press, Tucson, 2007; p
  655\relax
\mciteBstWouldAddEndPuncttrue
\mciteSetBstMidEndSepPunct{\mcitedefaultmidpunct}
{\mcitedefaultendpunct}{\mcitedefaultseppunct}\relax
\EndOfBibitem
\bibitem[{Boss}(1997)]{Boss_97}
{Boss},~A.~P. \emph{Science} \textbf{1997}, \emph{276}, 1836\relax
\mciteBstWouldAddEndPuncttrue
\mciteSetBstMidEndSepPunct{\mcitedefaultmidpunct}
{\mcitedefaultendpunct}{\mcitedefaultseppunct}\relax
\EndOfBibitem
\bibitem[{Carson} et~al.(2013){Carson}, {Thalmann}, {Janson}, {Kozakis},
  {Bonnefoy}, {Biller}, {Schlieder}, {Currie}, {McElwain}, {Goto}, {Henning},
  {Brandner}, {Feldt}, {Kandori}, {Kuzuhara}, {Stevens}, {Wong}, {Gainey},
  {Fukagawa}, {Kuwada}, {Brandt}, {Kwon}, {Abe}, {Egner}, {Grady}, {Guyon},
  {Hashimoto}, {Hayano}, {Hayashi}, {Hayashi}, {Hodapp}, {Ishii}, {Iye},
  {Knapp}, {Kudo}, {Kusakabe}, {Matsuo}, {Miyama}, {Morino}, {Moro-Martin},
  {Nishimura}, {Pyo}, {Serabyn}, {Suto}, {Suzuki}, {Takami}, {Takato},
  {Terada}, {Tomono}, {Turner}, {Watanabe}, {Wisniewski}, {Yamada}, {Takami},
  {Usuda}, and {Tamura}]{Carson_ea13a}
{Carson},~J. et~al.  \emph{Astrophys.~J.~Lett.,} \textbf{2013}, \emph{763},
  L32\relax
\mciteBstWouldAddEndPuncttrue
\mciteSetBstMidEndSepPunct{\mcitedefaultmidpunct}
{\mcitedefaultendpunct}{\mcitedefaultseppunct}\relax
\EndOfBibitem
\bibitem[{Janson} et~al.(2012){Janson}, {Bonavita}, {Klahr}, and
  {Lafreni{\`e}re}]{Janson_ea12a}
{Janson},~M.; {Bonavita},~M.; {Klahr},~H.; {Lafreni{\`e}re},~D.
  \emph{Astrophys.~J.,} \textbf{2012}, \emph{745}, 4\relax
\mciteBstWouldAddEndPuncttrue
\mciteSetBstMidEndSepPunct{\mcitedefaultmidpunct}
{\mcitedefaultendpunct}{\mcitedefaultseppunct}\relax
\EndOfBibitem
\bibitem[{Carmona} et~al.(2008){Carmona}, {van den Ancker}, {Henning},
  {Pavlyuchenkov}, {Dullemond}, {Goto}, {Thi}, {Bouwman}, and
  {Waters}]{Carmona_ea08}
{Carmona},~A.; {van den Ancker},~M.~E.; {Henning},~T.; {Pavlyuchenkov},~Y.;
  {Dullemond},~C.~P.; {Goto},~M.; {Thi},~W.~F.; {Bouwman},~J.;
  {Waters},~L.~B.~F.~M. \emph{Astron.~Astrophys.,} \textbf{2008}, \emph{477},
  839\relax
\mciteBstWouldAddEndPuncttrue
\mciteSetBstMidEndSepPunct{\mcitedefaultmidpunct}
{\mcitedefaultendpunct}{\mcitedefaultseppunct}\relax
\EndOfBibitem
\bibitem[{Tielens} et~al.(1991){Tielens}, {Tokunaga}, {Geballe}, and
  {Baas}]{Tielens_ea91}
{Tielens},~A.~G.~G.~M.; {Tokunaga},~A.~T.; {Geballe},~T.~R.; {Baas},~F.
  \emph{Astrophys.~J.,} \textbf{1991}, \emph{381}, 181\relax
\mciteBstWouldAddEndPuncttrue
\mciteSetBstMidEndSepPunct{\mcitedefaultmidpunct}
{\mcitedefaultendpunct}{\mcitedefaultseppunct}\relax
\EndOfBibitem
\bibitem[{van Dishoeck}(1988)]{vD88}
{van Dishoeck},~E.~F. In \emph{ASSL Vol. 146: Rate Coefficients in
  Astrochemistry}; {Millar},~T., {Williams},~D., Eds.; Kluwer Academic
  Publishers, Dordrecht, 1988; p~49\relax
\mciteBstWouldAddEndPuncttrue
\mciteSetBstMidEndSepPunct{\mcitedefaultmidpunct}
{\mcitedefaultendpunct}{\mcitedefaultseppunct}\relax
\EndOfBibitem
\bibitem[{Clayton}(2002)]{Clayton2002}
{Clayton},~R.~N. \emph{Nature} \textbf{2002}, \emph{415}, 860\relax
\mciteBstWouldAddEndPuncttrue
\mciteSetBstMidEndSepPunct{\mcitedefaultmidpunct}
{\mcitedefaultendpunct}{\mcitedefaultseppunct}\relax
\EndOfBibitem
\bibitem[{Lyons} et~al.(2007){Lyons}, {Boney}, and {Marcus}]{Lyons2007}
{Lyons},~J.~R.; {Boney},~E.; {Marcus},~R.~A. \emph{Lunar and Planetary
  Institute Conference Abstracts}; Lunar and Planetary Inst. Technical Report;
  Lunar and Planetary Institute, Houston, 2007; Vol.~38; p 2382\relax
\mciteBstWouldAddEndPuncttrue
\mciteSetBstMidEndSepPunct{\mcitedefaultmidpunct}
{\mcitedefaultendpunct}{\mcitedefaultseppunct}\relax
\EndOfBibitem
\bibitem[{Schreyer} et~al.(2006){Schreyer}, {Semenov}, {Henning}, and
  {Forbrich}]{2006ApJ...637L.129S}
{Schreyer},~K.; {Semenov},~D.; {Henning},~T.; {Forbrich},~J.
  \emph{Astrophys.~J.,} \textbf{2006}, \emph{637}, L129\relax
\mciteBstWouldAddEndPuncttrue
\mciteSetBstMidEndSepPunct{\mcitedefaultmidpunct}
{\mcitedefaultendpunct}{\mcitedefaultseppunct}\relax
\EndOfBibitem
\bibitem[{Bergin} et~al.(2013){Bergin}, {Cleeves}, {Gorti}, {Zhang}, {Blake},
  {Green}, {Andrews}, {Evans}, {Henning}, {{\"O}berg}, {Pontoppidan}, {Qi},
  {Salyk}, and {van Dishoeck}]{2013Natur.493..644B}
{Bergin},~E.~A.; {Cleeves},~L.~I.; {Gorti},~U.; {Zhang},~K.; {Blake},~G.~A.;
  {Green},~J.~D.; {Andrews},~S.~M.; {Evans},~N.~J.,~II; {Henning},~T.;
  {{\"O}berg},~K.; {Pontoppidan},~K.; {Qi},~C.; {Salyk},~C.; {van Dishoeck},~E.
  \emph{Nature} \textbf{2013}, \emph{493}, 644\relax
\mciteBstWouldAddEndPuncttrue
\mciteSetBstMidEndSepPunct{\mcitedefaultmidpunct}
{\mcitedefaultendpunct}{\mcitedefaultseppunct}\relax
\EndOfBibitem
\bibitem[{Andrews} and {Williams}(2005){Andrews}, and
  {Williams}]{Andrews_Williams05}
{Andrews},~S.~M.; {Williams},~J.~P. \emph{Astrophys.~J.,} \textbf{2005},
  \emph{631}, 1134\relax
\mciteBstWouldAddEndPuncttrue
\mciteSetBstMidEndSepPunct{\mcitedefaultmidpunct}
{\mcitedefaultendpunct}{\mcitedefaultseppunct}\relax
\EndOfBibitem
\bibitem[{Andrews} and {Williams}(2007){Andrews}, and
  {Williams}]{2007ApJ...671.1800A}
{Andrews},~S.~M.; {Williams},~J.~P. \emph{Astrophys.~J.,} \textbf{2007},
  \emph{671}, 1800\relax
\mciteBstWouldAddEndPuncttrue
\mciteSetBstMidEndSepPunct{\mcitedefaultmidpunct}
{\mcitedefaultendpunct}{\mcitedefaultseppunct}\relax
\EndOfBibitem
\bibitem[{Guilloteau} et~al.(2011){Guilloteau}, {Dutrey}, {Pi{\'e}tu}, and
  {Boehler}]{Guilloteau_ea11a}
{Guilloteau},~S.; {Dutrey},~A.; {Pi{\'e}tu},~V.; {Boehler},~Y.
  \emph{Astron.~Astrophys.,} \textbf{2011}, \emph{529}, A105\relax
\mciteBstWouldAddEndPuncttrue
\mciteSetBstMidEndSepPunct{\mcitedefaultmidpunct}
{\mcitedefaultendpunct}{\mcitedefaultseppunct}\relax
\EndOfBibitem
\bibitem[{Andrews} et~al.(2013){Andrews}, {Rosenfeld}, {Kraus}, and
  {Wilner}]{Andrews_ea13a}
{Andrews},~S.~M.; {Rosenfeld},~K.~A.; {Kraus},~A.~L.; {Wilner},~D.~J.
  \emph{Astrophys.~J.,} \textbf{2013}, \emph{771}, 129\relax
\mciteBstWouldAddEndPuncttrue
\mciteSetBstMidEndSepPunct{\mcitedefaultmidpunct}
{\mcitedefaultendpunct}{\mcitedefaultseppunct}\relax
\EndOfBibitem
\bibitem[{Muzerolle} et~al.(2003){Muzerolle}, {Hillenbrand}, {Calvet},
  {Brice{\~n}o}, and {Hartmann}]{Muzerolle_ea03}
{Muzerolle},~J.; {Hillenbrand},~L.; {Calvet},~N.; {Brice{\~n}o},~C.;
  {Hartmann},~L. \emph{Astrophys.~J.,} \textbf{2003}, \emph{592}, 266\relax
\mciteBstWouldAddEndPuncttrue
\mciteSetBstMidEndSepPunct{\mcitedefaultmidpunct}
{\mcitedefaultendpunct}{\mcitedefaultseppunct}\relax
\EndOfBibitem
\bibitem[{Natta} et~al.(2006){Natta}, {Testi}, and {Randich}]{Natta_ea06}
{Natta},~A.; {Testi},~L.; {Randich},~S. \emph{Astron.~Astrophys.,}
  \textbf{2006}, \emph{452}, 245\relax
\mciteBstWouldAddEndPuncttrue
\mciteSetBstMidEndSepPunct{\mcitedefaultmidpunct}
{\mcitedefaultendpunct}{\mcitedefaultseppunct}\relax
\EndOfBibitem
\bibitem[{Fang} et~al.(2009){Fang}, {van Boekel}, {Wang}, {Carmona},
  {Sicilia-Aguilar}, and {Henning}]{Fang_ea09}
{Fang},~M.; {van Boekel},~R.; {Wang},~W.; {Carmona},~A.; {Sicilia-Aguilar},~A.;
  {Henning},~T. \emph{Astron.~Astrophys.,} \textbf{2009}, \emph{504}, 461\relax
\mciteBstWouldAddEndPuncttrue
\mciteSetBstMidEndSepPunct{\mcitedefaultmidpunct}
{\mcitedefaultendpunct}{\mcitedefaultseppunct}\relax
\EndOfBibitem
\bibitem[{Fedele} et~al.(2010){Fedele}, {van den Ancker}, {Henning},
  {Jayawardhana}, and {Oliveira}]{2010A&A...510A..72F}
{Fedele},~D.; {van den Ancker},~M.~E.; {Henning},~T.; {Jayawardhana},~R.;
  {Oliveira},~J.~M. \emph{Astron.~Astrophys.,} \textbf{2010}, \emph{510},
  A72\relax
\mciteBstWouldAddEndPuncttrue
\mciteSetBstMidEndSepPunct{\mcitedefaultmidpunct}
{\mcitedefaultendpunct}{\mcitedefaultseppunct}\relax
\EndOfBibitem
\bibitem[{Andrews} et~al.(2012){Andrews}, {Wilner}, {Hughes}, {Qi},
  {Rosenfeld}, {{\"O}berg}, {Birnstiel}, {Espaillat}, {Cieza}, {Williams},
  {Lin}, and {Ho}]{Andrews_Wilner_ea12a}
{Andrews},~S.~M.; {Wilner},~D.~J.; {Hughes},~A.~M.; {Qi},~C.;
  {Rosenfeld},~K.~A.; {{\"O}berg},~K.~I.; {Birnstiel},~T.; {Espaillat},~C.;
  {Cieza},~L.~A.; {Williams},~J.~P.; {Lin},~S.-Y.; {Ho},~P.~T.~P.
  \emph{Astrophys.~J.,} \textbf{2012}, \emph{744}, 162\relax
\mciteBstWouldAddEndPuncttrue
\mciteSetBstMidEndSepPunct{\mcitedefaultmidpunct}
{\mcitedefaultendpunct}{\mcitedefaultseppunct}\relax
\EndOfBibitem
\bibitem[{Guilloteau} and {Dutrey}(1998){Guilloteau}, and {Dutrey}]{GD98}
{Guilloteau},~S.; {Dutrey},~A. \emph{Astron.~Astrophys.,} \textbf{1998},
  \emph{339}, 467\relax
\mciteBstWouldAddEndPuncttrue
\mciteSetBstMidEndSepPunct{\mcitedefaultmidpunct}
{\mcitedefaultendpunct}{\mcitedefaultseppunct}\relax
\EndOfBibitem
\bibitem[{Pani{\'c}} et~al.(2008){Pani{\'c}}, {Hogerheijde}, {Wilner}, and
  {Qi}]{2008A&A...491..219P}
{Pani{\'c}},~O.; {Hogerheijde},~M.~R.; {Wilner},~D.; {Qi},~C.
  \emph{Astron.~Astrophys.,} \textbf{2008}, \emph{491}, 219\relax
\mciteBstWouldAddEndPuncttrue
\mciteSetBstMidEndSepPunct{\mcitedefaultmidpunct}
{\mcitedefaultendpunct}{\mcitedefaultseppunct}\relax
\EndOfBibitem
\bibitem[{Qi} et~al.(2006){Qi}, {Wilner}, {Calvet}, {Bourke}, {Blake},
  {Hogerheijde}, {Ho}, and {Bergin}]{Qi_ea06}
{Qi},~C.; {Wilner},~D.~J.; {Calvet},~N.; {Bourke},~T.~L.; {Blake},~G.~A.;
  {Hogerheijde},~M.~R.; {Ho},~P.~T.~P.; {Bergin},~E. \emph{Astrophys.~J.,}
  \textbf{2006}, \emph{636}, L157\relax
\mciteBstWouldAddEndPuncttrue
\mciteSetBstMidEndSepPunct{\mcitedefaultmidpunct}
{\mcitedefaultendpunct}{\mcitedefaultseppunct}\relax
\EndOfBibitem
\bibitem[{Akiyama} et~al.(2012){Akiyama}, {Momose}, {Hayashi}, and
  {Kitamura}]{2012arXiv1205.6573A}
{Akiyama},~E.; {Momose},~M.; {Hayashi},~H.; {Kitamura},~Y.
  \emph{arXiv:astro-ph/1205.6573,} \textbf{2012}, \relax
\mciteBstWouldAddEndPunctfalse
\mciteSetBstMidEndSepPunct{\mcitedefaultmidpunct}
{}{\mcitedefaultseppunct}\relax
\EndOfBibitem
\bibitem[{Dartois} et~al.(2003){Dartois}, {Dutrey}, and {Guilloteau}]{DDG03}
{Dartois},~E.; {Dutrey},~A.; {Guilloteau},~S. \emph{Astron.~Astrophys.,}
  \textbf{2003}, \emph{399}, 773\relax
\mciteBstWouldAddEndPuncttrue
\mciteSetBstMidEndSepPunct{\mcitedefaultmidpunct}
{\mcitedefaultendpunct}{\mcitedefaultseppunct}\relax
\EndOfBibitem
\bibitem[{Haisch} et~al.(2001){Haisch}, {Lada}, and {Lada}]{disc_fraction}
{Haisch},~K.~E.; {Lada},~E.~A.; {Lada},~C.~J. \emph{Astrophys.~J.,}
  \textbf{2001}, \emph{553}, L153\relax
\mciteBstWouldAddEndPuncttrue
\mciteSetBstMidEndSepPunct{\mcitedefaultmidpunct}
{\mcitedefaultendpunct}{\mcitedefaultseppunct}\relax
\EndOfBibitem
\bibitem[{Mo{\'o}r} et~al.(2011){Mo{\'o}r}, {{\'A}brah{\'a}m}, {Juh{\'a}sz},
  {Kiss}, {Pascucci}, {K{\'o}sp{\'a}l}, {Apai}, {Henning}, {Csengeri}, and
  {Grady}]{2011ApJ...740L...7M}
{Mo{\'o}r},~A.; {{\'A}brah{\'a}m},~P.; {Juh{\'a}sz},~A.; {Kiss},~C.;
  {Pascucci},~I.; {K{\'o}sp{\'a}l},~{\'A}.; {Apai},~D.; {Henning},~T.;
  {Csengeri},~T.; {Grady},~C. \emph{Astrophys.~J.,} \textbf{2011}, \emph{740},
  L7\relax
\mciteBstWouldAddEndPuncttrue
\mciteSetBstMidEndSepPunct{\mcitedefaultmidpunct}
{\mcitedefaultendpunct}{\mcitedefaultseppunct}\relax
\EndOfBibitem
\bibitem[{Amelin} and {Krot}(2007){Amelin}, and {Krot}]{2007M&PS...42.1321A}
{Amelin},~Y.; {Krot},~A. \emph{Meteorit. Planet. Sci.,} \textbf{2007},
  \emph{42}, 1321\relax
\mciteBstWouldAddEndPuncttrue
\mciteSetBstMidEndSepPunct{\mcitedefaultmidpunct}
{\mcitedefaultendpunct}{\mcitedefaultseppunct}\relax
\EndOfBibitem
\bibitem[{Dutrey} et~al.(2007){Dutrey}, {Henning}, {Guilloteau}, {Semenov},
  {Pi{\'e}tu}, {Schreyer}, {Bacmann}, {Launhardt}, {Pety}, and
  {Gueth}]{Dutrey_ea07}
{Dutrey},~A.; {Henning},~T.; {Guilloteau},~S.; {Semenov},~D.; {Pi{\'e}tu},~V.;
  {Schreyer},~K.; {Bacmann},~A.; {Launhardt},~R.; {Pety},~J.; {Gueth},~F.
  \emph{Astron.~Astrophys.,} \textbf{2007}, \emph{464}, 615\relax
\mciteBstWouldAddEndPuncttrue
\mciteSetBstMidEndSepPunct{\mcitedefaultmidpunct}
{\mcitedefaultendpunct}{\mcitedefaultseppunct}\relax
\EndOfBibitem
\bibitem[{Dutrey} et~al.(2011){Dutrey}, {Wakelam}, {Boehler}, {Guilloteau},
  {Hersant}, {Semenov}, {Chapillon}, {Henning}, {Pi{\'e}tu}, {Launhardt},
  {Gueth}, and {Schreyer}]{Dutrey_ea11a}
{Dutrey},~A.; {Wakelam},~V.; {Boehler},~Y.; {Guilloteau},~S.; {Hersant},~F.;
  {Semenov},~D.; {Chapillon},~E.; {Henning},~T.; {Pi{\'e}tu},~V.;
  {Launhardt},~R.; {Gueth},~F.; {Schreyer},~K. \emph{Astron.~Astrophys.,}
  \textbf{2011}, \emph{535}, A104\relax
\mciteBstWouldAddEndPuncttrue
\mciteSetBstMidEndSepPunct{\mcitedefaultmidpunct}
{\mcitedefaultendpunct}{\mcitedefaultseppunct}\relax
\EndOfBibitem
\bibitem[{Chapillon} et~al.(2012){Chapillon}, {Guilloteau}, {Dutrey},
  {Pi{\'e}tu}, and {Gu{\'e}lin}]{Chapillon_ea12a}
{Chapillon},~E.; {Guilloteau},~S.; {Dutrey},~A.; {Pi{\'e}tu},~V.;
  {Gu{\'e}lin},~M. \emph{Astron.~Astrophys.,} \textbf{2012}, \emph{537},
  A60\relax
\mciteBstWouldAddEndPuncttrue
\mciteSetBstMidEndSepPunct{\mcitedefaultmidpunct}
{\mcitedefaultendpunct}{\mcitedefaultseppunct}\relax
\EndOfBibitem
\bibitem[{Chapillon} et~al.(2012){Chapillon}, {Dutrey}, {Guilloteau},
  {Pi{\'e}tu}, {Wakelam}, {Hersant}, {Gueth}, {Henning}, {Launhardt},
  {Schreyer}, and {Semenov}]{Chapillon_ea12b}
{Chapillon},~E.; {Dutrey},~A.; {Guilloteau},~S.; {Pi{\'e}tu},~V.;
  {Wakelam},~V.; {Hersant},~F.; {Gueth},~F.; {Henning},~T.; {Launhardt},~R.;
  {Schreyer},~K.; {Semenov},~D. \emph{Astrophys.~J.,} \textbf{2012},
  \emph{756}, 58\relax
\mciteBstWouldAddEndPuncttrue
\mciteSetBstMidEndSepPunct{\mcitedefaultmidpunct}
{\mcitedefaultendpunct}{\mcitedefaultseppunct}\relax
\EndOfBibitem
\bibitem[{Guilloteau} et~al.(2012){Guilloteau}, {Dutrey}, {Wakelam}, {Hersant},
  {Semenov}, {Chapillon}, {Henning}, and {Pi{\'e}tu}]{Guilloteau_ea12a}
{Guilloteau},~S.; {Dutrey},~A.; {Wakelam},~V.; {Hersant},~F.; {Semenov},~D.;
  {Chapillon},~E.; {Henning},~T.; {Pi{\'e}tu},~V. \emph{Astron.~Astrophys.,}
  \textbf{2012}, \emph{548}, A70\relax
\mciteBstWouldAddEndPuncttrue
\mciteSetBstMidEndSepPunct{\mcitedefaultmidpunct}
{\mcitedefaultendpunct}{\mcitedefaultseppunct}\relax
\EndOfBibitem
\bibitem[{{\"O}berg} et~al.(2011){{\"O}berg}, {Qi}, {Fogel}, {Bergin},
  {Andrews}, {Espaillat}, {Wilner}, {Pascucci}, and {Kastner}]{Oeberg_ea11a}
{{\"O}berg},~K.~I.; {Qi},~C.; {Fogel},~J.~K.~J.; {Bergin},~E.~A.;
  {Andrews},~S.~M.; {Espaillat},~C.; {Wilner},~D.~J.; {Pascucci},~I.;
  {Kastner},~J.~H. \emph{Astrophys.~J.,} \textbf{2011}, \emph{734}, 98\relax
\mciteBstWouldAddEndPuncttrue
\mciteSetBstMidEndSepPunct{\mcitedefaultmidpunct}
{\mcitedefaultendpunct}{\mcitedefaultseppunct}\relax
\EndOfBibitem
\bibitem[{Qi} et~al.(2013){Qi}, {{\"O}berg}, {Wilner}, and
  {Rosenfeld}]{2013arXiv1302.0251Q}
{Qi},~C.; {{\"O}berg},~K.~I.; {Wilner},~D.~J.; {Rosenfeld},~K.~A.
  \emph{Astrophys. J. Lett.,} \textbf{2013}, \emph{765}, L14\relax
\mciteBstWouldAddEndPuncttrue
\mciteSetBstMidEndSepPunct{\mcitedefaultmidpunct}
{\mcitedefaultendpunct}{\mcitedefaultseppunct}\relax
\EndOfBibitem
\bibitem[{Kastner} et~al.(1997){Kastner}, {Zuckerman}, {Weintraub}, and
  {Forveille}]{Kastner_ea97}
{Kastner},~J.~H.; {Zuckerman},~B.; {Weintraub},~D.~A.; {Forveille},~T.
  \emph{Science} \textbf{1997}, \emph{277}, 67\relax
\mciteBstWouldAddEndPuncttrue
\mciteSetBstMidEndSepPunct{\mcitedefaultmidpunct}
{\mcitedefaultendpunct}{\mcitedefaultseppunct}\relax
\EndOfBibitem
\bibitem[{Fedele} et~al.(2012){Fedele}, {Bruderer}, {van Dishoeck}, {Herczeg},
  {Evans}, {Bouwman}, {Henning}, and {Green}]{2012A&A...544L...9F}
{Fedele},~D.; {Bruderer},~S.; {van Dishoeck},~E.~F.; {Herczeg},~G.~J.;
  {Evans},~N.~J.; {Bouwman},~J.; {Henning},~T.; {Green},~J.
  \emph{Astron.~Astrophys.,} \textbf{2012}, \emph{544}, L9\relax
\mciteBstWouldAddEndPuncttrue
\mciteSetBstMidEndSepPunct{\mcitedefaultmidpunct}
{\mcitedefaultendpunct}{\mcitedefaultseppunct}\relax
\EndOfBibitem
\bibitem[{Meeus} et~al.(2012){Meeus}, {Montesinos}, {Mendigut{\'{\i}}a},
  {Kamp}, {Thi}, {Eiroa}, {Grady}, {Mathews}, {Sandell}, {Martin-Za{\"\i}di},
  {Brittain}, {Dent}, {Howard}, {M{\'e}nard}, {Pinte}, {Roberge},
  {Vandenbussche}, and {Williams}]{Meeus_ea12}
{Meeus},~G. et~al.  \emph{Astron.~Astrophys.,} \textbf{2012}, \emph{544},
  A78\relax
\mciteBstWouldAddEndPuncttrue
\mciteSetBstMidEndSepPunct{\mcitedefaultmidpunct}
{\mcitedefaultendpunct}{\mcitedefaultseppunct}\relax
\EndOfBibitem
\bibitem[{Riviere-Marichalar} et~al.(2012){Riviere-Marichalar}, {M{\'e}nard},
  {Thi}, {Kamp}, {Montesinos}, {Meeus}, {Woitke}, {Howard}, {Sandell}, {Podio},
  {Dent}, {Mendigut{\'{\i}}a}, {Pinte}, {White}, and
  {Barrado}]{2012A&A...538L...3R}
{Riviere-Marichalar},~P.; {M{\'e}nard},~F.; {Thi},~W.~F.; {Kamp},~I.;
  {Montesinos},~B.; {Meeus},~G.; {Woitke},~P.; {Howard},~C.; {Sandell},~G.;
  {Podio},~L.; {Dent},~W.~R.~F.; {Mendigut{\'{\i}}a},~I.; {Pinte},~C.;
  {White},~G.~J.; {Barrado},~D. \emph{Astron.~Astrophys.,} \textbf{2012},
  \emph{538}, L3\relax
\mciteBstWouldAddEndPuncttrue
\mciteSetBstMidEndSepPunct{\mcitedefaultmidpunct}
{\mcitedefaultendpunct}{\mcitedefaultseppunct}\relax
\EndOfBibitem
\bibitem[{Fedele} et~al.(2013){Fedele}, {Bruderer}, {van Dishoeck}, {Carr},
  {Herczeg}, {Salyk}, {Evans}, {Bouwman}, {Meeus}, {Henning}, {Green},
  {Najita}, and {Guedel}]{Fedele_ea13a}
{Fedele},~D.; {Bruderer},~S.; {van Dishoeck},~E.~F.; {Carr},~J.;
  {Herczeg},~G.~J.; {Salyk},~C.; {Evans},~N.~J.,~II; {Bouwman},~J.;
  {Meeus},~G.; {Henning},~T.; {Green},~J.; {Najita},~J.~R.; {Guedel},~M.
  \emph{ArXiv:astro-ph/1308.1578,} \textbf{2013}, \relax
\mciteBstWouldAddEndPunctfalse
\mciteSetBstMidEndSepPunct{\mcitedefaultmidpunct}
{}{\mcitedefaultseppunct}\relax
\EndOfBibitem
\bibitem[{Hogerheijde} et~al.(2011){Hogerheijde}, {Bergin}, {Brinch},
  {Cleeves}, {Fogel}, {Blake}, {Dominik}, {Lis}, {Melnick}, {Neufeld},
  {Pani{\'c}}, {Pearson}, {Kristensen}, {Y{\i}ld{\i}z}, and {van
  Dishoeck}]{2011Sci...334..338H}
{Hogerheijde},~M.~R.; {Bergin},~E.~A.; {Brinch},~C.; {Cleeves},~L.~I.;
  {Fogel},~J.~K.~J.; {Blake},~G.~A.; {Dominik},~C.; {Lis},~D.~C.;
  {Melnick},~G.; {Neufeld},~D.; {Pani{\'c}},~O.; {Pearson},~J.~C.;
  {Kristensen},~L.; {Y{\i}ld{\i}z},~U.~A.; {van Dishoeck},~E.~F. \emph{Science}
  \textbf{2011}, \emph{334}, 338\relax
\mciteBstWouldAddEndPuncttrue
\mciteSetBstMidEndSepPunct{\mcitedefaultmidpunct}
{\mcitedefaultendpunct}{\mcitedefaultseppunct}\relax
\EndOfBibitem
\bibitem[{Sturm} et~al.(2010){Sturm}, {Bouwman}, {Henning}, {Evans}, {Acke},
  {Mulders}, {Waters}, {van Dishoeck}, {Meeus}, {Green}, {Augereau},
  {Olofsson}, {Salyk}, {Najita}, {Herczeg}, {van Kempen}, {Kristensen},
  {Dominik}, {Carr}, {Waelkens}, {Bergin}, {Blake}, {Brown}, {Chen}, {Cieza},
  {Dunham}, {Glassgold}, {G{\"u}del}, {Harvey}, {Hogerheijde}, {Jaffe},
  {J{\o}rgensen}, {Kim}, {Knez}, {Lacy}, {Lee}, {Maret}, {Meijerink},
  {Mer{\'{\i}}n}, {Mundy}, {Pontoppidan}, {Visser}, and
  {Y{\i}ld{\i}z}]{2010A&A...518L.129S}
{Sturm},~B. et~al.  \emph{Astron.~Astrophys.,} \textbf{2010}, \emph{518},
  L129\relax
\mciteBstWouldAddEndPuncttrue
\mciteSetBstMidEndSepPunct{\mcitedefaultmidpunct}
{\mcitedefaultendpunct}{\mcitedefaultseppunct}\relax
\EndOfBibitem
\bibitem[{Meeus} et~al.(2013){Meeus}, {Salyk}, {Bruderer}, {Fedele},
  {Maaskant}, {Evans}, {van Dishoeck}, {Montesinos}, {Herczeg}, {Bouwman},
  {Green}, {Dominik}, {Henning}, {Vicente}, and {the DIGIT team}]{Meeus_ea13a}
{Meeus},~G.; {Salyk},~C.; {Bruderer},~S.; {Fedele},~D.; {Maaskant},~K.;
  {Evans},~N.~J.,~II; {van Dishoeck},~E.~F.; {Montesinos},~B.; {Herczeg},~G.;
  {Bouwman},~J.; {Green},~J.~D.; {Dominik},~C.; {Henning},~T.; {Vicente},~S.;
  {the DIGIT team}, \emph{ArXiv:astro-ph/1308.4160,} \textbf{2013}, \relax
\mciteBstWouldAddEndPunctfalse
\mciteSetBstMidEndSepPunct{\mcitedefaultmidpunct}
{}{\mcitedefaultseppunct}\relax
\EndOfBibitem
\bibitem[{Bruderer} et~al.(2012){Bruderer}, {van Dishoeck}, {Doty}, and
  {Herczeg}]{Bruderer_ea12}
{Bruderer},~S.; {van Dishoeck},~E.~F.; {Doty},~S.~D.; {Herczeg},~G.~J.
  \emph{Astron.~Astrophys.,} \textbf{2012}, \emph{541}, A91\relax
\mciteBstWouldAddEndPuncttrue
\mciteSetBstMidEndSepPunct{\mcitedefaultmidpunct}
{\mcitedefaultendpunct}{\mcitedefaultseppunct}\relax
\EndOfBibitem
\bibitem[{Thi} et~al.(2011){Thi}, {M{\'e}nard}, {Meeus}, {Martin-Za{\"\i}di},
  {Woitke}, {Tatulli}, {Benisty}, {Kamp}, {Pascucci}, {Pinte}, {Grady},
  {Brittain}, {White}, {Howard}, {Sandell}, and {Eiroa}]{Thi_ea11a}
{Thi},~W.-F. et~al.  \emph{Astron.~Astrophys.,} \textbf{2011}, \emph{530},
  L2\relax
\mciteBstWouldAddEndPuncttrue
\mciteSetBstMidEndSepPunct{\mcitedefaultmidpunct}
{\mcitedefaultendpunct}{\mcitedefaultseppunct}\relax
\EndOfBibitem
\bibitem[{Carr} and {Najita}(2008){Carr}, and {Najita}]{Carr_Najita08}
{Carr},~J.~S.; {Najita},~J.~R. \emph{Science} \textbf{2008}, \emph{319},
  1504\relax
\mciteBstWouldAddEndPuncttrue
\mciteSetBstMidEndSepPunct{\mcitedefaultmidpunct}
{\mcitedefaultendpunct}{\mcitedefaultseppunct}\relax
\EndOfBibitem
\bibitem[{Salyk} et~al.(2008){Salyk}, {Pontoppidan}, {Blake}, {Lahuis}, {van
  Dishoeck}, and {Evans}]{Salyk_ea08}
{Salyk},~C.; {Pontoppidan},~K.~M.; {Blake},~G.~A.; {Lahuis},~F.; {van
  Dishoeck},~E.~F.; {Evans},~N.~J.,~II \emph{Astrophys.~J.,} \textbf{2008},
  \emph{676}, L49\relax
\mciteBstWouldAddEndPuncttrue
\mciteSetBstMidEndSepPunct{\mcitedefaultmidpunct}
{\mcitedefaultendpunct}{\mcitedefaultseppunct}\relax
\EndOfBibitem
\bibitem[{Pascucci} et~al.(2009){Pascucci}, {Apai}, {Luhman}, {Henning},
  {Bouwman}, {Meyer}, {Lahuis}, and {Natta}]{2009ApJ...696..143P}
{Pascucci},~I.; {Apai},~D.; {Luhman},~K.; {Henning},~T.; {Bouwman},~J.;
  {Meyer},~M.~R.; {Lahuis},~F.; {Natta},~A. \emph{Astrophys.~J.,}
  \textbf{2009}, \emph{696}, 143\relax
\mciteBstWouldAddEndPuncttrue
\mciteSetBstMidEndSepPunct{\mcitedefaultmidpunct}
{\mcitedefaultendpunct}{\mcitedefaultseppunct}\relax
\EndOfBibitem
\bibitem[{Pontoppidan} et~al.(2010){Pontoppidan}, {Salyk}, {Blake}, and
  {K{\"a}ufl}]{2010ApJ...722L.173P}
{Pontoppidan},~K.~M.; {Salyk},~C.; {Blake},~G.~A.; {K{\"a}ufl},~H.~U.
  \emph{Astrophys.~J.,} \textbf{2010}, \emph{722}, L173\relax
\mciteBstWouldAddEndPuncttrue
\mciteSetBstMidEndSepPunct{\mcitedefaultmidpunct}
{\mcitedefaultendpunct}{\mcitedefaultseppunct}\relax
\EndOfBibitem
\bibitem[{Carr} and {Najita}(2011){Carr}, and {Najita}]{2011ApJ...733..102C}
{Carr},~J.~S.; {Najita},~J.~R. \emph{Astrophys.~J.,} \textbf{2011}, \emph{733},
  102\relax
\mciteBstWouldAddEndPuncttrue
\mciteSetBstMidEndSepPunct{\mcitedefaultmidpunct}
{\mcitedefaultendpunct}{\mcitedefaultseppunct}\relax
\EndOfBibitem
\bibitem[{Salyk} et~al.(2011){Salyk}, {Pontoppidan}, {Blake}, {Najita}, and
  {Carr}]{Salyk_ea11a}
{Salyk},~C.; {Pontoppidan},~K.~M.; {Blake},~G.~A.; {Najita},~J.~R.;
  {Carr},~J.~S. \emph{Astrophys.~J.,} \textbf{2011}, \emph{731}, 130\relax
\mciteBstWouldAddEndPuncttrue
\mciteSetBstMidEndSepPunct{\mcitedefaultmidpunct}
{\mcitedefaultendpunct}{\mcitedefaultseppunct}\relax
\EndOfBibitem
\bibitem[{Mandell} et~al.(2012){Mandell}, {Bast}, {van Dishoeck}, {Blake},
  {Salyk}, {Mumma}, and {Villanueva}]{2012ApJ...747...92M}
{Mandell},~A.~M.; {Bast},~J.; {van Dishoeck},~E.~F.; {Blake},~G.~A.;
  {Salyk},~C.; {Mumma},~M.~J.; {Villanueva},~G. \emph{Astrophys.~J.,}
  \textbf{2012}, \emph{747}, 92\relax
\mciteBstWouldAddEndPuncttrue
\mciteSetBstMidEndSepPunct{\mcitedefaultmidpunct}
{\mcitedefaultendpunct}{\mcitedefaultseppunct}\relax
\EndOfBibitem
\bibitem[{Lahuis} et~al.(2006){Lahuis}, {van Dishoeck}, {Boogert},
  {Pontoppidan}, {Blake}, {Dullemond}, {Evans}, {Hogerheijde}, {J{\o}rgensen},
  {Kessler-Silacci}, and {Knez}]{Lahuis_ea06}
{Lahuis},~F.; {van Dishoeck},~E.~F.; {Boogert},~A.~C.~A.; {Pontoppidan},~K.~M.;
  {Blake},~G.~A.; {Dullemond},~C.~P.; {Evans},~N.~J.,~II; {Hogerheijde},~M.~R.;
  {J{\o}rgensen},~J.~K.; {Kessler-Silacci},~J.~E.; {Knez},~C.
  \emph{Astrophys.~J.,} \textbf{2006}, \emph{636}, L145\relax
\mciteBstWouldAddEndPuncttrue
\mciteSetBstMidEndSepPunct{\mcitedefaultmidpunct}
{\mcitedefaultendpunct}{\mcitedefaultseppunct}\relax
\EndOfBibitem
\bibitem[{Gibb} et~al.(2007){Gibb}, {Van Brunt}, {Brittain}, and
  {Rettig}]{2007ApJ...660.1572G}
{Gibb},~E.~L.; {Van Brunt},~K.~A.; {Brittain},~S.~D.; {Rettig},~T.~W.
  \emph{Astrophys.~J.,} \textbf{2007}, \emph{660}, 1572\relax
\mciteBstWouldAddEndPuncttrue
\mciteSetBstMidEndSepPunct{\mcitedefaultmidpunct}
{\mcitedefaultendpunct}{\mcitedefaultseppunct}\relax
\EndOfBibitem
\bibitem[{Pontoppidan} et~al.(2010){Pontoppidan}, {Salyk}, {Blake},
  {Meijerink}, {Carr}, and {Najita}]{2010ApJ...720..887P}
{Pontoppidan},~K.~M.; {Salyk},~C.; {Blake},~G.~A.; {Meijerink},~R.;
  {Carr},~J.~S.; {Najita},~J. \emph{Astrophys.~J.,} \textbf{2010}, \emph{720},
  887\relax
\mciteBstWouldAddEndPuncttrue
\mciteSetBstMidEndSepPunct{\mcitedefaultmidpunct}
{\mcitedefaultendpunct}{\mcitedefaultseppunct}\relax
\EndOfBibitem
\bibitem[{Brittain} et~al.(2003){Brittain}, {Rettig}, {Simon}, {Kulesa},
  {DiSanti}, and {Dello Russo}]{Bea03}
{Brittain},~S.~D.; {Rettig},~T.~W.; {Simon},~T.; {Kulesa},~C.;
  {DiSanti},~M.~A.; {Dello Russo},~N. \emph{Astrophys.~J.,} \textbf{2003},
  \emph{588}, 535\relax
\mciteBstWouldAddEndPuncttrue
\mciteSetBstMidEndSepPunct{\mcitedefaultmidpunct}
{\mcitedefaultendpunct}{\mcitedefaultseppunct}\relax
\EndOfBibitem
\bibitem[{Pontoppidan} et~al.(2008){Pontoppidan}, {Blake}, {van Dishoeck},
  {Smette}, {Ireland}, and {Brown}]{Pontoppidan_ea08}
{Pontoppidan},~K.~M.; {Blake},~G.~A.; {van Dishoeck},~E.~F.; {Smette},~A.;
  {Ireland},~M.~J.; {Brown},~J. \emph{Astrophys.~J.,} \textbf{2008},
  \emph{684}, 1323\relax
\mciteBstWouldAddEndPuncttrue
\mciteSetBstMidEndSepPunct{\mcitedefaultmidpunct}
{\mcitedefaultendpunct}{\mcitedefaultseppunct}\relax
\EndOfBibitem
\bibitem[{Najita} et~al.(2003){Najita}, {Carr}, and
  {Mathieu}]{2003ApJ...589..931N}
{Najita},~J.; {Carr},~J.~S.; {Mathieu},~R.~D. \emph{Astrophys.~J.,}
  \textbf{2003}, \emph{589}, 931\relax
\mciteBstWouldAddEndPuncttrue
\mciteSetBstMidEndSepPunct{\mcitedefaultmidpunct}
{\mcitedefaultendpunct}{\mcitedefaultseppunct}\relax
\EndOfBibitem
\bibitem[{Najita} et~al.(2007){Najita}, {Carr}, {Glassgold}, and
  {Valenti}]{2007prpl.conf..507N}
{Najita},~J.~R.; {Carr},~J.~S.; {Glassgold},~A.~E.; {Valenti},~J.~A. In
  \emph{Protostars and Planets V}; {Reipurth},~B., {Jewitt},~D., {Keil},~K.,
  Eds.; University of Arizona Press, Tucson, 2007; p 507\relax
\mciteBstWouldAddEndPuncttrue
\mciteSetBstMidEndSepPunct{\mcitedefaultmidpunct}
{\mcitedefaultendpunct}{\mcitedefaultseppunct}\relax
\EndOfBibitem
\bibitem[{Thrower} et~al.(2012){Thrower}, {J{\o}rgensen}, {Friis}, {Baouche},
  {Mennella}, {Luntz}, {Andersen}, {Hammer}, and
  {Hornek{\ae}r}]{2012ApJ...752....3T}
{Thrower},~J.~D.; {J{\o}rgensen},~B.; {Friis},~E.~E.; {Baouche},~S.;
  {Mennella},~V.; {Luntz},~A.~C.; {Andersen},~M.; {Hammer},~B.;
  {Hornek{\ae}r},~L. \emph{Astrophys.~J.,} \textbf{2012}, \emph{752}, 3\relax
\mciteBstWouldAddEndPuncttrue
\mciteSetBstMidEndSepPunct{\mcitedefaultmidpunct}
{\mcitedefaultendpunct}{\mcitedefaultseppunct}\relax
\EndOfBibitem
\bibitem[{Acke} et~al.(2010){Acke}, {Bouwman}, {Juh{\'a}sz}, {Henning}, {van
  den Ancker}, {Meeus}, {Tielens}, and {Waters}]{2010ApJ...718..558A}
{Acke},~B.; {Bouwman},~J.; {Juh{\'a}sz},~A.; {Henning},~T.; {van den
  Ancker},~M.~E.; {Meeus},~G.; {Tielens},~A.~G.~G.~M.; {Waters},~L.~B.~F.~M.
  \emph{Astrophys.~J.,} \textbf{2010}, \emph{718}, 558\relax
\mciteBstWouldAddEndPuncttrue
\mciteSetBstMidEndSepPunct{\mcitedefaultmidpunct}
{\mcitedefaultendpunct}{\mcitedefaultseppunct}\relax
\EndOfBibitem
\bibitem[{Geers} et~al.(2007){Geers}, {van Dishoeck}, {Visser}, {Pontoppidan},
  {Augereau}, {Habart}, and {Lagrange}]{2007A&A...476..279G}
{Geers},~V.~C.; {van Dishoeck},~E.~F.; {Visser},~R.; {Pontoppidan},~K.~M.;
  {Augereau},~J.-C.; {Habart},~E.; {Lagrange},~A.~M. \emph{Astron.~Astrophys.,}
  \textbf{2007}, \emph{476}, 279\relax
\mciteBstWouldAddEndPuncttrue
\mciteSetBstMidEndSepPunct{\mcitedefaultmidpunct}
{\mcitedefaultendpunct}{\mcitedefaultseppunct}\relax
\EndOfBibitem
\bibitem[{Siebenmorgen} and {Kr{\"u}gel}(2010){Siebenmorgen}, and
  {Kr{\"u}gel}]{2010A&A...511A...6S}
{Siebenmorgen},~R.; {Kr{\"u}gel},~E. \emph{Astron.~Astrophys.,} \textbf{2010},
  \emph{511}, A6\relax
\mciteBstWouldAddEndPuncttrue
\mciteSetBstMidEndSepPunct{\mcitedefaultmidpunct}
{\mcitedefaultendpunct}{\mcitedefaultseppunct}\relax
\EndOfBibitem
\bibitem[{Tin{\'e}} et~al.(2000){Tin{\'e}}, {Roueff}, {Falgarone}, {Gerin}, and
  {Pineau des For{\^e}ts}]{2000A&A...356.1039T}
{Tin{\'e}},~S.; {Roueff},~E.; {Falgarone},~E.; {Gerin},~M.; {Pineau des
  For{\^e}ts},~G. \emph{Astron.~Astrophys.,} \textbf{2000}, \emph{356},
  1039\relax
\mciteBstWouldAddEndPuncttrue
\mciteSetBstMidEndSepPunct{\mcitedefaultmidpunct}
{\mcitedefaultendpunct}{\mcitedefaultseppunct}\relax
\EndOfBibitem
\bibitem[{Roberts} et~al.(2002){Roberts}, {Fuller}, {Millar}, {Hatchell}, and
  {Buckle}]{2002A&A...381.1026R}
{Roberts},~H.; {Fuller},~G.~A.; {Millar},~T.~J.; {Hatchell},~J.;
  {Buckle},~J.~V. \emph{Astron.~Astrophys.,} \textbf{2002}, \emph{381},
  1026\relax
\mciteBstWouldAddEndPuncttrue
\mciteSetBstMidEndSepPunct{\mcitedefaultmidpunct}
{\mcitedefaultendpunct}{\mcitedefaultseppunct}\relax
\EndOfBibitem
\bibitem[{Guilloteau} et~al.(2006){Guilloteau}, {Pi{\'e}tu}, {Dutrey}, and
  {Gu{\'e}lin}]{2006A&A...448L...5G}
{Guilloteau},~S.; {Pi{\'e}tu},~V.; {Dutrey},~A.; {Gu{\'e}lin},~M.
  \emph{Astron.~Astrophys.,} \textbf{2006}, \emph{448}, L5\relax
\mciteBstWouldAddEndPuncttrue
\mciteSetBstMidEndSepPunct{\mcitedefaultmidpunct}
{\mcitedefaultendpunct}{\mcitedefaultseppunct}\relax
\EndOfBibitem
\bibitem[{Caselli} et~al.(2008){Caselli}, {Vastel}, {Ceccarelli}, {van der
  Tak}, {Crapsi}, and {Bacmann}]{2008A&A...492..703C}
{Caselli},~P.; {Vastel},~C.; {Ceccarelli},~C.; {van der Tak},~F.~F.~S.;
  {Crapsi},~A.; {Bacmann},~A. \emph{Astron.~Astrophys.,} \textbf{2008},
  \emph{492}, 703\relax
\mciteBstWouldAddEndPuncttrue
\mciteSetBstMidEndSepPunct{\mcitedefaultmidpunct}
{\mcitedefaultendpunct}{\mcitedefaultseppunct}\relax
\EndOfBibitem
\bibitem[{Chapillon} et~al.(2011){Chapillon}, {Parise}, {Guilloteau}, and
  {Du}]{Chapillon_ea11}
{Chapillon},~E.; {Parise},~B.; {Guilloteau},~S.; {Du},~F.
  \emph{Astron.~Astrophys.,} \textbf{2011}, \emph{533}, 143\relax
\mciteBstWouldAddEndPuncttrue
\mciteSetBstMidEndSepPunct{\mcitedefaultmidpunct}
{\mcitedefaultendpunct}{\mcitedefaultseppunct}\relax
\EndOfBibitem
\bibitem[{{\"O}berg} et~al.(2012){{\"O}berg}, {Qi}, {Wilner}, and
  {Hogerheijde}]{Oeberg_ea12a}
{{\"O}berg},~K.~I.; {Qi},~C.; {Wilner},~D.~J.; {Hogerheijde},~M.~R.
  \emph{Astrophys.~J.,} \textbf{2012}, \emph{749}, 162\relax
\mciteBstWouldAddEndPuncttrue
\mciteSetBstMidEndSepPunct{\mcitedefaultmidpunct}
{\mcitedefaultendpunct}{\mcitedefaultseppunct}\relax
\EndOfBibitem
\bibitem[{Natta} et~al.(2007){Natta}, {Testi}, {Calvet}, {Henning}, {Waters},
  and {Wilner}]{Natta_ea07}
{Natta},~A.; {Testi},~L.; {Calvet},~N.; {Henning},~T.; {Waters},~R.;
  {Wilner},~D. In \emph{Protostars and Planets V}; {Reipurth},~B.,
  {Jewitt},~D., {Keil},~K., Eds.; University of Arizona Press, Tucson, 2007; p
  767\relax
\mciteBstWouldAddEndPuncttrue
\mciteSetBstMidEndSepPunct{\mcitedefaultmidpunct}
{\mcitedefaultendpunct}{\mcitedefaultseppunct}\relax
\EndOfBibitem
\bibitem[{Juh{\'a}sz} et~al.(2010){Juh{\'a}sz}, {Bouwman}, {Henning}, {Acke},
  {van den Ancker}, {Meeus}, {Dominik}, {Min}, {Tielens}, and
  {Waters}]{Juhasz_ea10a}
{Juh{\'a}sz},~A.; {Bouwman},~J.; {Henning},~T.; {Acke},~B.; {van den
  Ancker},~M.~E.; {Meeus},~G.; {Dominik},~C.; {Min},~M.;
  {Tielens},~A.~G.~G.~M.; {Waters},~L.~B.~F.~M. \emph{Astrophys.~J.,}
  \textbf{2010}, \emph{721}, 431\relax
\mciteBstWouldAddEndPuncttrue
\mciteSetBstMidEndSepPunct{\mcitedefaultmidpunct}
{\mcitedefaultendpunct}{\mcitedefaultseppunct}\relax
\EndOfBibitem
\bibitem[{van Boekel} et~al.(2005){van Boekel}, {Min}, {Waters}, {de Koter},
  {Dominik}, {van den Ancker}, and {Bouwman}]{2005A&A...437..189V}
{van Boekel},~R.; {Min},~M.; {Waters},~L.~B.~F.~M.; {de Koter},~A.;
  {Dominik},~C.; {van den Ancker},~M.~E.; {Bouwman},~J.
  \emph{Astron.~Astrophys.,} \textbf{2005}, \emph{437}, 189\relax
\mciteBstWouldAddEndPuncttrue
\mciteSetBstMidEndSepPunct{\mcitedefaultmidpunct}
{\mcitedefaultendpunct}{\mcitedefaultseppunct}\relax
\EndOfBibitem
\bibitem[{Bouwman} et~al.(2008){Bouwman}, {Henning}, {Hillenbrand}, {Meyer},
  {Pascucci}, {Carpenter}, {Hines}, {Kim}, {Silverstone}, {Hollenbach}, and
  {Wolf}]{Bouwman_ea08}
{Bouwman},~J.; {Henning},~T.; {Hillenbrand},~L.~A.; {Meyer},~M.~R.;
  {Pascucci},~I.; {Carpenter},~J.; {Hines},~D.; {Kim},~J.~S.;
  {Silverstone},~M.~D.; {Hollenbach},~D.; {Wolf},~S. \emph{Astrophys.~J.,}
  \textbf{2008}, \emph{683}, 479\relax
\mciteBstWouldAddEndPuncttrue
\mciteSetBstMidEndSepPunct{\mcitedefaultmidpunct}
{\mcitedefaultendpunct}{\mcitedefaultseppunct}\relax
\EndOfBibitem
\bibitem[{Meeus} et~al.(2009){Meeus}, {Juh{\'a}sz}, {Henning}, {Bouwman},
  {Chen}, {Lawson}, {Apai}, {Pascucci}, and
  {Sicilia-Aguilar}]{2009A&A...497..379M}
{Meeus},~G.; {Juh{\'a}sz},~A.; {Henning},~T.; {Bouwman},~J.; {Chen},~C.;
  {Lawson},~W.; {Apai},~D.; {Pascucci},~I.; {Sicilia-Aguilar},~A.
  \emph{Astron.~Astrophys.,} \textbf{2009}, \emph{497}, 379\relax
\mciteBstWouldAddEndPuncttrue
\mciteSetBstMidEndSepPunct{\mcitedefaultmidpunct}
{\mcitedefaultendpunct}{\mcitedefaultseppunct}\relax
\EndOfBibitem
\bibitem[{Testi} et~al.(2003){Testi}, {Natta}, {Shepherd}, and
  {Wilner}]{Testi_ea03}
{Testi},~L.; {Natta},~A.; {Shepherd},~D.~S.; {Wilner},~D.~J.
  \emph{Astron.~Astrophys.,} \textbf{2003}, \emph{403}, 323\relax
\mciteBstWouldAddEndPuncttrue
\mciteSetBstMidEndSepPunct{\mcitedefaultmidpunct}
{\mcitedefaultendpunct}{\mcitedefaultseppunct}\relax
\EndOfBibitem
\bibitem[{Wilner} et~al.(2005){Wilner}, {D'Alessio}, {Calvet}, {Claussen}, and
  {Hartmann}]{Wilner_ea05}
{Wilner},~D.~J.; {D'Alessio},~P.; {Calvet},~N.; {Claussen},~M.~J.;
  {Hartmann},~L. \emph{Astrophys. J. Lett.,} \textbf{2005}, \emph{626},
  L109\relax
\mciteBstWouldAddEndPuncttrue
\mciteSetBstMidEndSepPunct{\mcitedefaultmidpunct}
{\mcitedefaultendpunct}{\mcitedefaultseppunct}\relax
\EndOfBibitem
\bibitem[{Rodmann} et~al.(2006){Rodmann}, {Henning}, {Chandler}, {Mundy}, and
  {Wilner}]{Rodmann_ea06}
{Rodmann},~J.; {Henning},~T.; {Chandler},~C.~J.; {Mundy},~L.~G.;
  {Wilner},~D.~J. \emph{Astron.~Astrophys.,} \textbf{2006}, \emph{446},
  211\relax
\mciteBstWouldAddEndPuncttrue
\mciteSetBstMidEndSepPunct{\mcitedefaultmidpunct}
{\mcitedefaultendpunct}{\mcitedefaultseppunct}\relax
\EndOfBibitem
\bibitem[{P{\'e}rez} et~al.(2012){P{\'e}rez}, {Carpenter}, {Chandler},
  {Isella}, {Andrews}, {Ricci}, {Calvet}, {Corder}, {Deller}, {Dullemond},
  {Greaves}, {Harris}, {Henning}, {Kwon}, {Lazio}, {Linz}, {Mundy}, {Sargent},
  {Storm}, {Testi}, and {Wilner}]{Perez_ea12}
{P{\'e}rez},~L.~M. et~al.  \emph{Astrophys.~J.~Lett.,} \textbf{2012},
  \emph{760}, L17\relax
\mciteBstWouldAddEndPuncttrue
\mciteSetBstMidEndSepPunct{\mcitedefaultmidpunct}
{\mcitedefaultendpunct}{\mcitedefaultseppunct}\relax
\EndOfBibitem
\bibitem[{Sturm} et~al.(2013){Sturm}, {Bouwman}, {Henning}, {Evans}, {Waters},
  {van Dishoeck}, {Green}, {Olofsson}, {Meeus}, {Maaskant}, {Dominik},
  {Augereau}, {Mulders}, {Acke}, {Merin}, and {Herczeg}]{Sturm_ea13a}
{Sturm},~B. et~al.  \emph{Astron.~Astrophys.,} \textbf{2013}, \emph{553},
  A5\relax
\mciteBstWouldAddEndPuncttrue
\mciteSetBstMidEndSepPunct{\mcitedefaultmidpunct}
{\mcitedefaultendpunct}{\mcitedefaultseppunct}\relax
\EndOfBibitem
\bibitem[{Gail}(2004)]{Gail_04}
{Gail},~H.-P. \emph{Astron.~Astrophys.,} \textbf{2004}, \emph{413}, 571\relax
\mciteBstWouldAddEndPuncttrue
\mciteSetBstMidEndSepPunct{\mcitedefaultmidpunct}
{\mcitedefaultendpunct}{\mcitedefaultseppunct}\relax
\EndOfBibitem
\bibitem[{Harker} and {Desch}(2002){Harker}, and {Desch}]{HD_02}
{Harker},~D.~E.; {Desch},~S.~J. \emph{Astrophys.~J.,} \textbf{2002},
  \emph{565}, L109\relax
\mciteBstWouldAddEndPuncttrue
\mciteSetBstMidEndSepPunct{\mcitedefaultmidpunct}
{\mcitedefaultendpunct}{\mcitedefaultseppunct}\relax
\EndOfBibitem
\bibitem[{{\'A}brah{\'a}m} et~al.(2009){{\'A}brah{\'a}m}, {Juh{\'a}sz},
  {Dullemond}, {K{\'o}sp{\'a}l}, {van Boekel}, {Bouwman}, {Henning},
  {Mo{\'o}r}, {Mosoni}, {Sicilia-Aguilar}, and {Sipos}]{Abraham_ea09}
{{\'A}brah{\'a}m},~P.; {Juh{\'a}sz},~A.; {Dullemond},~C.~P.;
  {K{\'o}sp{\'a}l},~{\'A}.; {van Boekel},~R.; {Bouwman},~J.; {Henning},~T.;
  {Mo{\'o}r},~A.; {Mosoni},~L.; {Sicilia-Aguilar},~A.; {Sipos},~N.
  \emph{Nature} \textbf{2009}, \emph{459}, 224\relax
\mciteBstWouldAddEndPuncttrue
\mciteSetBstMidEndSepPunct{\mcitedefaultmidpunct}
{\mcitedefaultendpunct}{\mcitedefaultseppunct}\relax
\EndOfBibitem
\bibitem[{Juh{\'a}sz} et~al.(2012){Juh{\'a}sz}, {Dullemond}, {van Boekel},
  {Bouwman}, {{\'A}brah{\'a}m}, {Acosta-Pulido}, {Henning}, {K{\'o}sp{\'a}l},
  {Sicilia-Aguilar}, {Jones}, {Mo{\'o}r}, {Mosoni}, {Reg{\'a}ly}, {Szokoly},
  and {Sipos}]{2012ApJ...744..118J}
{Juh{\'a}sz},~A.; {Dullemond},~C.~P.; {van Boekel},~R.; {Bouwman},~J.;
  {{\'A}brah{\'a}m},~P.; {Acosta-Pulido},~J.~A.; {Henning},~T.;
  {K{\'o}sp{\'a}l},~A.; {Sicilia-Aguilar},~A.; {Jones},~A.; {Mo{\'o}r},~A.;
  {Mosoni},~L.; {Reg{\'a}ly},~Z.; {Szokoly},~G.; {Sipos},~N.
  \emph{Astrophys.~J.,} \textbf{2012}, \emph{744}, 118\relax
\mciteBstWouldAddEndPuncttrue
\mciteSetBstMidEndSepPunct{\mcitedefaultmidpunct}
{\mcitedefaultendpunct}{\mcitedefaultseppunct}\relax
\EndOfBibitem
\bibitem[{Pollack} et~al.(1994){Pollack}, {Hollenbach}, {Beckwith},
  {Simonelli}, {Roush}, and {Fong}]{Pea94}
{Pollack},~J.~B.; {Hollenbach},~D.; {Beckwith},~S.; {Simonelli},~D.~P.;
  {Roush},~T.; {Fong},~W. \emph{Astrophys.~J.,} \textbf{1994}, \emph{421},
  615\relax
\mciteBstWouldAddEndPuncttrue
\mciteSetBstMidEndSepPunct{\mcitedefaultmidpunct}
{\mcitedefaultendpunct}{\mcitedefaultseppunct}\relax
\EndOfBibitem
\bibitem[{Semenov} et~al.(2003){Semenov}, {Henning}, {Helling}, {Ilgner}, and
  {Sedlmayr}]{RP_opacities}
{Semenov},~D.; {Henning},~T.; {Helling},~C.; {Ilgner},~M.; {Sedlmayr},~E.
  \emph{Astron.~Astrophys.,} \textbf{2003}, \emph{410}, 611\relax
\mciteBstWouldAddEndPuncttrue
\mciteSetBstMidEndSepPunct{\mcitedefaultmidpunct}
{\mcitedefaultendpunct}{\mcitedefaultseppunct}\relax
\EndOfBibitem
\bibitem[{Acke} and {van den Ancker}(2006){Acke}, and {van den
  Ancker}]{2006A&A...457..171A}
{Acke},~B.; {van den Ancker},~M.~E. \emph{Astron.~Astrophys.,} \textbf{2006},
  \emph{457}, 171\relax
\mciteBstWouldAddEndPuncttrue
\mciteSetBstMidEndSepPunct{\mcitedefaultmidpunct}
{\mcitedefaultendpunct}{\mcitedefaultseppunct}\relax
\EndOfBibitem
\bibitem[{Guillois} et~al.(1999){Guillois}, {Ledoux}, and
  {Reynaud}]{1999ApJ...521L.133G}
{Guillois},~O.; {Ledoux},~G.; {Reynaud},~C. \emph{Astrophys.~J.,}
  \textbf{1999}, \emph{521}, L133\relax
\mciteBstWouldAddEndPuncttrue
\mciteSetBstMidEndSepPunct{\mcitedefaultmidpunct}
{\mcitedefaultendpunct}{\mcitedefaultseppunct}\relax
\EndOfBibitem
\bibitem[{Terada} et~al.(2007){Terada}, {Tokunaga}, {Kobayashi}, {Takato},
  {Hayano}, and {Takami}]{Terada_ea07a}
{Terada},~H.; {Tokunaga},~A.~T.; {Kobayashi},~N.; {Takato},~N.; {Hayano},~Y.;
  {Takami},~H. \emph{Astrophys.~J.,} \textbf{2007}, \emph{667}, 303\relax
\mciteBstWouldAddEndPuncttrue
\mciteSetBstMidEndSepPunct{\mcitedefaultmidpunct}
{\mcitedefaultendpunct}{\mcitedefaultseppunct}\relax
\EndOfBibitem
\bibitem[{Aikawa} et~al.(2012){Aikawa}, {Kamuro}, {Sakon}, {Itoh}, {Terada},
  {Noble}, {Pontoppidan}, {Fraser}, {Tamura}, {Kandori}, {Kawamura}, and
  {Ueno}]{2012A&A...538A..57A}
{Aikawa},~Y.; {Kamuro},~D.; {Sakon},~I.; {Itoh},~Y.; {Terada},~H.;
  {Noble},~J.~A.; {Pontoppidan},~K.~M.; {Fraser},~H.~J.; {Tamura},~M.;
  {Kandori},~R.; {Kawamura},~A.; {Ueno},~M. \emph{Astron.~Astrophys.,}
  \textbf{2012}, \emph{538}, A57\relax
\mciteBstWouldAddEndPuncttrue
\mciteSetBstMidEndSepPunct{\mcitedefaultmidpunct}
{\mcitedefaultendpunct}{\mcitedefaultseppunct}\relax
\EndOfBibitem
\bibitem[{Thi} et~al.(2002){Thi}, {Pontoppidan}, {van Dishoeck}, {Dartois}, and
  {d'Hendecourt}]{2002A&A...394L..27T}
{Thi},~W.~F.; {Pontoppidan},~K.~M.; {van Dishoeck},~E.~F.; {Dartois},~E.;
  {d'Hendecourt},~L. \emph{Astron.~Astrophys.,} \textbf{2002}, \emph{394},
  L27\relax
\mciteBstWouldAddEndPuncttrue
\mciteSetBstMidEndSepPunct{\mcitedefaultmidpunct}
{\mcitedefaultendpunct}{\mcitedefaultseppunct}\relax
\EndOfBibitem
\bibitem[{Pontoppidan} et~al.(2005){Pontoppidan}, {Dullemond}, {van Dishoeck},
  {Blake}, {Boogert}, {Evans}, {Kessler-Silacci}, and
  {Lahuis}]{Pontoppidan_ea05}
{Pontoppidan},~K.~M.; {Dullemond},~C.~P.; {van Dishoeck},~E.~F.;
  {Blake},~G.~A.; {Boogert},~A.~C.~A.; {Evans},~N.~J.,~II;
  {Kessler-Silacci},~J.~E.; {Lahuis},~F. \emph{Astrophys.~J.,} \textbf{2005},
  \emph{622}, 463\relax
\mciteBstWouldAddEndPuncttrue
\mciteSetBstMidEndSepPunct{\mcitedefaultmidpunct}
{\mcitedefaultendpunct}{\mcitedefaultseppunct}\relax
\EndOfBibitem
\bibitem[{Malfait} et~al.(1999){Malfait}, {Waelkens}, {Bouwman}, {de Koter},
  and {Waters}]{1999A&A...345..181M}
{Malfait},~K.; {Waelkens},~C.; {Bouwman},~J.; {de Koter},~A.;
  {Waters},~L.~B.~F.~M. \emph{Astron.~Astrophys.,} \textbf{1999}, \emph{345},
  181\relax
\mciteBstWouldAddEndPuncttrue
\mciteSetBstMidEndSepPunct{\mcitedefaultmidpunct}
{\mcitedefaultendpunct}{\mcitedefaultseppunct}\relax
\EndOfBibitem
\bibitem[{McClure} et~al.(2012){McClure}, {Manoj}, {Calvet}, {Adame},
  {Espaillat}, {Watson}, {Sargent}, {Forrest}, and {D'Alessio}]{McClure_ea12}
{McClure},~M.~K.; {Manoj},~P.; {Calvet},~N.; {Adame},~L.; {Espaillat},~C.;
  {Watson},~D.~M.; {Sargent},~B.; {Forrest},~W.~J.; {D'Alessio},~P.
  \emph{Astrophys. J. Lett.,} \textbf{2012}, \emph{759}, 10\relax
\mciteBstWouldAddEndPuncttrue
\mciteSetBstMidEndSepPunct{\mcitedefaultmidpunct}
{\mcitedefaultendpunct}{\mcitedefaultseppunct}\relax
\EndOfBibitem
\bibitem[{Harada} et~al.(2010){Harada}, {Herbst}, and {Wakelam}]{Harada_ea10}
{Harada},~N.; {Herbst},~E.; {Wakelam},~V. \emph{Astrophys.~J.,} \textbf{2010},
  \emph{721}, 1570\relax
\mciteBstWouldAddEndPuncttrue
\mciteSetBstMidEndSepPunct{\mcitedefaultmidpunct}
{\mcitedefaultendpunct}{\mcitedefaultseppunct}\relax
\EndOfBibitem
\bibitem[{Aikawa} et~al.(1999){Aikawa}, {Umebayashi}, {Nakano}, and
  {Miyama}]{Aea99}
{Aikawa},~Y.; {Umebayashi},~T.; {Nakano},~T.; {Miyama},~S.~M.
  \emph{Astrophys.~J.,} \textbf{1999}, \emph{519}, 705\relax
\mciteBstWouldAddEndPuncttrue
\mciteSetBstMidEndSepPunct{\mcitedefaultmidpunct}
{\mcitedefaultendpunct}{\mcitedefaultseppunct}\relax
\EndOfBibitem
\bibitem[{Willacy} et~al.(1998){Willacy}, {Klahr}, {Millar}, and
  {Henning}]{WKMH98}
{Willacy},~K.; {Klahr},~H.~H.; {Millar},~T.~J.; {Henning},~T.
  \emph{Astron.~Astrophys.,} \textbf{1998}, \emph{338}, 995\relax
\mciteBstWouldAddEndPuncttrue
\mciteSetBstMidEndSepPunct{\mcitedefaultmidpunct}
{\mcitedefaultendpunct}{\mcitedefaultseppunct}\relax
\EndOfBibitem
\bibitem[{Markwick} et~al.(2002){Markwick}, {Ilgner}, {Millar}, and
  {Henning}]{Mea02}
{Markwick},~A.~J.; {Ilgner},~M.; {Millar},~T.~J.; {Henning},~T.
  \emph{Astron.~Astrophys.,} \textbf{2002}, \emph{385}, 632\relax
\mciteBstWouldAddEndPuncttrue
\mciteSetBstMidEndSepPunct{\mcitedefaultmidpunct}
{\mcitedefaultendpunct}{\mcitedefaultseppunct}\relax
\EndOfBibitem
\bibitem[{Woods} and {Willacy}(2007){Woods}, and {Willacy}]{Woods_Willacy07}
{Woods},~P.~M.; {Willacy},~K. \emph{Astrophys.~J.,} \textbf{2007}, \emph{655},
  L49\relax
\mciteBstWouldAddEndPuncttrue
\mciteSetBstMidEndSepPunct{\mcitedefaultmidpunct}
{\mcitedefaultendpunct}{\mcitedefaultseppunct}\relax
\EndOfBibitem
\bibitem[{Bethell} and {Bergin}(2009){Bethell}, and {Bergin}]{Bethell_Bergin09}
{Bethell},~T.; {Bergin},~E. \emph{Science} \textbf{2009}, \emph{326},
  1675\relax
\mciteBstWouldAddEndPuncttrue
\mciteSetBstMidEndSepPunct{\mcitedefaultmidpunct}
{\mcitedefaultendpunct}{\mcitedefaultseppunct}\relax
\EndOfBibitem
\bibitem[{Najita} et~al.(2011){Najita}, {{\'A}d{\'a}mkovics}, and
  {Glassgold}]{2011ApJ...743..147N}
{Najita},~J.~R.; {{\'A}d{\'a}mkovics},~M.; {Glassgold},~A.~E.
  \emph{Astrophys.~J.,} \textbf{2011}, \emph{743}, 147\relax
\mciteBstWouldAddEndPuncttrue
\mciteSetBstMidEndSepPunct{\mcitedefaultmidpunct}
{\mcitedefaultendpunct}{\mcitedefaultseppunct}\relax
\EndOfBibitem
\bibitem[{Fedele} et~al.(2011){Fedele}, {Pascucci}, {Brittain}, {Kamp},
  {Woitke}, {Williams}, {Dent}, and {Thi}]{2011ApJ...732..106F}
{Fedele},~D.; {Pascucci},~I.; {Brittain},~S.; {Kamp},~I.; {Woitke},~P.;
  {Williams},~J.~P.; {Dent},~W.~R.~F.; {Thi},~W.-F. \emph{Astrophys.~J.,}
  \textbf{2011}, \emph{732}, 106\relax
\mciteBstWouldAddEndPuncttrue
\mciteSetBstMidEndSepPunct{\mcitedefaultmidpunct}
{\mcitedefaultendpunct}{\mcitedefaultseppunct}\relax
\EndOfBibitem
\bibitem[{Aikawa} and {Herbst}(1999){Aikawa}, and {Herbst}]{ah1999}
{Aikawa},~Y.; {Herbst},~E. \emph{Astron.~Astrophys.,} \textbf{1999},
  \emph{351}, 233\relax
\mciteBstWouldAddEndPuncttrue
\mciteSetBstMidEndSepPunct{\mcitedefaultmidpunct}
{\mcitedefaultendpunct}{\mcitedefaultseppunct}\relax
\EndOfBibitem
\bibitem[{Aikawa} et~al.(2002){Aikawa}, {van Zadelhoff}, {van Dishoeck}, and
  {Herbst}]{Aea02}
{Aikawa},~Y.; {van Zadelhoff},~G.~J.; {van Dishoeck},~E.~F.; {Herbst},~E.
  \emph{Astron.~Astrophys.,} \textbf{2002}, \emph{386}, 622\relax
\mciteBstWouldAddEndPuncttrue
\mciteSetBstMidEndSepPunct{\mcitedefaultmidpunct}
{\mcitedefaultendpunct}{\mcitedefaultseppunct}\relax
\EndOfBibitem
\bibitem[{Fogel} et~al.(2011){Fogel}, {Bethell}, {Bergin}, {Calvet}, and
  {Semenov}]{Fogel_ea11}
{Fogel},~J.~K.~J.; {Bethell},~T.~J.; {Bergin},~E.~A.; {Calvet},~N.;
  {Semenov},~D. \emph{Astrophys.~J.,} \textbf{2011}, \emph{726}, 29\relax
\mciteBstWouldAddEndPuncttrue
\mciteSetBstMidEndSepPunct{\mcitedefaultmidpunct}
{\mcitedefaultendpunct}{\mcitedefaultseppunct}\relax
\EndOfBibitem
\bibitem[{Herbst} and {Klemperer}(1973){Herbst}, and
  {Klemperer}]{HerbstKlemperer73}
{Herbst},~E.; {Klemperer},~W. \emph{Astrophys.~J.,} \textbf{1973}, \emph{185},
  505\relax
\mciteBstWouldAddEndPuncttrue
\mciteSetBstMidEndSepPunct{\mcitedefaultmidpunct}
{\mcitedefaultendpunct}{\mcitedefaultseppunct}\relax
\EndOfBibitem
\bibitem[{Watson}(1974)]{Watson_74a}
{Watson},~W.~D. \emph{Astrophys.~J.,} \textbf{1974}, \emph{188}, 35\relax
\mciteBstWouldAddEndPuncttrue
\mciteSetBstMidEndSepPunct{\mcitedefaultmidpunct}
{\mcitedefaultendpunct}{\mcitedefaultseppunct}\relax
\EndOfBibitem
\bibitem[{Woodall} et~al.(2007){Woodall}, {Ag{\'u}ndez}, {Markwick-Kemper}, and
  {Millar}]{Woodall_ea07}
{Woodall},~J.; {Ag{\'u}ndez},~M.; {Markwick-Kemper},~A.~J.; {Millar},~T.~J.
  \emph{Astron.~Astrophys.,} \textbf{2007}, \emph{466}, 1197\relax
\mciteBstWouldAddEndPuncttrue
\mciteSetBstMidEndSepPunct{\mcitedefaultmidpunct}
{\mcitedefaultendpunct}{\mcitedefaultseppunct}\relax
\EndOfBibitem
\bibitem[{Wakelam} et~al.(2012){Wakelam}, {Herbst}, {Loison}, {Smith},
  {Chandrasekaran}, {Pavone}, {Adams}, {Bacchus-Montabonel}, {Bergeat},
  {B{\'e}roff}, {Bierbaum}, {Chabot}, {Dalgarno}, {van Dishoeck}, {Faure},
  {Geppert}, {Gerlich}, {Galli}, {H{\'e}brard}, {Hersant}, {Hickson},
  {Honvault}, {Klippenstein}, {Le Picard}, {Nyman}, {Pernot}, {Schlemmer},
  {Selsis}, {Sims}, {Talbi}, {Tennyson}, {Troe}, {Wester}, and
  {Wiesenfeld}]{KIDA}
{Wakelam},~V. et~al.  \emph{Astrophys.~J.,~Suppl.~Ser.,} \textbf{2012},
  \emph{199}, 21\relax
\mciteBstWouldAddEndPuncttrue
\mciteSetBstMidEndSepPunct{\mcitedefaultmidpunct}
{\mcitedefaultendpunct}{\mcitedefaultseppunct}\relax
\EndOfBibitem
\bibitem[{Morbidelli} et~al.(2000){Morbidelli}, {Chambers}, {Lunine}, {Petit},
  {Robert}, {Valsecchi}, and {Cyr}]{2000M&PS...35.1309M}
{Morbidelli},~A.; {Chambers},~J.; {Lunine},~J.~I.; {Petit},~J.~M.;
  {Robert},~F.; {Valsecchi},~G.~B.; {Cyr},~K.~E. \emph{Meteorit. Planet. Sci.,}
  \textbf{2000}, \emph{35}, 1309\relax
\mciteBstWouldAddEndPuncttrue
\mciteSetBstMidEndSepPunct{\mcitedefaultmidpunct}
{\mcitedefaultendpunct}{\mcitedefaultseppunct}\relax
\EndOfBibitem
\bibitem[{Lecar} et~al.(2006){Lecar}, {Podolak}, {Sasselov}, and
  {Chiang}]{Lecar_ea06}
{Lecar},~M.; {Podolak},~M.; {Sasselov},~D.; {Chiang},~E. \emph{Astrophys.~J.,}
  \textbf{2006}, \emph{640}, 1115\relax
\mciteBstWouldAddEndPuncttrue
\mciteSetBstMidEndSepPunct{\mcitedefaultmidpunct}
{\mcitedefaultendpunct}{\mcitedefaultseppunct}\relax
\EndOfBibitem
\bibitem[{Encrenaz}(2008)]{Encrenaz_08}
{Encrenaz},~T. \emph{Ann.~Rev.~Astron.~Astrophys.,} \textbf{2008}, \emph{46},
  57\relax
\mciteBstWouldAddEndPuncttrue
\mciteSetBstMidEndSepPunct{\mcitedefaultmidpunct}
{\mcitedefaultendpunct}{\mcitedefaultseppunct}\relax
\EndOfBibitem
\bibitem[{Min} et~al.(2011){Min}, {Dullemond}, {Kama}, and {Dominik}]{Min_ea11}
{Min},~M.; {Dullemond},~C.~P.; {Kama},~M.; {Dominik},~C. \emph{Icarus}
  \textbf{2011}, \emph{212}, 416\relax
\mciteBstWouldAddEndPuncttrue
\mciteSetBstMidEndSepPunct{\mcitedefaultmidpunct}
{\mcitedefaultendpunct}{\mcitedefaultseppunct}\relax
\EndOfBibitem
\bibitem[{Zhang} et~al.(2013){Zhang}, {Pontoppidan}, {Salyk}, and
  {Blake}]{Zhang_ea13a}
{Zhang},~K.; {Pontoppidan},~K.~M.; {Salyk},~C.; {Blake},~G.~A.
  \emph{Astrophys.~J.,} \textbf{2013}, \emph{766}, 82\relax
\mciteBstWouldAddEndPuncttrue
\mciteSetBstMidEndSepPunct{\mcitedefaultmidpunct}
{\mcitedefaultendpunct}{\mcitedefaultseppunct}\relax
\EndOfBibitem
\bibitem[{Herczeg} et~al.(2002){Herczeg}, {Linsky}, {Valenti}, {Johns-Krull},
  and {Wood}]{Herczeg_ea02}
{Herczeg},~G.~J.; {Linsky},~J.~L.; {Valenti},~J.~A.; {Johns-Krull},~C.~M.;
  {Wood},~B.~E. \emph{Astrophys.~J.,} \textbf{2002}, \emph{572}, 310\relax
\mciteBstWouldAddEndPuncttrue
\mciteSetBstMidEndSepPunct{\mcitedefaultmidpunct}
{\mcitedefaultendpunct}{\mcitedefaultseppunct}\relax
\EndOfBibitem
\bibitem[{Schindhelm} et~al.(2012){Schindhelm}, {France}, {Herczeg}, {Bergin},
  {Yang}, {Brown}, {Brown}, {Linsky}, and {Valenti}]{Schindhelm_ea12}
{Schindhelm},~E.; {France},~K.; {Herczeg},~G.~J.; {Bergin},~E.; {Yang},~H.;
  {Brown},~A.; {Brown},~J.~M.; {Linsky},~J.~L.; {Valenti},~J.
  \emph{Astrophys.~J.~Lett.,} \textbf{2012}, \emph{756}, L23\relax
\mciteBstWouldAddEndPuncttrue
\mciteSetBstMidEndSepPunct{\mcitedefaultmidpunct}
{\mcitedefaultendpunct}{\mcitedefaultseppunct}\relax
\EndOfBibitem
\bibitem[{Semenov} et~al.(2005){Semenov}, {Pavlyuchenkov}, {Schreyer},
  {Henning}, {Dullemond}, and {Bacmann}]{Semenov_ea05}
{Semenov},~D.; {Pavlyuchenkov},~Y.; {Schreyer},~K.; {Henning},~T.;
  {Dullemond},~C.; {Bacmann},~A. \emph{Astrophys.~J.,} \textbf{2005},
  \emph{621}, 853\relax
\mciteBstWouldAddEndPuncttrue
\mciteSetBstMidEndSepPunct{\mcitedefaultmidpunct}
{\mcitedefaultendpunct}{\mcitedefaultseppunct}\relax
\EndOfBibitem
\bibitem[{Draine}(1978)]{G}
{Draine},~B.~T. \emph{Astrophys.~J.,} \textbf{1978}, \emph{36}, 595\relax
\mciteBstWouldAddEndPuncttrue
\mciteSetBstMidEndSepPunct{\mcitedefaultmidpunct}
{\mcitedefaultendpunct}{\mcitedefaultseppunct}\relax
\EndOfBibitem
\bibitem[{van Dishoeck} and {Black}(1988){van Dishoeck}, and
  {Black}]{1988ApJ...334..771V}
{van Dishoeck},~E.~F.; {Black},~J.~H. \emph{Astrophys.~J.,} \textbf{1988},
  \emph{334}, 771\relax
\mciteBstWouldAddEndPuncttrue
\mciteSetBstMidEndSepPunct{\mcitedefaultmidpunct}
{\mcitedefaultendpunct}{\mcitedefaultseppunct}\relax
\EndOfBibitem
\bibitem[{Draine} and {Bertoldi}(1996){Draine}, and {Bertoldi}]{DB96}
{Draine},~B.~T.; {Bertoldi},~F. \emph{Astrophys.~J.,} \textbf{1996},
  \emph{468}, 269\relax
\mciteBstWouldAddEndPuncttrue
\mciteSetBstMidEndSepPunct{\mcitedefaultmidpunct}
{\mcitedefaultendpunct}{\mcitedefaultseppunct}\relax
\EndOfBibitem
\bibitem[{Getman} et~al.(2009){Getman}, {Feigelson}, {Luhman},
  {Sicilia-Aguilar}, {Wang}, and {Garmire}]{Getman_ea09}
{Getman},~K.~V.; {Feigelson},~E.~D.; {Luhman},~K.~L.; {Sicilia-Aguilar},~A.;
  {Wang},~J.; {Garmire},~G.~P. \emph{Astrophys.~J.,} \textbf{2009}, \emph{699},
  1454\relax
\mciteBstWouldAddEndPuncttrue
\mciteSetBstMidEndSepPunct{\mcitedefaultmidpunct}
{\mcitedefaultendpunct}{\mcitedefaultseppunct}\relax
\EndOfBibitem
\bibitem[{{\"O}berg} et~al.(2007){{\"O}berg}, {Fuchs}, {Awad}, {Fraser},
  {Schlemmer}, {van Dishoeck}, and {Linnartz}]{Oeberg_ea07}
{{\"O}berg},~K.~I.; {Fuchs},~G.~W.; {Awad},~Z.; {Fraser},~H.~J.;
  {Schlemmer},~S.; {van Dishoeck},~E.~F.; {Linnartz},~H. \emph{Astrophys.~J.,}
  \textbf{2007}, \emph{662}, L23\relax
\mciteBstWouldAddEndPuncttrue
\mciteSetBstMidEndSepPunct{\mcitedefaultmidpunct}
{\mcitedefaultendpunct}{\mcitedefaultseppunct}\relax
\EndOfBibitem
\bibitem[{{\"O}berg} et~al.(2009){{\"O}berg}, {van Dishoeck}, and
  {Linnartz}]{Oeberg_ea09a}
{{\"O}berg},~K.~I.; {van Dishoeck},~E.~F.; {Linnartz},~H.
  \emph{Astron.~Astrophys.,} \textbf{2009}, \emph{496}, 281\relax
\mciteBstWouldAddEndPuncttrue
\mciteSetBstMidEndSepPunct{\mcitedefaultmidpunct}
{\mcitedefaultendpunct}{\mcitedefaultseppunct}\relax
\EndOfBibitem
\bibitem[{{\"O}berg} et~al.(2009){{\"O}berg}, {Linnartz}, {Visser}, and {van
  Dishoeck}]{Oeberg_ea09b}
{{\"O}berg},~K.~I.; {Linnartz},~H.; {Visser},~R.; {van Dishoeck},~E.~F.
  \emph{Astrophys.~J.,} \textbf{2009}, \emph{693}, 1209\relax
\mciteBstWouldAddEndPuncttrue
\mciteSetBstMidEndSepPunct{\mcitedefaultmidpunct}
{\mcitedefaultendpunct}{\mcitedefaultseppunct}\relax
\EndOfBibitem
\bibitem[{Fayolle} et~al.(2011){Fayolle}, {Bertin}, {Romanzin}, {Michaut},
  {{\"O}berg}, {Linnartz}, and {Fillion}]{Fayolle_ea11a}
{Fayolle},~E.~C.; {Bertin},~M.; {Romanzin},~C.; {Michaut},~X.;
  {{\"O}berg},~K.~I.; {Linnartz},~H.; {Fillion},~J.-H. \emph{Astrophys.~J.,}
  \textbf{2011}, \emph{739}, L36\relax
\mciteBstWouldAddEndPuncttrue
\mciteSetBstMidEndSepPunct{\mcitedefaultmidpunct}
{\mcitedefaultendpunct}{\mcitedefaultseppunct}\relax
\EndOfBibitem
\bibitem[{Leger} et~al.(1985){Leger}, {Jura}, and {Omont}]{Leger_ea85}
{Leger},~A.; {Jura},~M.; {Omont},~A. \emph{Astron.~Astrophys.,} \textbf{1985},
  \emph{144}, 147\relax
\mciteBstWouldAddEndPuncttrue
\mciteSetBstMidEndSepPunct{\mcitedefaultmidpunct}
{\mcitedefaultendpunct}{\mcitedefaultseppunct}\relax
\EndOfBibitem
\bibitem[{Hasegawa} and {Herbst}(1993){Hasegawa}, and {Herbst}]{HH93}
{Hasegawa},~T.~I.; {Herbst},~E. \emph{Mon.~Not.~R.~Astron.~Soc,} \textbf{1993},
  \emph{263}, 589\relax
\mciteBstWouldAddEndPuncttrue
\mciteSetBstMidEndSepPunct{\mcitedefaultmidpunct}
{\mcitedefaultendpunct}{\mcitedefaultseppunct}\relax
\EndOfBibitem
\bibitem[{Najita} et~al.(2001){Najita}, {Bergin}, and {Ullom}]{Najita_ea01}
{Najita},~J.; {Bergin},~E.~A.; {Ullom},~J.~N. \emph{Astrophys.~J.,}
  \textbf{2001}, \emph{561}, 880\relax
\mciteBstWouldAddEndPuncttrue
\mciteSetBstMidEndSepPunct{\mcitedefaultmidpunct}
{\mcitedefaultendpunct}{\mcitedefaultseppunct}\relax
\EndOfBibitem
\bibitem[{Cameron}(1995)]{cameron1995}
{Cameron},~A.~G.~W. \emph{Meteoritics} \textbf{1995}, \emph{30}, 133\relax
\mciteBstWouldAddEndPuncttrue
\mciteSetBstMidEndSepPunct{\mcitedefaultmidpunct}
{\mcitedefaultendpunct}{\mcitedefaultseppunct}\relax
\EndOfBibitem
\bibitem[{Grossman}(1972)]{Grossman1972}
{Grossman},~L. \emph{Geochim.~Cosmochim.~Acta,} \textbf{1972}, \emph{36},
  597\relax
\mciteBstWouldAddEndPuncttrue
\mciteSetBstMidEndSepPunct{\mcitedefaultmidpunct}
{\mcitedefaultendpunct}{\mcitedefaultseppunct}\relax
\EndOfBibitem
\bibitem[{Morfill} and {V{\"o}lk}(1984){Morfill}, and
  {V{\"o}lk}]{Morfill_Voelk84}
{Morfill},~G.~E.; {V{\"o}lk},~H.~J. \emph{Astrophys.~J.,} \textbf{1984},
  \emph{287}, 371\relax
\mciteBstWouldAddEndPuncttrue
\mciteSetBstMidEndSepPunct{\mcitedefaultmidpunct}
{\mcitedefaultendpunct}{\mcitedefaultseppunct}\relax
\EndOfBibitem
\bibitem[{Wood} and {Hashimoto}(1993){Wood}, and {Hashimoto}]{Wood_Hashimoto93}
{Wood},~J.~A.; {Hashimoto},~A. \emph{Geochim.~Cosmochim.~Acta,} \textbf{1993},
  \emph{57}, 2377\relax
\mciteBstWouldAddEndPuncttrue
\mciteSetBstMidEndSepPunct{\mcitedefaultmidpunct}
{\mcitedefaultendpunct}{\mcitedefaultseppunct}\relax
\EndOfBibitem
\bibitem[{Prinn}(1993)]{Prinn_93}
{Prinn},~R.~G. In \emph{Protostars and Planets III}; {Levy},~E.~H.,
  {Lunine},~J.~I., Eds.; University of Arizona Press, Tucson, 1993; p
  1005\relax
\mciteBstWouldAddEndPuncttrue
\mciteSetBstMidEndSepPunct{\mcitedefaultmidpunct}
{\mcitedefaultendpunct}{\mcitedefaultseppunct}\relax
\EndOfBibitem
\bibitem[{Duschl} et~al.(1996){Duschl}, {Gail}, and {Tscharnuter}]{Duschl_ea96}
{Duschl},~W.~J.; {Gail},~H.; {Tscharnuter},~W.~M. \emph{Astron.~Astrophys.,}
  \textbf{1996}, \emph{312}, 624\relax
\mciteBstWouldAddEndPuncttrue
\mciteSetBstMidEndSepPunct{\mcitedefaultmidpunct}
{\mcitedefaultendpunct}{\mcitedefaultseppunct}\relax
\EndOfBibitem
\bibitem[{Bauer} et~al.(1997){Bauer}, {Finocchi}, {Duschl}, {Gail}, and
  {Schloeder}]{bauer}
{Bauer},~I.; {Finocchi},~F.; {Duschl},~W.~J.; {Gail},~H.-P.; {Schloeder},~J.~P.
  \emph{Astron.~Astrophys.,} \textbf{1997}, \emph{317}, 273\relax
\mciteBstWouldAddEndPuncttrue
\mciteSetBstMidEndSepPunct{\mcitedefaultmidpunct}
{\mcitedefaultendpunct}{\mcitedefaultseppunct}\relax
\EndOfBibitem
\bibitem[{Finocchi} et~al.(1997){Finocchi}, {Gail}, and {Duschl}]{FG97}
{Finocchi},~F.; {Gail},~H.-P.; {Duschl},~W.~J. \emph{Astron.~Astrophys.,}
  \textbf{1997}, \emph{325}, 1264\relax
\mciteBstWouldAddEndPuncttrue
\mciteSetBstMidEndSepPunct{\mcitedefaultmidpunct}
{\mcitedefaultendpunct}{\mcitedefaultseppunct}\relax
\EndOfBibitem
\bibitem[{Finocchi} and {Gail}(1997){Finocchi}, and {Gail}]{Finocchi_Gail97}
{Finocchi},~F.; {Gail},~H.-P. \emph{Astron.~Astrophys.,} \textbf{1997},
  \emph{327}, 825\relax
\mciteBstWouldAddEndPuncttrue
\mciteSetBstMidEndSepPunct{\mcitedefaultmidpunct}
{\mcitedefaultendpunct}{\mcitedefaultseppunct}\relax
\EndOfBibitem
\bibitem[{Gail}(1998)]{Gail98}
{Gail},~H.-P. \emph{Astron.~Astrophys.,} \textbf{1998}, \emph{332}, 1099\relax
\mciteBstWouldAddEndPuncttrue
\mciteSetBstMidEndSepPunct{\mcitedefaultmidpunct}
{\mcitedefaultendpunct}{\mcitedefaultseppunct}\relax
\EndOfBibitem
\bibitem[{Gail}(2001)]{G01}
{Gail},~H.-P. \emph{Astron.~Astrophys.,} \textbf{2001}, \emph{378}, 192\relax
\mciteBstWouldAddEndPuncttrue
\mciteSetBstMidEndSepPunct{\mcitedefaultmidpunct}
{\mcitedefaultendpunct}{\mcitedefaultseppunct}\relax
\EndOfBibitem
\bibitem[{Keller} and {Gail}(2004){Keller}, and {Gail}]{Keller_Gail04}
{Keller},~C.; {Gail},~H. \emph{Astron.~Astrophys.,} \textbf{2004}, \emph{415},
  1177\relax
\mciteBstWouldAddEndPuncttrue
\mciteSetBstMidEndSepPunct{\mcitedefaultmidpunct}
{\mcitedefaultendpunct}{\mcitedefaultseppunct}\relax
\EndOfBibitem
\bibitem[{Lyons} and {Young}(2005){Lyons}, and {Young}]{Lyons_Young05}
{Lyons},~J.~R.; {Young},~E.~D. \emph{Nature} \textbf{2005}, \emph{435},
  317\relax
\mciteBstWouldAddEndPuncttrue
\mciteSetBstMidEndSepPunct{\mcitedefaultmidpunct}
{\mcitedefaultendpunct}{\mcitedefaultseppunct}\relax
\EndOfBibitem
\bibitem[{Ciesla} and {Cuzzi}(2006){Ciesla}, and {Cuzzi}]{Ciesla_Cuzzi06}
{Ciesla},~F.~J.; {Cuzzi},~J.~N. \emph{Icarus} \textbf{2006}, \emph{181},
  178\relax
\mciteBstWouldAddEndPuncttrue
\mciteSetBstMidEndSepPunct{\mcitedefaultmidpunct}
{\mcitedefaultendpunct}{\mcitedefaultseppunct}\relax
\EndOfBibitem
\bibitem[{Ciesla}(2009)]{Ciesla_09}
{Ciesla},~F.~J. \emph{Icarus} \textbf{2009}, \emph{200}, 655\relax
\mciteBstWouldAddEndPuncttrue
\mciteSetBstMidEndSepPunct{\mcitedefaultmidpunct}
{\mcitedefaultendpunct}{\mcitedefaultseppunct}\relax
\EndOfBibitem
\bibitem[{Geiss} and {Reeves}(1981){Geiss}, and {Reeves}]{Geiss_Reeves81}
{Geiss},~J.; {Reeves},~H. \emph{Astron.~Astrophys.,} \textbf{1981}, \emph{93},
  189\relax
\mciteBstWouldAddEndPuncttrue
\mciteSetBstMidEndSepPunct{\mcitedefaultmidpunct}
{\mcitedefaultendpunct}{\mcitedefaultseppunct}\relax
\EndOfBibitem
\bibitem[{Petaev} and {Wood}(1998){Petaev}, and {Wood}]{Petaev_Wood98}
{Petaev},~M.~I.; {Wood},~J.~A. \emph{Meteorit. Planet. Sci.,} \textbf{1998},
  \emph{33}, 1123\relax
\mciteBstWouldAddEndPuncttrue
\mciteSetBstMidEndSepPunct{\mcitedefaultmidpunct}
{\mcitedefaultendpunct}{\mcitedefaultseppunct}\relax
\EndOfBibitem
\bibitem[{Fegley}(2000)]{Fegley00}
{Fegley},~B.,~Jr. \emph{Space~Sci.~Rev.,} \textbf{2000}, \emph{92}, 177\relax
\mciteBstWouldAddEndPuncttrue
\mciteSetBstMidEndSepPunct{\mcitedefaultmidpunct}
{\mcitedefaultendpunct}{\mcitedefaultseppunct}\relax
\EndOfBibitem
\bibitem[{Aikawa} et~al.(1996){Aikawa}, {Miyama}, {Nakano}, and
  {Umebayashi}]{Aea96}
{Aikawa},~Y.; {Miyama},~S.~M.; {Nakano},~T.; {Umebayashi},~T.
  \emph{Astrophys.~J.,} \textbf{1996}, \emph{467}, 684\relax
\mciteBstWouldAddEndPuncttrue
\mciteSetBstMidEndSepPunct{\mcitedefaultmidpunct}
{\mcitedefaultendpunct}{\mcitedefaultseppunct}\relax
\EndOfBibitem
\bibitem[{Aikawa} et~al.(1997){Aikawa}, {Umebayashi}, {Nakano}, and
  {Miyama}]{Aikawa_ea97a}
{Aikawa},~Y.; {Umebayashi},~T.; {Nakano},~T.; {Miyama},~S.~M.
  \emph{Astrophys.~J.,~Lett.,} \textbf{1997}, \emph{486}, L51\relax
\mciteBstWouldAddEndPuncttrue
\mciteSetBstMidEndSepPunct{\mcitedefaultmidpunct}
{\mcitedefaultendpunct}{\mcitedefaultseppunct}\relax
\EndOfBibitem
\bibitem[{Aikawa} and {Nomura}(2006){Aikawa}, and {Nomura}]{Aikawa_ea06}
{Aikawa},~Y.; {Nomura},~H. \emph{Astrophys.~J.,} \textbf{2006}, \emph{642},
  1152\relax
\mciteBstWouldAddEndPuncttrue
\mciteSetBstMidEndSepPunct{\mcitedefaultmidpunct}
{\mcitedefaultendpunct}{\mcitedefaultseppunct}\relax
\EndOfBibitem
\bibitem[{Ag{\'u}ndez} et~al.(2008){Ag{\'u}ndez}, {Cernicharo}, and
  {Goicoechea}]{Agundez_ea08}
{Ag{\'u}ndez},~M.; {Cernicharo},~J.; {Goicoechea},~J.~R.
  \emph{Astron.~Astrophys.,} \textbf{2008}, \emph{483}, 831\relax
\mciteBstWouldAddEndPuncttrue
\mciteSetBstMidEndSepPunct{\mcitedefaultmidpunct}
{\mcitedefaultendpunct}{\mcitedefaultseppunct}\relax
\EndOfBibitem
\bibitem[{Walsh} et~al.(2010){Walsh}, {Millar}, and {Nomura}]{Walsh_ea10}
{Walsh},~C.; {Millar},~T.~J.; {Nomura},~H. \emph{Astrophys.~J.,} \textbf{2010},
  \emph{722}, 1607\relax
\mciteBstWouldAddEndPuncttrue
\mciteSetBstMidEndSepPunct{\mcitedefaultmidpunct}
{\mcitedefaultendpunct}{\mcitedefaultseppunct}\relax
\EndOfBibitem
\bibitem[{D'Alessio} et~al.(1998){D'Alessio}, {Canto}, {Calvet}, and
  {Lizano}]{DAea98}
{D'Alessio},~P.; {Canto},~J.; {Calvet},~N.; {Lizano},~S. \emph{Astrophys.~J.,}
  \textbf{1998}, \emph{500}, 411\relax
\mciteBstWouldAddEndPuncttrue
\mciteSetBstMidEndSepPunct{\mcitedefaultmidpunct}
{\mcitedefaultendpunct}{\mcitedefaultseppunct}\relax
\EndOfBibitem
\bibitem[{Weingartner} and {Draine}(2001){Weingartner}, and
  {Draine}]{Weingartner_Draine01}
{Weingartner},~J.~C.; {Draine},~B.~T. \emph{Astrophys.~J.,} \textbf{2001},
  \emph{548}, 296\relax
\mciteBstWouldAddEndPuncttrue
\mciteSetBstMidEndSepPunct{\mcitedefaultmidpunct}
{\mcitedefaultendpunct}{\mcitedefaultseppunct}\relax
\EndOfBibitem
\bibitem[{Woitke} et~al.(2009){Woitke}, {Kamp}, and {Thi}]{Woitke_ea09}
{Woitke},~P.; {Kamp},~I.; {Thi},~W. \emph{Astron.~Astrophys.,} \textbf{2009},
  \emph{501}, 383\relax
\mciteBstWouldAddEndPuncttrue
\mciteSetBstMidEndSepPunct{\mcitedefaultmidpunct}
{\mcitedefaultendpunct}{\mcitedefaultseppunct}\relax
\EndOfBibitem
\bibitem[{Nomura} et~al.(2007){Nomura}, {Aikawa}, {Tsujimoto}, {Nakagawa}, and
  {Millar}]{Nomura_ea07a}
{Nomura},~H.; {Aikawa},~Y.; {Tsujimoto},~M.; {Nakagawa},~Y.; {Millar},~T.~J.
  \emph{Astrophys.~J.,} \textbf{2007}, \emph{661}, 334\relax
\mciteBstWouldAddEndPuncttrue
\mciteSetBstMidEndSepPunct{\mcitedefaultmidpunct}
{\mcitedefaultendpunct}{\mcitedefaultseppunct}\relax
\EndOfBibitem
\bibitem[{Glassgold} et~al.(2005){Glassgold}, {Feigelson}, {Montmerle}, and
  {Wolk}]{Glassgold_ea05}
{Glassgold},~A.~E.; {Feigelson},~E.~D.; {Montmerle},~T.; {Wolk},~S. In
  \emph{Chondrites and the Protoplanetary Disk}; {A.~N.~Krot, E.~R.~D.~Scott,
  \& B.~Reipurth},, Ed.; Astronomical Society of the Pacific Conference Series;
  Astronomical Society of the Pacific, San Francisco, 2005; Vol. 341; p
  165\relax
\mciteBstWouldAddEndPuncttrue
\mciteSetBstMidEndSepPunct{\mcitedefaultmidpunct}
{\mcitedefaultendpunct}{\mcitedefaultseppunct}\relax
\EndOfBibitem
\bibitem[{Aresu} et~al.(2011){Aresu}, {Kamp}, {Meijerink}, {Woitke}, {Thi}, and
  {Spaans}]{Aresu_ea10a}
{Aresu},~G.; {Kamp},~I.; {Meijerink},~R.; {Woitke},~P.; {Thi},~W.; {Spaans},~M.
  \emph{Astron.~Astrophys.,} \textbf{2011}, \emph{526}, 163\relax
\mciteBstWouldAddEndPuncttrue
\mciteSetBstMidEndSepPunct{\mcitedefaultmidpunct}
{\mcitedefaultendpunct}{\mcitedefaultseppunct}\relax
\EndOfBibitem
\bibitem[{Aikawa} and {Herbst}(1999){Aikawa}, and {Herbst}]{AH99b}
{Aikawa},~Y.; {Herbst},~E. \emph{Astrophys.~J.,} \textbf{1999}, \emph{526},
  314\relax
\mciteBstWouldAddEndPuncttrue
\mciteSetBstMidEndSepPunct{\mcitedefaultmidpunct}
{\mcitedefaultendpunct}{\mcitedefaultseppunct}\relax
\EndOfBibitem
\bibitem[{Aikawa} and {Herbst}(2001){Aikawa}, and {Herbst}]{ah2001}
{Aikawa},~Y.; {Herbst},~E. \emph{Astron.~Astrophys.,} \textbf{2001},
  \emph{371}, 1107\relax
\mciteBstWouldAddEndPuncttrue
\mciteSetBstMidEndSepPunct{\mcitedefaultmidpunct}
{\mcitedefaultendpunct}{\mcitedefaultseppunct}\relax
\EndOfBibitem
\bibitem[{Ceccarelli} and {Dominik}(2005){Ceccarelli}, and
  {Dominik}]{Ceccarelli_Dominik05}
{Ceccarelli},~C.; {Dominik},~C. \emph{Astron.~Astrophys.,} \textbf{2005},
  \emph{440}, 583\relax
\mciteBstWouldAddEndPuncttrue
\mciteSetBstMidEndSepPunct{\mcitedefaultmidpunct}
{\mcitedefaultendpunct}{\mcitedefaultseppunct}\relax
\EndOfBibitem
\bibitem[{Willacy}(2007)]{Willacy_07}
{Willacy},~K. \emph{Astrophys.~J.,} \textbf{2007}, \emph{660}, 441\relax
\mciteBstWouldAddEndPuncttrue
\mciteSetBstMidEndSepPunct{\mcitedefaultmidpunct}
{\mcitedefaultendpunct}{\mcitedefaultseppunct}\relax
\EndOfBibitem
\bibitem[{Willacy} and {Woods}(2009){Willacy}, and {Woods}]{Willacy_Woods09}
{Willacy},~K.; {Woods},~P.~M. \emph{Astrophys.~J.,} \textbf{2009}, \emph{703},
  479\relax
\mciteBstWouldAddEndPuncttrue
\mciteSetBstMidEndSepPunct{\mcitedefaultmidpunct}
{\mcitedefaultendpunct}{\mcitedefaultseppunct}\relax
\EndOfBibitem
\bibitem[{Woods} and {Willacy}(2009){Woods}, and {Willacy}]{Woods_Willacy08}
{Woods},~P.~M.; {Willacy},~K. \emph{Astrophys.~J.,} \textbf{2009}, \emph{693},
  1360\relax
\mciteBstWouldAddEndPuncttrue
\mciteSetBstMidEndSepPunct{\mcitedefaultmidpunct}
{\mcitedefaultendpunct}{\mcitedefaultseppunct}\relax
\EndOfBibitem
\bibitem[{Nomura} et~al.(2009){Nomura}, {Aikawa}, {Nakagawa}, and
  {Millar}]{Nomura_ea09}
{Nomura},~H.; {Aikawa},~Y.; {Nakagawa},~Y.; {Millar},~T.~J.
  \emph{Astron.~Astrophys.,} \textbf{2009}, \emph{495}, 183\relax
\mciteBstWouldAddEndPuncttrue
\mciteSetBstMidEndSepPunct{\mcitedefaultmidpunct}
{\mcitedefaultendpunct}{\mcitedefaultseppunct}\relax
\EndOfBibitem
\bibitem[{Visser} et~al.(2011){Visser}, {Doty}, and {van
  Dishoeck}]{Visser_ea11a}
{Visser},~R.; {Doty},~S.~D.; {van Dishoeck},~E.~F. \emph{Astron.~Astrophys.,}
  \textbf{2011}, \emph{534}, 132\relax
\mciteBstWouldAddEndPuncttrue
\mciteSetBstMidEndSepPunct{\mcitedefaultmidpunct}
{\mcitedefaultendpunct}{\mcitedefaultseppunct}\relax
\EndOfBibitem
\bibitem[{Ilgner} and {Nelson}(2008){Ilgner}, and {Nelson}]{Ilgner_ea08}
{Ilgner},~M.; {Nelson},~R.~P. \emph{Astron.~Astrophys.,} \textbf{2008},
  \emph{483}, 815\relax
\mciteBstWouldAddEndPuncttrue
\mciteSetBstMidEndSepPunct{\mcitedefaultmidpunct}
{\mcitedefaultendpunct}{\mcitedefaultseppunct}\relax
\EndOfBibitem
\bibitem[{Semenov} et~al.(2006){Semenov}, {Wiebe}, and {Henning}]{Semenov_ea06}
{Semenov},~D.; {Wiebe},~D.; {Henning},~T. \emph{Astrophys.~J.,} \textbf{2006},
  \emph{647}, L57\relax
\mciteBstWouldAddEndPuncttrue
\mciteSetBstMidEndSepPunct{\mcitedefaultmidpunct}
{\mcitedefaultendpunct}{\mcitedefaultseppunct}\relax
\EndOfBibitem
\bibitem[{Turner} et~al.(2006){Turner}, {Willacy}, {Bryden}, and
  {Yorke}]{Turner_ea06}
{Turner},~N.~J.; {Willacy},~K.; {Bryden},~G.; {Yorke},~H.~W.
  \emph{Astrophys.~J.,} \textbf{2006}, \emph{639}, 1218\relax
\mciteBstWouldAddEndPuncttrue
\mciteSetBstMidEndSepPunct{\mcitedefaultmidpunct}
{\mcitedefaultendpunct}{\mcitedefaultseppunct}\relax
\EndOfBibitem
\bibitem[{Vorobyov} and {Basu}(2006){Vorobyov}, and {Basu}]{Vorobyov_Basu06}
{Vorobyov},~E.~I.; {Basu},~S. \emph{Astrophys.~J.,} \textbf{2006}, \emph{650},
  956\relax
\mciteBstWouldAddEndPuncttrue
\mciteSetBstMidEndSepPunct{\mcitedefaultmidpunct}
{\mcitedefaultendpunct}{\mcitedefaultseppunct}\relax
\EndOfBibitem
\bibitem[{Vorobyov} and {Basu}(2010){Vorobyov}, and {Basu}]{Vorobyov_Basu10a}
{Vorobyov},~E.~I.; {Basu},~S. \emph{Astrophys.~J.,} \textbf{2010}, \emph{719},
  1896\relax
\mciteBstWouldAddEndPuncttrue
\mciteSetBstMidEndSepPunct{\mcitedefaultmidpunct}
{\mcitedefaultendpunct}{\mcitedefaultseppunct}\relax
\EndOfBibitem
\bibitem[{Ilee} et~al.(2011){Ilee}, {Boley}, {Caselli}, {Durisen}, {Hartquist},
  and {Rawlings}]{Ilee_ea11a}
{Ilee},~J.~D.; {Boley},~A.~C.; {Caselli},~P.; {Durisen},~R.~H.;
  {Hartquist},~T.~W.; {Rawlings},~J.~M.~C. \emph{Mon.~Not.~R.~Astron.~Soc,}
  \textbf{2011}, \emph{417}, 2950\relax
\mciteBstWouldAddEndPuncttrue
\mciteSetBstMidEndSepPunct{\mcitedefaultmidpunct}
{\mcitedefaultendpunct}{\mcitedefaultseppunct}\relax
\EndOfBibitem
\bibitem[{Pavlyuchenkov} et~al.(2007){Pavlyuchenkov}, {Semenov}, {Henning},
  {Guilloteau}, {Pi{\'e}tu}, {Launhardt}, and {Dutrey}]{Pavlyuchenkov_ea07a}
{Pavlyuchenkov},~Y.; {Semenov},~D.; {Henning},~T.; {Guilloteau},~S.;
  {Pi{\'e}tu},~V.; {Launhardt},~R.; {Dutrey},~A. \emph{Astrophys.~J.,}
  \textbf{2007}, \emph{669}, 1262\relax
\mciteBstWouldAddEndPuncttrue
\mciteSetBstMidEndSepPunct{\mcitedefaultmidpunct}
{\mcitedefaultendpunct}{\mcitedefaultseppunct}\relax
\EndOfBibitem
\bibitem[{Semenov} et~al.(2008){Semenov}, {Pavlyuchenkov}, {Henning}, {Wolf},
  and {Launhardt}]{Semenov_ea08}
{Semenov},~D.; {Pavlyuchenkov},~Y.; {Henning},~T.; {Wolf},~S.; {Launhardt},~R.
  \emph{Astrophys.~J.~Lett.,} \textbf{2008}, \emph{673}, L195\relax
\mciteBstWouldAddEndPuncttrue
\mciteSetBstMidEndSepPunct{\mcitedefaultmidpunct}
{\mcitedefaultendpunct}{\mcitedefaultseppunct}\relax
\EndOfBibitem
\bibitem[{Kamp} et~al.(2011){Kamp}, {Woitke}, {Pinte}, {Tilling}, {Thi},
  {Menard}, {Duchene}, and {Augereau}]{Kamp_ea11a}
{Kamp},~I.; {Woitke},~P.; {Pinte},~C.; {Tilling},~I.; {Thi},~W.-F.;
  {Menard},~F.; {Duchene},~G.; {Augereau},~J.-C. \emph{Astron.~Astrophys.,}
  \textbf{2011}, \emph{532}, A85\relax
\mciteBstWouldAddEndPuncttrue
\mciteSetBstMidEndSepPunct{\mcitedefaultmidpunct}
{\mcitedefaultendpunct}{\mcitedefaultseppunct}\relax
\EndOfBibitem
\bibitem[{Igea} and {Glassgold}(1999){Igea}, and {Glassgold}]{IG99}
{Igea},~J.; {Glassgold},~A.~E. \emph{Astrophys.~J.,} \textbf{1999}, \emph{518},
  848\relax
\mciteBstWouldAddEndPuncttrue
\mciteSetBstMidEndSepPunct{\mcitedefaultmidpunct}
{\mcitedefaultendpunct}{\mcitedefaultseppunct}\relax
\EndOfBibitem
\bibitem[{Pierce} and {A'Hearn}(2010){Pierce}, and {A'Hearn}]{Pierce_AHearn10a}
{Pierce},~D.~M.; {A'Hearn},~M.~F. \emph{Astrophys.~J.,} \textbf{2010},
  \emph{718}, 340\relax
\mciteBstWouldAddEndPuncttrue
\mciteSetBstMidEndSepPunct{\mcitedefaultmidpunct}
{\mcitedefaultendpunct}{\mcitedefaultseppunct}\relax
\EndOfBibitem
\bibitem[{Hayashi}(1981)]{Hayashi_81}
{Hayashi},~C. \emph{Prog. Theor. Phys. Suppl.,} \textbf{1981}, \emph{70},
  35\relax
\mciteBstWouldAddEndPuncttrue
\mciteSetBstMidEndSepPunct{\mcitedefaultmidpunct}
{\mcitedefaultendpunct}{\mcitedefaultseppunct}\relax
\EndOfBibitem
\bibitem[{Aikawa} et~al.(1998){Aikawa}, {Umebayashi}, {Nakano}, and
  {Miyama}]{Aikawa_ea98}
{Aikawa},~Y.; {Umebayashi},~T.; {Nakano},~T.; {Miyama},~S. \emph{Faraday
  Discussions}; Royal Society of Chemistry, Cambridge, 1998; Vol. 109; p
  281\relax
\mciteBstWouldAddEndPuncttrue
\mciteSetBstMidEndSepPunct{\mcitedefaultmidpunct}
{\mcitedefaultendpunct}{\mcitedefaultseppunct}\relax
\EndOfBibitem
\bibitem[{Chiang} and {Goldreich}(1997){Chiang}, and {Goldreich}]{CG97}
{Chiang},~E.~I.; {Goldreich},~P. \emph{Astrophys.~J.,} \textbf{1997},
  \emph{490}, 368\relax
\mciteBstWouldAddEndPuncttrue
\mciteSetBstMidEndSepPunct{\mcitedefaultmidpunct}
{\mcitedefaultendpunct}{\mcitedefaultseppunct}\relax
\EndOfBibitem
\bibitem[{Kamp} and {van Zadelhoff}(2001){Kamp}, and {van
  Zadelhoff}]{Kamp_Zadelhoff01}
{Kamp},~I.; {van Zadelhoff},~G.-J. \emph{Astron.~Astrophys.,} \textbf{2001},
  \emph{373}, 641\relax
\mciteBstWouldAddEndPuncttrue
\mciteSetBstMidEndSepPunct{\mcitedefaultmidpunct}
{\mcitedefaultendpunct}{\mcitedefaultseppunct}\relax
\EndOfBibitem
\bibitem[{Kamp} and {Bertoldi}(2000){Kamp}, and {Bertoldi}]{Kamp_Bertoldi00}
{Kamp},~I.; {Bertoldi},~F. \emph{Astron.~Astrophys.,} \textbf{2000},
  \emph{353}, 276\relax
\mciteBstWouldAddEndPuncttrue
\mciteSetBstMidEndSepPunct{\mcitedefaultmidpunct}
{\mcitedefaultendpunct}{\mcitedefaultseppunct}\relax
\EndOfBibitem
\bibitem[{Millar} et~al.(2003){Millar}, {Nomura}, and {Markwick}]{Millar_ea03}
{Millar},~T.~J.; {Nomura},~H.; {Markwick},~A.~J. \emph{Astrophys.~Space.~Sci,}
  \textbf{2003}, \emph{285}, 761\relax
\mciteBstWouldAddEndPuncttrue
\mciteSetBstMidEndSepPunct{\mcitedefaultmidpunct}
{\mcitedefaultendpunct}{\mcitedefaultseppunct}\relax
\EndOfBibitem
\bibitem[{Glassgold} et~al.(2007){Glassgold}, {Najita}, and
  {Igea}]{Glassgold_ea07}
{Glassgold},~A.~E.; {Najita},~J.~R.; {Igea},~J. \emph{Astrophys.~J.,}
  \textbf{2007}, \emph{656}, 515\relax
\mciteBstWouldAddEndPuncttrue
\mciteSetBstMidEndSepPunct{\mcitedefaultmidpunct}
{\mcitedefaultendpunct}{\mcitedefaultseppunct}\relax
\EndOfBibitem
\bibitem[{Meijerink} et~al.(2008){Meijerink}, {Glassgold}, and
  {Najita}]{Meijerink_ea08a}
{Meijerink},~R.; {Glassgold},~A.~E.; {Najita},~J.~R. \emph{Astrophys.~J.,}
  \textbf{2008}, \emph{676}, 518\relax
\mciteBstWouldAddEndPuncttrue
\mciteSetBstMidEndSepPunct{\mcitedefaultmidpunct}
{\mcitedefaultendpunct}{\mcitedefaultseppunct}\relax
\EndOfBibitem
\bibitem[{Glassgold} et~al.(2009){Glassgold}, {Meijerink}, and
  {Najita}]{Glassgold_ea09}
{Glassgold},~A.~E.; {Meijerink},~R.; {Najita},~J.~R. \emph{Astrophys.~J.,}
  \textbf{2009}, \emph{701}, 142x\relax
\mciteBstWouldAddEndPuncttrue
\mciteSetBstMidEndSepPunct{\mcitedefaultmidpunct}
{\mcitedefaultendpunct}{\mcitedefaultseppunct}\relax
\EndOfBibitem
\bibitem[{Jonkheid} et~al.(2004){Jonkheid}, {Faas}, {van Zadelhoff}, and {van
  Dishoeck}]{Jonkheid_ea04}
{Jonkheid},~B.; {Faas},~F.~G.~A.; {van Zadelhoff},~G.-J.; {van Dishoeck},~E.~F.
  \emph{Astron.~Astrophys.,} \textbf{2004}, \emph{428}, 511\relax
\mciteBstWouldAddEndPuncttrue
\mciteSetBstMidEndSepPunct{\mcitedefaultmidpunct}
{\mcitedefaultendpunct}{\mcitedefaultseppunct}\relax
\EndOfBibitem
\bibitem[{Dullemond} et~al.(2002){Dullemond}, {van Zadelhoff}, and
  {Natta}]{Dullemond_ea02}
{Dullemond},~C.~P.; {van Zadelhoff},~G.~J.; {Natta},~A.
  \emph{Astron.~Astrophys.,} \textbf{2002}, \emph{389}, 464\relax
\mciteBstWouldAddEndPuncttrue
\mciteSetBstMidEndSepPunct{\mcitedefaultmidpunct}
{\mcitedefaultendpunct}{\mcitedefaultseppunct}\relax
\EndOfBibitem
\bibitem[{Dominik} et~al.(2005){Dominik}, {Ceccarelli}, {Hollenbach}, and
  {Kaufman}]{Dominik_ea05a}
{Dominik},~C.; {Ceccarelli},~C.; {Hollenbach},~D.; {Kaufman},~M.
  \emph{Astrophys.~J.,} \textbf{2005}, \emph{635}, L85\relax
\mciteBstWouldAddEndPuncttrue
\mciteSetBstMidEndSepPunct{\mcitedefaultmidpunct}
{\mcitedefaultendpunct}{\mcitedefaultseppunct}\relax
\EndOfBibitem
\bibitem[{Dullemond} and {Dominik}(2004){Dullemond}, and {Dominik}]{DD04}
{Dullemond},~C.~P.; {Dominik},~C. \emph{Astron.~Astrophys.,} \textbf{2004},
  \emph{417}, 159\relax
\mciteBstWouldAddEndPuncttrue
\mciteSetBstMidEndSepPunct{\mcitedefaultmidpunct}
{\mcitedefaultendpunct}{\mcitedefaultseppunct}\relax
\EndOfBibitem
\bibitem[{Nomura} and {Millar}(2005){Nomura}, and {Millar}]{Nomura_Millar05}
{Nomura},~H.; {Millar},~T.~J. \emph{Astron.~Astrophys.,} \textbf{2005},
  \emph{438}, 923\relax
\mciteBstWouldAddEndPuncttrue
\mciteSetBstMidEndSepPunct{\mcitedefaultmidpunct}
{\mcitedefaultendpunct}{\mcitedefaultseppunct}\relax
\EndOfBibitem
\bibitem[{Mathis} et~al.(1977){Mathis}, {Rumpl}, and {Nordsieck}]{MRN}
{Mathis},~J.~S.; {Rumpl},~W.; {Nordsieck},~K.~H. \emph{Astrophys.~J.,}
  \textbf{1977}, \emph{217}, 425\relax
\mciteBstWouldAddEndPuncttrue
\mciteSetBstMidEndSepPunct{\mcitedefaultmidpunct}
{\mcitedefaultendpunct}{\mcitedefaultseppunct}\relax
\EndOfBibitem
\bibitem[{Semenov} et~al.(2010){Semenov}, {Hersant}, {Wakelam}, {Dutrey},
  {Chapillon}, {Guilloteau}, {Henning}, {Launhardt}, {Pi{\'e}tu}, and
  {Schreyer}]{Semenov_ea10}
{Semenov},~D.; {Hersant},~F.; {Wakelam},~V.; {Dutrey},~A.; {Chapillon},~E.;
  {Guilloteau},~S.; {Henning},~T.; {Launhardt},~R.; {Pi{\'e}tu},~V.;
  {Schreyer},~K. \emph{Astron.~Astrophys.,} \textbf{2010}, \emph{522},
  A42\relax
\mciteBstWouldAddEndPuncttrue
\mciteSetBstMidEndSepPunct{\mcitedefaultmidpunct}
{\mcitedefaultendpunct}{\mcitedefaultseppunct}\relax
\EndOfBibitem
\bibitem[{Jonkheid} et~al.(2007){Jonkheid}, {Dullemond}, {Hogerheijde}, and
  {van Dishoeck}]{Jonkheid_ea07}
{Jonkheid},~B.; {Dullemond},~C.~P.; {Hogerheijde},~M.~R.; {van Dishoeck},~E.~F.
  \emph{Astron.~Astrophys.,} \textbf{2007}, \emph{463}, 203\relax
\mciteBstWouldAddEndPuncttrue
\mciteSetBstMidEndSepPunct{\mcitedefaultmidpunct}
{\mcitedefaultendpunct}{\mcitedefaultseppunct}\relax
\EndOfBibitem
\bibitem[{Oppenheimer} and {Dalgarno}(1974){Oppenheimer}, and {Dalgarno}]{OD74}
{Oppenheimer},~M.; {Dalgarno},~A. \emph{Astrophys.~J.,} \textbf{1974},
  \emph{192}, 29\relax
\mciteBstWouldAddEndPuncttrue
\mciteSetBstMidEndSepPunct{\mcitedefaultmidpunct}
{\mcitedefaultendpunct}{\mcitedefaultseppunct}\relax
\EndOfBibitem
\bibitem[{D'Alessio} et~al.(2001){D'Alessio}, {Calvet}, and {Hartmann}]{DCH01}
{D'Alessio},~P.; {Calvet},~N.; {Hartmann},~L. \emph{Astrophys.~J.,}
  \textbf{2001}, \emph{553}, 321\relax
\mciteBstWouldAddEndPuncttrue
\mciteSetBstMidEndSepPunct{\mcitedefaultmidpunct}
{\mcitedefaultendpunct}{\mcitedefaultseppunct}\relax
\EndOfBibitem
\bibitem[{Chapillon} et~al.(2008){Chapillon}, {Guilloteau}, {Dutrey}, and
  {Pi{\'e}tu}]{Chapillon_ea08}
{Chapillon},~E.; {Guilloteau},~S.; {Dutrey},~A.; {Pi{\'e}tu},~V.
  \emph{Astron.~Astrophys.,} \textbf{2008}, \emph{488}, 565\relax
\mciteBstWouldAddEndPuncttrue
\mciteSetBstMidEndSepPunct{\mcitedefaultmidpunct}
{\mcitedefaultendpunct}{\mcitedefaultseppunct}\relax
\EndOfBibitem
\bibitem[{Chapillon} et~al.(2010){Chapillon}, {Parise}, {Guilloteau}, {Dutrey},
  and {Wakelam}]{Chapillon_ea10}
{Chapillon},~E.; {Parise},~B.; {Guilloteau},~S.; {Dutrey},~A.; {Wakelam},~V.
  \emph{Astron.~Astrophys.,} \textbf{2010}, \emph{520}, 61\relax
\mciteBstWouldAddEndPuncttrue
\mciteSetBstMidEndSepPunct{\mcitedefaultmidpunct}
{\mcitedefaultendpunct}{\mcitedefaultseppunct}\relax
\EndOfBibitem
\bibitem[{Vasyunin} et~al.(2008){Vasyunin}, {Semenov}, {Henning}, {Wakelam},
  {Herbst}, and {Sobolev}]{Vasyunin_ea08}
{Vasyunin},~A.~I.; {Semenov},~D.; {Henning},~T.; {Wakelam},~V.; {Herbst},~E.;
  {Sobolev},~A.~M. \emph{Astrophys.~J.,} \textbf{2008}, \emph{672}, 629\relax
\mciteBstWouldAddEndPuncttrue
\mciteSetBstMidEndSepPunct{\mcitedefaultmidpunct}
{\mcitedefaultendpunct}{\mcitedefaultseppunct}\relax
\EndOfBibitem
\bibitem[{Visser} et~al.(2009){Visser}, {van Dishoeck}, {Doty}, and
  {Dullemond}]{Visser_ea09}
{Visser},~R.; {van Dishoeck},~E.~F.; {Doty},~S.~D.; {Dullemond},~C.~P.
  \emph{Astron.~Astrophys.,} \textbf{2009}, \emph{495}, 881\relax
\mciteBstWouldAddEndPuncttrue
\mciteSetBstMidEndSepPunct{\mcitedefaultmidpunct}
{\mcitedefaultendpunct}{\mcitedefaultseppunct}\relax
\EndOfBibitem
\bibitem[{Woitke} et~al.(2011){Woitke}, {Riaz}, {Duch{\^e}ne}, {Pascucci},
  {Lyo}, {Dent}, {Phillips}, {Thi}, {M{\'e}nard}, {Herczeg}, {Bergin}, {Brown},
  {Mora}, {Kamp}, {Aresu}, {Brittain}, {de Gregorio-Monsalvo}, and
  {Sandell}]{Woitke_ea11a}
{Woitke},~P. et~al.  \emph{Astron.~Astrophys.,} \textbf{2011}, \emph{534},
  44\relax
\mciteBstWouldAddEndPuncttrue
\mciteSetBstMidEndSepPunct{\mcitedefaultmidpunct}
{\mcitedefaultendpunct}{\mcitedefaultseppunct}\relax
\EndOfBibitem
\bibitem[{Meijerink} et~al.(2012){Meijerink}, {Aresu}, {Kamp}, {Spaans}, {Thi},
  and {Woitke}]{Meijerink_ea12a}
{Meijerink},~R.; {Aresu},~G.; {Kamp},~I.; {Spaans},~M.; {Thi},~W.-F.;
  {Woitke},~P. \emph{Astron.~Astrophys.,} \textbf{2012}, \emph{547}, A68\relax
\mciteBstWouldAddEndPuncttrue
\mciteSetBstMidEndSepPunct{\mcitedefaultmidpunct}
{\mcitedefaultendpunct}{\mcitedefaultseppunct}\relax
\EndOfBibitem
\bibitem[{Pinte} et~al.(2006){Pinte}, {M{\'e}nard}, {Duch{\^e}ne}, and
  {Bastien}]{Pinte_ea06}
{Pinte},~C.; {M{\'e}nard},~F.; {Duch{\^e}ne},~G.; {Bastien},~P.
  \emph{Astron.~Astrophys.,} \textbf{2006}, \emph{459}, 797\relax
\mciteBstWouldAddEndPuncttrue
\mciteSetBstMidEndSepPunct{\mcitedefaultmidpunct}
{\mcitedefaultendpunct}{\mcitedefaultseppunct}\relax
\EndOfBibitem
\bibitem[{Cleeves} et~al.(2011){Cleeves}, {Bergin}, {Bethell}, {Calvet},
  {Fogel}, {Sauter}, and {Wolf}]{Cleeves_ea11a}
{Cleeves},~L.~I.; {Bergin},~E.~A.; {Bethell},~T.~J.; {Calvet},~N.;
  {Fogel},~J.~K.~J.; {Sauter},~J.; {Wolf},~S. \emph{Astrophys.~J.~Lett.,}
  \textbf{2011}, \emph{743}, L2\relax
\mciteBstWouldAddEndPuncttrue
\mciteSetBstMidEndSepPunct{\mcitedefaultmidpunct}
{\mcitedefaultendpunct}{\mcitedefaultseppunct}\relax
\EndOfBibitem
\bibitem[{Bjorkman} and {Wood}(2001){Bjorkman}, and {Wood}]{Bjorkman_Wood01}
{Bjorkman},~J.~E.; {Wood},~K. \emph{Astrophys.~J.,} \textbf{2001}, \emph{554},
  615\relax
\mciteBstWouldAddEndPuncttrue
\mciteSetBstMidEndSepPunct{\mcitedefaultmidpunct}
{\mcitedefaultendpunct}{\mcitedefaultseppunct}\relax
\EndOfBibitem
\bibitem[{Bruderer} et~al.(2009){Bruderer}, {Benz}, {Doty}, {van Dishoeck}, and
  {Bourke}]{Bruderer_ea09a}
{Bruderer},~S.; {Benz},~A.~O.; {Doty},~S.~D.; {van Dishoeck},~E.~F.;
  {Bourke},~T.~L. \emph{Astrophys.~J.,} \textbf{2009}, \emph{700}, 872\relax
\mciteBstWouldAddEndPuncttrue
\mciteSetBstMidEndSepPunct{\mcitedefaultmidpunct}
{\mcitedefaultendpunct}{\mcitedefaultseppunct}\relax
\EndOfBibitem
\bibitem[{Birnstiel} et~al.(2012){Birnstiel}, {Klahr}, and
  {Ercolano}]{Birnstiel_ea12}
{Birnstiel},~T.; {Klahr},~H.; {Ercolano},~B. \emph{Astron.~Astrophys.,}
  \textbf{2012}, \emph{539}, 148\relax
\mciteBstWouldAddEndPuncttrue
\mciteSetBstMidEndSepPunct{\mcitedefaultmidpunct}
{\mcitedefaultendpunct}{\mcitedefaultseppunct}\relax
\EndOfBibitem
\bibitem[{Yurimoto} et~al.(2007){Yurimoto}, {Kuramoto}, {Krot}, {Scott},
  {Cuzzi}, {Thiemens}, and {Lyons}]{Yurimoto_ea07}
{Yurimoto},~H.; {Kuramoto},~K.; {Krot},~A.~N.; {Scott},~E.~R.~D.;
  {Cuzzi},~J.~N.; {Thiemens},~M.~H.; {Lyons},~J.~R. In \emph{Protostars and
  Planets V}; {Reipurth},~B., {Jewitt},~D., {Keil},~K., Eds.; University of
  Arizona Press, Tucson, 2007; p 849\relax
\mciteBstWouldAddEndPuncttrue
\mciteSetBstMidEndSepPunct{\mcitedefaultmidpunct}
{\mcitedefaultendpunct}{\mcitedefaultseppunct}\relax
\EndOfBibitem
\bibitem[{Draine}(2003)]{2003ARA&A..41..241D}
{Draine},~B.~T. \emph{Ann.~Rev.~Astron.~Astrophys.,} \textbf{2003}, \emph{41},
  241\relax
\mciteBstWouldAddEndPuncttrue
\mciteSetBstMidEndSepPunct{\mcitedefaultmidpunct}
{\mcitedefaultendpunct}{\mcitedefaultseppunct}\relax
\EndOfBibitem
\bibitem[{Palme}(2001)]{Palme_01}
{Palme},~H. \emph{Philos. Trans. R. Soc., A} \textbf{2001}, \emph{359},
  2061\relax
\mciteBstWouldAddEndPuncttrue
\mciteSetBstMidEndSepPunct{\mcitedefaultmidpunct}
{\mcitedefaultendpunct}{\mcitedefaultseppunct}\relax
\EndOfBibitem
\bibitem[{Trieloff} and {Palme}(2006){Trieloff}, and {Palme}]{Trieloff_Palme06}
{Trieloff},~M.; {Palme},~H. In \emph{Planet Formation}; {Klahr},~H.,
  {Brandner},~W., Eds.; Cambridge University Press, Cambridge, 2006; p~64\relax
\mciteBstWouldAddEndPuncttrue
\mciteSetBstMidEndSepPunct{\mcitedefaultmidpunct}
{\mcitedefaultendpunct}{\mcitedefaultseppunct}\relax
\EndOfBibitem
\bibitem[{Ehrenfreund} and {Charnley}(2000){Ehrenfreund}, and
  {Charnley}]{Ehrenfreund_Charnley00}
{Ehrenfreund},~P.; {Charnley},~S.~B. \emph{Ann.~Rev.~Astron.~Astrophys.,}
  \textbf{2000}, \emph{38}, 427\relax
\mciteBstWouldAddEndPuncttrue
\mciteSetBstMidEndSepPunct{\mcitedefaultmidpunct}
{\mcitedefaultendpunct}{\mcitedefaultseppunct}\relax
\EndOfBibitem
\bibitem[{Busemann} et~al.(2006){Busemann}, {Young}, {O'D.~Alexander}, {Hoppe},
  {Mukhopadhyay}, and {Nittler}]{Busemann_ea06}
{Busemann},~H.; {Young},~A.~F.; {O'D.~Alexander},~C.~M.; {Hoppe},~P.;
  {Mukhopadhyay},~S.; {Nittler},~L.~R. \emph{Science} \textbf{2006},
  \emph{312}, 727\relax
\mciteBstWouldAddEndPuncttrue
\mciteSetBstMidEndSepPunct{\mcitedefaultmidpunct}
{\mcitedefaultendpunct}{\mcitedefaultseppunct}\relax
\EndOfBibitem
\bibitem[{Pizzarello} et~al.(2006){Pizzarello}, {Cooper}, and
  {Flynn}]{Pizzarello_ea06}
{Pizzarello},~S.; {Cooper},~G.~W.; {Flynn},~G.~J. In \emph{Meteorites and the
  Early Solar System II}; {Lauretta},~D.~S., {McSween},~H.~Y., Eds.; University
  of Arizona Press, Tucson, 2006; p 625\relax
\mciteBstWouldAddEndPuncttrue
\mciteSetBstMidEndSepPunct{\mcitedefaultmidpunct}
{\mcitedefaultendpunct}{\mcitedefaultseppunct}\relax
\EndOfBibitem
\bibitem[{Herbst} and {van Dishoeck}(2009){Herbst}, and {van
  Dishoeck}]{Herbst_vanDishoeck09}
{Herbst},~E.; {van Dishoeck},~E.~F. \emph{Ann.~Rev.~Astron.~Astrophys.,}
  \textbf{2009}, \emph{47}, 427\relax
\mciteBstWouldAddEndPuncttrue
\mciteSetBstMidEndSepPunct{\mcitedefaultmidpunct}
{\mcitedefaultendpunct}{\mcitedefaultseppunct}\relax
\EndOfBibitem
\bibitem[{Morgan} et~al.(1991){Morgan}, {Feigelson}, {Wang}, and
  {Frenklach}]{Morgan_ea91}
{Morgan},~W.~A.,~Jr.; {Feigelson},~E.~D.; {Wang},~H.; {Frenklach},~M.
  \emph{Science} \textbf{1991}, \emph{252}, 109\relax
\mciteBstWouldAddEndPuncttrue
\mciteSetBstMidEndSepPunct{\mcitedefaultmidpunct}
{\mcitedefaultendpunct}{\mcitedefaultseppunct}\relax
\EndOfBibitem
\bibitem[{Wozniakiewicz} et~al.(2012){Wozniakiewicz}, {Kearsley}, {Ishii},
  {Burchell}, {Bradley}, {Teslich}, {Cole}, and {Price}]{2012M&PS...47..660W}
{Wozniakiewicz},~P.~J.; {Kearsley},~A.~T.; {Ishii},~H.~A.; {Burchell},~M.~J.;
  {Bradley},~J.~P.; {Teslich},~N.; {Cole},~M.~J.; {Price},~M.~C.
  \emph{Meteorit. Planet. Sci.,} \textbf{2012}, \emph{47}, 660\relax
\mciteBstWouldAddEndPuncttrue
\mciteSetBstMidEndSepPunct{\mcitedefaultmidpunct}
{\mcitedefaultendpunct}{\mcitedefaultseppunct}\relax
\EndOfBibitem
\bibitem[{Crovisier} et~al.(1997){Crovisier}, {Leech}, {Bockel{\'e}e-Morvan},
  {Brooke}, {Hanner}, {Altieri}, {Keller}, and {Lellouch}]{Crovisier1997}
{Crovisier},~J.; {Leech},~K.; {Bockel{\'e}e-Morvan},~D.; {Brooke},~T.~Y.;
  {Hanner},~M.~S.; {Altieri},~B.; {Keller},~H.~U.; {Lellouch},~E.
  \emph{Science} \textbf{1997}, \emph{275}, 1904\relax
\mciteBstWouldAddEndPuncttrue
\mciteSetBstMidEndSepPunct{\mcitedefaultmidpunct}
{\mcitedefaultendpunct}{\mcitedefaultseppunct}\relax
\EndOfBibitem
\bibitem[{Kelley} et~al.(2006){Kelley}, {Woodward}, {Harker}, {Wooden},
  {Gehrz}, {Campins}, {Hanner}, {Lederer}, {Osip}, {Pittichov{\'a}}, and
  {Polomski}]{2006ApJ...651.1256K}
{Kelley},~M.~S.; {Woodward},~C.~E.; {Harker},~D.~E.; {Wooden},~D.~H.;
  {Gehrz},~R.~D.; {Campins},~H.; {Hanner},~M.~S.; {Lederer},~S.~M.;
  {Osip},~D.~J.; {Pittichov{\'a}},~J.; {Polomski},~E. \emph{Astrophys.~J.,}
  \textbf{2006}, \emph{651}, 1256\relax
\mciteBstWouldAddEndPuncttrue
\mciteSetBstMidEndSepPunct{\mcitedefaultmidpunct}
{\mcitedefaultendpunct}{\mcitedefaultseppunct}\relax
\EndOfBibitem
\bibitem[{Mousis} et~al.(2007){Mousis}, {Petit}, {Wurm}, {Krauss}, {Alibert},
  and {Horner}]{2007A&A...466L...9M}
{Mousis},~O.; {Petit},~J.-M.; {Wurm},~G.; {Krauss},~O.; {Alibert},~Y.;
  {Horner},~J. \emph{Astron.~Astrophys.,} \textbf{2007}, \emph{466}, L9\relax
\mciteBstWouldAddEndPuncttrue
\mciteSetBstMidEndSepPunct{\mcitedefaultmidpunct}
{\mcitedefaultendpunct}{\mcitedefaultseppunct}\relax
\EndOfBibitem
\bibitem[{Wooden} et~al.(2007){Wooden}, {Desch}, {Harker}, {Gail}, and
  {Keller}]{2007prpl.conf..815W}
{Wooden},~D.; {Desch},~S.; {Harker},~D.; {Gail},~H.-P.; {Keller},~L. In
  \emph{Protostars and Planets V}; {Reipurth},~B., {Jewitt},~D., {Keil},~K.,
  Eds.; University of Arizona Press, Tucson, 2007; p 815\relax
\mciteBstWouldAddEndPuncttrue
\mciteSetBstMidEndSepPunct{\mcitedefaultmidpunct}
{\mcitedefaultendpunct}{\mcitedefaultseppunct}\relax
\EndOfBibitem
\bibitem[{Ciesla}(2009)]{2009M&PS...44.1663C}
{Ciesla},~F.~J. \emph{Meteorit. Planet. Sci.,} \textbf{2009}, \emph{44},
  1663\relax
\mciteBstWouldAddEndPuncttrue
\mciteSetBstMidEndSepPunct{\mcitedefaultmidpunct}
{\mcitedefaultendpunct}{\mcitedefaultseppunct}\relax
\EndOfBibitem
\bibitem[{Wozniakiewicz} et~al.(2012){Wozniakiewicz}, {Bradley}, {Ishii},
  {Brownlee}, {Kearsley}, {Burchell}, and {Price}]{2012ApJ...760L..23W}
{Wozniakiewicz},~P.~J.; {Bradley},~J.~P.; {Ishii},~H.~A.; {Brownlee},~D.~E.;
  {Kearsley},~A.~T.; {Burchell},~M.~J.; {Price},~M.~C. \emph{Astrophys.~J.,}
  \textbf{2012}, \emph{760}, L23\relax
\mciteBstWouldAddEndPuncttrue
\mciteSetBstMidEndSepPunct{\mcitedefaultmidpunct}
{\mcitedefaultendpunct}{\mcitedefaultseppunct}\relax
\EndOfBibitem
\bibitem[{Hubbard} et~al.(2012){Hubbard}, {McNally}, and {Mac
  Low}]{2012ApJ...761...58H}
{Hubbard},~A.; {McNally},~C.~P.; {Mac Low},~M.-M. \emph{Astrophys.~J.,}
  \textbf{2012}, \emph{761}, 58\relax
\mciteBstWouldAddEndPuncttrue
\mciteSetBstMidEndSepPunct{\mcitedefaultmidpunct}
{\mcitedefaultendpunct}{\mcitedefaultseppunct}\relax
\EndOfBibitem
\bibitem[{Oliveira} et~al.(2011){Oliveira}, {Olofsson}, {Pontoppidan}, {van
  Dishoeck}, {Augereau}, and {Mer{\'{\i}}n}]{2011ApJ...734...51O}
{Oliveira},~I.; {Olofsson},~J.; {Pontoppidan},~K.~M.; {van Dishoeck},~E.~F.;
  {Augereau},~J.-C.; {Mer{\'{\i}}n},~B. \emph{Astrophys.~J.,} \textbf{2011},
  \emph{734}, 51\relax
\mciteBstWouldAddEndPuncttrue
\mciteSetBstMidEndSepPunct{\mcitedefaultmidpunct}
{\mcitedefaultendpunct}{\mcitedefaultseppunct}\relax
\EndOfBibitem
\bibitem[{Riaz} et~al.(2012){Riaz}, {Honda}, {Campins}, {Micela}, {Guarcello},
  {Gledhill}, {Hough}, and {Mart{\'{\i}}n}]{2012MNRAS.420.2603R}
{Riaz},~B.; {Honda},~M.; {Campins},~H.; {Micela},~G.; {Guarcello},~M.~G.;
  {Gledhill},~T.; {Hough},~J.; {Mart{\'{\i}}n},~E.~L.
  \emph{Mon.~Not.~R.~Astron.~Soc,} \textbf{2012}, \emph{420}, 2603\relax
\mciteBstWouldAddEndPuncttrue
\mciteSetBstMidEndSepPunct{\mcitedefaultmidpunct}
{\mcitedefaultendpunct}{\mcitedefaultseppunct}\relax
\EndOfBibitem
\bibitem[{Hughes} et~al.(2011){Hughes}, {Wilner}, {Andrews}, {Qi}, and
  {Hogerheijde}]{Hughes_ea11a}
{Hughes},~A.~M.; {Wilner},~D.~J.; {Andrews},~S.~M.; {Qi},~C.;
  {Hogerheijde},~M.~R. \emph{Astrophys.~J.,} \textbf{2011}, \emph{727},
  85\relax
\mciteBstWouldAddEndPuncttrue
\mciteSetBstMidEndSepPunct{\mcitedefaultmidpunct}
{\mcitedefaultendpunct}{\mcitedefaultseppunct}\relax
\EndOfBibitem
\bibitem[{Cyr} et~al.(1998){Cyr}, {Sears}, and {Lunine}]{Cyr_ea98}
{Cyr},~K.~E.; {Sears},~W.~D.; {Lunine},~J.~I. \emph{Icarus} \textbf{1998},
  \emph{135}, 537\relax
\mciteBstWouldAddEndPuncttrue
\mciteSetBstMidEndSepPunct{\mcitedefaultmidpunct}
{\mcitedefaultendpunct}{\mcitedefaultseppunct}\relax
\EndOfBibitem
\bibitem[{Xie} et~al.(1995){Xie}, {Allen}, and {Langer}]{Xie_ea95}
{Xie},~T.; {Allen},~M.; {Langer},~W.~D. \emph{Astrophys.~J.,} \textbf{1995},
  \emph{440}, 674\relax
\mciteBstWouldAddEndPuncttrue
\mciteSetBstMidEndSepPunct{\mcitedefaultmidpunct}
{\mcitedefaultendpunct}{\mcitedefaultseppunct}\relax
\EndOfBibitem
\bibitem[{Schr{\"a}pler} and {Henning}(2004){Schr{\"a}pler}, and
  {Henning}]{SchraeplerHenning04}
{Schr{\"a}pler},~R.; {Henning},~T. \emph{Astrophys.~J.,} \textbf{2004},
  \emph{614}, 960\relax
\mciteBstWouldAddEndPuncttrue
\mciteSetBstMidEndSepPunct{\mcitedefaultmidpunct}
{\mcitedefaultendpunct}{\mcitedefaultseppunct}\relax
\EndOfBibitem
\bibitem[{Lissauer} and {Stevenson}(2007){Lissauer}, and
  {Stevenson}]{2007prpl.conf..591L}
{Lissauer},~J.~J.; {Stevenson},~D.~J. In \emph{Protostars and Planets V};
  {Reipurth},~B., {Jewitt},~D., {Keil},~K., Eds.; University of Arizona Press,
  Tucson, 2007; p 591\relax
\mciteBstWouldAddEndPuncttrue
\mciteSetBstMidEndSepPunct{\mcitedefaultmidpunct}
{\mcitedefaultendpunct}{\mcitedefaultseppunct}\relax
\EndOfBibitem
\bibitem[{Kennedy} and {Kenyon}(2008){Kennedy}, and {Kenyon}]{Kennedy_Kenyon08}
{Kennedy},~G.~M.; {Kenyon},~S.~J. \emph{Astrophys.~J.,} \textbf{2008},
  \emph{673}, 502\relax
\mciteBstWouldAddEndPuncttrue
\mciteSetBstMidEndSepPunct{\mcitedefaultmidpunct}
{\mcitedefaultendpunct}{\mcitedefaultseppunct}\relax
\EndOfBibitem
\bibitem[{Bockel{\'e}e-Morvan}(2011)]{2011IAUS..280..261B}
{Bockel{\'e}e-Morvan},~D. In \emph{IAU Symposium 280}; {Cernicharo},~J.,
  {Bachiller},~R., Eds.; Cambridge University Press, Cambridge, 2011; Vol. 280;
  p 261\relax
\mciteBstWouldAddEndPuncttrue
\mciteSetBstMidEndSepPunct{\mcitedefaultmidpunct}
{\mcitedefaultendpunct}{\mcitedefaultseppunct}\relax
\EndOfBibitem
\bibitem[{Drake}(2005)]{2005M&PS...40..519D}
{Drake},~M.~J. \emph{Meteorit. Planet. Sci.,} \textbf{2005}, \emph{40},
  519\relax
\mciteBstWouldAddEndPuncttrue
\mciteSetBstMidEndSepPunct{\mcitedefaultmidpunct}
{\mcitedefaultendpunct}{\mcitedefaultseppunct}\relax
\EndOfBibitem
\bibitem[{Messenger} et~al.(2003){Messenger}, {Stadermann}, {Floss}, {Nittler},
  and {Mukhopadhyay}]{2003SSRv..106..155M}
{Messenger},~S.; {Stadermann},~F.~J.; {Floss},~C.; {Nittler},~L.~R.;
  {Mukhopadhyay},~S. \emph{Space~Sci.~Rev,} \textbf{2003}, \emph{106},
  155\relax
\mciteBstWouldAddEndPuncttrue
\mciteSetBstMidEndSepPunct{\mcitedefaultmidpunct}
{\mcitedefaultendpunct}{\mcitedefaultseppunct}\relax
\EndOfBibitem
\bibitem[{Hartogh} et~al.(2011){Hartogh}, {Lis}, {Bockel{\'e}e-Morvan}, {de
  Val-Borro}, {Biver}, {K{\"u}ppers}, {Emprechtinger}, {Bergin}, {Crovisier},
  {Rengel}, {Moreno}, {Szutowicz}, and {Blake}]{2011Natur.478..218H}
{Hartogh},~P.; {Lis},~D.~C.; {Bockel{\'e}e-Morvan},~D.; {de Val-Borro},~M.;
  {Biver},~N.; {K{\"u}ppers},~M.; {Emprechtinger},~M.; {Bergin},~E.~A.;
  {Crovisier},~J.; {Rengel},~M.; {Moreno},~R.; {Szutowicz},~S.; {Blake},~G.~A.
  \emph{Nature} \textbf{2011}, \emph{478}, 218\relax
\mciteBstWouldAddEndPuncttrue
\mciteSetBstMidEndSepPunct{\mcitedefaultmidpunct}
{\mcitedefaultendpunct}{\mcitedefaultseppunct}\relax
\EndOfBibitem
\bibitem[{Bergin} et~al.(2010){Bergin}, {Hogerheijde}, {Brinch}, {Fogel},
  {Y{\i}ld{\i}z}, {Kristensen}, {van Dishoeck}, {Bell}, {Blake}, {Cernicharo},
  {Dominik}, {Lis}, {Melnick}, {Neufeld}, {Pani{\'c}}, {Pearson}, {Bachiller},
  {Baudry}, {Benedettini}, {Benz}, {Bjerkeli}, {Bontemps}, {Braine},
  {Bruderer}, {Caselli}, {Codella}, {Daniel}, {di Giorgio}, {Doty}, {Encrenaz},
  {Fich}, {Fuente}, {Giannini}, {Goicoechea}, {de Graauw}, {Helmich},
  {Herczeg}, {Herpin}, {Jacq}, {Johnstone}, {J{\o}rgensen}, {Larsson},
  {Liseau}, {Marseille}, {McCoey}, {Nisini}, {Olberg}, {Parise}, {Plume},
  {Risacher}, {Santiago-Garc{\'{\i}}a}, {Saraceno}, { Shipman}, {Tafalla}, {van
  Kempen}, {Visser}, {Wampfler}, {Wyrowski}, {van der Tak}, {Jellema},
  {Tielens}, {Hartogh}, {St{\"u}tzki}, and {Szczerba}]{Bergin_ea10a}
{Bergin},~E.~A. et~al.  \emph{Astron.~Astrophys.,} \textbf{2010}, \emph{521},
  L33\relax
\mciteBstWouldAddEndPuncttrue
\mciteSetBstMidEndSepPunct{\mcitedefaultmidpunct}
{\mcitedefaultendpunct}{\mcitedefaultseppunct}\relax
\EndOfBibitem
\bibitem[{Teske} et~al.(2011){Teske}, {Najita}, {Carr}, {Pascucci}, {Apai}, and
  {Henning}]{2011ApJ...734...27T}
{Teske},~J.~K.; {Najita},~J.~R.; {Carr},~J.~S.; {Pascucci},~I.; {Apai},~D.;
  {Henning},~T. \emph{Astrophys.~J.,} \textbf{2011}, \emph{734}, 27\relax
\mciteBstWouldAddEndPuncttrue
\mciteSetBstMidEndSepPunct{\mcitedefaultmidpunct}
{\mcitedefaultendpunct}{\mcitedefaultseppunct}\relax
\EndOfBibitem
\bibitem[{Banzatti} et~al.(2012){Banzatti}, {Meyer}, {Bruderer}, {Geers},
  {Pascucci}, {Lahuis}, {Juh{\'a}sz}, {Henning}, and
  {{\'A}brah{\'a}m}]{2012ApJ...745...90B}
{Banzatti},~A.; {Meyer},~M.~R.; {Bruderer},~S.; {Geers},~V.; {Pascucci},~I.;
  {Lahuis},~F.; {Juh{\'a}sz},~A.; {Henning},~T.; {{\'A}brah{\'a}m},~P.
  \emph{Astrophys.~J.,} \textbf{2012}, \emph{745}, 90\relax
\mciteBstWouldAddEndPuncttrue
\mciteSetBstMidEndSepPunct{\mcitedefaultmidpunct}
{\mcitedefaultendpunct}{\mcitedefaultseppunct}\relax
\EndOfBibitem
\bibitem[{Elitzur} and {de Jong}(1978){Elitzur}, and {de
  Jong}]{1978A&A....67..323E}
{Elitzur},~M.; {de Jong},~T. \emph{Astron.~Astrophys.,} \textbf{1978},
  \emph{67}, 323\relax
\mciteBstWouldAddEndPuncttrue
\mciteSetBstMidEndSepPunct{\mcitedefaultmidpunct}
{\mcitedefaultendpunct}{\mcitedefaultseppunct}\relax
\EndOfBibitem
\bibitem[{Kaufman} and {Neufeld}(1996){Kaufman}, and
  {Neufeld}]{1996ApJ...456..250K}
{Kaufman},~M.~J.; {Neufeld},~D.~A. \emph{Astrophys.~J.,} \textbf{1996},
  \emph{456}, 250\relax
\mciteBstWouldAddEndPuncttrue
\mciteSetBstMidEndSepPunct{\mcitedefaultmidpunct}
{\mcitedefaultendpunct}{\mcitedefaultseppunct}\relax
\EndOfBibitem
\bibitem[{Wagner} and {Graff}(1987){Wagner}, and {Graff}]{1987ApJ...317..423W}
{Wagner},~A.~F.; {Graff},~M.~M. \emph{Astrophys.~J.,} \textbf{1987},
  \emph{317}, 423\relax
\mciteBstWouldAddEndPuncttrue
\mciteSetBstMidEndSepPunct{\mcitedefaultmidpunct}
{\mcitedefaultendpunct}{\mcitedefaultseppunct}\relax
\EndOfBibitem
\bibitem[{Kamp} et~al.(2013){Kamp}, {Thi}, {Meeus}, {Woitke}, {Pinte},
  {Meijerink}, {Spaans}, {Pascucci}, {Aresu}, and {Dent}]{2013arXiv1308.1772K}
{Kamp},~I.; {Thi},~W.-F.; {Meeus},~G.; {Woitke},~P.; {Pinte},~C.;
  {Meijerink},~R.; {Spaans},~M.; {Pascucci},~I.; {Aresu},~G.; {Dent},~W.~R.~F.
  \emph{ArXiv e-prints} \textbf{2013}, \relax
\mciteBstWouldAddEndPunctfalse
\mciteSetBstMidEndSepPunct{\mcitedefaultmidpunct}
{}{\mcitedefaultseppunct}\relax
\EndOfBibitem
\bibitem[{Mumma} et~al.(1987){Mumma}, {Weaver}, and {Larson}]{Mumma_ea87}
{Mumma},~M.~J.; {Weaver},~H.~A.; {Larson},~H.~P. \emph{Astron.~Astrophys.,}
  \textbf{1987}, \emph{187}, 419\relax
\mciteBstWouldAddEndPuncttrue
\mciteSetBstMidEndSepPunct{\mcitedefaultmidpunct}
{\mcitedefaultendpunct}{\mcitedefaultseppunct}\relax
\EndOfBibitem
\bibitem[{Woodward} et~al.(2007){Woodward}, {Kelley}, {Bockel{\'e}e-Morvan},
  and {Gehrz}]{Woodward_ea07}
{Woodward},~C.~E.; {Kelley},~M.~S.; {Bockel{\'e}e-Morvan},~D.; {Gehrz},~R.~D.
  \emph{Astrophys.~J.,} \textbf{2007}, \emph{671}, 1065\relax
\mciteBstWouldAddEndPuncttrue
\mciteSetBstMidEndSepPunct{\mcitedefaultmidpunct}
{\mcitedefaultendpunct}{\mcitedefaultseppunct}\relax
\EndOfBibitem
\bibitem[{Bockel{\'e}e-Morvan} et~al.(2009){Bockel{\'e}e-Morvan}, {Woodward},
  {Kelley}, and {Wooden}]{BM_ea09}
{Bockel{\'e}e-Morvan},~D.; {Woodward},~C.~E.; {Kelley},~M.~S.; {Wooden},~D.~H.
  \emph{Astrophys.~J.,} \textbf{2009}, \emph{696}, 1075\relax
\mciteBstWouldAddEndPuncttrue
\mciteSetBstMidEndSepPunct{\mcitedefaultmidpunct}
{\mcitedefaultendpunct}{\mcitedefaultseppunct}\relax
\EndOfBibitem
\bibitem[{Shinnaka} et~al.(2012){Shinnaka}, {Kawakita}, {Kobayashi}, {Boice},
  and {Martinez}]{Shinnaka_ea12}
{Shinnaka},~Y.; {Kawakita},~H.; {Kobayashi},~H.; {Boice},~D.~C.;
  {Martinez},~S.~E. \emph{Astrophys.~J.,} \textbf{2012}, \emph{749}, 101\relax
\mciteBstWouldAddEndPuncttrue
\mciteSetBstMidEndSepPunct{\mcitedefaultmidpunct}
{\mcitedefaultendpunct}{\mcitedefaultseppunct}\relax
\EndOfBibitem
\bibitem[{Bonev} et~al.(2013){Bonev}, {Villanueva}, {Paganini}, {DiSanti},
  {Gibb}, {Keane}, {Meech}, and {Mumma}]{Bonev_ea13a}
{Bonev},~B.~P.; {Villanueva},~G.~L.; {Paganini},~L.; {DiSanti},~M.~A.;
  {Gibb},~E.~L.; {Keane},~J.~V.; {Meech},~K.~J.; {Mumma},~M.~J. \emph{Icarus}
  \textbf{2013}, \emph{222}, 740\relax
\mciteBstWouldAddEndPuncttrue
\mciteSetBstMidEndSepPunct{\mcitedefaultmidpunct}
{\mcitedefaultendpunct}{\mcitedefaultseppunct}\relax
\EndOfBibitem
\bibitem[{Alexander} et~al.(1998){Alexander}, {Russell}, {Arden}, {Ash},
  {Grady}, and {Pillinger}]{1998M&PS...33..603A}
{Alexander},~C.~M.~O.; {Russell},~S.~S.; {Arden},~J.~W.; {Ash},~R.~D.;
  {Grady},~M.~M.; {Pillinger},~C.~T. \emph{Meteorit. Planet. Sci.,}
  \textbf{1998}, \emph{33}, 603\relax
\mciteBstWouldAddEndPuncttrue
\mciteSetBstMidEndSepPunct{\mcitedefaultmidpunct}
{\mcitedefaultendpunct}{\mcitedefaultseppunct}\relax
\EndOfBibitem
\bibitem[{Clayton}(1993)]{Clayton_93}
{Clayton},~R.~N. \emph{Annu. Rev. Earth Planet. Sci.,} \textbf{1993},
  \emph{21}, 115\relax
\mciteBstWouldAddEndPuncttrue
\mciteSetBstMidEndSepPunct{\mcitedefaultmidpunct}
{\mcitedefaultendpunct}{\mcitedefaultseppunct}\relax
\EndOfBibitem
\bibitem[{Clayton}(2007)]{Clayton2007}
{Clayton},~R.~N. \emph{Annu. Rev. Earth Planet. Sci.,} \textbf{2007},
  \emph{35}, 1\relax
\mciteBstWouldAddEndPuncttrue
\mciteSetBstMidEndSepPunct{\mcitedefaultmidpunct}
{\mcitedefaultendpunct}{\mcitedefaultseppunct}\relax
\EndOfBibitem
\bibitem[{Glavin} et~al.(2010){Glavin}, {Aubrey}, {Callahan}, {Dworkin},
  {Elsila}, {Parker}, {Bada}, {Jenniskens}, and {Shaddad}]{Glavin_ea10}
{Glavin},~D.~P.; {Aubrey},~A.~D.; {Callahan},~M.~P.; {Dworkin},~J.~P.;
  {Elsila},~J.~E.; {Parker},~E.~T.; {Bada},~J.~L.; {Jenniskens},~P.;
  {Shaddad},~M.~H. \emph{Meteorit. Planet. Sci.,} \textbf{2010}, \emph{45},
  1695\relax
\mciteBstWouldAddEndPuncttrue
\mciteSetBstMidEndSepPunct{\mcitedefaultmidpunct}
{\mcitedefaultendpunct}{\mcitedefaultseppunct}\relax
\EndOfBibitem
\bibitem[{Elsila} et~al.(2012){Elsila}, {Charnley}, {Burton}, {Glavin}, and
  {Dworkin}]{2012M&PS..tmp..201E}
{Elsila},~J.~E.; {Charnley},~S.~B.; {Burton},~A.~S.; {Glavin},~D.~P.;
  {Dworkin},~J.~P. \emph{Meteorit. Planet. Sci.,} \textbf{2012}, 201\relax
\mciteBstWouldAddEndPuncttrue
\mciteSetBstMidEndSepPunct{\mcitedefaultmidpunct}
{\mcitedefaultendpunct}{\mcitedefaultseppunct}\relax
\EndOfBibitem
\bibitem[{Elsila} et~al.(2009){Elsila}, {Glavin}, and {Dworkin}]{Elsila_ea09}
{Elsila},~J.~E.; {Glavin},~D.~P.; {Dworkin},~J.~P. \emph{Meteorit. Planet.
  Sci.,} \textbf{2009}, \emph{44}, 1323\relax
\mciteBstWouldAddEndPuncttrue
\mciteSetBstMidEndSepPunct{\mcitedefaultmidpunct}
{\mcitedefaultendpunct}{\mcitedefaultseppunct}\relax
\EndOfBibitem
\bibitem[{Pettini} and {Cooke}(2012){Pettini}, and
  {Cooke}]{2012MNRAS.425.2477P}
{Pettini},~M.; {Cooke},~R. \emph{Mon.~Not.~R.~Astron.~Soc,} \textbf{2012},
  \emph{425}, 2477\relax
\mciteBstWouldAddEndPuncttrue
\mciteSetBstMidEndSepPunct{\mcitedefaultmidpunct}
{\mcitedefaultendpunct}{\mcitedefaultseppunct}\relax
\EndOfBibitem
\bibitem[{Linsky} et~al.(2006){Linsky}, {Draine}, {Moos}, {Jenkins}, {Wood},
  {Oliveira}, {Blair}, {Friedman}, {Gry}, {Knauth}, {Kruk}, {Lacour}, {Lehner},
  {Redfield}, {Shull}, {Sonneborn}, and {Williger}]{2006ApJ...647.1106L}
{Linsky},~J.~L. et~al.  \emph{Astrophys.~J.,} \textbf{2006}, \emph{647},
  1106\relax
\mciteBstWouldAddEndPuncttrue
\mciteSetBstMidEndSepPunct{\mcitedefaultmidpunct}
{\mcitedefaultendpunct}{\mcitedefaultseppunct}\relax
\EndOfBibitem
\bibitem[{Roberts} et~al.(2002){Roberts}, {Fuller}, {Millar}, {Hatchell}, and
  {Buckle}]{2002P&SS...50.1173R}
{Roberts},~H.; {Fuller},~G.~A.; {Millar},~T.~J.; {Hatchell},~J.;
  {Buckle},~J.~V. \emph{Planet.~Space~Sci.,} \textbf{2002}, \emph{50},
  1173\relax
\mciteBstWouldAddEndPuncttrue
\mciteSetBstMidEndSepPunct{\mcitedefaultmidpunct}
{\mcitedefaultendpunct}{\mcitedefaultseppunct}\relax
\EndOfBibitem
\bibitem[{Bacmann} et~al.(2003){Bacmann}, {Lefloch}, {Ceccarelli},
  {Steinacker}, {Castets}, and {Loinard}]{2003ApJ...585L..55B}
{Bacmann},~A.; {Lefloch},~B.; {Ceccarelli},~C.; {Steinacker},~J.;
  {Castets},~A.; {Loinard},~L. \emph{Astrophys.~J.,} \textbf{2003}, \emph{585},
  L55\relax
\mciteBstWouldAddEndPuncttrue
\mciteSetBstMidEndSepPunct{\mcitedefaultmidpunct}
{\mcitedefaultendpunct}{\mcitedefaultseppunct}\relax
\EndOfBibitem
\bibitem[{Vastel} et~al.(2006){Vastel}, {Caselli}, {Ceccarelli}, {Phillips},
  {Wiedner}, {Peng}, {Houde}, and {Dominik}]{2006ApJ...645.1198V}
{Vastel},~C.; {Caselli},~P.; {Ceccarelli},~C.; {Phillips},~T.;
  {Wiedner},~M.~C.; {Peng},~R.; {Houde},~M.; {Dominik},~C.
  \emph{Astrophys.~J.,} \textbf{2006}, \emph{645}, 1198\relax
\mciteBstWouldAddEndPuncttrue
\mciteSetBstMidEndSepPunct{\mcitedefaultmidpunct}
{\mcitedefaultendpunct}{\mcitedefaultseppunct}\relax
\EndOfBibitem
\bibitem[{Irvine} et~al.(2000){Irvine}, {Schloerb}, {Crovisier}, {Fegley}, and
  {Mumma}]{2000prpl.conf.1159I}
{Irvine},~W.~M.; {Schloerb},~F.~P.; {Crovisier},~J.; {Fegley},~B.,~Jr.;
  {Mumma},~M.~J. In \emph{Protostars and Planets IV}; {Mannings},~V.,
  {Boss},~A.~P., {Russell},~S.~S., Eds.; University of Arizona Press, Tucson,
  2000; p 1159\relax
\mciteBstWouldAddEndPuncttrue
\mciteSetBstMidEndSepPunct{\mcitedefaultmidpunct}
{\mcitedefaultendpunct}{\mcitedefaultseppunct}\relax
\EndOfBibitem
\bibitem[{Qi} et~al.(2008){Qi}, {Wilner}, {Aikawa}, {Blake}, and
  {Hogerheijde}]{Qi_ea08}
{Qi},~C.; {Wilner},~D.~J.; {Aikawa},~Y.; {Blake},~G.~A.; {Hogerheijde},~M.~R.
  \emph{Astrophys.~J.,} \textbf{2008}, \emph{681}, 1396\relax
\mciteBstWouldAddEndPuncttrue
\mciteSetBstMidEndSepPunct{\mcitedefaultmidpunct}
{\mcitedefaultendpunct}{\mcitedefaultseppunct}\relax
\EndOfBibitem
\bibitem[{Millar} et~al.(1989){Millar}, {Bennett}, and {Herbst}]{Millarea89}
{Millar},~T.-J.; {Bennett},~A.; {Herbst},~E. \emph{Astrophys.~J.,}
  \textbf{1989}, \emph{340}, 906\relax
\mciteBstWouldAddEndPuncttrue
\mciteSetBstMidEndSepPunct{\mcitedefaultmidpunct}
{\mcitedefaultendpunct}{\mcitedefaultseppunct}\relax
\EndOfBibitem
\bibitem[{Gerlich} et~al.(2002){Gerlich}, {Herbst}, and {Roueff}]{GHR_02}
{Gerlich},~D.; {Herbst},~E.; {Roueff},~E. \emph{Planet.~Space~Sci.,}
  \textbf{2002}, \emph{50}, 1275\relax
\mciteBstWouldAddEndPuncttrue
\mciteSetBstMidEndSepPunct{\mcitedefaultmidpunct}
{\mcitedefaultendpunct}{\mcitedefaultseppunct}\relax
\EndOfBibitem
\bibitem[{Roberts} and {Millar}(2000){Roberts}, and {Millar}]{RobertsMillar00}
{Roberts},~H.; {Millar},~T.~J. \emph{Astron.~Astrophys.,} \textbf{2000},
  \emph{361}, 388\relax
\mciteBstWouldAddEndPuncttrue
\mciteSetBstMidEndSepPunct{\mcitedefaultmidpunct}
{\mcitedefaultendpunct}{\mcitedefaultseppunct}\relax
\EndOfBibitem
\bibitem[{Roberts} et~al.(2003){Roberts}, {Herbst}, and {Millar}]{Robertsea03}
{Roberts},~H.; {Herbst},~E.; {Millar},~T.~J. \emph{Astrophys.~J.,}
  \textbf{2003}, \emph{591}, L41\relax
\mciteBstWouldAddEndPuncttrue
\mciteSetBstMidEndSepPunct{\mcitedefaultmidpunct}
{\mcitedefaultendpunct}{\mcitedefaultseppunct}\relax
\EndOfBibitem
\bibitem[{Albertsson} et~al.(2013){Albertsson}, {Semenov}, {Vasyunin},
  {Henning}, and {Herbst}]{2011arXiv1110.2644A}
{Albertsson},~T.; {Semenov},~D.~A.; {Vasyunin},~A.~I.; {Henning},~T.;
  {Herbst},~E. \emph{Astrophys.~J.,~Suppl.~Ser.,} \textbf{2013}, \emph{207},
  27\relax
\mciteBstWouldAddEndPuncttrue
\mciteSetBstMidEndSepPunct{\mcitedefaultmidpunct}
{\mcitedefaultendpunct}{\mcitedefaultseppunct}\relax
\EndOfBibitem
\bibitem[{Nagaoka} et~al.(2005){Nagaoka}, {Watanabe}, and
  {Kouchi}]{Nagaoka_ea05}
{Nagaoka},~A.; {Watanabe},~N.; {Kouchi},~A. \emph{Astrophys.~J.,~Lett.,}
  \textbf{2005}, \emph{624}, L29\relax
\mciteBstWouldAddEndPuncttrue
\mciteSetBstMidEndSepPunct{\mcitedefaultmidpunct}
{\mcitedefaultendpunct}{\mcitedefaultseppunct}\relax
\EndOfBibitem
\bibitem[{Watanabe}(2005)]{2005IAUS..231..415W}
{Watanabe},~N. In \emph{Astrochemistry: Recent Successes and Current
  Challenges}; {D.~C.~Lis, G.~A.~Blake, \& E.~Herbst},, Ed.; IAU Symposium 231;
  Cambridge University Press, Cambridge, 2005; Vol. 231; p 415\relax
\mciteBstWouldAddEndPuncttrue
\mciteSetBstMidEndSepPunct{\mcitedefaultmidpunct}
{\mcitedefaultendpunct}{\mcitedefaultseppunct}\relax
\EndOfBibitem
\bibitem[{Hiraoka} et~al.(2006){Hiraoka}, {Mochizuki}, and
  {Wada}]{2006AIPC..855...86H}
{Hiraoka},~K.; {Mochizuki},~N.; {Wada},~A. In \emph{Astrochemistry - From
  Laboratory Studies to Astronomical Observations}; {R.~I.~Kaiser, P.~Bernath,
  Y.~Osamura, S.~Petrie, \& A.~M.~Mebel},, Ed.; American Institute of Physics
  Conference Series; American Institute of Physics, Melville, New York, 2006;
  Vol. 855; p~86\relax
\mciteBstWouldAddEndPuncttrue
\mciteSetBstMidEndSepPunct{\mcitedefaultmidpunct}
{\mcitedefaultendpunct}{\mcitedefaultseppunct}\relax
\EndOfBibitem
\bibitem[{Hidaka} et~al.(2006){Hidaka}, {Watanabe}, and
  {Kouchi}]{2006AIPC..855..107H}
{Hidaka},~H.; {Watanabe},~N.; {Kouchi},~A. In \emph{Astrochemistry - From
  Laboratory Studies to Astronomical Observations}; {R.~I.~Kaiser, P.~Bernath,
  Y.~Osamura, S.~Petrie, \& A.~M.~Mebel},, Ed.; American Institute of Physics
  Conference Series; American Institute of Physics, Melville, New York, 2006;
  Vol. 855; p 107\relax
\mciteBstWouldAddEndPuncttrue
\mciteSetBstMidEndSepPunct{\mcitedefaultmidpunct}
{\mcitedefaultendpunct}{\mcitedefaultseppunct}\relax
\EndOfBibitem
\bibitem[{Flower} et~al.(2006){Flower}, {Pineau Des For{\^e}ts}, and
  {Walmsley}]{2006A&A...449..621F}
{Flower},~D.~R.; {Pineau Des For{\^e}ts},~G.; {Walmsley},~C.~M.
  \emph{Astron.~Astrophys.,} \textbf{2006}, \emph{449}, 621\relax
\mciteBstWouldAddEndPuncttrue
\mciteSetBstMidEndSepPunct{\mcitedefaultmidpunct}
{\mcitedefaultendpunct}{\mcitedefaultseppunct}\relax
\EndOfBibitem
\bibitem[{Pagani} et~al.(2009){Pagani}, {Vastel}, {Hugo}, {Kokoouline},
  {Greene}, {Bacmann}, {Bayet}, {Ceccarelli}, {Peng}, and
  {Schlemmer}]{2009A&A...494..623P}
{Pagani},~L.; {Vastel},~C.; {Hugo},~E.; {Kokoouline},~V.; {Greene},~C.~H.;
  {Bacmann},~A.; {Bayet},~E.; {Ceccarelli},~C.; {Peng},~R.; {Schlemmer},~S.
  \emph{Astron.~Astrophys.,} \textbf{2009}, \emph{494}, 623\relax
\mciteBstWouldAddEndPuncttrue
\mciteSetBstMidEndSepPunct{\mcitedefaultmidpunct}
{\mcitedefaultendpunct}{\mcitedefaultseppunct}\relax
\EndOfBibitem
\bibitem[{Asvany} et~al.(2004){Asvany}, {Schlemmer}, and
  {Gerlich}]{Asvany_ea04}
{Asvany},~O.; {Schlemmer},~S.; {Gerlich},~D. \emph{Astrophys.~J.,}
  \textbf{2004}, \emph{617}, 685\relax
\mciteBstWouldAddEndPuncttrue
\mciteSetBstMidEndSepPunct{\mcitedefaultmidpunct}
{\mcitedefaultendpunct}{\mcitedefaultseppunct}\relax
\EndOfBibitem
\bibitem[{Herbst} et~al.(1987){Herbst}, {Adams}, {Smith}, and
  {Defrees}]{Herbst_ea87}
{Herbst},~E.; {Adams},~N.~G.; {Smith},~D.; {Defrees},~D.~J.
  \emph{Astrophys.~J.,} \textbf{1987}, \emph{312}, 351\relax
\mciteBstWouldAddEndPuncttrue
\mciteSetBstMidEndSepPunct{\mcitedefaultmidpunct}
{\mcitedefaultendpunct}{\mcitedefaultseppunct}\relax
\EndOfBibitem
\bibitem[{Roueff} et~al.(2007){Roueff}, {Parise}, and {Herbst}]{Roueff_ea07}
{Roueff},~E.; {Parise},~B.; {Herbst},~E. \emph{Astron.~Astrophys.,}
  \textbf{2007}, \emph{464}, 245\relax
\mciteBstWouldAddEndPuncttrue
\mciteSetBstMidEndSepPunct{\mcitedefaultmidpunct}
{\mcitedefaultendpunct}{\mcitedefaultseppunct}\relax
\EndOfBibitem
\bibitem[{Henning} and {Salama}(1998){Henning}, and
  {Salama}]{1998Sci...282.2204H}
{Henning},~T.; {Salama},~F. \emph{Science} \textbf{1998}, \emph{282},
  2204\relax
\mciteBstWouldAddEndPuncttrue
\mciteSetBstMidEndSepPunct{\mcitedefaultmidpunct}
{\mcitedefaultendpunct}{\mcitedefaultseppunct}\relax
\EndOfBibitem
\bibitem[{Alexander} et~al.(2007){Alexander}, {Boss}, {Keller}, {Nuth}, and
  {Weinberger}]{2007prpl.conf..801A}
{Alexander},~C.~M.~O.; {Boss},~A.~P.; {Keller},~L.~P.; {Nuth},~J.~A.;
  {Weinberger},~A. In \emph{Protostars and Planets V}; {Reipurth},~B.,
  {Jewitt},~D., {Keil},~K., Eds.; University of Arizona Press, Tucson, 2007; p
  801\relax
\mciteBstWouldAddEndPuncttrue
\mciteSetBstMidEndSepPunct{\mcitedefaultmidpunct}
{\mcitedefaultendpunct}{\mcitedefaultseppunct}\relax
\EndOfBibitem
\bibitem[{Tielens}(2010)]{2010pcim.book.....T}
{Tielens},~A.~G.~G.~M. \emph{The Physics and Chemistry of the Interstellar
  Medium}; Cambridge University Press, Cambridge, 2010\relax
\mciteBstWouldAddEndPuncttrue
\mciteSetBstMidEndSepPunct{\mcitedefaultmidpunct}
{\mcitedefaultendpunct}{\mcitedefaultseppunct}\relax
\EndOfBibitem
\bibitem[{Schnaiter} et~al.(1999){Schnaiter}, {Henning}, {Mutschke}, {Kohn},
  {Ehbrecht}, and {Huisken}]{1999ApJ...519..687S}
{Schnaiter},~M.; {Henning},~T.; {Mutschke},~H.; {Kohn},~B.; {Ehbrecht},~M.;
  {Huisken},~F. \emph{Astrophys.~J.,} \textbf{1999}, \emph{519}, 687\relax
\mciteBstWouldAddEndPuncttrue
\mciteSetBstMidEndSepPunct{\mcitedefaultmidpunct}
{\mcitedefaultendpunct}{\mcitedefaultseppunct}\relax
\EndOfBibitem
\bibitem[{Fitzpatrick} and {Massa}(1988){Fitzpatrick}, and
  {Massa}]{1988ApJ...328..734F}
{Fitzpatrick},~E.~L.; {Massa},~D. \emph{Astrophys.~J.,} \textbf{1988},
  \emph{328}, 734\relax
\mciteBstWouldAddEndPuncttrue
\mciteSetBstMidEndSepPunct{\mcitedefaultmidpunct}
{\mcitedefaultendpunct}{\mcitedefaultseppunct}\relax
\EndOfBibitem
\bibitem[{Schnaiter} et~al.(1998){Schnaiter}, {Mutschke}, {Dorschner},
  {Henning}, and {Salama}]{1998ApJ...498..486S}
{Schnaiter},~M.; {Mutschke},~H.; {Dorschner},~J.; {Henning},~T.; {Salama},~F.
  \emph{Astrophys.~J.,} \textbf{1998}, \emph{498}, 486\relax
\mciteBstWouldAddEndPuncttrue
\mciteSetBstMidEndSepPunct{\mcitedefaultmidpunct}
{\mcitedefaultendpunct}{\mcitedefaultseppunct}\relax
\EndOfBibitem
\bibitem[{Steglich} et~al.(2011){Steglich}, {Bouwman}, {Huisken}, and
  {Henning}]{2011ApJ...742....2S}
{Steglich},~M.; {Bouwman},~J.; {Huisken},~F.; {Henning},~T.
  \emph{Astrophys.~J.,} \textbf{2011}, \emph{742}, 2\relax
\mciteBstWouldAddEndPuncttrue
\mciteSetBstMidEndSepPunct{\mcitedefaultmidpunct}
{\mcitedefaultendpunct}{\mcitedefaultseppunct}\relax
\EndOfBibitem
\bibitem[{Duley} and {Hu}(2012){Duley}, and {Hu}]{2012ApJ...761..115D}
{Duley},~W.~W.; {Hu},~A. \emph{Astrophys.~J.,} \textbf{2012}, \emph{761},
  115\relax
\mciteBstWouldAddEndPuncttrue
\mciteSetBstMidEndSepPunct{\mcitedefaultmidpunct}
{\mcitedefaultendpunct}{\mcitedefaultseppunct}\relax
\EndOfBibitem
\bibitem[{Allamandola} et~al.(1989){Allamandola}, {Tielens}, and
  {Barker}]{1989ApJS...71..733A}
{Allamandola},~L.~J.; {Tielens},~A.~G.~G.~M.; {Barker},~J.~R.
  \emph{Astrophys.~J.,~Suppl.~Ser.,} \textbf{1989}, \emph{71}, 733\relax
\mciteBstWouldAddEndPuncttrue
\mciteSetBstMidEndSepPunct{\mcitedefaultmidpunct}
{\mcitedefaultendpunct}{\mcitedefaultseppunct}\relax
\EndOfBibitem
\bibitem[{Tielens}(2008)]{2008ARA&A..46..289T}
{Tielens},~A.~G.~G.~M. \emph{Ann.~Rev.~Astron.~Astrophys.,} \textbf{2008},
  \emph{46}, 289\relax
\mciteBstWouldAddEndPuncttrue
\mciteSetBstMidEndSepPunct{\mcitedefaultmidpunct}
{\mcitedefaultendpunct}{\mcitedefaultseppunct}\relax
\EndOfBibitem
\bibitem[{Bernatowicz} et~al.(1987){Bernatowicz}, {Fraundorf}, {Ming},
  {Anders}, {Wopenka}, {Zinner}, and {Fraundorf}]{1987Natur.330..728B}
{Bernatowicz},~T.; {Fraundorf},~G.; {Ming},~T.; {Anders},~E.; {Wopenka},~B.;
  {Zinner},~E.; {Fraundorf},~P. \emph{Nature} \textbf{1987}, \emph{330},
  728\relax
\mciteBstWouldAddEndPuncttrue
\mciteSetBstMidEndSepPunct{\mcitedefaultmidpunct}
{\mcitedefaultendpunct}{\mcitedefaultseppunct}\relax
\EndOfBibitem
\bibitem[{Henning} and {Mutschke}(2001){Henning}, and
  {Mutschke}]{2001AcSpe..57..815H}
{Henning},~T.; {Mutschke},~H. \emph{Spectrochim. Acta} \textbf{2001},
  \emph{57}, 815\relax
\mciteBstWouldAddEndPuncttrue
\mciteSetBstMidEndSepPunct{\mcitedefaultmidpunct}
{\mcitedefaultendpunct}{\mcitedefaultseppunct}\relax
\EndOfBibitem
\bibitem[{Matrajt} et~al.(2012){Matrajt}, {Messenger}, {Brownlee}, and
  {Joswiak}]{2012M&PS...47..525M}
{Matrajt},~G.; {Messenger},~S.; {Brownlee},~D.; {Joswiak},~D. \emph{Meteorit.
  Planet. Sci.,} \textbf{2012}, \emph{47}, 525\relax
\mciteBstWouldAddEndPuncttrue
\mciteSetBstMidEndSepPunct{\mcitedefaultmidpunct}
{\mcitedefaultendpunct}{\mcitedefaultseppunct}\relax
\EndOfBibitem
\bibitem[{Jessberger} et~al.(1988){Jessberger}, {Christoforidis}, and
  {Kissel}]{JCK88}
{Jessberger},~E.~K.; {Christoforidis},~A.; {Kissel},~J. \emph{Nature}
  \textbf{1988}, \emph{332}, 691\relax
\mciteBstWouldAddEndPuncttrue
\mciteSetBstMidEndSepPunct{\mcitedefaultmidpunct}
{\mcitedefaultendpunct}{\mcitedefaultseppunct}\relax
\EndOfBibitem
\bibitem[{Pizzarello} and {Huang}(2005){Pizzarello}, and
  {Huang}]{2005GeCoA..69..599P}
{Pizzarello},~S.; {Huang},~Y. \emph{Geochim.~Cosmochim.~Acta,} \textbf{2005},
  \emph{69}, 599\relax
\mciteBstWouldAddEndPuncttrue
\mciteSetBstMidEndSepPunct{\mcitedefaultmidpunct}
{\mcitedefaultendpunct}{\mcitedefaultseppunct}\relax
\EndOfBibitem
\bibitem[{Caselli} and {Ceccarelli}(2012){Caselli}, and
  {Ceccarelli}]{2012A&ARv..20...56C}
{Caselli},~P.; {Ceccarelli},~C. \emph{Astron.~Astrophys.~Rev,} \textbf{2012},
  \emph{20}, 56\relax
\mciteBstWouldAddEndPuncttrue
\mciteSetBstMidEndSepPunct{\mcitedefaultmidpunct}
{\mcitedefaultendpunct}{\mcitedefaultseppunct}\relax
\EndOfBibitem
\bibitem[{Parise} et~al.(2004){Parise}, {Castets}, {Herbst}, {Caux},
  {Ceccarelli}, {Mukhopadhyay}, and {Tielens}]{CD3OH}
{Parise},~B.; {Castets},~A.; {Herbst},~E.; {Caux},~E.; {Ceccarelli},~C.;
  {Mukhopadhyay},~I.; {Tielens},~A.~G.~G.~M. \emph{Astron.~Astrophys.,}
  \textbf{2004}, \emph{416}, 159\relax
\mciteBstWouldAddEndPuncttrue
\mciteSetBstMidEndSepPunct{\mcitedefaultmidpunct}
{\mcitedefaultendpunct}{\mcitedefaultseppunct}\relax
\EndOfBibitem
\bibitem[{Bacmann} et~al.(2012){Bacmann}, {Taquet}, {Faure}, {Kahane}, and
  {Ceccarelli}]{2012A&A...541L..12B}
{Bacmann},~A.; {Taquet},~V.; {Faure},~A.; {Kahane},~C.; {Ceccarelli},~C.
  \emph{Astron.~Astrophys.,} \textbf{2012}, \emph{541}, L12\relax
\mciteBstWouldAddEndPuncttrue
\mciteSetBstMidEndSepPunct{\mcitedefaultmidpunct}
{\mcitedefaultendpunct}{\mcitedefaultseppunct}\relax
\EndOfBibitem
\bibitem[{Hollis} et~al.(2004){Hollis}, {Jewell}, {Lovas}, and
  {Remijan}]{Hollis_ea04}
{Hollis},~J.~M.; {Jewell},~P.~R.; {Lovas},~F.~J.; {Remijan},~A.
  \emph{Astrophys.~J.,} \textbf{2004}, \emph{613}, L45\relax
\mciteBstWouldAddEndPuncttrue
\mciteSetBstMidEndSepPunct{\mcitedefaultmidpunct}
{\mcitedefaultendpunct}{\mcitedefaultseppunct}\relax
\EndOfBibitem
\bibitem[{J{\o}rgensen} et~al.(2012){J{\o}rgensen}, {Favre}, {Bisschop},
  {Bourke}, {van Dishoeck}, and {Schmalzl}]{2012ApJ...757L...4J}
{J{\o}rgensen},~J.~K.; {Favre},~C.; {Bisschop},~S.~E.; {Bourke},~T.~L.; {van
  Dishoeck},~E.~F.; {Schmalzl},~M. \emph{Astrophys.~J.~Lett.,} \textbf{2012},
  \emph{757}, L4\relax
\mciteBstWouldAddEndPuncttrue
\mciteSetBstMidEndSepPunct{\mcitedefaultmidpunct}
{\mcitedefaultendpunct}{\mcitedefaultseppunct}\relax
\EndOfBibitem
\bibitem[{Kahane} et~al.(2013){Kahane}, {Ceccarelli}, {Faure}, and
  {Caux}]{2013ApJ...763L..38K}
{Kahane},~C.; {Ceccarelli},~C.; {Faure},~A.; {Caux},~E.
  \emph{Astrophys.~J.~Lett.,} \textbf{2013}, \emph{763}, L38\relax
\mciteBstWouldAddEndPuncttrue
\mciteSetBstMidEndSepPunct{\mcitedefaultmidpunct}
{\mcitedefaultendpunct}{\mcitedefaultseppunct}\relax
\EndOfBibitem
\bibitem[{Hassel}(2004)]{Hassel_04}
{Hassel},~G.~E.,~Jr. {Shock processing of icy grain mantles in protoplanetary
  disks}. Ph.D.\ thesis, Rensselaer Polytechnic Institute, 110 8th St Troy, NY
  12180, USA, 2004\relax
\mciteBstWouldAddEndPuncttrue
\mciteSetBstMidEndSepPunct{\mcitedefaultmidpunct}
{\mcitedefaultendpunct}{\mcitedefaultseppunct}\relax
\EndOfBibitem
\bibitem[{d'Hendecourt} et~al.(1982){d'Hendecourt}, {Allamandola}, {Baas}, and
  {Greenberg}]{dHendecourtea82}
{d'Hendecourt},~L.~B.; {Allamandola},~L.~J.; {Baas},~F.; {Greenberg},~J.~M.
  \emph{Astron.~Astrophys.,} \textbf{1982}, \emph{109}, L12\relax
\mciteBstWouldAddEndPuncttrue
\mciteSetBstMidEndSepPunct{\mcitedefaultmidpunct}
{\mcitedefaultendpunct}{\mcitedefaultseppunct}\relax
\EndOfBibitem
\bibitem[{Shalabiea} and {Greenberg}(1994){Shalabiea}, and
  {Greenberg}]{ShalabeiaGreenberg94}
{Shalabiea},~O.~M.; {Greenberg},~J.~M. \emph{Astron.~Astrophys.,}
  \textbf{1994}, \emph{290}, 266\relax
\mciteBstWouldAddEndPuncttrue
\mciteSetBstMidEndSepPunct{\mcitedefaultmidpunct}
{\mcitedefaultendpunct}{\mcitedefaultseppunct}\relax
\EndOfBibitem
\bibitem[{Garrod} et~al.(2007){Garrod}, {Wakelam}, and {Herbst}]{Garrod_ea07}
{Garrod},~R.~T.; {Wakelam},~V.; {Herbst},~E. \emph{Astron.~Astrophys.,}
  \textbf{2007}, \emph{467}, 1103\relax
\mciteBstWouldAddEndPuncttrue
\mciteSetBstMidEndSepPunct{\mcitedefaultmidpunct}
{\mcitedefaultendpunct}{\mcitedefaultseppunct}\relax
\EndOfBibitem
\bibitem[{Hill} and {Nuth}(2003){Hill}, and {Nuth}]{2003AsBio...3..291H}
{Hill},~H.~G.~M.; {Nuth},~J.~A. \emph{Astrobiology} \textbf{2003}, \emph{3},
  291\relax
\mciteBstWouldAddEndPuncttrue
\mciteSetBstMidEndSepPunct{\mcitedefaultmidpunct}
{\mcitedefaultendpunct}{\mcitedefaultseppunct}\relax
\EndOfBibitem
\bibitem[{Nuth} and {Berg}(1994){Nuth}, and {Berg}]{NB94}
{Nuth},~J.~A.; {Berg},~O. \emph{Lunar and Planetary Institute Conference
  Abstracts}; Lunar and Planetary Institute, Houston, 1994; p 1011\relax
\mciteBstWouldAddEndPuncttrue
\mciteSetBstMidEndSepPunct{\mcitedefaultmidpunct}
{\mcitedefaultendpunct}{\mcitedefaultseppunct}\relax
\EndOfBibitem
\bibitem[{Kress} et~al.(2010){Kress}, {Tielens}, and {Frenklach}]{Kress_ea10a}
{Kress},~M.~E.; {Tielens},~A.~G.~G.~M.; {Frenklach},~M. \emph{Adv. Space Res.,}
  \textbf{2010}, \emph{46}, 44\relax
\mciteBstWouldAddEndPuncttrue
\mciteSetBstMidEndSepPunct{\mcitedefaultmidpunct}
{\mcitedefaultendpunct}{\mcitedefaultseppunct}\relax
\EndOfBibitem
\bibitem[{Qi} et~al.(2003){Qi}, {Kessler}, {Koerner}, {Sargent}, and
  {Blake}]{Qi_ea03}
{Qi},~C.; {Kessler},~J.~E.; {Koerner},~D.~W.; {Sargent},~A.~I.; {Blake},~G.~A.
  \emph{Astrophys.~J.,} \textbf{2003}, \emph{597}, 986\relax
\mciteBstWouldAddEndPuncttrue
\mciteSetBstMidEndSepPunct{\mcitedefaultmidpunct}
{\mcitedefaultendpunct}{\mcitedefaultseppunct}\relax
\EndOfBibitem
\bibitem[{Zasowski} et~al.(2009){Zasowski}, {Kemper}, {Watson}, {Furlan},
  {Bohac}, {Hull}, and {Green}]{Zasowski_ea08}
{Zasowski},~G.; {Kemper},~F.; {Watson},~D.~M.; {Furlan},~E.; {Bohac},~C.~J.;
  {Hull},~C.; {Green},~J.~D. \emph{Astrophys.~J.,} \textbf{2009}, \emph{694},
  459\relax
\mciteBstWouldAddEndPuncttrue
\mciteSetBstMidEndSepPunct{\mcitedefaultmidpunct}
{\mcitedefaultendpunct}{\mcitedefaultseppunct}\relax
\EndOfBibitem
\bibitem[{Garrod} and {Herbst}(2006){Garrod}, and {Herbst}]{Garrod_ea06}
{Garrod},~R.~T.; {Herbst},~E. \emph{Astron.~Astrophys.,} \textbf{2006},
  \emph{457}, 927\relax
\mciteBstWouldAddEndPuncttrue
\mciteSetBstMidEndSepPunct{\mcitedefaultmidpunct}
{\mcitedefaultendpunct}{\mcitedefaultseppunct}\relax
\EndOfBibitem
\bibitem[{Evans} et~al.(2011){Evans}, {Bennett}, {Ullrich}, and
  {Kaiser}]{2011ApJ...730...69E}
{Evans},~N.~L.; {Bennett},~C.~J.; {Ullrich},~S.; {Kaiser},~R.~I.
  \emph{Astrophys.~J.,} \textbf{2011}, \emph{730}, 69\relax
\mciteBstWouldAddEndPuncttrue
\mciteSetBstMidEndSepPunct{\mcitedefaultmidpunct}
{\mcitedefaultendpunct}{\mcitedefaultseppunct}\relax
\EndOfBibitem
\bibitem[{Jones} et~al.(2011){Jones}, {Bennett}, and
  {Kaiser}]{2011ApJ...734...78J}
{Jones},~B.~M.; {Bennett},~C.~J.; {Kaiser},~R.~I. \emph{Astrophys.~J.,}
  \textbf{2011}, \emph{734}, 78\relax
\mciteBstWouldAddEndPuncttrue
\mciteSetBstMidEndSepPunct{\mcitedefaultmidpunct}
{\mcitedefaultendpunct}{\mcitedefaultseppunct}\relax
\EndOfBibitem
\bibitem[{Kim} and {Kaiser}(2012){Kim}, and {Kaiser}]{2012ApJ...758...37K}
{Kim},~Y.~S.; {Kaiser},~R.~I. \emph{Astrophys.~J.,} \textbf{2012}, \emph{758},
  37\relax
\mciteBstWouldAddEndPuncttrue
\mciteSetBstMidEndSepPunct{\mcitedefaultmidpunct}
{\mcitedefaultendpunct}{\mcitedefaultseppunct}\relax
\EndOfBibitem
\bibitem[{van Broekhuizen} et~al.(2004){van Broekhuizen}, {Keane}, and
  {Schutte}]{vanBroekhuizen_ea2004}
{van Broekhuizen},~F.~A.; {Keane},~J.~V.; {Schutte},~W.~A.
  \emph{Astron.~Astrophys.,} \textbf{2004}, \emph{415}, 425\relax
\mciteBstWouldAddEndPuncttrue
\mciteSetBstMidEndSepPunct{\mcitedefaultmidpunct}
{\mcitedefaultendpunct}{\mcitedefaultseppunct}\relax
\EndOfBibitem
\bibitem[{Tideswell} et~al.(2010){Tideswell}, {Fuller}, {Millar}, and
  {Markwick}]{Tideswell_ea10}
{Tideswell},~D.~M.; {Fuller},~G.~A.; {Millar},~T.~J.; {Markwick},~A.~J.
  \emph{Astron.~Astrophys.,} \textbf{2010}, \emph{510}, 85\relax
\mciteBstWouldAddEndPuncttrue
\mciteSetBstMidEndSepPunct{\mcitedefaultmidpunct}
{\mcitedefaultendpunct}{\mcitedefaultseppunct}\relax
\EndOfBibitem
\end{mcitethebibliography}

\end{document}